\documentclass[a4paper,12pt]{article}
\pdfoutput=1
\usepackage{amssymb}
\usepackage{amsmath}
\usepackage{graphicx}
\usepackage{epsfig}
\usepackage{slashed}
\usepackage{braket}
\usepackage{lineno}
\usepackage{setspace}

\usepackage{xcolor,colortbl}
\newcommand{\dbl}{}

\usepackage[utf8]{inputenc}
\usepackage[top=15mm,bottom=20mm,left=15mm,right=15mm]{geometry}
\usepackage[sorting=none,style=nature,backend=biber]{biblatex}
\usepackage[colorlinks=true,hypertexnames=false]{hyperref}
\usepackage{color}
\AtEveryBibitem{\clearfield{doi}}
\AtEveryBibitem{\clearfield{url}}
\usepackage{amstext}
\usepackage{multirow,bigdelim}
\usepackage{array}
\newcolumntype{C}{>{$}c<{$}}
\newcolumntype{L}{>{$}l<{$}}
\newcolumntype{R}{>{$}r<{$}}
\usepackage{arydshln}

\newcommand{\lsim}{ {\
\lower-1.2pt\vbox{\hbox{\rlap{$<$}\lower5pt\vbox{\hbox{$\sim$}}}}\ } }
\newcommand{\gsim}{ {\
\lower-1.2pt\vbox{\hbox{\rlap{$>$}\lower5pt\vbox{\hbox{$\sim$}}}}\ } }

\DeclareMathOperator*{\sumint}{%
\mathchoice%
  {\ooalign{$\displaystyle\sum$\cr\hidewidth$\displaystyle\int$\hidewidth\cr}}
  {\ooalign{\raisebox{.14\height}{\scalebox{.7}{$\textstyle\sum$}}\cr\hidewidth$\textstyle\int$\hidewidth\cr}}
  {\ooalign{\raisebox{.2\height}{\scalebox{.6}{$\scriptstyle\sum$}}\cr$\scriptstyle\int$\cr}}
  {\ooalign{\raisebox{.2\height}{\scalebox{.6}{$\scriptstyle\sum$}}\cr$\scriptstyle\int$\cr}}
}

\newcommand{\authors}
{
    \noindent
    Sz.\ Borsanyi$^{1}$,
    Z.\ Fodor$^{1,2,3,4,5,*}$,
    J.\ N.\ Guenther$^{6}$,
    C.\ Hoelbling$^{1}$,
    S.\ D.\ Katz$^{4}$,
    L.\ Lellouch$^{7}$,
    T.\ Lippert$^{1,2}$,
    K.\ Miura$^{7,8,9}$,
    L.\ Parato$^{7}$,
    K.\ K.\ Szabo$^{1,2}$,
    F.\ Stokes$^{2}$,
    B.\ C.\ Toth$^{1}$,
    Cs.\ Torok$^{2}$,
    L.\ Varnhorst$^{1}$
    \vspace*{1cm}

    \noindent
    $^{1}$\ Department of Physics, University of Wuppertal, D-42119 Wuppertal, Germany\\
    $^{2}$\ J\"ulich Supercomputing Centre, Forschungszentrum J\"ulich, D-52428 J\"ulich, Germany\\
    $^{3}$\ Department of Physics, Pennsylvania State University, University Park, PA 16802, USA\\
    $^{4}$\ Institute for Theoretical Physics, E\"otv\"os University, H-1117 Budapest, Hungary\\
    $^{5}$\ University of California, San Diego, 9500 Gilman Drive, La Jolla, CA 92093, USA\\
    $^{6}$\ Department of Physics, University of Regensburg, Regensburg D-93053, Germany\\
    $^{7}$\ Aix Marseille Univ, Universit\'e de Toulon, CNRS, CPT, IPhU, Marseille, France\\
    $^{8}$\ Helmholtz Institute Mainz, D-55099 Mainz, Germany\\
    $^{9}$\ Kobayashi-Maskawa Institute for the Origin of Particles and the Universe, Nagoya University, Nagoya 464-8602, Japan
    \vspace*{1cm}
}

\begin{document}

\dbl

\clearpage
\setcounter{page}{1}
\begin{refsegment}
    \vspace*{1cm}
    \begin{center}
	\huge\bf
	Leading hadronic contribution to the muon
	magnetic moment from lattice QCD
    \end{center}
    \vspace*{1cm}
    \authors
    \newpage
    {\bf The standard model of particle physics describes the vast majority of
experiments and observations involving elementary particles. Any deviation from
its predictions would be a sign of new, fundamental physics. One long-standing
discrepancy concerns the anomalous magnetic moment of the muon, $(g_\mu-2)$, a
measure of the magnetic field surrounding that particle. Indeed, standard model
predictions for $(g_\mu-2)$, reviewed in~\cite{Tanabashi:2018oca}, exhibit
disagreement with the measurement~\cite{Bennett:2006fi} that is tightly
scattered around $3.7$ standard deviations. Today, theory and measurement
errors are comparable.  However, a new experiment is underway at Fermilab and
another is planned at J-PARC, aiming to reduce the measurement’s error by a
factor of four. On the theory side, the dominant source of error is the
leading-order, hadronic vacuum polarization (LO-HVP) contribution. To fully
leverage the upcoming measurements, it is critical to check the prediction for
this contribution with independent methods and to reduce its uncertainties. The
most precise, model-independent determinations currently rely on dispersive
techniques, combined with measurements of the cross-section for
electron-positron annihilation into
hadrons~\cite{Davier:2019can,Keshavarzi:2019abf,Colangelo:2018mtw,Hoferichter:2019mqg}.
Here we use {\it ab initio} simulations in quantum chromodynamics and quantum
electrodynamics to compute the LO-HVP contribution with sufficient precision to
discriminate between the measurement of $(g_\mu-2)$ and the dispersive
predictions. Our result, $[(g_\mu-2)/2]_\mathrm{LO-HVP}=707.5[5.5]\times
10^{-10}$, favors the experimentally measured value of $(g_\mu-2)$ over the
results based on the dispersion relation. Moreover, the methods used and
developed here will allow further increases in precision, as more powerful
computers become available.}

The muon is an ephemeral sibling of the electron. It is 207 times more
massive, but has the same electric charge and spin. Similarly to the
electron, it behaves like a tiny magnet, characterized by a magnetic
moment. This quantity is proportional to the spin and charge of the
muon, and inversely proportional to twice its mass. Dirac's
relativistic quantum mechanics predicts that the constant of
proportionality, $g_\mu$, should be 2. However, in a relativistic
quantum field theory such as the standard model, this prediction
receives small corrections due to quantum, vacuum fluctuations. These
corrections are called the anomalous magnetic moment and are
quantified by $(g_\mu-2)/2$. They were measured to an exquisite
0.54~ppm at the Brookhaven National Laboratory in the early
2000s~\cite{Bennett:2006fi}, and have been calculated with a comparable
precision (see~\cite{Aoyama:2020ynm} for a recent review).

At this level of precision, all of the interactions of the standard
model contribute. The leading contributions are electromagnetic and
described by quantum electrodynamics (QED), but the one that dominates
the theory error is induced by the strong interaction and requires
solving the highly non-linear equations of quantum chromodynamics
(QCD) at low energies. This contribution is determined by the
leading-order, hadronic vacuum polarization (LO-HVP), which describes
how the propagation of a virtual photon is modified by the presence of
quark and gluon fluctuations in the vacuum. Here we compute this
LO-HVP contribution to $(g_\mu-2)/2$, denoted by
$a_\mu^\mathrm{LO-HVP}$, using {\it ab initio} simulations in QCD and
QED.

QCD is a generalized version of QED. The Euclidean Lagrangian for this
theory is ${\cal L} = 1/(4e^2) F_{\mu\nu}F_{\mu\nu}
+1/(2g^2)\text{Tr}G_{\mu\nu}G_{\mu\nu} +
\sum_f{\bar \psi_f}[\gamma_\mu (\partial_\mu + i q_f A_\mu + i B_\mu) + m_f ]
\psi_f$, where $\gamma_\mu$ are the Dirac-matrices, $f$ runs over the flavors
of quarks, the $m_f$ are their masses and the $q_f$ are their charges in units
of the electron charge $e$. Moreover, $F_{\mu\nu}=\partial_\mu A_\nu -
\partial_\nu A_\mu$ and $G_{\mu\nu}=\partial_\mu B_\nu - \partial_\nu
B_\mu+[B_\mu,B_\nu]$ and $g$ is the QCD coupling constant.  In electrodynamics,
the gauge potential $A_\mu$ is a real valued field, whereas in QCD, $B_\mu$ is
a $3\times3$ Hermitian matrix field. The different ``flavors'' of quarks are
represented by independent fermionic fields, $\psi_f$. These fields have an
additional ``color'' index in QCD, which runs from 1 to 3.  In the present
work, we include both QED and QCD, as well as four non-degenerate quark flavors
(up, down, strange and charm), in a lattice formulation taking into account all
dynamical effects.  We also consider the tiny contributions of the bottom and
top quarks, as discussed in the Supplementary Information.

We compute $a_\mu^\mathrm{LO-HVP}$ in the so-called time-momentum
representation \cite{Bernecker:2011gh}, which relies on the following, zero
three-momentum, two-point function in Euclidean time $t$:
\begin{linenomath*}
\begin{equation}
    \label{eq:main_gdef}
    G(t)= \frac{1}{3e^2}\sum_{\mu=1,2,3}\int d^3x
    \langle J_\mu(\vec{x},t) J_\mu(0)\rangle\ ,
\end{equation}
\end{linenomath*}
where $J_\mu$ is the quark electromagnetic current with $J_\mu/e= \frac{2}{3}
\bar{u}\gamma_\mu u - \frac{1}{3} \bar{d}\gamma_\mu d - \frac{1}{3}
\bar{s}\gamma_\mu s + \frac{2}{3} \bar{c} \gamma_\mu c$. $u,d,s$ and $c$ are
the up, down, strange and charm quark fields and the angle brackets stand for
the QCD+QED expectation value to order $e^2$. It is convenient to decompose
$G(t)$ into light, strange, charm and disconnected components, which have very
different statistical and systematic uncertainties. Integrating the
one-photon-irreducible ($\mathrm{1\gamma I}$) part of the two-point function
\eqref{eq:main_gdef} yields the LO-HVP contribution to the magnetic moment of the
muon \cite{Lautrup:1971jf,deRafael:1993za,Blum:2002ii,Bernecker:2011gh}:
\begin{linenomath*}
\begin{equation}
    \label{eq:intkg}
    a_{\mu}^\mathrm{LO-HVP}=\alpha^2
    \int_0^\infty dt\ K(t)\ G_\mathrm{1\gamma I}(t)\ ,
\end{equation}
\end{linenomath*}
with the weight function,
\begin{linenomath*}
\begin{equation}
    \label{eq:main_kdef}
    K(t)=\int_0^\infty \frac{dQ^2}{m_\mu^2}\ 
      \omega\left(\frac{Q^2}{m_\mu^2}\right)
      \left[t^2-\frac4{Q^2}\sin^2\left(\frac{Qt}2\right) \right]
\ ,
\end{equation}
\end{linenomath*}
and where $\omega(r)=[r+2-\sqrt{r(r+4)}]^2/\sqrt{r(r+4)}$, $\alpha$ is the fine
structure constant in the Thomson limit and $m_\mu$ is the muon mass. Since we
consider only the LO-HVP contribution, for brevity we drop the superscript and
multiply the result by $10^{10}$, ie. $a_\mu$ stands for
$a_{\mu}^\mathrm{LO-HVP} \times 10^{10}$ throughout this work.

The subpercent precision, that we are aiming for, represents a huge challenge
for lattice QCD. To reach that goal, we have to address four critical issues:
scale determination; noise reduction; QED and strong-isospin breaking;
infinite-volume and continuum extrapolations. We discuss these one by one.

The first issue is the scale determination. The quantity $a_\mu$ depends on the muon mass. When computing
\eqref{eq:intkg} on the lattice, $m_\mu$ has to be converted into lattice
units, $am_\mu$, where $a$ is the lattice spacing. A relative error of the
lattice spacing propagates into about a twice as large a relative error on
$a_\mu$, so that $a$ has to be determined with a few permil precision. We use
the mass of the $\Omega$ baryon, $M_\Omega=1672.45(29)$~MeV
\cite{Tanabashi:2018oca}, to set the lattice spacing. We also use the
$w_0$-scale from \cite{Borsanyi:2012zs}, in order to define an isospin
decomposition of our observables.  Though $w_0$ can be determined with
sub-permil precision on the lattice, it is inaccessible experimentally.  In
this work we determine the physical value of $w_0$ including QED and
strong-isospin-breaking effects: $w_0=0.17236(29)(63)[70]$~fm, where the first
error is statistical, the second is systematic and the third is the total
error.  In total we reach a relative accuracy of four permil, which is better
than the error of the previous best determination of \cite{Dowdall:2013rya},
whose value agrees with ours. There the pion-decay-constant was used as
experimental input and the isospin-breaking effects were only included as an
estimate.

The second issue is the noise reduction. Our result for $a_\mu$ is obtained as
an integral over the conserved current-current correlation function, from zero
to infinite time separation, as shown in Equation \eqref{eq:intkg}. For large
separations the correlator is quite noisy. This noise manifests itself as a
statistical error in $a_\mu$. To reach the desired accuracy on $a_\mu$, one
needs high-precision at every step. Over 20,000 configurations were accumulated
for our 27 ensembles on $L\approx 6$~fm lattices. In addition, we also include
a lattice with $L\approx 11$~fm. The most important improvement
over our earlier $a_\mu$ determination in \cite{Borsanyi:2017zdw} is the
extensive use of analysis techniques based on the lowest eigenmodes of the
Dirac operator, see eg.
\cite{Neff:2001zr,Giusti:2004yp,DeGrand:2004qw,Shintani:2014vja}.  About an
order of magnitude accuracy-gain can be reached using this technique for
$a_\mu$ \cite{Blum:2018mom,Aubin:2019usy}.

\begin{figure}[p]
    \centering
    \includegraphics[width=\textwidth]{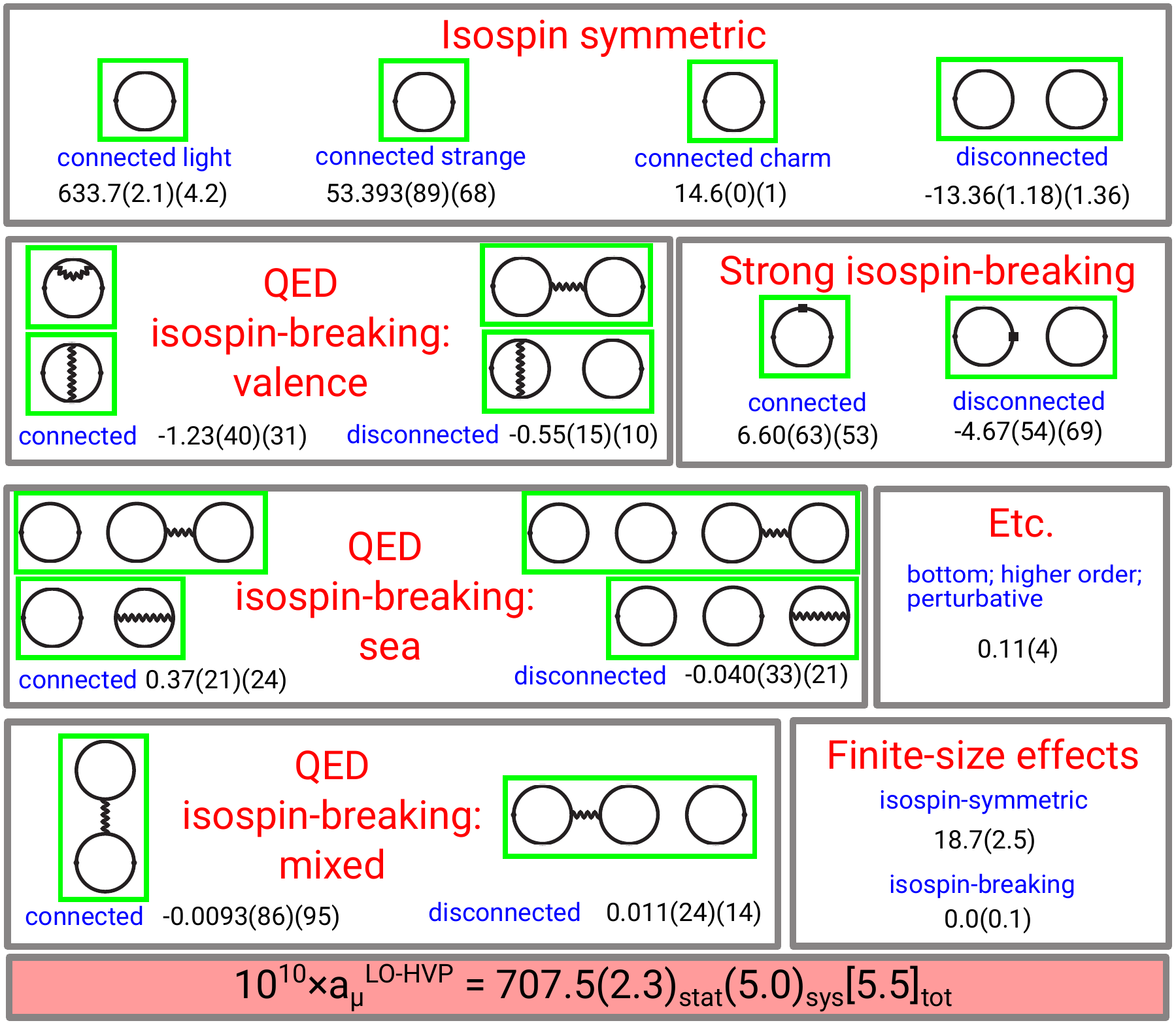}
    \caption
    {
	\dbl \label{fi:main_feynman} List of the contributions to $a_\mu$,
	including examples of the corresponding Feynman diagrams.  Solid lines
	are quarks and curly lines are photons. Gluons are not shown
	explicitly, and internal quark loops, only if they are attached to
	photons.  Dots represent coordinates in position space, a box indicates
	the mass insertion relevant for strong-isospin breaking. The numbers
	give our result for each contribution, they correspond to our
	``reference'' system size given by $L_\mathrm{ref}=6.272$~fm spatial
	and $T_\mathrm{ref}=9.408$~fm temporal lattice extents.  We also
	explicitly compute the finite-size corrections that must be added to
	these results, these are given separately in the lower right panel. The
	first error is the statistical and the second is the systematic
	uncertainty; except for the contributions where only a single, total
	error is given. Errors are s.e.m.
    }
\end{figure}

The third issue is the isospin-breaking. The precision needed cannot be reached
with pure, isospin-symmetric QCD. Thus, we include QED effects and allow the up
and down quarks to have different masses. These effects are included both in
the scale determination and in the current-current correlators. Note that the
separation of isospin symmetric and isospin breaking contributions requires a
convention, which we discuss in detail in the Supplementary Information.
Strong-isospin breaking is implemented by taking derivatives of QCD+QED
expectation values with respect to up/down quark masses and computing the
resulting observables on isospin-symmetric configurations
\cite{deDivitiis:2011eh}. Note that the first derivative of the fermionic
determinant vanishes. We also implement derivatives with respect to the
electric charge \cite{deDivitiis:2013xla}. It is useful to distinguish between
the electric charge in the fermionic determinant, $e_s$ or sea electric-charge,
and in the observables, $e_v$ or valence electric-charge.  The complete list of
graphs that should be evaluated are shown in Figure~\ref{fi:main_feynman} with
our numerical results for them.

The final observable is given as a Taylor-expansion around the
isospin-symmetric, physical-mass point with zero sea and valence charges.
Instead of the quark masses, we use the pseudoscalar meson masses of pions and
kaons, which can be determined with high precision. With the expansion
coefficients, we extrapolate in the charges, in the strong-isospin breaking
parameter and in the lattice spacing and interpolate in the quark masses to the
physical point. Thus, we obtain $a_\mu$ and its statistical and systematic
uncertainties.

The fourth issue is the extrapolation to the infinite-volume and
continuum-limit. The standard wisdom for lattice calculations is that $M_\pi
L>4$ should be taken, where $M_\pi$ is the mass of the pion and $L$ is the
spatial extent of the lattice. Unfortunately, this is not satisfactory in the
present case: $a_\mu$ is far more sensitive to $L$ than other quantities, such
as hadron masses, and large volumes are needed to reach permil accuracy. For
less volume-sensitive quantities in this work, we use well-established results
to determine the finite-volume corrections on the pion-decay constant
\cite{Colangelo:2005gd} and on charged hadron masses
\cite{Davoudi:2014qua,Borsanyi:2014jba,Fodor:2015pna}. Leading-order chiral
perturbation theory \cite{Aubin:2015rzx} or two-loop, partially-quenched chiral
perturbation theory \cite{Bijnens:2017esv,Aubin:2019usy} for $a_\mu$ help, but
the non-perturbative, leading-order, large-$L$ expansion of
\cite{Hansen:2019rbh} indicates that those approaches still lead to systematic
effects which are larger than the accuracy that we are aiming for. In addition
to the infinite-volume extrapolation, the continuum extrapolation is also
difficult.  This is connected to the taste-symmetry breaking of staggered
fermions, which we use in this work.

We correct for finite-volume effects on $a_\mu$ by computing them directly,
thanks to lattice simulations that we performed for that purpose on $L\approx
11$~fm lattices, with highly-suppressed taste violations and with physical,
taste-averaged pion masses. These corrections are cross-checked against three
models that describe the relevant long-distance physics, in turn validating the
use of these models for the residual, sub-permil extrapolation to infinite
volume. These models include: i.\ the full two-loop, finite-volume, chiral
perturbation theory corrections for $a_\mu$; ii.\ the
Meyer-Lellouch-L\"uscher-Gounaris-Sakurai technique described in the
Supplementary Information; iii.\ the rho-pion-gamma model of Jegerlehner and
Szafron \cite{Jegerlehner:2011ti}, already used in a lattice context by HPQCD
\cite{Chakraborty:2016mwy}. Moreover, to reduce discretization errors in the
light-quark contributions to $a_\mu$, before extrapolating those contributions
to the continuum, we apply a taste-improvement procedure that reduces lattice
artefacts due to taste-symmetry breaking. The procedure is built upon the three
models of pion-rho physics mentioned above. We provide evidence validating this
procedure in the Supplementary Information.

\begin{figure}[p]
    \centering
    \includegraphics[width=0.68\textwidth]{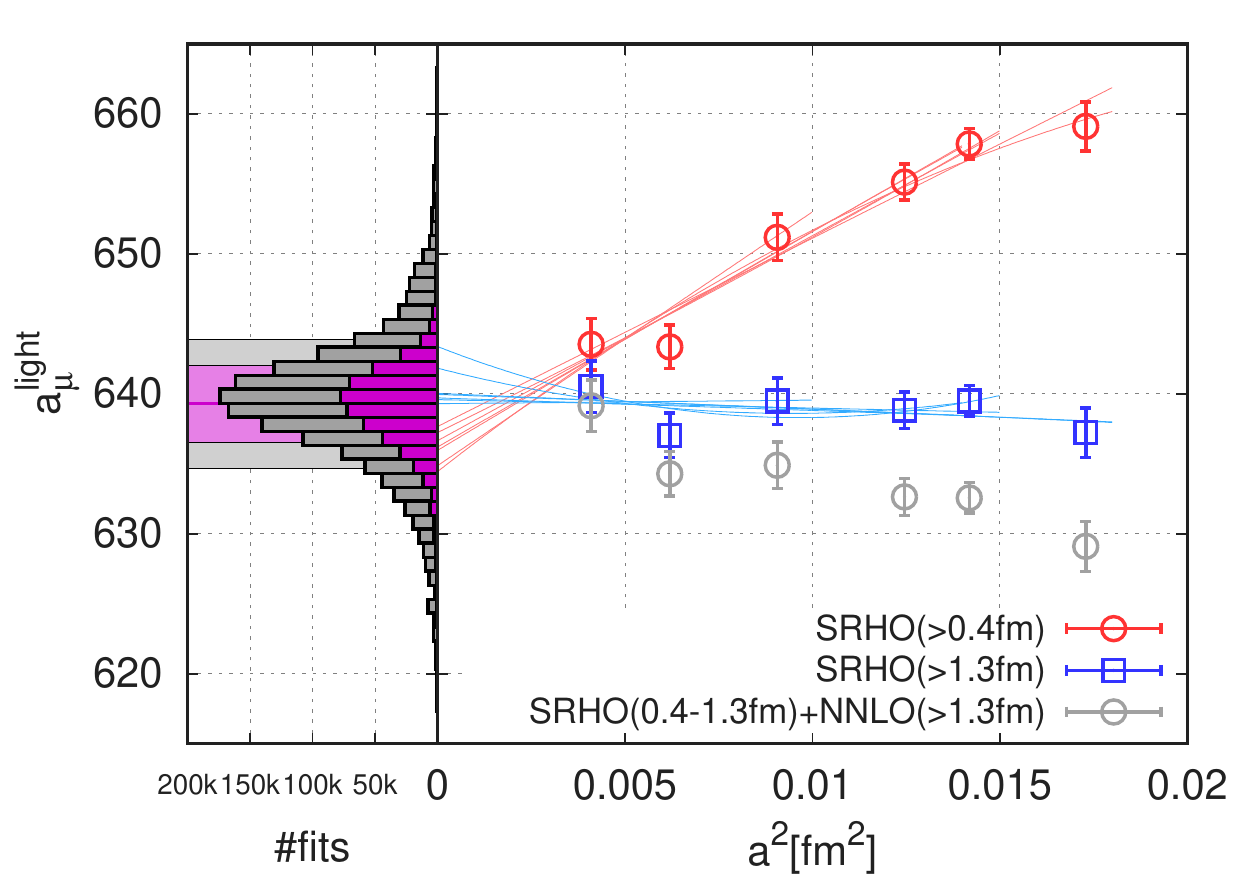}
    \caption
    {
	\dbl \label{fi:main_cont} Continuum extrapolation of the light
	connected component of $a_\mu$, denoted by $a_\mu^\mathrm{light}$.
	Before extrapolation we apply a taste-improvement procedure on the
	correlator starting at some distance $t_\mathrm{sep}$. (See the
	Supplementary Information for details on the improvement ``SRHO''.)
	Data sets are shown for two choices of $t_\mathrm{sep}$, $0.4$ and
	$1.3$~fm, with red and blue. The corresponding lines show fits using
	linear and quadratic terms in $a^2$ with varying number of lattice
	spacings in the fit.  Our final analysis involves about 500,000
	different continuum extrapolations, a histogram of which is shown on
	the left in purple. The purple line in the left panel shows the central
	value of the final result.  To estimate the error related to the taste
	improvement procedure we use next-to-next-to-leading-order staggered
	chiral perturbation theory (NNLO) in the long distance part of the
	correlator ($t>1.3$~fm). The corresponding data are shown with grey
	points, together with a histogram, from which the systematic error
	related to the taste improvement is obtained. The total error of the
	final result is given by the grey band in the left panel. Errors are
	s.e.m. The results are obtained on lattices of sizes $L\approx6$~fm.
    }
\end{figure}

Combining all of these ingredients we obtain, as a final result,
$a_\mu=707.5(2.3)(5.0)[5.5]$. The first, statistical error comes mostly from
the noisy, large-distance region of the current-current correlator. The second,
systematic error is dominated by the continuum extrapolation and the
finite-size effect computation. The third, total error is obtained by adding
the first two in quadrature. In total we reach a relative accuracy of $0.8$\%.
In Figure \ref{fi:main_cont} we show the continuum extrapolation of the light,
connected component of $a_\mu$, which gives the dominant contribution to
$a_\mu$. 

\begin{figure}[p]
    \centering
    \includegraphics[width=0.68\textwidth]{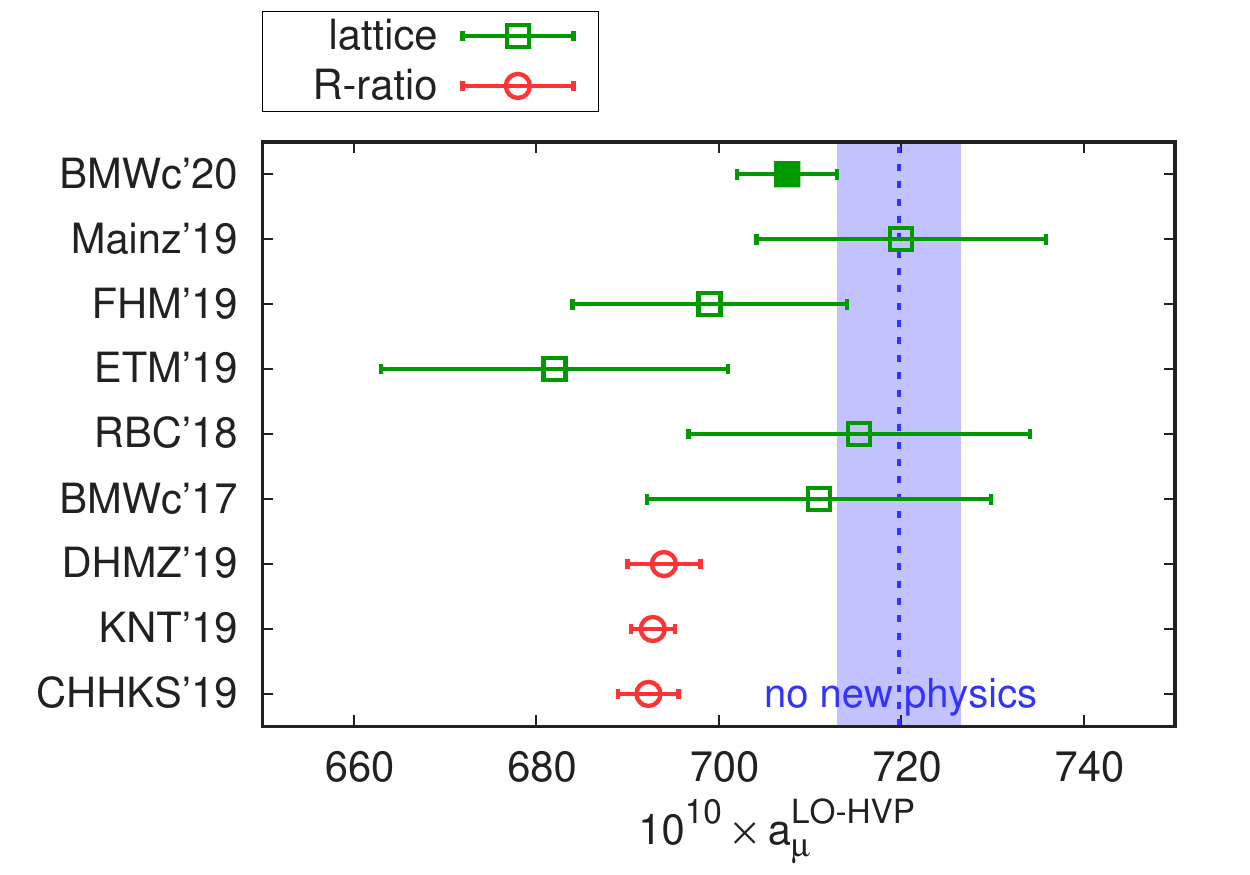}
    \caption
    {
	\dbl \label{fi:main_cmp} Comparison of recent results for the
	leading-order, hadronic vacuum polarization contribution to the
	anomalous magnetic moment of the muon. See \cite{Aoyama:2020ynm} for a
	recent review. Green squares are lattice results: this work's result,
	denoted by BMWc'20 and represented by a filled symbol at the top of the
	figure, is followed by Mainz'19 \cite{Gerardin:2019rua}, FHM'19
	\cite{Davies:2019efs}, ETM'19 \cite{Giusti:2019xct}, RBC'18
	\cite{Blum:2018mom} and our earlier work BMWc'17
	\cite{Borsanyi:2017zdw}. Errorbars are s.e.m. Compared to BMWc'17, the
	present work has increased the accuracy of the scale-setting from the
	per-cent to the per-mill level; has decreased the statistical error
	from 7.5 to 2.3; has computed all isospin-breaking contributions as
	opposed to estimating it, the corresponding error is 1.4 down from 5.1;
	has made a dedicated finite-size study to decrease the finite-size
	error from 13.5 to 2.5; has decreased the continuum extrapolation error
	from 8.0 to 4.1 by having much more statistics on our finest lattice
	and applying taste improvement. Red circles were obtained using the
	R-ratio method by DHMZ'19 \cite{Davier:2019can}, KNT'19
	\cite{Keshavarzi:2019abf} and CHHKS'19
	\cite{Colangelo:2018mtw,Hoferichter:2019mqg}; these results use the
	same experimental data as input.  The blue shaded region is the value
	that $a_\mu^\mathrm{LO-HVP}$ would have to have to explain the
	experimental measurement of $(g_\mu-2)$, assuming no new physics.
    }
\end{figure}

Figure \ref{fi:main_cmp} compares our result with previous lattice computations
and also with results from the R-ratio method, which have recently been
reviewed in \cite{Aoyama:2020ynm}. In principle, one can reduce the uncertainty
of our result by combining our lattice correlator, $G(t)$, with the one
obtained from the R-ratio method, in regions of Euclidean time where the latter
is more precise~\cite{Blum:2018mom}.  We do not do so here because there is a
tension between our result and those obtained by the R-ratio method, as can be
seen in Figure \ref{fi:main_cmp}. For the total, LO-HVP contribution to
$a_\mu$, our result is $2.0\sigma$, $2.5\sigma$, $2.4\sigma$ and $2.2\sigma$
larger than the R-ratio results of $a_\mu=694.0(4.0)$ \cite{Davier:2019can},
$a_\mu=692.78(2.42)$ \cite{Keshavarzi:2019abf}, $a_\mu=692.3(3.3)$
\cite{Colangelo:2018mtw,Hoferichter:2019mqg} and the combined result
$a_\mu=693.1(4.0)$ of \cite{Aoyama:2020ynm}, respectively.  It is worth noting
that the R-ratio determinations are based on the same experimental data sets
and are therefore strongly correlated, though these data sets were obtained in
several different and independent experiments that we have no reason to believe
are collectively biased. Clearly, these comparisons need further investigation,
though it should also be kept in mind that the tensions observed here are
smaller, for instance, than what is usually considered experimental evidence
for a new phenomenon ($3\sigma$) and much smaller than what is needed to claim
an experimental discovery ($5\sigma$).

\begin{figure}[p]
    \centering
    \includegraphics[width=0.68\textwidth]{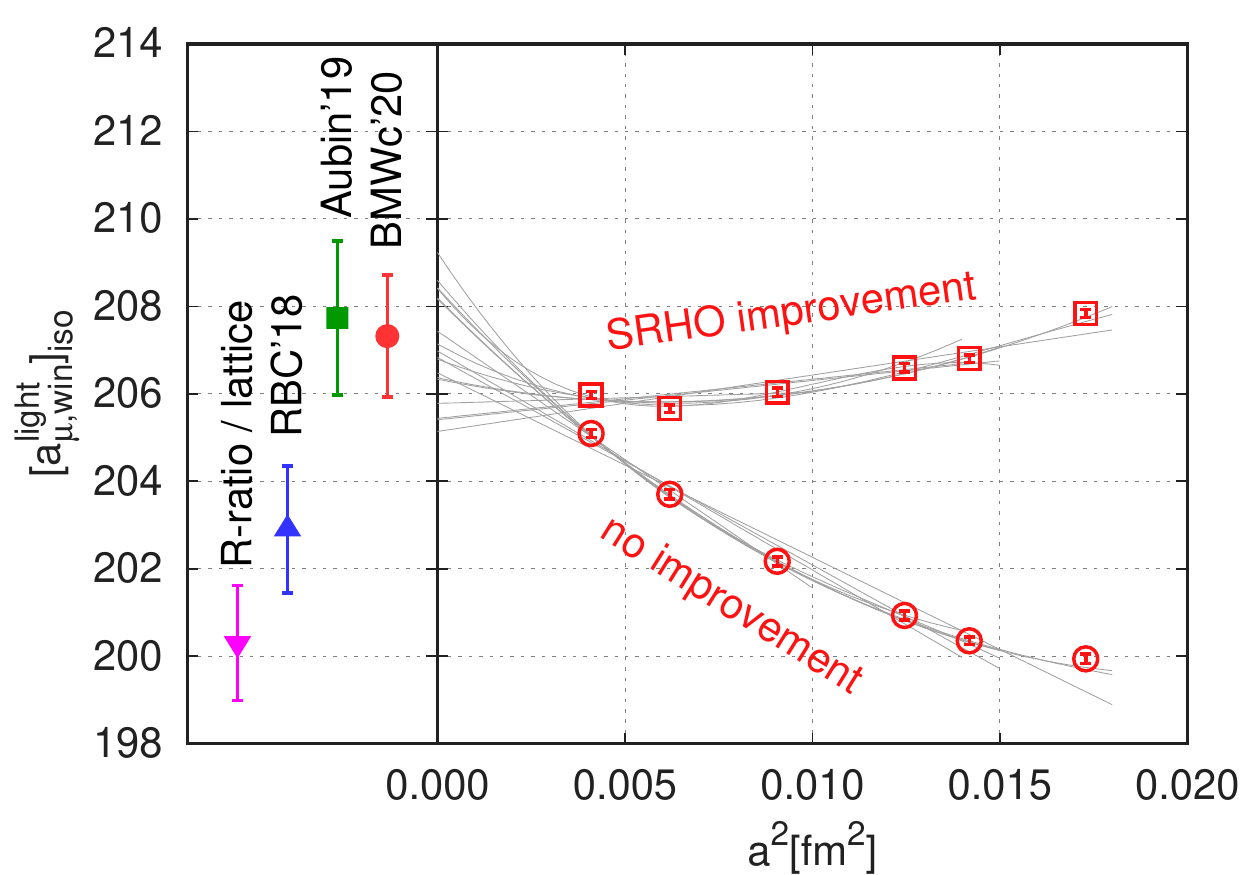}
    \caption
    {
	\dbl \label{fi:main_amuw} Continuum extrapolation of the
	isospin-symmetric, light, connected component of the window observable
	$a_{\mu,\mathrm{win}}$, denoted by
	$[a_{\mu,\mathrm{win}}^\mathrm{light}]_\mathrm{iso}$. The data points
	are extrapolated to the infinite-volume limit. Errorbars are s.e.m. Two
	different ways to perform the continuum extrapolations are shown: one
	without improvement, and another with corrections from a model
	involving the $\rho$-meson (SRHO).  In both cases the lines show
	linear, quadratic and cubic fits in $a^2$ with varying number of
	lattice spacings in the fit. The continuum extrapolated result is shown
	with the results from other lattice groups, RBC'18 \cite{Blum:2018mom}
	and Aubin'19 \cite{Aubin:2019usy}.  Also plotted is our R-ratio-based
	determination, obtained using the experimental data compiled by the
	authors of \cite{Keshavarzi:2019abf} and our lattice results for the
	non light connected contributions. This plot is convenient for
	comparing different lattice results with each other. Regarding the
	total $a_{\mu,\mathrm{win}}$, for which we also have to include the
	contributions of other-than-light flavors and isospin-breaking effects,
	we obtain $236.7[1.4]$ on the lattice and $229.7[1.3]$ from the
	R-ratio, the latter is $3.7\sigma$ or $3.1\%$ smaller than the lattice
	result.
    }
\end{figure}

As a first step in that direction, it is instructive to consider a modified
observable, where the correlator $G(t)$ is restricted to a finite interval by a
smooth window function \cite{Blum:2018mom}.  This observable, which we denote
by $a_{\mu,\mathrm{win}}$, is obtained much more readily than $a_\mu$ on the
lattice. Its shorter-distance nature makes it significantly less susceptible to
statistical noise and to finite-volume effects. Moreover, in the case of
staggered fermions, it has reduced discretization artefacts. This is shown in
Figure~\ref{fi:main_amuw}, where the light, connected component of
$a_{\mu,\mathrm{win}}$ is plotted as a function of $a^2$. Because the
determination of this quantity does not require overcoming many of the
challenges described above, other lattice groups have obtained it with errors
comparable to ours \cite{Blum:2018mom,Aubin:2019usy}. This allows for a sharper
benchmarking of our calculation of this challenging, light-quark contribution
that dominates $a_\mu$.  Our $a_{\mu,\mathrm{win}}^\mathrm{light}$ differs by
$0.2\sigma$ and $2.2\sigma$ from the lattice results of \cite{Aubin:2019usy}
and \cite{Blum:2018mom}, respectively.  Moreover, $a_{\mu,\mathrm{win}}$ can be
computed in the R-ratio approach, and we do so using the data set courteously
given to us by the authors of \cite{Keshavarzi:2019abf}. However, here we find
a $3.7\sigma$ tension with our lattice result.

To conclude, when combined with the other standard model contributions (see eg.
\cite{Davier:2019can,Keshavarzi:2019abf}), our result for the leading-order
hadronic contribution to the anomalous magnetic moment of the muon,
$a_\mu^\mathrm{LO-HVP}= 707.5[5.5]\times 10^{-10}$, weakens the longstanding
discrepancy between experiment and theory.  However, as discussed above and can
be seen in Figure \ref{fi:main_cmp}, our lattice result shows some tension with
the R-ratio determinations of
\cite{Keshavarzi:2019abf,Davier:2019can,Colangelo:2018mtw,Hoferichter:2019mqg}.
Obviously, our findings should be confirmed --or refuted-- by other
collaborations using other discretizations of QCD. Those investigations are
underway.

{\bf Supplementary Information} is available for this paper.

{\bf Acknowledgments.} We thank J.~Charles, A.~El-Khadra, M.~Hoferichter,
F.~Jegerlehner, C.~Lehner, M.~Knecht, A.~Kronfeld, E.~de Rafael and
participants of the online workshop, ``The hadronic vacuum polarization from
lattice QCD at high precision'' (16-20/11/2020), for informative discussions.
We thank J.~Bailey, W.~Lee and S.~Sharpe for correspondence on staggered chiral
perturbation theory. Special thanks go to A. Keshavarzi for providing us
cross-section data and useful discussions, and to the referees for their
constructive criticism. The computations were performed on JUQUEEN, JURECA,
JUWELS and QPACE at Forschungszentrum J\"ulich, on SuperMUC and SuperMUC-NG at
Leibniz Supercomputing Centre in M\"unchen, on Hazel Hen and HAWK at the High
Performance Computing Center in Stuttgart, on Turing and Jean Zay at CNRS'
IDRIS, on Joliot-Curie at CEA's TGCC, on Marconi in Roma and on GPU clusters in
Wuppertal and Budapest. We thank the Gauss Centre for Supercomputing, PRACE and
GENCI (grant 52275) for awarding us computer time on these machines. This
project was partially funded by the DFG grant SFB/TR55, by the BMBF Grant No.
05P18PXFCA, by the Hungarian National Research, Development and Innovation
Office grant KKP126769 and by the Excellence Initiative of Aix-Marseille
University - A*MIDEX, a French “Investissements d’Avenir” program, through
grants AMX-18-ACE-005, AMX-19-IET-008 - IPhU and ANR-11-LABX-0060.

{\bf Data availability.} The datasets for the figures and tables are available
from the corresponding author on request.

{\bf Code availability.} A CPU-code for configuration production and
measurements can be obtained from the corresponding author upon request. The
Wilson flow evolution code, which was used to determine $w_0$, can be
downloaded from {\texttt https://arxiv.org/abs/1203.4469}.

    \printbibliography[segment=1]
\end{refsegment}

\clearpage
\setcounter{page}{1}
\begin{refsegment}
    \renewcommand{\figurename}{Extended Data Figure}
    \renewcommand{\tablename}{Extended Data Table}
    \setcounter{figure}{0}
    \setcounter{table}{0}
    \begin{center}
	\huge\bf
	Methods
    \end{center}
    \section*{Finite-size effects}

Finite-size effects on $a_\mu$ were the largest source of uncertainty in our
previous work \cite{Borsanyi:2017zdw}. We compute these effects in a systematic
way, which includes dedicated lattice simulations, chiral perturbation theory
and phenomenological models.  The concrete goal is to provide a single number
that is to be added to the continuum-extrapolated lattice result obtained in a
reference box, which is defined by a spatial extent of
$L_\mathrm{ref}=6.272$~fm and a temporal extent of
$T_\mathrm{ref}=\tfrac{3}{2}L_\mathrm{ref}$.  Here we summarize our findings on
the finite-size effect of the isospin-symmetric part. More details and a
discussion of the isospin-breaking part can be found in the Supplementary
Information.

We perform dedicated lattice simulations with two different lattice geometries:
one is a $56\times84$ lattice with the reference box size and the other is a
large $96\times96$ lattice with box size $L=L_\mathrm{big}=10.752$~fm and
$T=T_\mathrm{big}=L_\mathrm{big}$.  Since taste violations distort the
finite-size effects, we designed a new action with highly-suppressed taste
breaking, which we call {\tt 4HEX}. Our
strategy is then to compute the finite-size correction as the following sum:
\begin{linenomath*}
\begin{equation}
    \label{exteq:telescope}
    \begin{gathered}
	a_\mu(\infty,\infty)-a_\mu(L_\mathrm{ref},T_\mathrm{ref})=\\
	=[a_\mu(L_\mathrm{big},T_\mathrm{big})-a_\mu(L_\mathrm{ref},T_\mathrm{ref})]_{\tt 4HEX} +
	[a_\mu(\infty,\infty)-a_\mu(L_\mathrm{big},T_\mathrm{big})]_\mathrm{XPT}.
    \end{gathered}
\end{equation}
\end{linenomath*}
The first difference on the right hand side is taken from the dedicated {\tt
4HEX} simulations.  The second difference is expected to be much smaller than
the first and is taken from a non-lattice approach: two-loop chiral
perturbation theory.

We consider four non-lattice approaches to compute both differences on the
right hand side of Equation \eqref{exteq:telescope}. In the case of the first
difference, the results obtained are compared to our {\tt 4HEX} simulations.
The first approach is chiral perturbation theory (XPT) to next-to-leading and
next-to-next-to-leading orders (NLO and NNLO), the second is the
Meyer-Lellouch-Luscher-Gounaris-Sakurai model (MLLGS), the third approach is
that of Hansen and Patella (HP) \cite{Hansen:2019rbh} and the fourth is the
rho-pion-gamma model of \cite{Chakraborty:2016mwy}, which we abbreviate as RHO
here.

We compute the first difference in Equation \eqref{exteq:telescope} using
dedicated simulations with the {\tt 4HEX} action. We
use the
harmonic-mean-square (HMS) to set the physical point:
\begin{linenomath*}
\begin{equation*}
    M^{-2}_{\pi,\mathrm{HMS}}\equiv \frac{1}{16} \sum_\alpha M^{-2}_{\pi,\alpha}\ ,
\end{equation*}
\end{linenomath*}
defined as an average over the masses of the 16 pion tastes, $M_{\pi,\alpha}$.
We set $M_{\pi,\mathrm{HMS}}$ to the physical value of the pion mass, which
requires lowering the Goldstone-pion mass to $110$~MeV. This way of fixing the
physical point results in much smaller lattice artefacts than the usual setting
with the Goldstone-pion, at least for an observable like the finite-size
effect.

To generate the {\tt 4HEX} data set, we performed simulations with two
different Goldstone pion masses: $M_\pi=104$~MeV and $121$~MeV. To set the
physical point as described above, we perform an interpolation from these two
pion masses to $M_\pi=110$~MeV. 

\begin{figure}
    \centering
    \includegraphics[width=0.7\textwidth]{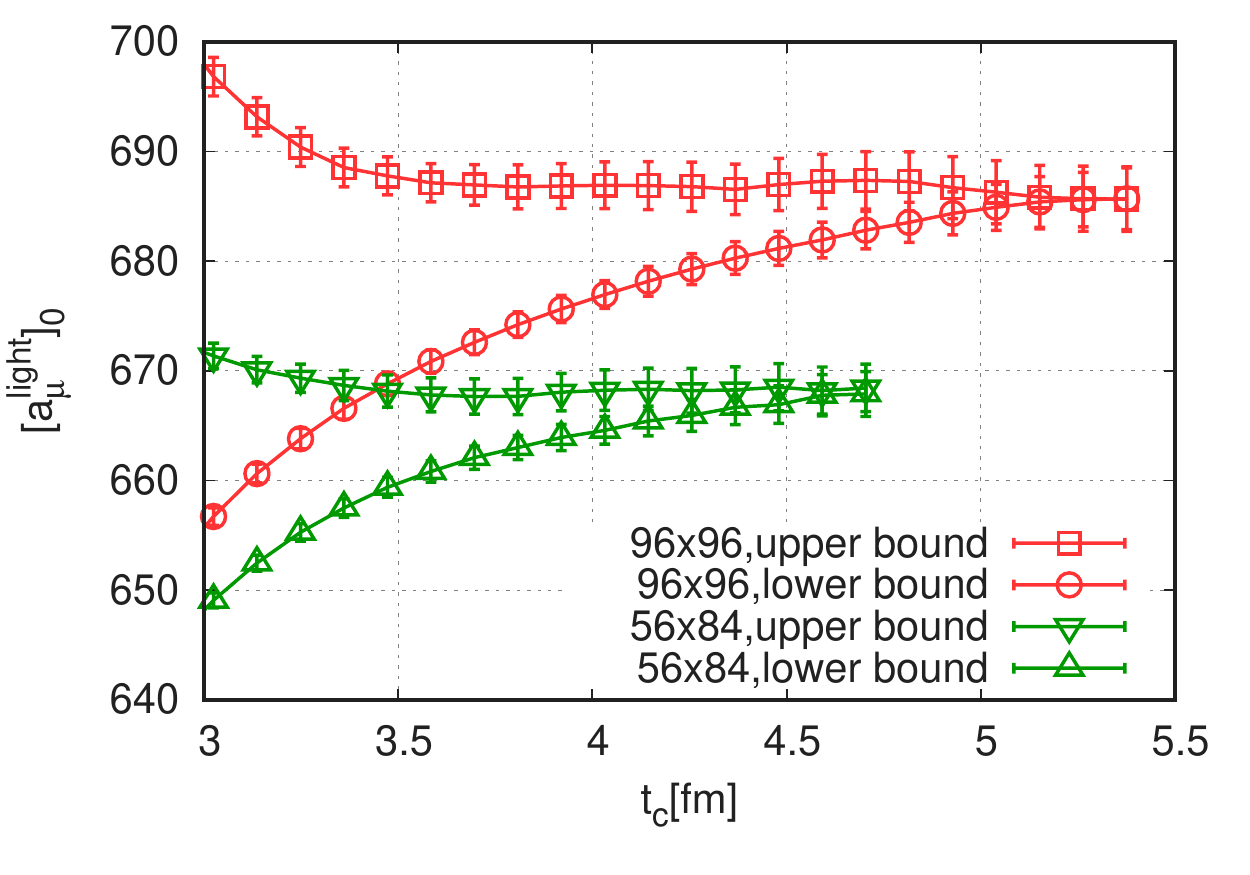}
    \caption
    {
	\dbl \label{extfi:fv4hex}Upper and lower bounds on the light
	isospin-symmetric component of $a_\mu$. The results shown here are
	obtained with the {\tt 4HEX} action on two different volumes at
	$a=0.112$~fm lattice spacing and $M_\pi=121$~MeV Goldstone-pion mass.
	We also have another simulation with $M_\pi=104$~MeV mass. From these
	two we interpolate to $M_\pi=110$~MeV. This value ensures that a
	particular average of pion tastes is fixed to the physical value of the
	pion mass (see text). Errorbars are statistical errors (s.e.m.).
    }
\end{figure}

To compute $a_{\mu}^\mathrm{light}$ from the current propagator in our {\tt
4HEX} simulations we use the upper and lower bound technique described in the
Supplementary Information.  Results for the $M_\pi=121$~MeV simulation point
are plotted in Extended Data Figure \ref{extfi:fv4hex}.  The bounds meet at
around $4.2$~fm and $4.7$~fm on the small and large volumes, respectively. At
these distances we take the average of the two bounds as an estimate for
$a_{\mu}^\mathrm{light}$.  The results are given in the Extended Data Table
\ref{extta:4hex}.

We only have one lattice spacing with the {\tt 4HEX} action, so the finite-size
effects cannot be extrapolated to the continuum limit. We estimate the cutoff
effect of the result by comparing the total $a_\mu$ with the {\tt 4HEX} action
at this single lattice spacing to the continuum extrapolated {\tt 4stout}
lattice result, both in the $L_\mathrm{ref}$ volume. The {\tt 4HEX} result is
about 7\% larger than the continuum value. Therefore we reduce the measured
finite-size effect by 7\%, and assign a 7\% uncertainty to this correction
step. For the difference we get
\begin{linenomath*}
\begin{equation}
    \label{exteq:fvref}
    a_\mu(L_\mathrm{big},T_\mathrm{big})-a_\mu(L_\mathrm{ref},T_\mathrm{ref})= 18.1(2.0)_\mathrm{stat}(1.4)_\mathrm{cont}\ .
\end{equation}
\end{linenomath*}
The result is obtained from the $a_\mu^\mathrm{light}$ numbers of Extended Data
Table \ref{extta:4hex}, including a $\left(\frac{9}{10}\right)$ charge
factor.  The first error is statistical and the second is an estimate of the
cutoff effect.

Extended Data Table \ref{extta:xpt} collects the finite-size effect computed in
various non-lattice approaches.  The different models give finite-size effects
of similar size, which agree well with the lattice determination of Equation
\eqref{exteq:fvref}. Only the NLO result differs by about $3\sigma$. The fact
that NLO chiral perturbation theory underestimates the finite-size effect was
already shown in \cite{Giusti:2018mdh}, at a non-physical pion mass. Using the
physical pion mass, a dedicated finite-volume study was carried out in
\cite{Shintani:2019wai}. It reaches the same conclusion as we do, albeit with
larger errors.  We also see that, according to the models, the finite-$T$
effect is much smaller than the finite-$L$ effect.

The good agreement for the finite-size effect of the reference box, between the
models and the lattice, gives us confidence that the models can be used to
reliably compute the very small, residual, finite-size effect of the large box.
The corresponding model estimates can be found in Extended Data Table
\ref{extta:xpt}.  For an infinite-time extent the NNLO XPT, the HP and RHO
approaches agree nicely. As a final value for the large-box, finite-size effect
we take the NNLO XPT result including finite-$T$ effects:
\begin{linenomath*}
\begin{equation*}
    a_\mu(\infty,\infty)-a_\mu(L_\mathrm{big},T_\mathrm{big})= 0.6(0.3)_\mathrm{big}\ ,
\end{equation*}
\end{linenomath*}
where the uncertainty is an estimate of higher-order effects, given here by the
difference of the NNLO and NLO values.

For our final result for the finite-size effect of the reference box, we also
include the contribution of isoscalar channel and isospin-breaking effects
giving:
\begin{linenomath*}
\begin{equation*}
    a_\mu(\infty,\infty) - a_\mu(L_\mathrm{ref},T_\mathrm{ref})= 18.7(2.0)_\mathrm{stat}(1.4)_\mathrm{cont}(0.3)_\mathrm{big}(0.6)_{I=0}(0.1)_\mathrm{qed}[2.5]\ .
\end{equation*}
\end{linenomath*}
The first error is the statistical uncertainty of our {\tt 4HEX} computation,
the second is an estimate of the {\tt 4HEX} cutoff effects, the third is the
uncertainty of the residual finite-size effect of the ``big'' lattice, the
fourth is a XPT estimate of the $I=0$ finite size effect and the fifth is an
estimate of the isospin-breaking effects. The last, total error in the
square-brackets is the sum of the first five, added in quadrature. The vast
majority of the finite-size effect is obtained using the {\tt 4HEX} lattice
computation; for the rest we apply analytic methods. These methods have been
validated by the lattice computation: for the main contribution they give
values that are consistent with the lattice result.

\section*{Taste improvement}

As is well known, some of the most important cutoff effects of staggered
fermions are taste violations. At long distances, these violations distort the
pion spectrum. Since $a_\mu$ is predominantly a long-distance observable,
dominated by a two-pion contribution, including the $\rho$ resonance, we expect
these effects to be largest in the light-quark terms. 

We investigate various physically motivated models for reducing long-distance
taste violations in our lattice results. We consider three techniques:
next-to-next-to-leading order chiral perturbation theory (NNLO XPT), a
Meyer-Lellouch-L\"uscher-Gounaris-Sakurai model (MLLGS) and the rho-pion-gamma
model (RHO). For the definition of these models see the Supplementary
Information. We investigate and discuss the suitability of their staggered
versions for reducing the taste violations present in our lattice data. We call
the resulting corrections taste improvements, because they improve the
continuum extrapolation of our lattice data without, in principle, modifying
the continuum-limit value. Indeed, these corrections vanish in that limit, as
taste-breaking effects should.  These improvements are applied on light-quark
observables at the isospin-symmetric point, whose taste violations have the
largest impact on our final uncertainties.

\begin{figure}
    \centering
    \includegraphics[width=0.7\textwidth]{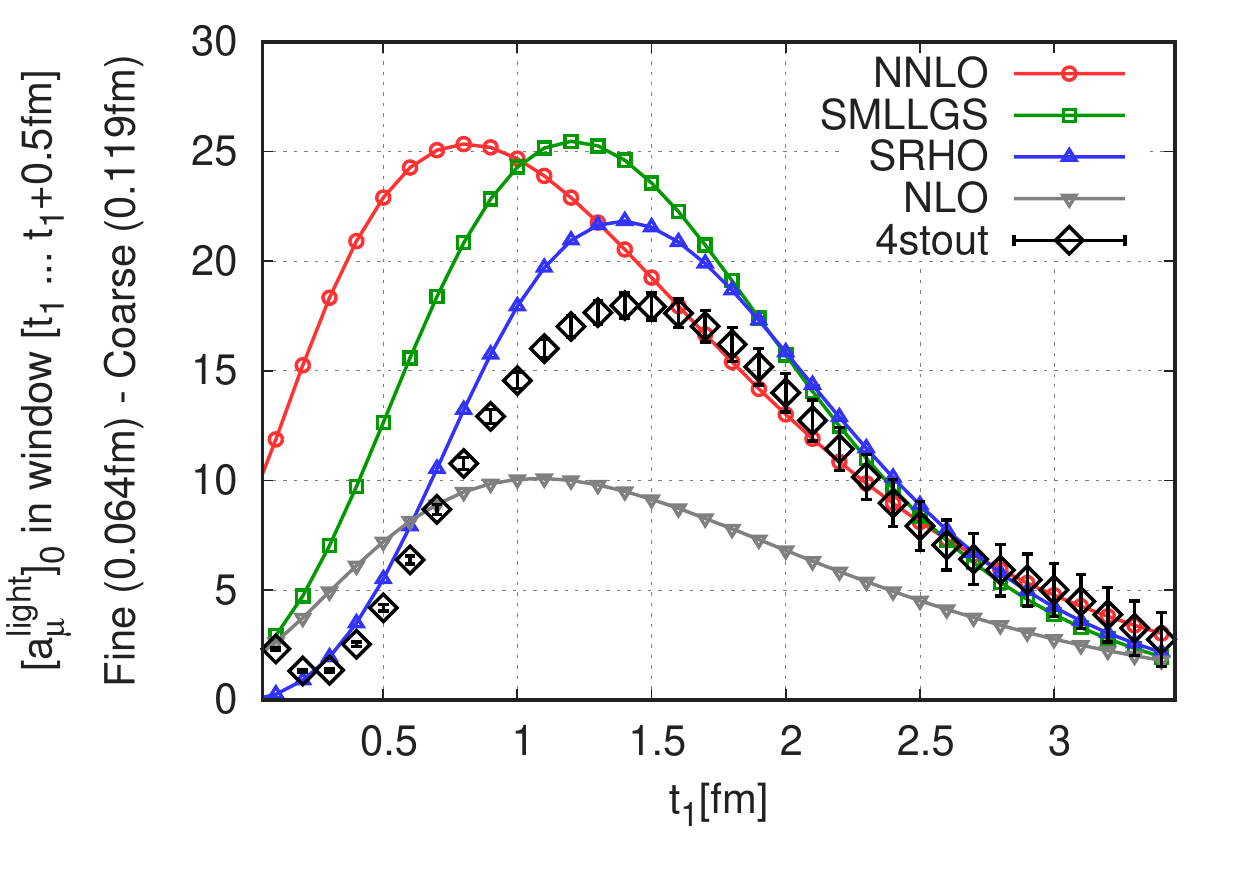}
    \caption
    {
	\dbl \label{extfi:amuwsxx} Isospin-symmetric component of
	$a_\mu^\mathrm{light}$ computed with a sliding window. The window
	starts at $t_1$ and ends $0.5\,\mathrm{fm}$ later.  The plot shows the
	difference between a fine and a coarse lattice, the volumes are
	$L=6.14\,\mathrm{fm}$ and $L=6.67\,\mathrm{fm}$. The black squares with
	errors are obtained from the simulation, errors are statistical
	(s.e.m.). The colored curves are the predictions of NLO and NNLO SXPT,
	the SRHO and the SMLLGS models.  They are computed at the parameters
	(pion mass, taste violation, volume) of the simulations. 
    }
\end{figure}

The models NNLO XPT, MLLGS and RHO describe the long-distance physics
associated with finite-volume effects, as measured in our simulations. One can
also define corresponding models describing the taste violations, they are
denoted NNLO SXPT, SMLLGS and SRHO.  We find that they describe the physics
associated with taste violations, at least at larger distances.  This is
illustrated in Extended Data Figure \ref{extfi:amuwsxx}, where cutoff effects in
the integrand of $a_\mu^\mathrm{light}$ are plotted as a function of Euclidean
time. More specifically, we define the physical observable, obtained by
convoluting the integrand of $a_\mu^\mathrm{light}$ with a smooth window
function $W(t;t_1)$ of a width of $0.5$~fm and starting at a time of $t_1$.
Then we consider the difference in the value of this observable, obtained on a
fine and a coarse lattice at a sequence of $t_1$ separated by
$0.1\,\mathrm{fm}$. These are compared to the NLO SXPT, NNLO SXPT, SRHO and
SMLLGS predictions for this quantity, evaluated at the exact parameters of the
ensembles.
    
The SMLLGS, the SRHO and the NNLO SXPT taste improvements describe the
numerical data very nicely for $t_1\gsim 2.0\,\mathrm{fm}$, fairly well for
$t_1\gsim 1.0\,\mathrm{fm}$ and all the way down to $t_1\simeq
0.4\,\mathrm{fm}$ in the case of SRHO. All three slightly overestimate the
observed cutoff effects, the rho-meson based approach performing best, whereas
NNLO displays a large deviation from the lattice results in the $t_1\le
0.8\,\mathrm{fm}$ region.  The lattice results have a maximum at
$t_1=1.4\,\mathrm{fm}$, as does the SRHO improvement, reinforcing our
confidence that this model captures the relevant physics.

These findings lead us to apply the following taste corrections to our
simulations results for $a_\mu^\mathrm{light}(L,T,a)$, obtained on an
$L^3\times T$ lattice with lattice spacing $a$, before performing continuum
extrapolations:
\begin{linenomath*}
\begin{equation*}
    \begin{aligned}[]
	a_\mu^\mathrm{light}(L,T,a) \to
	a_\mu^\mathrm{light}(L,T,a)
	&+\tfrac{10}{9}\left[a_{\mu,t \ge t_\mathrm{sep}}^\mathrm{RHO}(L_\mathrm{ref},T_\mathrm{ref})-a_{\mu,t \ge t_\mathrm{sep}}^\mathrm{SRHO}(L,T,a) \right],
    \end{aligned}
\end{equation*}
\end{linenomath*}
with $t_\mathrm{sep}=0.4,0.7,1.0,1.3\,\mathrm{fm}$, and where the factor
$(10/9)$ is related to the quark charges. Note that by using $L_\mathrm{ref}$
and $T_\mathrm{ref}$ in the above Equation, we are applying a very small volume
correction to interpolate all of our simulation results to the same reference,
four-volume so that they can be extrapolated to the continuum limit together.

The taste-improved data is then extrapolated to the continuum using our
standard fit procedure, in the course of which isospin-breaking effects are
also included.  For estimating the systematic error we use a histogram
technique. The central values and the detailed error budget of this analysis
can be found in the Supplementary Information.

The procedure described above does not yet take into account the systematic
uncertainty associated with our choice of SRHO for taste improvement for
$t>1.3\,\mathrm{fm}$. Since applying no taste improvement in that region is not
an option, because of the nonlinearities introduced by two-pion, taste
violations, we turn to NNLO SXPT, only as a means to estimate the uncertainty
associated with this choice. Thus, we define this systematic uncertainty as ERR
= (SRHO $-$ NNLO SXPT) for $t>1.3\,\mathrm{fm}$.  Then, we perform the same
histogram analysis but with SRHO, SRHO-ERR and SRHO+ERR improvements. From this
histogram we extract the contribution which comes from the variation in the
improvement model from SRHO-ERR to SRHO+ERR.  We assign this full spread to the
systematic uncertainty associated with the taste-improvement procedure. We add
this error in quadrature to the error given by the histogram technique
discussed in the previous paragraph.

\begin{figure}
    \centering
    \includegraphics[width=0.7\textwidth]{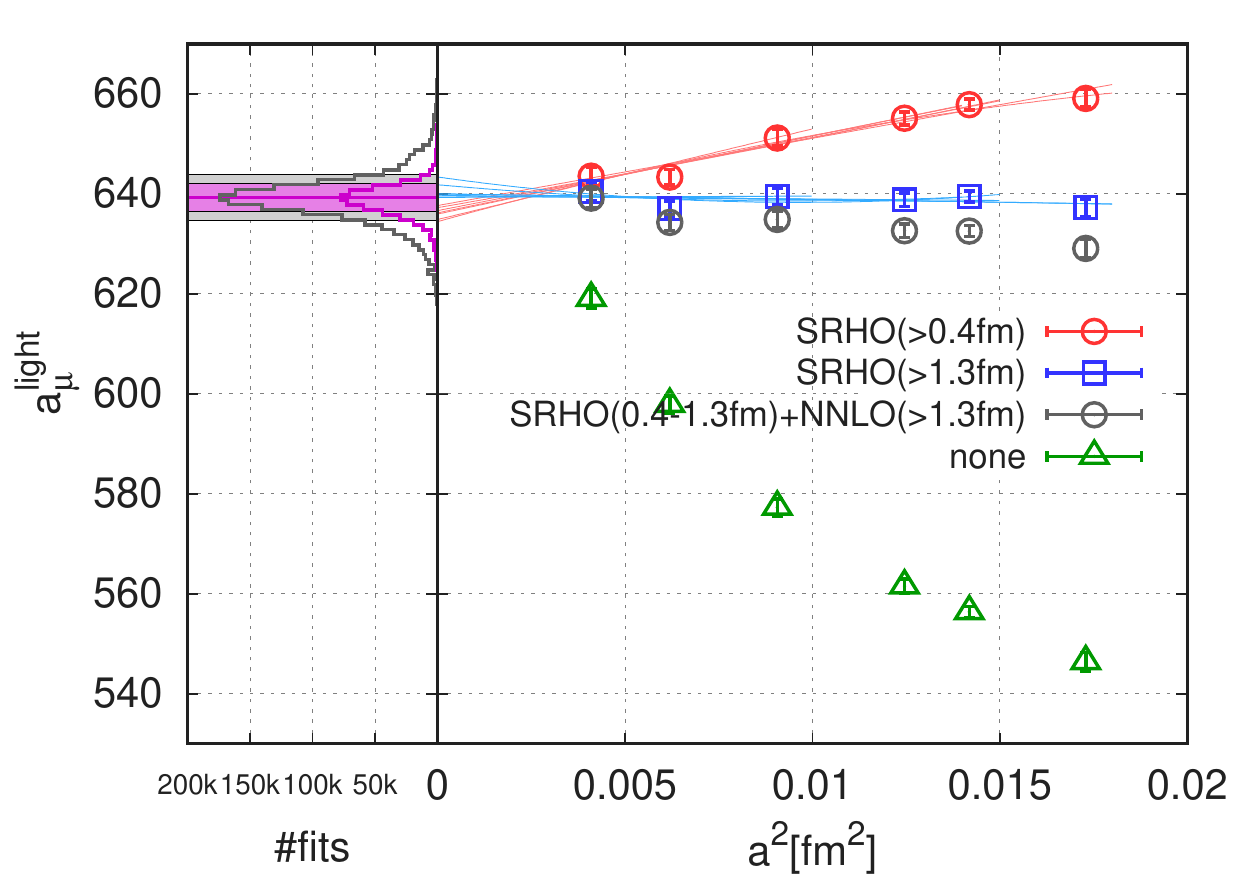}
    \caption
    {
	\dbl \label{extfi:cont} Example continuum limits of $a_\mu^\mathrm{light}$.
	The light green triangles labeled none correspond to our lattice
	results with no taste improvement. The blue squares have undergone no
	taste improvement for $t< 1.3\,\mathrm{fm}$ and SRHO improvement above.
	The blue curves correspond to example continuum extrapolations of those
	improved data to polynomials in $a^2$, up to and including $a^4$. Note
	that extrapolations in $a^2\alpha_s(1/a)^3$, with $\alpha_s(1/a)$ the
	strong coupling at the lattice scale, are also considered in our final
	result. The red circles and curves are the same as the blue points, but
	correspond to SRHO taste improvement for $t\ge 0.4\,\mathrm{fm}$ and
	none for smaller $t$.  The purple histogram results from the fits using
	the SRHO improvement, the corresponding central value and error is the
	purple band.  The darker grey circles correspond to results corrected
	with SRHO in the range $0.4{-}1.3\,\mathrm{fm}$, NNLO SXPT for larger
	$t$. These latter fits serve to estimate the systematic uncertainty of
	the SRHO improvement.  The grey band includes this uncertainty, the
	corresponding histogram is shown with grey. Errors are s.e.m.
    }
\end{figure}

The procedure is illustrated in Extended Data Figure~\ref{extfi:cont}, which shows the data
sets for $a_\mu^\mathrm{light}$ without and with taste improvements, as
functions of $a^2$. (See also Figure~\ref{fi:main_cont} of the main paper,
which zooms in on the taste-improved, continuum extrapolations.) The SRHO
improvement with $t_\mathrm{sep}=0.4\,\mathrm{fm}$ are shown as red points,
while blue points correspond to $t_\mathrm{sep}=1.3\,\mathrm{fm}$. These plots
already include isospin-breaking contributions.  An example of our lattice
results with SRHO improvement between $t=0.4\,\mathrm{fm}$ and
$t=1.3\,\mathrm{fm}$ and NNLO SXPT improvement above are shown as grey points
in Extended Data Figure~\ref{extfi:cont}. 

\begin{figure}
    \centering
    \includegraphics[width=0.7\textwidth]{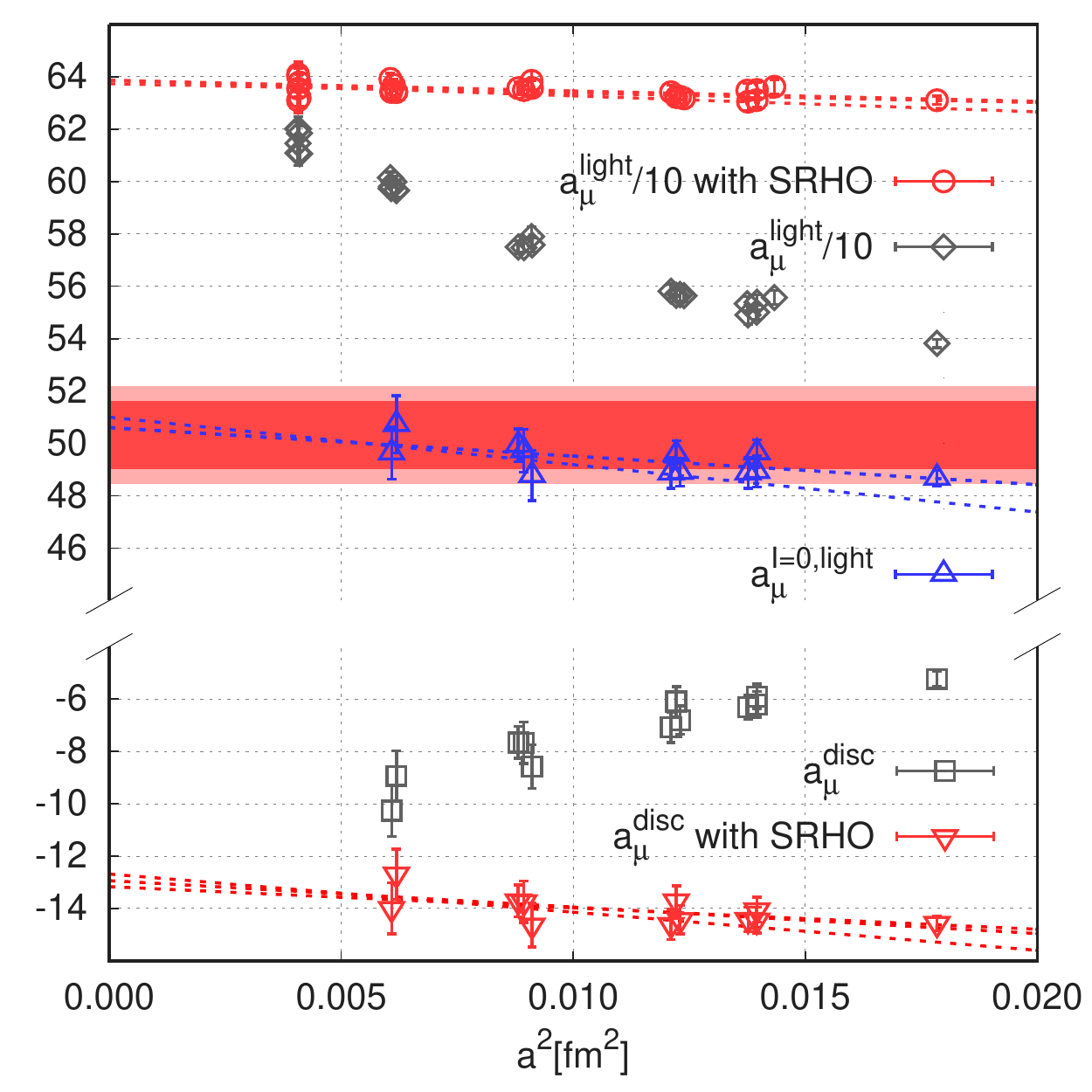}
    \caption
    {
	\dbl \label{extfi:ieq0} Comparison of the continuum extrapolation of
	$a_\mu^{I=0,\mathrm{light}}$ to those of $a_\mu^\mathrm{light}$ and
	$a_\mu^\mathrm{disc}$. The upper set of grey points corresponds to our
	uncorrected results for $\frac1{10}a_\mu^\mathrm{light}$. The upper red
	ones are these same results with our standard SRHO taste improvement.
	They have a much milder continuum limit that exhibits none of the
	nonlinear behavior of the grey points. The red curves show typical
	examples of illustrative continuum extrapolations of those
	points. The lower set of grey and red points and curves are the
	same quantities, but for $a_\mu^\mathrm{disc}$. Combining the results
	from the two, individual, continuum extrapolations of
	$\frac1{10}a_\mu^\mathrm{light}$ and $a_\mu^\mathrm{disc}$, according
	to Equation~\eqref{exteq:amuIeq0}, gives the result with statistical
	errors illustrated by the red band, and with combined statistical and
	systematic errors, by the broader pink band. The blue points correspond
	to our results for $a_\mu^{I=0,\mathrm{light}}$, for each of our
	simulations, and are obtained by combining the two sets of grey points,
	according to Equation~\eqref{exteq:amuIeq0}. As these blue points show,
	the resulting continuum-limit behavior of $a_\mu^{I=0,\mathrm{light}}$
	is much milder than that of either the uncorrected
	$a_\mu^\mathrm{light}$ or $a_\mu^\mathrm{disc}$, and shows none of the
	curvature exhibited by them. This behavior resembles much more that of
	the taste-improved, red points. Moreover, all of the blue points,
	including typical continuum extrapolations drawn as blue lines, lie
	within the bands.  This suggests that our taste improvements neither
	bias the central values of our continuum extrapolated
	$a_\mu^\mathrm{light}$ and $a_\mu^\mathrm{disc}$, nor do they lead to
	an underestimate of uncertainties. Errors are s.e.m.
    }
\end{figure}

An important check of our taste-improvement procedure is provided by the study
of the isoscalar or $I=0$ contribution to $a_\mu$, as suggested by arguments
made in \cite{Gerardin:2019rua}. Here we work with the isospin-symmetric data sets.
Then the $I=0$ contribution is defined as
\begin{linenomath*}
\begin{equation}
    \label{exteq:amuIeq0}
    a_\mu^{I=0}\equiv
    \frac1{10}a_\mu^\mathrm{light} +a_\mu^\mathrm{disc}+\cdots \ ,
\end{equation}
\end{linenomath*}
where the ellipsis stands for the quark-connected contributions of the more
massive $s$, $c$, \ldots\ quarks. This quantity receives no two-pion
contributions: it starts with three pions, whose taste-breaking effects should
be very small. Thus, if our understanding of discretization errors in
$a_\mu^\mathrm{light}$ and $a_\mu^\mathrm{disc}$ is correct, the taste-breaking
corrections observed in the light and disconnected quantities must be largely
absent from $a_\mu^{I=0}$. As a consequence, we expect the continuum
extrapolation of $a_\mu^{I=0}$ to be much milder. That is exactly what is shown
in Extended Data Figure \ref{extfi:ieq0} and explained in its caption.

\section*{Results for $w_0$, $M_{ss}$ and $\Delta M^2$}

Here we describe the details of the analyses that are used to obtain the
physical values of $w_0$, $M_{ss}$ and $\Delta M^2=M_{dd}^2-M_{uu}^2$ from the
experimental values of hadron masses, including the mass of the $\Omega$
baryon. The $M_{uu}$/$M_{dd}$/$M_{ss}$ denote the masses of mesons built from
an up/down/strange and an anti-up/down/strange quark without quark-disconnected
contributions. The results for these quantities are used to define the
isopin-symmetric point in this paper, as described in the Supplemental
Information.

In the case of $w_0$, the observable we fit is $w_0M_\Omega$.  To account for
the systematic error due to the different continuum extrapolations we apply
both linear and quadratic functions in the isospin-symmetric component. We also
remove zero/one/two/three of the coarsest lattice spacings in the linear and
zero/one/two lattice spacings in the quadratic fits. For the tiny valence QED
component only linear fits are applied, with zero/one/two points removed; for
the even smaller sea QED contributions, we have either a constant or a linear
fit with all lattice spacings.

The systematic error of the hadron mass fits is taken into account by 24
different combinations of the fit ranges: three for the $M_\Omega$ mass, two
for the pseudoscalars, two for the isospin breaking of the $M_\Omega$ and two
for the isospin breaking of the pseudoscalars.  To account for the experimental
error on $M_\Omega$ we carry out the analysis with two different experimental
values: one that corresponds to the central value plus the experimental error;
the other with this error subtracted. 

Altogether, these yield a total of 129024 fits. When the different analyses
are combined into a histogram to determine the systematic error, the results from
different fit functions or lattice spacing cuts are weighted with the Akaike
Information Criterion, the rest with flat weighting. We obtain
\begin{linenomath*}
\begin{equation}
    \label{exteq:w0}
    w_0 = 0.17236(29)(63)[70]\ \text{fm}\ ,
\end{equation}
\end{linenomath*}
where the first error is statistical, and the second is systematic and the
third is the total error; we reach a relative precision of $0.4$\%.  The split
up of the error into different sources can be found in the Supplementary
Information.  In Extended Data Figure \ref{extfi:w0} we show the various
isospin components of $w_0M_\Omega$ versus the lattice spacing squared,
together with the different continuum extrapolations.  Our result
\eqref{exteq:w0} is in good agreement with earlier four-flavor determinations:
$w_0=0.1715(9)$~fm of \cite{Dowdall:2013rya} and
$w_0=0.1714\left(\begin{smallmatrix}+15 \\-12\end{smallmatrix}\right)$~fm of
    \cite{Bazavov:2015yea}. In those studies the isospin-breaking effects were
    only estimated, whereas in our case they are fully accounted for.

The same procedure is used for $M_{ss}$. We actually work with
$(M_{ss}/M_{\Omega})^2$ instead of $M_{ss}/M_{\Omega}$, since the fit
qualities are much better in the first case.  The 129024 different fits give
\begin{linenomath*}
\begin{equation}
    M_{ss}= 689.89(28)(40)[49]\ \text{MeV}\ ,
\end{equation}
\end{linenomath*}
with statistical, systematic and total errors.

Finally we also carry out the analysis for $\Delta M^2/M_\Omega^2$ with $\Delta
M^2=M_{dd}^2-M_{uu}^2$, which is a measure of the strong-isospin breaking.
Altogether we have 3328 fits, which give the following central value with
statistical, systematic and total errors:
\begin{linenomath*}
\begin{equation}
    \Delta M^2= 13170(320)(270)[420]\ \text{MeV}^2\ .
\end{equation}
\end{linenomath*}

\printbibliography[segment=2]

\clearpage

\begin{table}
    \centering
    \begin{tabular}{C|C|C|C}
	M_\pi \text{ in {\tt 4HEX}}\to & 104\text{ MeV} & 121\text{ MeV} & 110\text{ MeV} \\
	\hline
	a_{\mu}^\mathrm{light}(56\times 84) &685.9(2.7)& 668.3(2.0)& 679.5(1.9)\\
	a_{\mu}^\mathrm{light}(96\times 96) &710.7(1.9)& 684.3(1.7)& 701.1(1.3)
    \end{tabular}
    \caption
    {
	\dbl \label{extta:4hex} Isospin-symmetric component of
	$a_\mu^\mathrm{light}$. The figures are obtained in simulations with
	the {\tt 4HEX} action on two different volumes and two different
	Goldston-pion masses.  In the last column we also give the interpolated
	value at the physical point, using the HMS averaged pion-mass
	prescription.
    }
\end{table}

\begin{table}
    \centering
    \begin{tabular}{L|C|C|C|C|C}
	& \text{NLO XPT} & \text{NNLO XPT} & \text{MLLGS} & \text{HP} & \text{RHO} \\
	\hline
	a_\mu(L_\mathrm{big},T_\mathrm{big})-a_\mu(L_\mathrm{ref},T_\mathrm{ref}) & 11.6 & 15.7 & 17.8 &    - &    - \\
	a_\mu(L_\mathrm{big},\infty)-a_\mu(L_\mathrm{ref},\infty)                 & 11.2 & 15.3 & 17.4 & 16.3 & 14.8 \\
	\hline
	a_\mu(\infty,\infty)-a_\mu(L_\mathrm{big},T_\mathrm{big}) & 0.3 & 0.6 & - &  - &   - \\
	a_\mu(\infty,\infty)-a_\mu(L_\mathrm{big},\infty)         & 1.2 & 1.4 & - &1.4 & 1.4 \\
    \end{tabular}
    \caption
    {
	\dbl \label{extta:xpt} Finite-size effect in the reference box of the
	isospin-symmetric component of $a_\mu$. The figures are obtained in
	various model approaches.
    }
\end{table}

\begin{figure}
    \centering
    \includegraphics[width=0.7\textwidth]{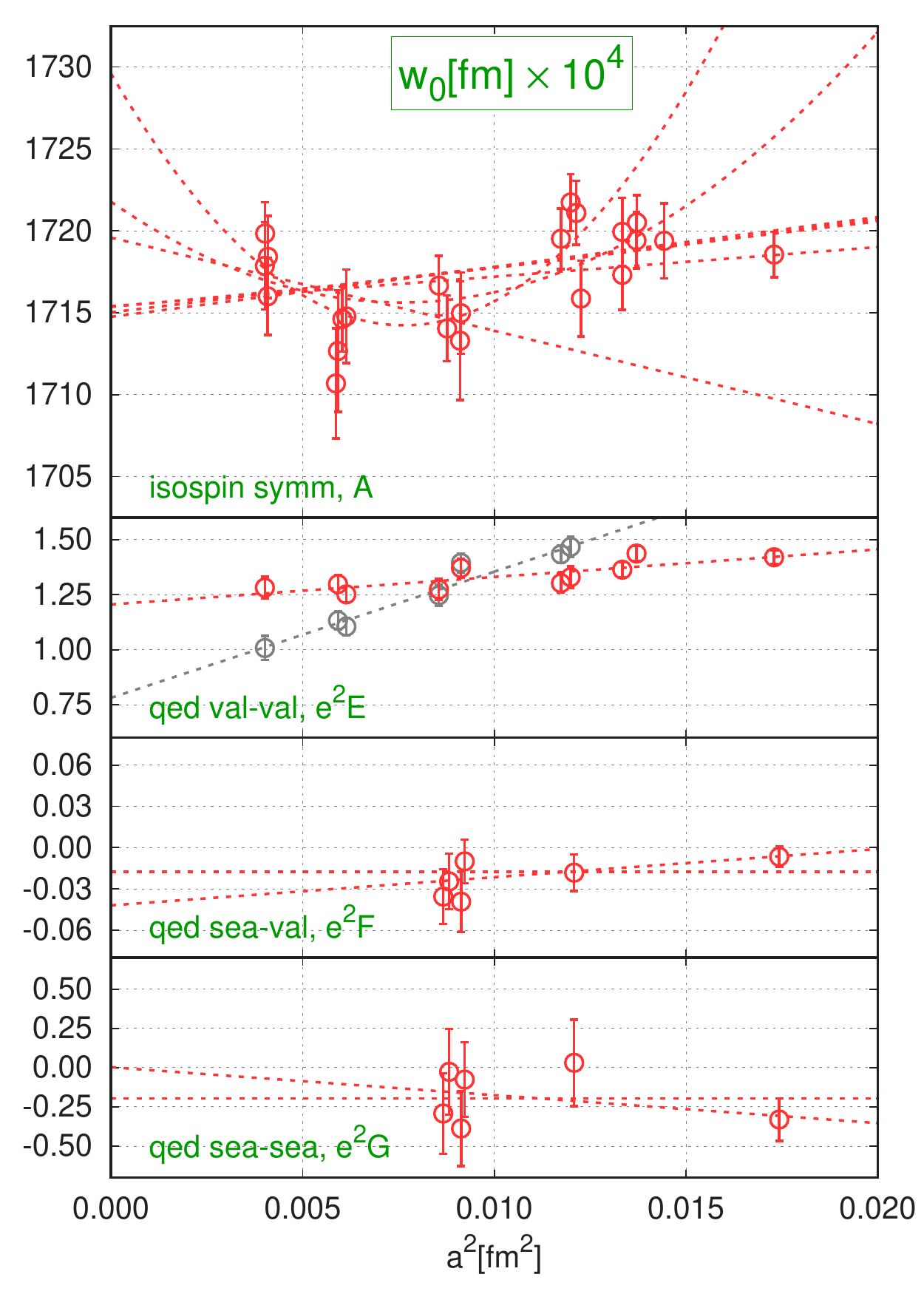}
    \caption
    {
	\dbl \label{extfi:w0} Continuum extrapolations of the contributions to $w_0
	M_\Omega$. From top to bottom: isospin-symmetric, electromagnetic
	valence-valence, sea-valence and sea-sea component. The results are
	multiplied by $10^4/[M_\Omega]_*$, the electric derivatives are
	multiplied by $e_*^2$; here the star subscript denotes the physical
	value. Errorbars show statistical errors (s.e.m.).  Dashed lines are
	continuum extrapolations, they are illustrative examples from our
	several thousand fits.  Only the lattice spacing dependence is shown:
	the data points are moved to the physical light and strange quark mass
	point.  This adjustment varies from fit to fit, the red datapoints are
	obtained in an $a^2$-linear fit to all ensembles. If in a fit the
	adjusted points differed significantly from the red points, we show
	them with grey color. The final result is obtained from a weighted
	histogram of the several thousand fits.
    }
\end{figure}

\end{refsegment}

\clearpage
\def\labelmaincont{2} 
\def\labelmaingdef{(1)} 
\def\labelmainkdef{(3)} 
\renewcommand{\figurename}{Figure}
\renewcommand{\tablename}{Table}
\setcounter{table}{0}
\setcounter{figure}{0}
\setcounter{page}{1}
\setcounter{section}{0}
\setcounter{equation}{0}

\begin{refsection}
    \begin{center}
	{\bf\Huge Supplementary Information}\\
	\vspace*{1.0cm}
	{\bf\Large Leading hadronic contribution to the muon
	magnetic moment from lattice QCD}
    \end{center}
    \authors
    \tableofcontents
    \newpage
    \section{The {\tt 4stout} action and gauge ensembles}
\label{se:act_4stout}

\begin{figure}[t]
    \centering
    \includegraphics[width=0.9\textwidth]{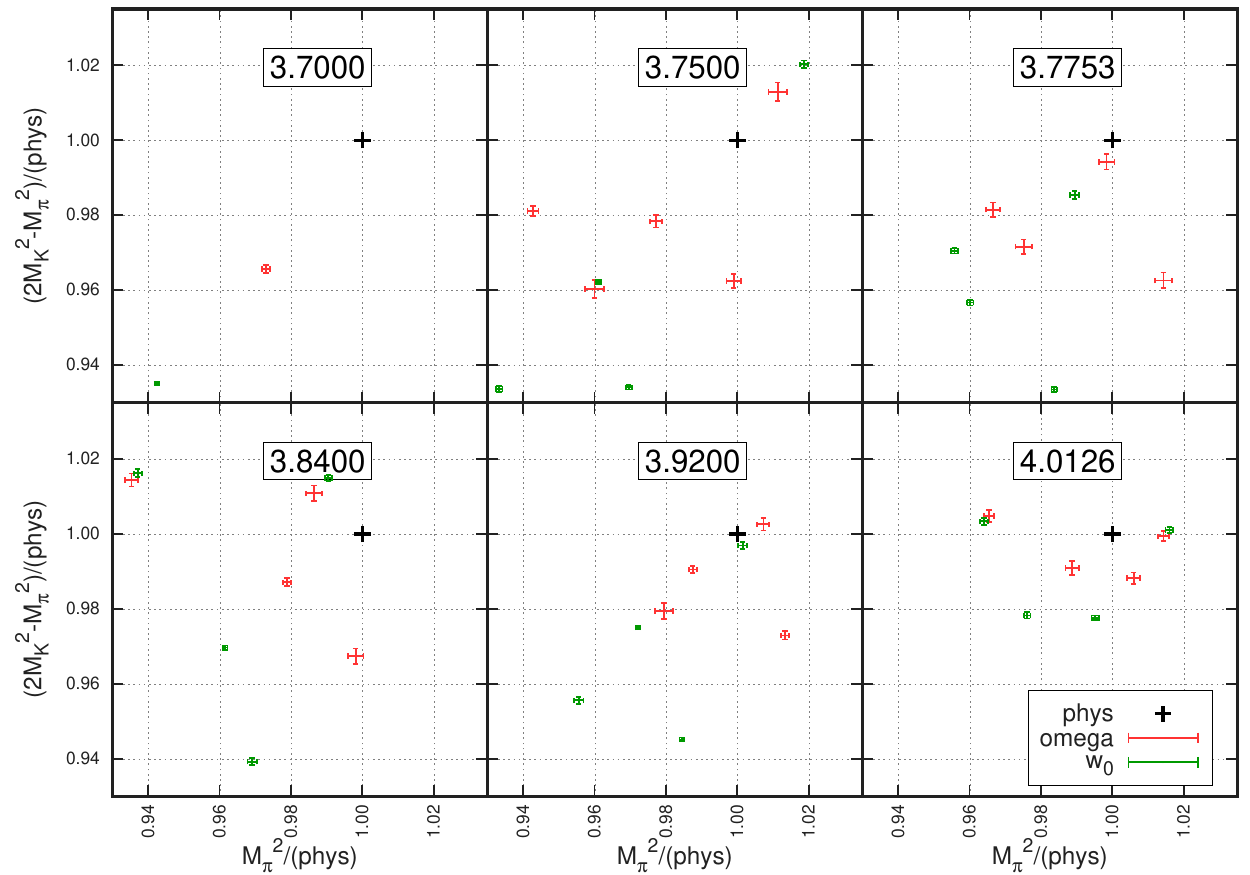}
    \caption{
	\label{fi:microlandscape} Position of the {\tt 4stout} ensembles in the
	plane of the hadron mass combinations of Equation
	\eqref{eq:mpimkratios}. These correspond
	approximately to the light and strange quark masses. The lattice
	spacings are $a=0.1315$, $0.1191$, $0.1116$, $0.0952$,
	$0.0787$ and $0.0640$~fm, respectively. The corresponding $\beta$
	gauge couplings are indicated at the top of each panel.
    }
\end{figure}

\begin{figure}[t]
    \centering
    \includegraphics[width=0.7\textwidth]{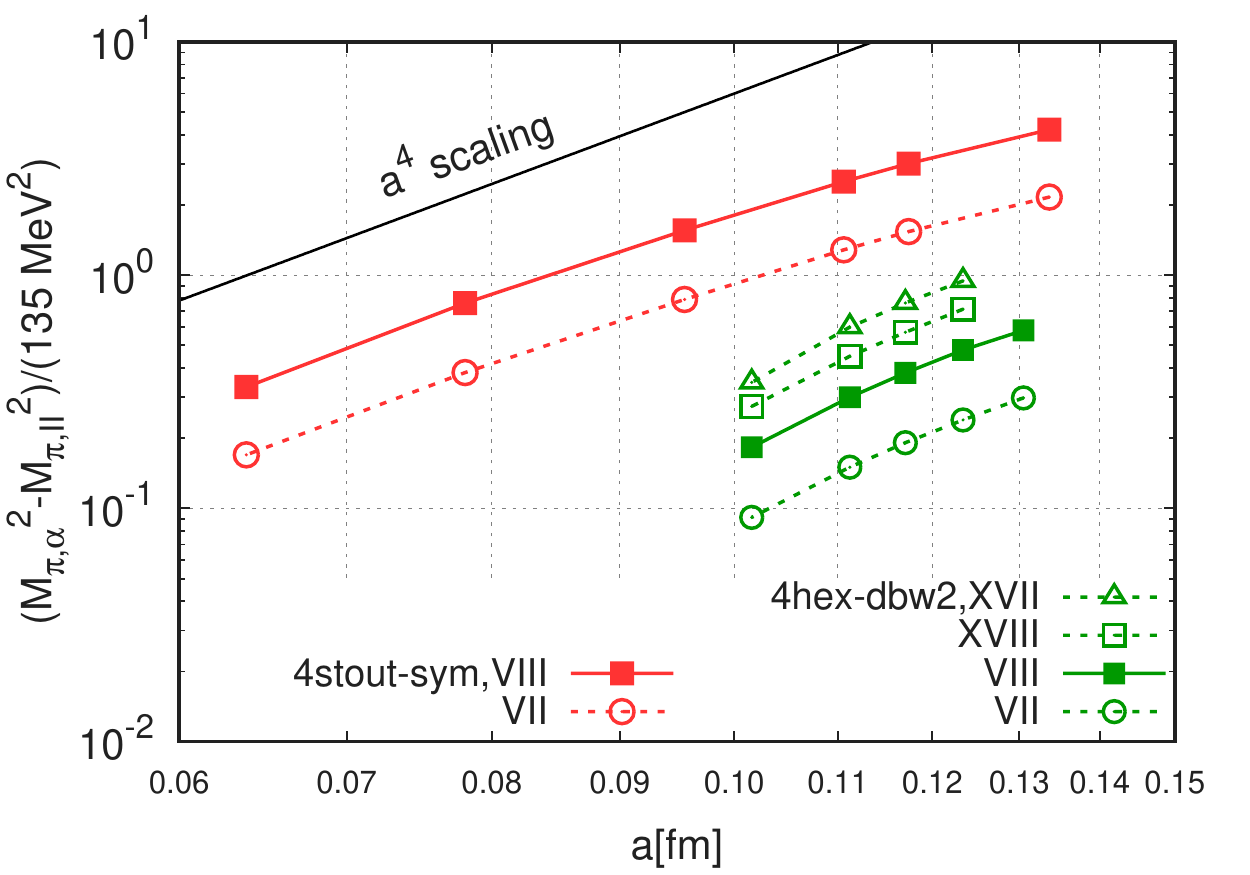}
    \caption
    {
	\label{fi:tavi}Taste multiplet of staggered pions as a function of
	lattice spacing, both for the {\tt 4stout} and the {\tt 4HEX} action.
	We label the meson operators by Roman numbers as in
	\cite{Ishizuka:1993mt}. The pseudo-Goldstone-boson is labeled by $\alpha$=II
	and the root-mean-square pion mass corresponds approximately to the
	operator $\alpha$=VIII, drawn with a solid line.
    }
\end{figure}

\begin{figure}[t]
    \centering
    \includegraphics[width=0.7\textwidth]{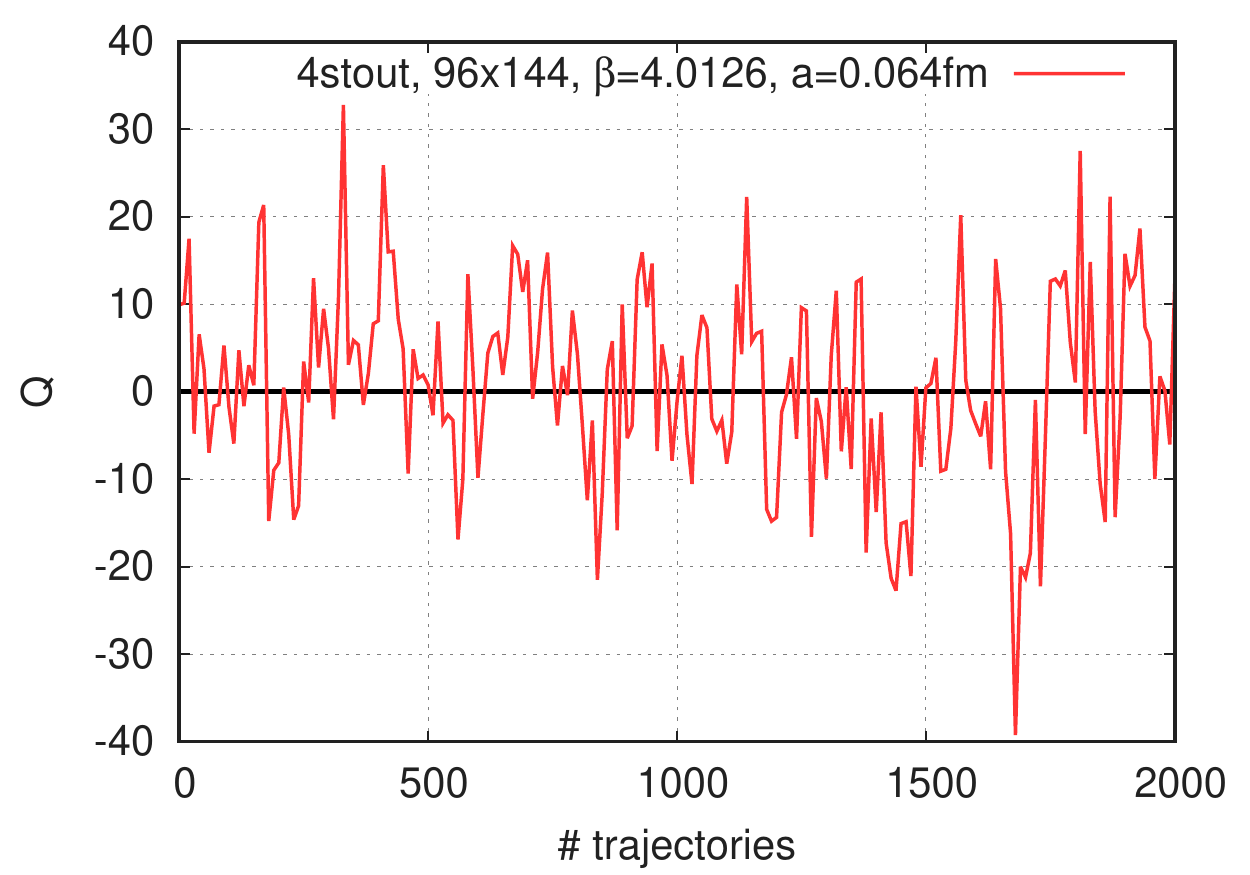}
    \caption
    {
	\label{fi:qhist2}History of topological charge $Q$, defined from the
	Wilson-flow in a {\tt 4stout} run at the physical point. The lattice
	spacing is about $a=0.064$~fm.
    }
\end{figure}

\begin{table}[p]
    \centering
    \begin{tabular}{C|C|C|C|C|R}
	\beta & a[\text{fm}] & L\times T & m_s & m_s/m_l & \# \text{conf} \\
	\hline
	\hline
	3.7000 & 0.1315 & 48\times  64 & 0.057291 & 27.899 &  904 \\ 
	\hline
	3.7500 & 0.1191 & 56\times  96 & 0.049593 & 28.038 &  315 \\ 
	       &        &              & 0.049593 & 26.939 &  516 \\ 
	       &        &              & 0.051617 & 29.183 &  504 \\ 
	       &        &              & 0.051617 & 28.038 &  522 \\ 
	       &        &              & 0.055666 & 28.038 &  215 \\ 
	\hline
	3.7753 & 0.1116 & 56\times  84 & 0.047615 & 27.843 &  510 \\ 
	       &        &              & 0.048567 & 28.400 &  505 \\ 
	       &        &              & 0.046186 & 26.479 &  507 \\ 
	       &        &              & 0.049520 & 27.852 &  385 \\ 
	\hline
	3.8400 & 0.0952 & 64\times  96 & 0.043194 & 28.500 &  510 \\ 
	       &        &              & 0.043194 & 30.205 &  190 \\ 
	       &        &              & 0.043194 & 30.205 &  436 \\ 
	       &        &              & 0.040750 & 28.007 & 1503 \\ 
	       &        &              & 0.039130 & 26.893 &  500 \\ 
	\hline
	3.9200 & 0.0787 & 80\times 128 & 0.032440 & 27.679 &  506 \\ 
	       &        &              & 0.034240 & 27.502 &  512 \\ 
	       &        &              & 0.032000 & 26.512 & 1001 \\ 
	       &        &              & 0.032440 & 27.679 &  327 \\ 
	       &        &              & 0.033286 & 27.738 & 1450 \\ 
	       &        &              & 0.034240 & 27.502 &  500 \\ 
	\hline
	4.0126 & 0.0640 & 96\times 144 & 0.026500 & 27.634 &  446 \\ 
	       &        &              & 0.026500 & 27.124 &  551 \\ 
	       &        &              & 0.026500 & 27.634 & 2248 \\ 
	       &        &              & 0.026500 & 27.124 & 1000 \\ 
	       &        &              & 0.027318 & 27.263 &  985 \\ 
	       &        &              & 0.027318 & 28.695 & 1750 \\ 
    \end{tabular}
    \caption
    {
	\label{ta:4stout}List of {\tt 4stout} ensembles with gauge coupling,
	lattice spacing at the physical point, lattice size, quark masses and
	number of configurations. Two different lines with the same parameters
	means that the ensembles were generated in two different streams.
    }
\end{table}

The main part of the simulation effort was carried out using the
\texttt{4stout} lattice action. This discretization is defined through the use
of the tree-level Symanzik gauge action \cite{Luscher:1984xn} and a one-link
staggered fermion action with four steps of stout smearing
\cite{Morningstar:2003gk}. The smearing parameter was set to $\rho=0.125$.

We have chosen six gauge coupling parameters, $\beta=6/g^2$, as shown in Table
\ref{ta:4stout}. All of these ensembles were generated using 2+1+1 dynamical
flavors with no isospin breaking. The charm mass is set by its ratio to the
strange mass, $m_c/m_s=11.85$, which comes from the spectroscopy of the
pseudoscalar charmed mesons in the continuum limit worked out in
\cite{Davies:2009ih}. This value is within one per-cent of the latest FLAG
average \cite{Aoki:2019cca}.  The light and strange quark masses are chosen to
scatter around a ``physical point'' defined by the pseudoscalar masses $M_\pi$
and $M_K$ and the mass of the Omega baryon, $M_\Omega$, as follows:
\begin{gather}
    \frac{M_\pi^2}{M_\Omega^2} = \left[\frac{M_{\pi_0}^2}{M_{\Omega_-}^2}\right]_*,\qquad
    \frac{M_K^2-\tfrac{1}{2}M_\pi^2}{M_\Omega^2} = \left[ \frac{M_{K_\chi}^2}{M_{\Omega_-}^2}\right]_*.
    \label{eq:mpimkratios}
\end{gather}
where $*$ denotes the experimental value and
\begin{gather}
    M_{K_\chi}^2\equiv \tfrac{1}{2}\left( M_{K^+}^2 + M_{K^0}^2- M_{\pi^+}^2\right)\,.
    \label{eq:mkchi}
\end{gather}
The latter quantity is designed to be approximately proportional to the strange
quark mass with a vanishing leading order sensitivity to strong-isospin
breaking.

In Equation \eqref{eq:mpimkratios}, the mass of the Omega baryon plays the role
of the scale setting variable. It could, in principle, be replaced by any other
dimensionful quantity that satisfies the criteria: a) moderate quark mass
dependence, b) precisely determined in a lattice simulation, c) known
experimental value to an accuracy better than a permil level.  The mass
$M_\Omega$ satisfies all three criteria, see Section \ref{se:obs_mass} for more
details. In this work we also use the $w_0$-scale \cite{Borsanyi:2012zs}, which
is derived from the Wilson-flow of the gauge fields \cite{Luscher:2010iy}. The
main motivation for this scale setting is to define an isospin decomposition,
see Section \ref{se:obs_split}. The $w_0$-scale readily satisfies both the a)
and b) criteria but, alas, it is defined in terms of observables in Euclidean
space, not by any experiment. In order to use $w_0$ for scale setting, we first
determine its physical value from our simulations, using the accurate
$M_\Omega$ scale as an input. This is described in Section \ref{se:res_w0etal}.
Evidently whenever we use the $w_0$ scale setting, both the statistical and the
systematic error of $w_0$, as well as the statistical correlation, will be
accounted for. As a by-product of this procedure, we give a physical value for
$w_0$, including dynamical QED effects, for the first time in the literature.

Our main analysis is based on the 27 ensembles shown in Table \ref{ta:4stout}.
In Figure \ref{fi:microlandscape} we show the ``landscape'' for each of our
lattice spacings: we plot the ensembles in a plane where the $x$ and $y$ axes
give the relative deviation of the light and strange quark masses from their
physical value. These are defined by the hadronic observables and their
experimental values in Equation \eqref{eq:mpimkratios}. The simulation
parameters are chosen in a way that makes interpolation to the physical point
possible. This ``bracketing'' feature is not available for each lattice
spacing, but only if all lattice spacings are considered together. This is not
a problem, since in our analyses we apply global fits with all ensembles
included. In Figure \ref{fi:microlandscape} each ensemble is represented by two
points, corresponding to the $M_\Omega$ and $w_0$ scale settings.  Although we
determined the physical value of $w_0$ using $M_\Omega$ itself, the mass ratios
vary with the choice of scale setting. This is because there are discretization
effects in the $w_0 M_\Omega$ product on the lattice.  Notice that the finer
the lattices, the smaller the difference in the respective mass ratios.

We also measured the taste violation for all six lattice spacings. The result
is shown in Figure \ref{fi:tavi}. The plot shows the mass-squared difference
between a non-Goldstone pion and the Goldstone pion as a function of the
lattice spacing. The difference shows a behavior that resembles $a^4$ in the
range of our smallest three lattice spacings. This is much faster than the
$\alpha_s a^2$ \cite{Lepage:1998vj}, where $\alpha_s$ is the strong coupling
constant at the lattice cutoff scale. The faster falloff is most probably due
to higher order terms of the type $\alpha_s^n a^2$ with $n>1$. At the smallest
lattice spacing the root-mean-square pion mass is about
$m_{\pi,\mathrm{RMS}}=155$~MeV.

In Figure \ref{fi:qhist2} we show the topological-charge history in a run on
our finest lattice. The charge $Q$ was computed using the standard
discretization of the topological charge density at a Wilson-flow time of
$\tau$, which was set to have a smearing radius of about
$\sqrt{8\tau}\approx0.6$~fm.  The integrated autocorrelation time of $Q$ is
found to be $19(2)$ trajectories.

    \section{The {\tt 4HEX} action and gauge ensembles}
\label{se:act_4hex}

\begin{figure}[p]
    \centering
    \includegraphics[width=0.7\textwidth]{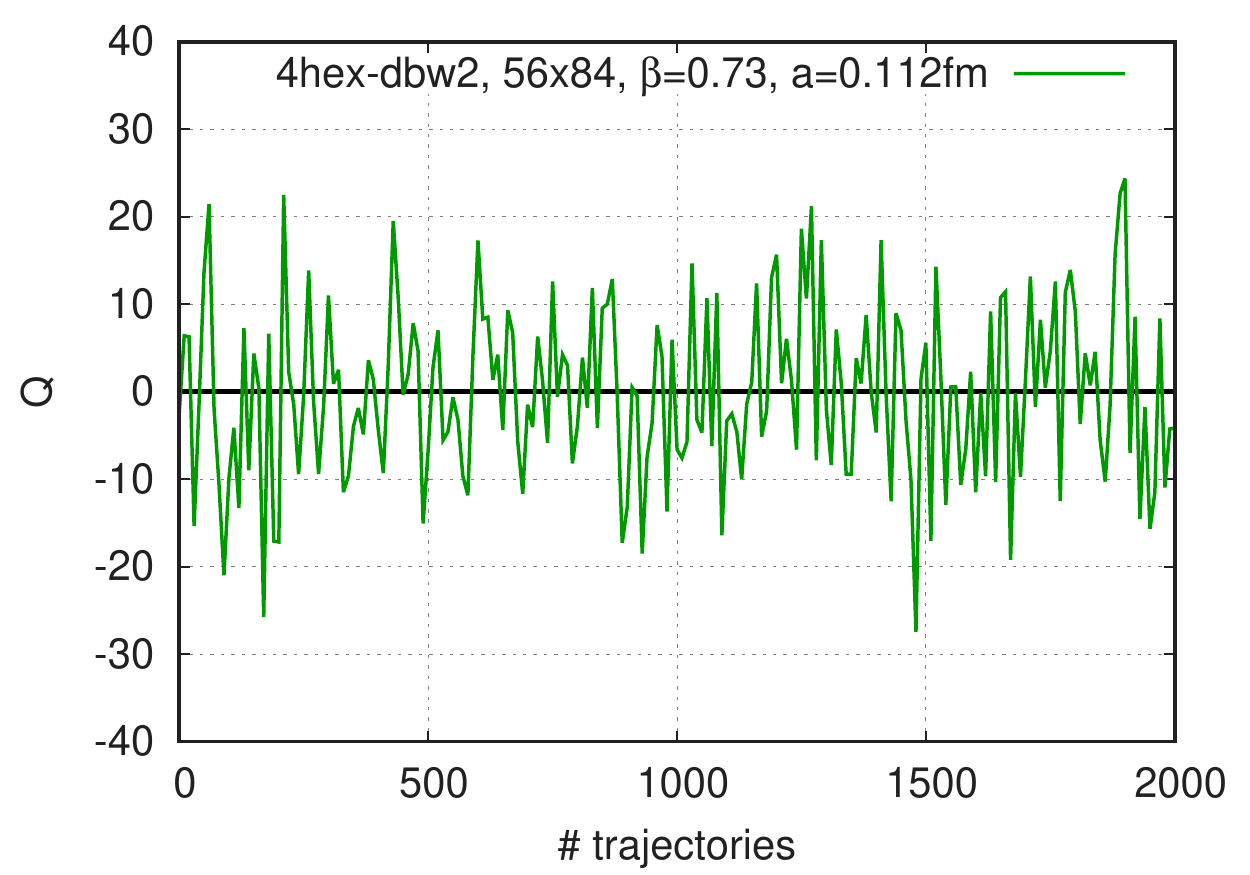}
    \caption
    {
	\label{fi:qhist}History of topological charge, defined from the
	Wilson-flow in a {\tt 4HEX} run at the physical point. The lattice
	spacing is about $a=0.112$~fm and the strange-to-light-quark mass ratio
	is $m_s/m_l=33.728$.
    }
\end{figure}

\begin{figure}[p]
    \centering
    \includegraphics[width=0.7\textwidth]{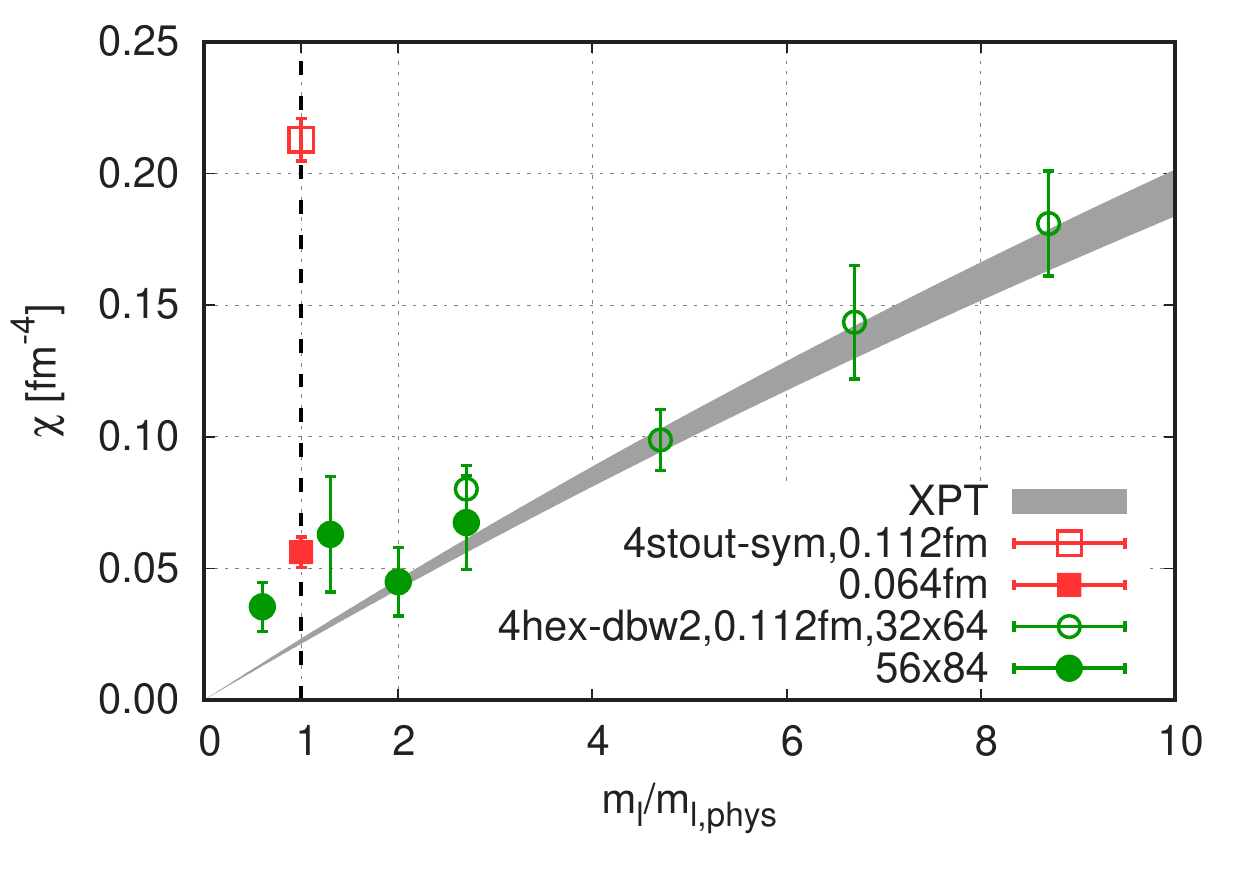}
    \caption
    {
	\label{fi:qsusc}Topological susceptibility as a function of the quark
	mass with the {\tt 4HEX} action at $a=0.112$~fm lattice spacing.  Also
	plotted are {\tt 4stout} results at the physical point and with two
	different lattice spacings.  The grey band is the prediction of leading
	order chiral perturbation theory, with parameters taken from
	\cite{Aoki:2016frl}. The {\tt 4HEX} simulation with the lightest quark
	mass ($m_l= 0.6\cdot m_\mathrm{phys}$) has a topological
	susceptibility, that is about the same as in the continuum at the
	physical point.
    }
\end{figure}

\begin{table}[t]
    \centering
    \begin{tabular}{C|C|C|C|C|R|C}
	\beta & a[\text{fm}] & m_s & m_s/m_l & L\times T & \#\text{ conf} & M_{\pi}[\text{MeV}]  \\
	\hline
	0.7300 & 0.112 & 0.060610 & 44.971 & 56\times84 & 7709 & 104 \\ 
	       &       &          &        & 96\times96 &  962 &     \\ 
	       &       &          & 33.728 & 56\times84 & 8173 & 121 \\ 
	       &       &          &        & 96\times96 &  813 &     \\ 
    \end{tabular}
    \caption
    {
	\label{ta:4hex}List of {\tt 4HEX} ensembles with gauge coupling,
	lattice spacing, quark masses, lattice size, number of configurations
	and Goldstone pion mass. These masses are chosen so that they bracket
	the point where a certain taste-average pion mass has the physical
	value of the pion mass (see text). At that point the Goldstone-pion
	mass is $M_\pi=110$~MeV.
    }
\end{table}

A major systematic effect of the {\tt 4stout} ensembles is related to their box
size, which is about $L\approx6$~fm. To obtain the infinite-volume result it is
desirable to extend the data set with a significantly larger box.  A large box
is computationally only affordable with a large lattice spacing. On coarse
lattices, however, taste violations make the pions too heavy, which completely
distorts the finite-volume behavior and makes a finite-size study pointless. 

We introduce here a new staggered action, called {\tt 4HEX}, to drastically
reduce the taste violation. This requires a gauge action that heavily
suppresses ultraviolet fluctuations and a fermion action with a more aggressive
smearing than {\tt 4stout}. As a result, taste splitting is reduced by an order
of magnitude. Additionally, we lower the Goldstone-pion mass below the physical
value, ensuring that the heavier tastes are closer to the physical pion mass.
The topological susceptibility is particularly sensitive to taste violations.
We show here that the observed reduced taste violation is paired with having
the topological susceptibility much closer to the continuum than in {\tt
4stout} at the same lattice spacing. We use this {\tt 4HEX} action to generate
the lattices for our finite-size study.

The {\tt 4HEX} gauge ensembles use the same action both for the valence and the
sea quarks. They have the following characteristics:
\begin{itemize}

    \item $n_f=2+1$ flavors of one-link staggered quarks with four steps of HEX
	smeared \cite{Capitani:2006ni} gauge links,

    \item DBW2 gauge action \cite{Takaishi:1996xj}, which differs from the
	Symanzik gauge action used in {\tt 4stout} only in the coefficient of
	the $1\times 2$ Wilson-loop.

\end{itemize}
The rationale for this choice is to drastically reduce the ultraviolet
fluctuations in the gauge configurations, since these directly impact the size
of the taste violation.

Four steps of HEX smearing suppresses the ultraviolet fluctuations much more
than four steps of stout smearing, and does so without increasing the locality
range of the smearing procedure. Though higher numbers of HEX smearing steps
would also have been possible, the increasing cost of the smearing and the
marginal improvement in the taste violations make an even higher number of
steps less practical.

The DBW2 gauge action suppresses taste violations even more than the Symanzik
gauge action, as shown eg.\ in \cite{DeGrand:2002vu}. However, it slows down
the decorrelation of the topological charge towards the continuum limit much
more dramatically than other gauge actions. We use the {\tt 4HEX} action only
at lattice spacings where sufficient tunnelings in the topological charge $Q$
are observed. This includes the lattice spacing where we carry out the
finite-volume study of the hadronic vacuum polarization. Figure \ref{fi:qhist}
shows the history of the charge $Q$ in one of these runs. Again, $Q$ is
computed using the standard discretization of the topological charge density at
a Wilson-flow time of $\tau$, which was set to have a smearing radius of about
$\sqrt{8\tau}\approx1.1$~fm. The integrated autocorrelation time of $Q$ is
found to be $6(1)$ trajectories.

In exploring the parameter space of the action we carried out $n_f=3$
simulations at five different $\beta$ values in the range $\beta= 0.70\dots
0.75$, corresponding to lattice spacings $a\approx 0.13\dots 0.10$~fm. Note
that such small $\beta$ values are typical with the DBW2 gauge action. The
quark mass was tuned to the vicinity of the three flavor symmetric point where
the quark mass equals the physical value of $\tfrac{1}{3}(2 m_l +m_s)$. The
lattice sizes were $32\times 64$. Members of the pion taste multiplet are shown
as a function of the lattice spacing in Figure \ref{fi:tavi}, together with
{\tt 4stout} data. The {\tt 4HEX} taste violations are an order of magnitude
smaller than that of the {\tt 4stout} action. At a lattice spacing of
$a=0.112$~fm, {\tt 4HEX} is as good as {\tt 4stout} at $a=0.064$~fm, which is
the finest lattice spacing available.

This reduced taste violation is also reflected in the topological
susceptibility. Non-chiral actions, including staggered fermions, typically
show large discretization errors in this quantity. Figure \ref{fi:qsusc} shows
$n_f=2+1$ {\tt 4HEX} simulations at $\beta=0.73$, $a= 0.112$~fm, with the
physical strange-quark mass. The light-quark mass was varied from $0.6$ to
$8.7$ times its physical value, set by the Goldstone-pion mass. The lattice
sizes are $32\times 64$ and $56\times84$. Results obtained with the {\tt
4stout} action at the physical point are also given in the plot. The {\tt
4stout} result at the same lattice spacing is off by an order of magnitude from
the continuum expectation.  On the other hand the {\tt 4HEX} data closely
follows the continuum curve almost down to the chiral limit, and is as good as
the {\tt 4stout} result at the finest lattice spacing available.

For the finite-size study of the hadronic vacuum polarization we work at
$\beta=0.73$ and $m_s=0.06061$. This choice corresponds to $a=0.112$~fm and
about a physical strange-quark mass.  The ensembles generated are listed in
Table \ref{ta:4hex}.  We have two different volumes with the same parameters.
We will refer to the smaller volume as the ``reference volume''.  It has a
spatial and time extent of
\begin{gather}
    L_\mathrm{ref}=6.272\text{ fm}
    \qquad\text{and}\qquad
    T_\mathrm{ref}=\frac{3}{2}L_\mathrm{ref}\ .
\end{gather}
This geometry corresponds approximately to the geometry of the lattices in the
{\tt 4stout} data set. The larger lattice extents will be denoted by
$L_\mathrm{big}$ and $T_\mathrm{big}$ and are given as
\begin{gather}
    L_\mathrm{big}=T_\mathrm{big}=10.752\text{ fm}\ .
\end{gather}
We also use two light-quark-mass values, so that we can bracket the physical
point. Here, differently from above and also from the {\tt 4stout} data set,
we set the physical point, not with the Goldstone-pion mass, rather with a
prescription that takes into account taste violations. Such a choice is
advantageous for studying finite-size effects, which depend strongly on the
masses of the pions.  The precise definition of the taste-average pion mass
will be given in the section on finite-size effects, Section \ref{se:obs_fv}.
For the Goldstone pion this prescription gives $M_{\pi}=110$~MeV. Let us note
here also that the topological susceptibility, with such a choice, is about the
same as in the continuum limit at the physical point, as shown in Figure
\ref{fi:qsusc}.  The number of configurations saved is also given in Table
\ref{ta:4hex}. They are separated by 10 unit-length RHMC trajectories. 

    \section{Overlap action}
\label{se:act_overlap}

\begin{table}[t]
    \centering
    \begin{tabular}{C|C|C|C|C|R}
	\beta & a[\text{fm}] & L\times T & m_s & m_s/m_l & \# \text{conf} \\
	\hline
	\hline
	3.7000 & 0.1315 & 24\times 48 & 0.057291 & 27.899 &  716 \\ 
	\hline
	3.7753 & 0.1116 & 28\times 56 & 0.047615 & 27.843 &  887 \\ 
	\hline
	3.8400 & 0.0952 & 32\times 64 & 0.043194 & 28.500 & 1110 \\ 
	\hline
	3.9200 & 0.0787 & 40\times 80 & 0.032000 & 26.512 &  559 \\ 
	       &        &             & 0.033286 & 27.738 &  364 \\ 
	\hline
	4.0126 & 0.0640 & 48\times 96 & 0.026500 & 27.634 &  339 \\ 
	       &        &             & 0.027318 & 27.263 &  264 \\ 
    \end{tabular}
    \caption
    {
	\label{ta:win3fm}List of {\tt 4stout} ensembles used in a crosscheck
	with valence overlap quarks. The columns are gauge coupling, lattice
	spacing at the physical point, lattice size, quark masses and number of
	configurations.
    }
\end{table}

\begin{table}[t]
    \centering
    \begin{tabular}{C|C|C|C}
	\beta & m_l & m_\mathrm{ov} & Z_V \\
	\hline
	3.7000 & 0.0021 & 0.0164(2) & 1.1474(3)\\
	3.7753 & 0.0017 & 0.0076(1) & 1.1162(3)\\
 	3.8400 & 0.0015 & 0.0041(1) & 1.0981(2)\\
	3.9200 & 0.0012 & 0.0021(1) & 1.0805(1)\\
    \end{tabular}
    \caption
    {
	\label{ta:match}Staggered light-quark mass, matched overlap quark mass
	and vector renormalization constant for different lattice spacings.
    }
\end{table}

\begin{figure}[t]
    \centering
    \includegraphics[width=0.7\textwidth]{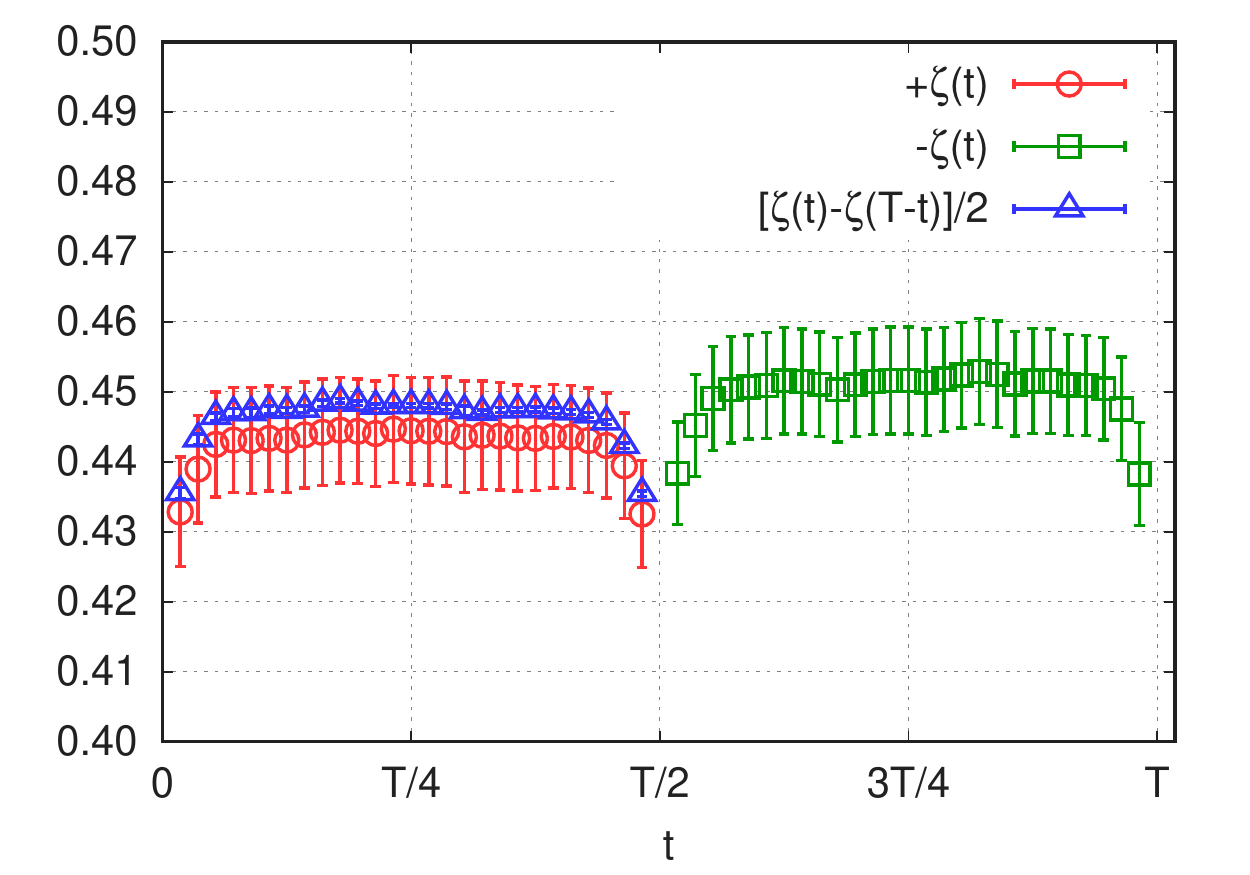}
    \caption
    {
	\label{fi:ovzv}Ratio of three-point and two-point functions $\zeta(t)$
	as function of the current insertion time $t$. This ratio is used to
	define the local vector current renormalization factor as
	$Z_V=[\zeta(T/4)-\zeta(3T/4)]^{-1}$. The plot shows data from a
	$\beta=3.7753$ ensemble, $T=56$.
    }
\end{figure}

In order to crosscheck our results for staggered valence quark artefacts,
including the normalization of the vector current, we compute
$a_{\mu,\mathrm{win}}^\mathrm{light}$ in a mixed action setup, with overlap
valence quarks on gauge backgrounds generated with the {\tt 4stout} staggered
action. We work at the isospin-symmetric point in this crosscheck.

For the sea quarks we use the {\tt 4stout} staggered action and generate
configurations with $L\approx 3$~fm box sizes and at five lattice
spacings, given in Table \ref{ta:win3fm}. The parameters are chosen to match
the parameters of a subset of the $L\approx6$~fm lattices of the {\tt 4stout}
data set given in Table \ref{ta:4stout}.  Though there are significant
finite-size effects in a box of $L\approx3$~fm for an observable like
$a_\mu^\mathrm{light}$, these are much less severe for the window observable
$a_{\mu,\mathrm{win}}^\mathrm{light}$, which is our target here. Our setup is
appropriate for crosschecking the continuum extrapolation and also the
normalization of the vector current.

For the valence quarks we use the overlap fermion formulation
\cite{Neuberger:1997fp}. In particular, the overlap Dirac operator
$D_\mathrm{ov}$ is constructed from the sign function of the Wilson Dirac
operator $D_\mathrm{W}$ as
\begin{gather}
    D_\mathrm{ov}= m_\mathrm{W}\left[ \mathrm{sgn}(\gamma_5 D_\mathrm{W}) + 1\right]\ ,
\end{gather}
where the Wilson operator has a mass of $-m_\mathrm{W}$. We choose
$m_\mathrm{W}=1.3$ and use the Zolotarev approximation of the sign function.
The gauge fields undergo two steps of HEX smearing \cite{Durr:2007cy}. The
overlap mass $m_\mathrm{ov}$ is introduced as
\begin{gather}
    D_\mathrm{ov}(m_\mathrm{ov})= \left( 1- \tfrac{1}{2}m_\mathrm{ov}\right) D_\mathrm{ov} + m_\mathrm{ov}\ .
\end{gather}
This version of the operator has been extensively used in previous thermodynamical
studies \cite{Borsanyi:2016ksw,Borsanyi:2012xf}. We apply $O(a)$
improvement to the overlap propagator by transforming each instance of
$D_\mathrm{ov}^{-1}(m_\mathrm{ov})$ as
\begin{gather}
    D_\mathrm{ov}^{-1}(m_\mathrm{ov}) \to \left( 1 - \tfrac{1}{2}D_\mathrm{ov}/m_\mathrm{W} \right) D_\mathrm{ov}^{-1}(m_\mathrm{ov})\ .
\end{gather}
At the same time we also compute the propagators with {\tt 4stout} staggered
quarks, using the noise reduction technique as on the large volume
ensembles.

We set the overlap quark mass by matching the staggered and overlap pion
masses. For this purpose, we compute overlap pion masses at four values of the
quark mass $m_\mathrm{ov}=0.002,0.005,0.010,0.020$ and interpolate the pion
mass squared using the form
\begin{gather}
    M_{\pi,\mathrm{ov}}^2(m_\mathrm{ov})= A m_\mathrm{ov}^B + C m_\mathrm{ov}^2
\end{gather}
with $A$, $B$ and $C$ fit parameters. This form can capture a possible quenched
chiral logarithm \cite{Sharpe:1992ft} typical in mixed action setups. Our
matching condition is to set the root-mean-square staggered pion mass equal to
the overlap pion mass.  Using the RMS pion is more advantageous than using the
Goldstone-pion mass, since in the latter case we face drastic increase in the
statistical error on our coarsest lattices. For the matched overlap quark
masses we get the values given in Table \ref{ta:match}.

Our determination of $a_{\mu,\mathrm{win}}^\mathrm{light}$ proceeds in a
similar way as in the staggered-on-staggered case, described in Sections
\ref{se:obs_jj} and \ref{se:obs_hvp}. A major difference is that we use the
local vector current in the overlap case. We thus need to compute the current
renormalization constant $Z_V$, which we get by measuring the electric charge
of the pion. For this we compute the ratio of three-point and two-point functions:
\begin{gather}
    \zeta(t)= \frac{\langle P(T/2) V_4(t) \bar{P}(0)\rangle}{\langle P(T/2) \bar{P}(0)\rangle}\ ,
\end{gather}
where the pseudoscalar density $P$ and the local vector current $V_\mu$ are given
in terms of valence overlap fermion fields $\psi_1$ and $\psi_2$ as:
\begin{gather}
    P(t)= \sum_{\vec x} (\bar{\psi}_2 \gamma_5 \psi_1)(\vec x,t)\ ,\quad
    \bar{P}(t)= \sum_{\vec x} (\bar{\psi}_1 \gamma_5 \psi_2)(\vec x,t)\ ,\quad
    V_\mu(t)=\sum_{\vec x} (\bar{\psi}_1 \gamma_\mu \psi_1)(\vec x,t)\ .
\end{gather}
In Figure \ref{fi:ovzv} we show the ratio $\zeta(t)$ as a function of the
timeslice of the current insertion $t$. In the case of a conserved current,
$\zeta(t)=\tfrac{1}{2}$ for $t<T/2$ and $\zeta(t)=-\tfrac{1}{2}$ otherwise. The
renormalization factor should be defined, so that the $\zeta$-ratio
for the renormalized current $Z_V V_4$ equals to $1$ at some physical distance,
for which we take $T/4$. We define the renormalization factor as
$Z_V=[\zeta(T/4) - \zeta(3T/4)]^{-1}$, which includes a trivial symmetrization
in time. The values for the different ensembles are given in Table
\ref{ta:match}.

    \newpage
    \section{Hadron mass measurements}
\label{se:obs_mass}

\begin{table}[t]
    \centering
    \begin{tabular}{L|C|C}
	& \text{range \#1 [fm]} & \text{range \#2 [fm]}\\
	\hline
	\text{pion} &  1.8 \dots 3.0  &  2.0 \dots 3.8 \\
	\text{kaon} &  2.1 \dots 3.3  &  2.4 \dots 3.6 \\
	\bar ss     &  2.1 \dots 3.3  &  2.4 \dots 3.6
    \end{tabular}
    \caption{\label{ta:meson}Fit ranges for extracting pseudoscalar masses on isospin
    symmetric ensembles.}
\end{table}

\begin{table}[t]
    \centering
    \begin{tabular}{C||C|C||C|C||C|C|C}
	& & &
	\multicolumn{2}{c||}{four-state fit} &
	\multicolumn{3}{c}{GEVP fit}\\
	\beta &
	N_\mathrm{Wptl} &
	N_\mathrm{3d} &
	\text{range \#1} & \text{range \#2} & t_a & t_b & \text{range} \\
	\hline
	3.7000& 24& 32&  7\dots 17 &  6\dots 14 & 4& 7&   6\dots14\\
	3.7500& 30& 40&  7\dots 19 &  6\dots 16 & 4& 7&   6\dots14\\
	3.7753& 34& 46&  8\dots 20 &  7\dots 17 & 4& 7&   7\dots17\\
	3.8400& 46& 62&  9\dots 23 &  8\dots 20 & 4& 9&   8\dots16\\
	3.9200& 67& 90& 11\dots 28 & 10\dots 24 & 6& 9&   9\dots17\\
	4.0126&101&135& 14\dots 30 & 12\dots 30 & 6&11&  11\dots19\\
    \end{tabular}
    \caption
    {
	\label{ta:omega} Parameters used for obtaining the $\Omega$ mass:
	number of Wuppertal and gauge-link smearing steps in the $\Omega$
	operator; fit ranges \#1 and \#2 for the four-state mass fit in
	Equation \eqref{eq:4state}; parameters and fit ranges for the GEVP
	based mass fit.
    }
\end{table}

\begin{figure}[t]
    \centering
    \includegraphics[width=0.7\textwidth]{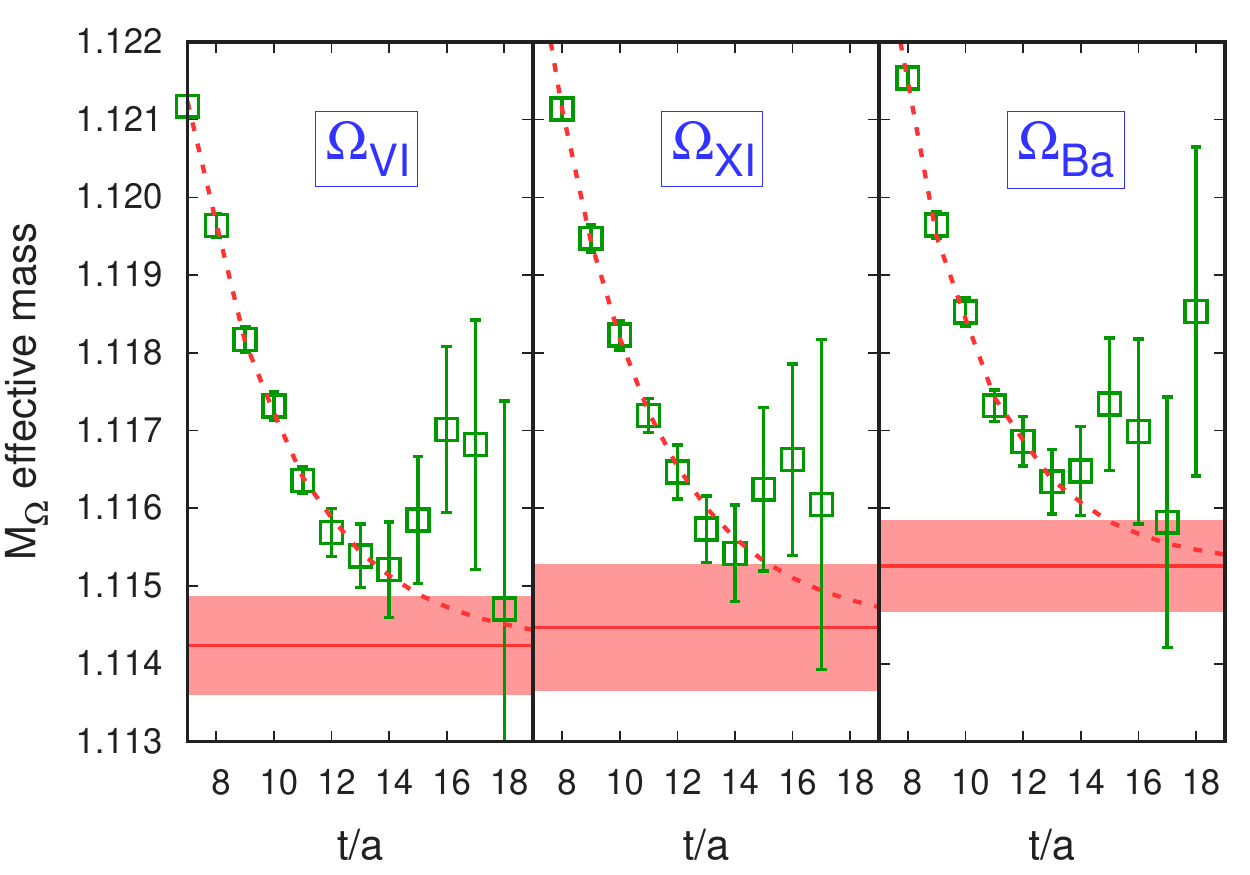}
    \caption
    {
	\label{fi:omefm} Effective mass of the ground state of the $\Omega$
	baryon in lattice units on our coarsest ensemble with $\beta=3.7000$.
	Results with three different staggered operators, $\Omega_\mathrm{VI}$,
	$\Omega_\mathrm{XI}$ and $\Omega_\mathrm{Ba}$ are shown. The horizontal
	lines and the shaded regions represent the fit values and the errors
	obtained with a four-state fit, Equation \eqref{eq:4state}, using
	range \#1 from Table \ref{ta:omega}. The respective $\chi^2$-values
	including the contribution of the priors are $2.1$, $1.6$ and $2.7$ for
	$6$ degrees of freedom. The dashed lines are the effective masses
	computed from the fitted functions.
    }
\end{figure}

\begin{figure}[t]
    \centering
    \includegraphics[width=0.7\textwidth]{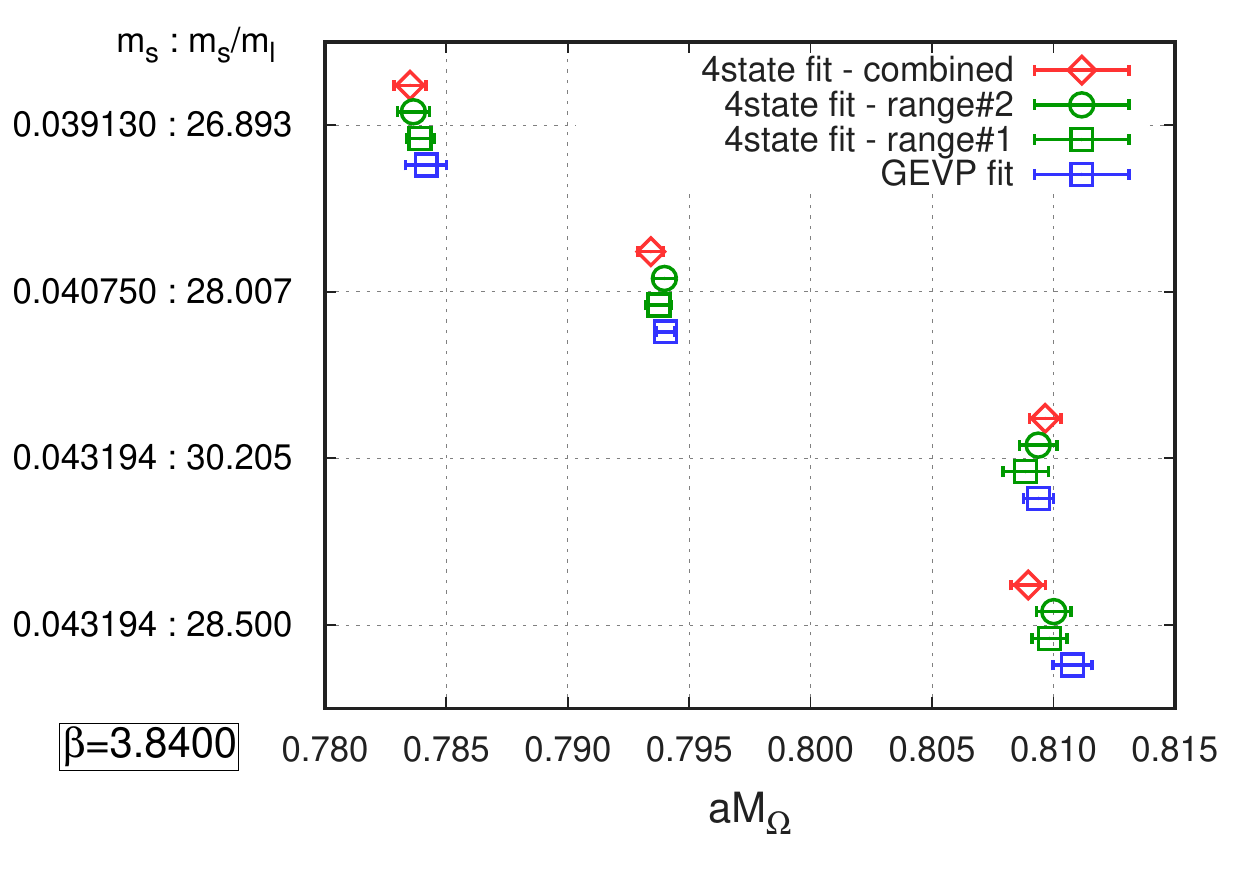}
    \caption
    {
	\label{fi:omega} The mass of the $\Omega$ baryon in lattice units
	extracted for four ensembles with $\beta=3.8400$, which bracket the
	physical point. We compare four methods. The deviation between them is
	used to estimate the systematic error (see text for details).
    }
\end{figure}

\subsection*{Pseudoscalar mass measurements}

The pseudoscalar propagators are computed with random wall sources and point
sinks, using an operator corresponding to the pseudo-Goldstone taste.  To
extract the mass and the decay constant we performed a correlated
$\cosh[M(t-T/2)]$ fit, using sufficiently late time slices to allow for this
simple form. To estimate systematic errors, we selected two fit windows which
are given for the different pseudoscalars in Table \ref{ta:meson}. For the
kaons we selected the even/odd slices for the first/second fit window,
respectively. These fit ranges were chosen by performing a Kolmogorov-Smirnov
test, which ascertains whether the fit qualities in all of the fits on the
ensembles of Table \ref{ta:4stout} follow a uniform distribution.

\subsection*{Omega propagators}

To extract the mass of the positive-parity, ground-state $\Omega$ baryon, a
number of different operators are available in the staggered formalism. First,
there are two operators from the pioneering work of Golterman and Smit
\cite{Golterman:1984dn}. To label these operators we use the convention of
\cite{Ishizuka:1993mt}:
\begin{align}
    \Omega_\mathrm{VI}(t)&=
    \sum_{x_k \mathrm{even}} \epsilon_{abc} \left[
	S_1 \chi_a S_{12} \chi_b S_{13} \chi_c -
	S_2 \chi_a S_{21} \chi_b S_{23} \chi_c +
	S_3 \chi_a S_{31} \chi_b S_{32} \chi_c\right](x),\\
    \Omega_\mathrm{XI}(t)&=
    \sum_{x_k \mathrm{even}} \epsilon_{abc} [S_1 \chi_a S_2 \chi_b S_3 \chi_c](x).
\end{align}
Here, $\chi_a(x)$ is the strange-quark field with color index $a$. The operator
$S_\mu$ performs a symmetric, gauge-covariant shift in direction $\mu$, while
$S_{\mu\nu}\equiv S_\mu S_\nu$. Both $\Omega_\mathrm{VI}$ and
$\Omega_\mathrm{XI}$ couple to two different tastes of the $\Omega$ baryon,
which become degenerate in the continuum limit. At finite lattice spacing
however, there is a splitting between the two tastes. In principle they could
be disentangled by carrying out an analysis involving the correlators of both
$\Omega_\mathrm{VI}$ and $\Omega_\mathrm{XI}$ and also their cross terms.
Later, Bailey successfully constructed an operator which only couples to a
single taste \cite{Bailey:2006zn}.  To achieve this, two additional (valence)
strange quarks are introduced. In other words, the strange-quark field gets an
additional ``flavor'' index: $\chi_{a\alpha}$ with $\alpha=1,2,3$. The operator
is then given as
\begin{align}
    \begin{aligned}
	\Omega_\mathrm{Ba}(t)&=
	\left[
	    2\delta_{\alpha1}\delta_{\beta2}\delta_{\gamma3} -
	    \delta_{\alpha3}\delta_{\beta1}\delta_{\gamma2} -
	    \delta_{\alpha2}\delta_{\beta3}\delta_{\gamma1} +
	    (\dots\beta \leftrightarrow\gamma\dots) \right]\cdot \\
	&\cdot\sum_{x_k \mathrm{even}} \epsilon_{abc} \left[
	    S_1 \chi_{a\alpha} S_{12} \chi_{b\beta} S_{13} \chi_{c\gamma} -
	    S_2 \chi_{a\alpha} S_{21} \chi_{b\beta} S_{23} \chi_{c\gamma} +
	    S_3 \chi_{a\alpha} S_{31} \chi_{b\beta} S_{32} \chi_{c\gamma}\right](x).
    \end{aligned}
\end{align}
The mass of this state becomes degenerate with the above two taste partners in
the continuum limit. We investigated the difference between these three
operators on an ensemble with large statistics. At $\beta=3.7000$,
corresponding to our coarsest lattice spacing, we generated about 3000
configurations in addition to the statistics listed in Table \ref{ta:4stout}.
Note that only the $\Omega$ operators were measured on these extra
configurations.  The effective masses for the above three operators are shown
in Figure \ref{fi:omefm}. In the asymptotic regime we see deviations below
0.1\%, which gives an estimate of the taste violation. We expect that these
will get smaller as we go to finer lattice spacings, as it does for pions.  In
this work we chose the $\Omega_\mathrm{VI}$ operator for our scale setting
measurements. This is justified, since typical statistical and systematic
errors on our ensembles are around 0.1\%, and thus cover the taste-violation
effects estimated here.

As is usual with staggered fermions, these propagators have an oscillating
contribution, corresponding to negative parity states.  There are also
excited states for both parities in the propagators. We suppress the excited
state by a number of Wuppertal smearing steps~\cite{Gusken:1989ad} applied
equally to the source and sink. To avoid mixing between time slices, we define
the smearing to act in the spatial directions only. Since staggered baryon
operators are defined on a coarse lattice with spacing $2a$, our implementation
of the Wuppertal smearing connects only sites that reside on the same
sublattice with $2a$ lattice spacing. In this way there is no interference with
the spin-taste structure of the operators. The action of the smearing operator
$\hat{W}$ on a vector $v_x$ with coefficient $\sigma$ is given as:
\begin{gather}
    \label{eq:wptl}
    [\hat{W}v]_x= (1-\sigma) v_x + \frac{\sigma}{6} \sum_{\mu=1,2,3}
    \left( U^\mathrm{3d}_{\mu,x} U^\mathrm{3d}_{\mu,x+\mu} v_{x+2\mu} + U^{\mathrm{3d},\dagger}_{\mu,x-\mu} U^{\mathrm{3d},\dagger}_{\mu,x-2\mu} v_{x-2\mu} \right),
\end{gather}
where the $U^\mathrm{3d}[U]$ parallel transporters are inserted to keep the
recipe gauge invariant: no gauge fixing is needed. These parallel
transporters are built from a smeared version of the underlying gauge field
configuration $U$, by applying a number of three dimensional stout smearing
steps $N_\mathrm{3d}$. For this we use the same $\rho=0.125$ parameter that we
also have in the link smearing of the Dirac operator. Note that the
$U^\mathrm{3d}$ links are also needed to build the baryon operator, as they
appear in the $S_\mu$ shifts. The smearing operation in Equation
\eqref{eq:wptl} is iterated $N_\mathrm{Wptl}$ times on both source and sink
sides on a point vector with a coefficient of $\sigma=0.5$. In Table
\ref{ta:omega} we list the number of smearing steps for each lattice spacing.
The smearing radii corresponding to the number of smearing steps approximately
follow the change in the lattice spacing. That way the smearing radius in
physical units is kept constant. Note that we are discussing a three dimensional
smearing here: this only effects the overlap of the mass eigenstates with the
operator and leaves the masses invariant.

To enhance the signal, we calculate the $\Omega$ propagator from 256 different
source fields on each gauge configuration listed in Table \ref{ta:4stout}. For
each source field we select a random time slice, which, in turn, is
populated with eight independent $Z_3$ random point sources at $(0,0,0)$,
$(L/2,0,0)$, \dots and $(L/2,L/2,L/2)$. This formation of sources is usually
called a grid source. We also randomize the center of the grid
source.

\subsection*{Omega mass determination: four-state fits}

Our model for the propagator is a four-state fit function $h$, with two
positive and two negative parity states:
\begin{gather}
    \label{eq:4state}
    h(t,A,M)= 
    A_0 h_+(M_0,t) + A_1 h_-(M_1,t) + A_2 h_+(M_2,t) + A_3 h_-(M_3,t)
\end{gather}
where the
\begin{gather}
    h_+(M,t)= e^{-Mt} + (-1)^{t-1} e^{-M(T-t)}\qquad
    \text{and}\qquad
    h_-(M,t)= -h_+(M,T-t)
\end{gather}
functions describe the time dependence of the positive and negative parity
states, see eg. Equation (123) of \cite{Bazavov:2009bb}.  Here $M_0$ and $A_0$
are the mass and amplitude of the ground state.  Our $\chi^2$ function is
defined as a sum of the correlated $\chi^2$ of the model $h$ and a prior term:
\begin{gather}
    \chi^{2}(A,M)=
    \sum_{i,k}
    \left[ h(t_i,A,M)-H_i \right]
    \mathrm{Cov}^{-1}_{ik}
    \left[ h(t_k,A,M)-H_k \right] + \chi^2_\mathrm{prior}( M ),
\end{gather}
where $H_i$ is the value of the hadron propagator on the time slice $t_i$
and $\mathrm{Cov}_{ik}$ stands for the covariance between $H_i$
and $H_k$. A prior term was introduced to stabilize the fit, containing
priors on the masses except for the ground state.  The concrete form is:
\begin{gather}
    \chi^2_\mathrm{prior}(M)= 
    \sum_{s=1}^{3} \left(\frac{M_s/M_0-\mu_s}{\delta \mu_s}\right)^2,
\end{gather}
where the prior parameters are set as follows:
\begin{center}
    \begin{tabular}{C|C|C}
	s & \mu_s \cdot 1672 \text{ MeV} & \delta \mu_s\\
	\hline
	1 & 2012\text{ MeV}& 0.10 \\
	2 & 2250\text{ MeV}& 0.10 \\
	3 & 2400\text{ MeV}& 0.15
    \end{tabular}
\end{center}
The prior for the negative parity ground state, $s=1$, is motivated by the
recent observation from the Belle collaboration \cite{Yelton:2018mag}. The
excited states, $s=2,3$, have not been discovered in experiments so far, so their
priors follow from the quark model \cite{Capstick:1986bm}. The existence of
these undiscovered states is also motivated by lattice thermodynamics
below the chiral transition \cite{Bazavov:2014xya,Alba:2017mqu}.

Beside these resonant states there are also non-resonant, scattering states
corresponding to $(\Omega,\pi,\pi)$, $(\Xi,K)$, $(\Xi,K,\pi)$ and other such
multi-hadron combinations. These are expected to couple very weakly to our
single hadron operator and we have not taken them into account in this
four-state analysis. In our alternative approach (GEVP), that we discuss later,
no priors are introduced on the excited states. In that approach the energy of
the first excited state is consistent with the mass of the first resonance, but
not with the expected energy levels of the above scattering states. This
indicates that the couplings of these states to our interpolating operator are
indeed small.

The range of time slices, that are included in the $\chi^2$, were chosen by an
optimization on the coarsest lattice, $\beta=3.7000$. As already mentioned,
we have around 4000 configurations there, which is about four times larger than
on the ensembles at other lattice spacings.  In the fit range $[7\dots 17]$ we
obtained the mass with a relative precision of $0.06\%$ and with fit quality of
$Q=0.55$: $M_{\Omega,\mathrm{VI}}=1.11424(63)$ in lattice units. The priors did not impose a
significant pull on the result, ie. the final values of the fit parameters
were well within the prior widths. A different fit range $[6\dots 14]$ resulted in a
change in $M_\Omega$ within a small fraction of the statistical error.  These
results reassure us that the excited state effects are smaller than the
statistical error with these two fit ranges. On the other ensembles, with
lesser statistics, we used these two fit ranges, keeping their values in
physical units approximately constant upon changing the lattice spacing. The
exact fit ranges used are given in Table~\ref{ta:omega} in lattice units.

In Figure \ref{fi:omega} we show a comparison of the $\Omega$ masses obtained
with the various fits on four ensembles at $\beta=3.8400$. Besides the
four-state fit with two different fit ranges, see Table \ref{ta:omega}, we also
show a value from a ``combined'' fit. Here we combine the correlators from all
of the ensembles at a given $\beta$ and apply to them a common four-state fit.
Assuming that all excited state masses are a linear function of the bare
strange mass for a given $\beta$, one can fit this linear dependence across the
ensembles along with the still independent ground state masses.  This reduces
the number of fit parameters and results in more stable fits.  The bare
light-mass dependence of the excited states can be ignored, since their
statistical precision is not sufficient to resolve this dependence. (Note also,
that all our ensembles are in close vicinity of the physical point.) The result
of this combined fit agrees well with the ones obtained from the individual fit
ranges. We do not use this combined fit in our final analyses though, it would
introduce correlations between ensembles, making the analysis procedure more
complicated.

\subsection*{Omega mass determination: GEVP method}

In addition to the above four-state fit to the $\Omega$ propagator we also used
a mass extraction procedure proposed in \cite{Aubin:2010jc}, which is based on
the Generalized Eigenvalue Problem (GEVP). The method has the advantage of not
using priors. We first apply a folding transformation to the original hadron
propagator $H_t$:
\begin{gather}
    H_{t}\to \left\{\begin{array}{ll}
	\frac{1}{2}\left[ H_t+ (-1)^{t+1} H_{T-t}\right]& 0<t<\frac{T}{2}\\
	H_t & t = 0 ~\mathrm{or}~t=\frac{T}{2}
    \end{array}\right.
\end{gather}
Then we construct a matrix for each time slice $t$:
\begin{gather}
    \mathcal{H}(t) = \left(\begin{array}{cccc}
	H_{t+0}&H_{t+1}&H_{t+2}&H_{t+3}\\
	H_{t+1}&H_{t+2}&H_{t+3}&H_{t+4}\\
	H_{t+2}&H_{t+3}&H_{t+4}&H_{t+5}\\
	H_{t+3}&H_{t+4}&H_{t+5}&H_{t+6}
\end{array}\right)
\end{gather}
For a given $t_a$ and $t_b$ let $\lambda(t_a,t_b)$ be an eigenvalue and $v(t_a,t_b)$
an eigenvector solution to this $4\times 4$ generalized eigenvalue problem:
\begin{gather}
    \mathcal{H}(t_a) v(t_a,t_b) = \lambda(t_a,t_b) \mathcal{H}(t_b) v(t_a,t_b).
\end{gather}
Here we select the smallest eigenvalue $\lambda$ and use the corresponding
eigenvector $v$ to project out the ground state:
\begin{gather}
    \label{eq:vHv}
    v^+(t_a,t_b)\mathcal{H}(t)v(t_a,t_b),
\end{gather}
which then can be fitted to a simple $\exp(-M t)$ type function. This assumes
that backward propagating states are negligible between $t_a$ and $t_b$ as well
as in the range used to fit Equation \eqref{eq:vHv}. The parity partner states
inherent in the staggered formulation appear as excited states, that give large
contributions to the correlation functions \eqref{eq:vHv} constructed with an
eigenvalue $\lambda$ with non-minimal absolute value. In that case, both the
correlation function \eqref{eq:vHv} and the eigenvalue $\lambda$ exhibit
oscillating signs. In the case of the correlation function, this oscillation
occurs as a function of $t$, and in the case of $\lambda$ as a function of
$t_b-t_a$. See \cite{DeTar:2014gla} for details of the variational method with
staggered fermions.

The tuneable parameters of the procedure are $t_a$ and $t_b$ for specifying the
GEVP, as well as the fit range for the $\exp$ fitting in the last step; they
are given in the last three columns of Table \ref{ta:omega}. Similar to the
pion mass analysis these parameters were chosen by performing a
Kolmogorov-Smirnov test across all the ensembles. In Figure \ref{fi:omega} we
show the fit results obtained with this GEVP procedure for ensembles at
$\beta=3.8400$. They are in good agreement with the four-state fit values.

The mass extracted using the GEVP gives a third $M_\Omega$ value for each
ensemble, beside the results with the four-state fit procedure with the two fit
ranges. We will use the deviation between these three values as a systematic
error in the $\Omega$ mass determination.

\subsection*{Finite size corrections}

In order to determine the finite-volume corrections for the pseudoscalar
masses, $M_\pi(L)-M_\pi(\infty)$, and decay constants,
$f_\pi(L)-f_\pi(\infty)$, we use the chiral perturbation theory based formulae
of Reference \cite{Colangelo:2005gd}. Our pion masses are very close to the
physical point, where one obtains a relative correction of $0.022$\% for the
mass and a relative correction of $0.077$\% for the decay constant.  The mass
of the kaon also receives a correction due to the finite volume.  However, this
correction is so small, and any uncertainty related to it is so subdominant,
that we ignore it.

We also take into account the effect of the finite time extent $T$ in the decay
constants, both for pions and kaons, assuming that they are free particles. This is
obtained by noting that the $T$-dependence of the free particle propagator is
given by $\cosh[M(t-T/2)]/\sinh(MT/2)$. Therefore, we fit our propagators to the
form $A\cosh[M(t-T/2)]/\sinh(MT/2)$ and extract the decay constant from the
amplitude $A$. 

The finite-size effects on the $\Omega$ mass is estimated from next-to-leading
order, three-flavor, heavy-baryon chiral perturbation theory. See
\cite{Ishikawa:2009vc} for the corresponding formulas. To this order the pions
give no contribution to the finite-size effects, but only the kaons and the eta
do. As a result, the finite-size correction is so tiny that it can be safely
neglected.

    \section{Path integral and expectation values}
\label{se:obs_gen}

Our staggered path integral includes four flavors of quarks, $f= \{ u, d, s,
c\}$, gluon fields $U$ and photon fields $A$ and is given by:
\begin{gather}
    \label{eq:z}
    Z= \int [dU] \exp(-S_g[U]) \int [dA] \exp(-S_\gamma[A])\ \prod_f \det M^{1/4}[ V_U\exp(ieq_fA), m_f ].
\end{gather}
The ensemble specific definition of the gauge action $S_g$ is given in Sections
\ref{se:act_4stout} and \ref{se:act_4hex}. The photon integral measure $[dA]$
and action $S_\gamma$ are defined in the $\mathrm{QED}_\mathrm{L}$ scheme
\cite{Hayakawa:2008an}. 
The one-hop staggered matrix in a background field $W_\mu$ can
be written as
\begin{gather}
    M[W,m]= D[W]+m= \sum_\mu D_\mu[W_\mu] + m,
\end{gather}
where $D_\mu$ is the covariant differentiation in the $\mu$ direction involving
$W$ and its adjoint $W^\dagger$ together with the obligatory staggered phases.
In the path integral the fermions are coupled to a gauge field that is a
product of the exponentiated photon field and of the smeared gluon gauge field
$V_U$.  Our smearing recipes are given in Sections \ref{se:act_4stout} and
\ref{se:act_4hex}. The photon field is not smeared.  $q_f\in \{ +\tfrac{2}{3},
-\tfrac{1}{3}, -\tfrac{1}{3}, +\tfrac{2}{3}\}$ stand for the quark 
electric charges in units of the positron charge $e$,
$m_f$ for the quark masses and $\alpha=e^2/(4\pi)$.
We use the notation $\delta m\equiv m_d-m_u$ for the difference in the up and
down quark masses and $m_l\equiv \tfrac{1}{2}(m_u+m_d)$ for their average.  To simplify
later formulas we also introduce the notations
\begin{gather}
    \label{eq:dets}
    M_f\equiv M[ V_U e^{ieq_fA}, m_f]
    \qquad\text{and}\qquad
    \mathrm{dets}[U,A;\{m_f\},\{q_f\},e]\equiv \prod_f \det M_f^{1/4}\ ,
\end{gather}
where the latter is the product of all fermion determinants.

In this work, isospin-breaking is implemented by taking derivatives with
respect to the isospin-breaking parameters and by measuring the so obtained
derivative operators on isospin-symmetric configurations
\cite{deDivitiis:2013xla}. A different approach would be to generate
configurations at non-zero values of the isospin breaking parameters and use
the same operators as at zero isospin breaking, see eg.
\cite{Borsanyi:2014jba}. We choose the former approach in this work, so as to
optimally distribute the computing resources among the various isospin-breaking
contributions.

We introduce a set of notations for isospin-symmetric observables and their
isospin-breaking derivatives. Consider the observable $X(e,\delta m)$, which is
a function of $e$ and $\delta m$. Then we define
\begin{gather}
    \label{eq:ibder1}
    X_0\equiv X(0,0),\qquad
    X'_m\equiv m_l\frac{\partial X}{\partial \delta m}(0,0),\qquad
    X'_1\equiv \frac{\partial X}{\partial e}(0,0),\qquad
    X''_2\equiv \frac{1}{2}\frac{\partial^2 X}{\partial e^2}(0,0).
\end{gather}
The isospin-breaking derivatives are denoted by prime(s) and an index. The
mass derivative has the index $m$, it requires no renormalization, since
$\delta m$ and $m_l$ have the same renormalization factor at zero
electromagnetic coupling.  The electric charge derivatives have a single digit
index: $1$ or $2$. Below, we also define electric charge derivatives with
two-digit indices.  We take into account only leading-order isospin-breaking in
this work, so no higher derivatives are needed.

In the case of the fermion determinant, the isospin-symmetric value is denoted
by $\mathrm{dets}_0$.  The strong-isospin-breaking of $\mathrm{dets}$ is zero
at leading order:
\begin{align}
    \mathrm{dets}'_m= 0,
\end{align}
since $\mathrm{dets}$ is symmetric under the exchange $u\leftrightarrow d$.
The electromagnetic derivatives are
\begin{align}
    \label{eq:detsd}
    \begin{split}
	\frac{\mathrm{dets}'_1}{\mathrm{dets}_0}&= \sum_f \frac{q_f}{4} \mathrm{Tr} \left( M_f^{-1} D[iAV_U] \right), \\
	\frac{\mathrm{dets}''_2}{\mathrm{dets}_0}&=\frac{1}{2}\left[
	\left( \frac{\mathrm{dets}'_1}{\mathrm{dets}_0} \right)^2
	- \sum_f \frac{q_f^2}{4} \mathrm{Tr}\left( M_f^{-1} D[A^2V_U] \right)
	- \sum_f \frac{q_f^2}{4} \mathrm{Tr}\left( M_f^{-1} D[iAV_U] M_f^{-1} D[iAV_U] \right)\right],
    \end{split}
\end{align}
where $\mathrm{Tr}$ is trace over color and spacetime indices and the
argument of the $D$ operator is a $3\times3$ complex matrix valued field, eg.
$A^2V_U$ has components $A_{\mu,x}^2 [V_U]_{\mu,x}$.  The implementation of these
derivatives is given in Section \ref{se:obs_dynqed}.

We also make a distinction between the electric charge in the fermion
determinant and in the operator that we measure. We call the former sea
electric charge and denote it by $e_s$, the latter is the valence electric charge
and is denoted by $e_v$.  For an observable $X$ that depends on both the
valence and sea charges, $X(e_v,e_s)$, the second order electric charge
derivatives are defined as follows:
\begin{gather}
    \label{eq:ibder2}
    X''_{20}\equiv \frac{1}{2}\frac{\partial^2 X}{\partial e_v^2}(0,0), \qquad
    X''_{11}\equiv \frac{\partial^2 X}{\partial e_v\partial e_s}(0,0), \qquad
    X''_{02}\equiv \frac{1}{2}\frac{\partial^2 X}{\partial e_s^2}(0,0).
\end{gather}
For functions that depend on either $e_v$ or $e_s$, but not on both, we use the
single digit notations of Equation \eqref{eq:ibder1}.

The expectation value of an operator $O$ is calculated by inserting $O[U,A]$
into the integrand of the path integral of Equation \eqref{eq:z} and
normalizing the integral by $Z$. Here we consider operators whose photon field
dependence arises entirely from the photon-quark interaction, ie. $O=O[U,e_v
A]$. The expectation value of this operator depends on $\delta m$, $e_v$
and $e_s$, and the isospin expansion can be written as:
\begin{gather}
    \label{eq:exp}
    \langle O \rangle =
    \left[ \langle O \rangle\right]_0 +
    e_v^2  \langle O \rangle''_{20} +
    e_ve_s \langle O \rangle''_{11} +
    e_s^2  \langle O \rangle''_{02} +
    \tfrac{\delta m}{m_l} \langle O \rangle'_m.
\end{gather}
Here, the individual terms can be expressed as expectation values obtained with
the isospin-symmetric path integral, which we denote by $\langle \dots
\rangle_0$. The concrete expressions are:
\begin{align}
    \label{eq:ixx}
    \begin{split}
	\text{isospin-symmetric:}\quad& \left[\langle O \rangle\right]_0=  \langle O_0 \rangle_0\\
	\text{qed valence-valence:}\quad& \langle O \rangle''_{20}=
	\langle O''_{2}\rangle_0\\
	\text{qed sea-valence:}\quad& \langle O \rangle''_{11}=
	\left\langle O'_1 \frac{\mathrm{dets}'_1}{\mathrm{dets}_0} \right\rangle_0\\
	\text{qed sea-sea:}\quad& \langle O \rangle''_{02}=
	\left\langle O_0  \frac{\mathrm{dets}''_2}{\mathrm{dets}_0} \right\rangle_0 - 
	\left\langle O_0\right\rangle_0 \left\langle \frac{\mathrm{dets}''_2}{\mathrm{dets}_0} \right\rangle_0\\
	\text{strong-isospin-breaking:}\quad& \langle O \rangle'_m=
	\langle O'_m \rangle_0
    \end{split}
\end{align}
In the derivation of these expressions we use $\left\langle
\tfrac{\mathrm{dets}'_1}{\mathrm{dets}_0} \right\rangle_0=0$.  In Table
\ref{ta:obs_gen} we give an overview of the isospin-breaking derivatives for
the observables that are computed in this paper.

Note that Equation \eqref{eq:exp} is an expansion in bare parameters and not
what we consider a decomposition into isospin-symmetric and isospin breaking
parts.  The latter involves derivatives with respect to renormalized
observables and our prescription for that is given in Section
\ref{se:obs_split}. There is no need to introduce a renormalized
electromagnetic coupling though: its running is an $O(e^4)$ effect, ie. beyond
the leading order isospin approximation that we consider here.

\begin{table}[t]
    \centering
    \begin{tabular}{l|c|c|c|c|c}
	$X$ & $X''_{20}$ & $X''_{11}$ & $X''_{02}$ & $X'_m$ & Section \\
	\hline
	$M_\Omega,M_{\pi_\chi},M_{K_\chi}$ & \checkmark & \checkmark & \checkmark & -          & \ref{se:obs_ibmass} \\
	$\Delta M_K^2,\Delta M^2$          & \checkmark & \checkmark & -          & \checkmark & \ref{se:obs_ibmass} \\
	\hline
	$w_0$                              & -          & -          & \checkmark & -          & \ref{se:obs_ibw0} \\
	\hline
	$\langle JJ \rangle$-light         & \checkmark & \checkmark & \checkmark & \checkmark & \ref{se:obs_ibjj} \\
	$\langle JJ \rangle$-strange       & \checkmark & \checkmark & \checkmark & -          & \ref{se:obs_ibjj} \\
	$\langle JJ \rangle$-disc.         & \checkmark & \checkmark & \checkmark & \checkmark & \ref{se:obs_ibjj}
    \end{tabular}
    \caption
    {
	\label{ta:obs_gen}Overview of isospin-breaking derivatives computed in
	this paper. For each observable, we specify the Section in which the
	implementation details can be found. A dash indicates, that the
	particular contribution vanishes.
    }
\end{table}

    \section{Isospin breaking: decomposition}
\label{se:obs_split}

For various purposes it is useful to decompose the observables into
isospin-symmetric and isospin-breaking parts. This requires a matching of the
isospin symmetric and full theories, in which we specify a set of observables
that must be equal in both theories. Of course, different sets will lead to
different decompositions, which is commonly referred to as scheme dependence.
Only the sum of the components, ie. the result in the full theory, is scheme
independent.

A possible choice for the observables are the Wilson-flow--based $w_0$ scale
and the masses of mesons built from an up/down/strange and an
anti-up/down/strange quark, $M_{uu}$/$M_{dd}$/$M_{ss}$. These mesons are
defined by taking into account only the quark-connected contributions in their
two-point functions \cite{Borsanyi:2013lga}. Their masses are practical
substitutes for the quark masses. Also, they are neutral and have no magnetic
moment, so they are a reasonable choice for an isospin decomposition.  These
masses cannot be measured in experiments, but have a well defined continuum
limit and thus a physical value can be associated to them.

According to partially-quenched chiral perturbation theory coupled to photons
\cite{Bijnens:2006mk}, the combination
\begin{gather}
    M_{\pi_\chi}^2\equiv \tfrac{1}{2}(M_{uu}^2+M_{dd}^2)
\end{gather}
equals the neutral pion mass, $M_{\pi_\chi}=M_{\pi_0}$, up to terms that are
beyond leading order in isospin breaking.  Since such terms are beyond the accuracy needed
in this work, we use the experimental value of the
neutral pion mass as the physical value of $M_{\pi_\chi}$.
Furthermore the difference,
\begin{gather}
    \Delta M^2\equiv M_{dd}^2- M_{uu}^2
\end{gather}
is a measure of strong-isospin-breaking
not affected by electromagnetism. According to
\cite{Bijnens:2006mk}, $\Delta M^2= 2B_2\delta m$ is valid up to effects that
are beyond leading order in isospin breaking, at least around the physical
point. Here, $B_2$ is the two-flavor chiral condensate parameter.
For the determination of the physical values of $w_0$, $M_{ss}$ and $\Delta M^2$,
see Section \ref{se:res_w0etal}.

For the decomposition we start with the QCD+QED theory and parameterize our
observable $\langle O \rangle$ with the quantities defined above:
\begin{gather}
    \langle O \rangle( M_{\pi_\chi}w_0, M_{ss}w_0, \tfrac{L}{w_0}, \Delta M w_0, e )\ .
\end{gather}
Here, the continuum limit is assumed. We can isolate the
electromagnetic part by switching off the electromagnetic coupling, while
keeping the other parameters fixed:
\begin{gather}
    \langle O \rangle_\mathrm{qed}\equiv
    e^2 \cdot \left.\frac{\partial \langle O \rangle}{\partial e^2}\right|_{M_{\pi_\chi}w_0, M_{ss}w_0, \tfrac{L}{w_0}, \Delta M w_0,e=0}\ .
\end{gather}
The strong-isospin-breaking part is given by the response to the $\Delta M$ parameter:
\begin{gather}
    \langle O \rangle_\mathrm{sib}\equiv
    (\Delta M w_0)^2 \cdot \left.\frac{\partial \langle O \rangle}{\partial (\Delta M w_0)^2}\right|_{M_{\pi_\chi}w_0, M_{ss}w_0, \tfrac{L}{w_0}, \Delta M w_0=0,e=0}\ ,
\end{gather}
and the isospin-symmetric part is just the remainder:
\begin{gather}
    \langle O \rangle_\mathrm{iso}\equiv
    \langle O \rangle( M_{\pi_\chi}w_0, M_{ss}w_0, \tfrac{L}{w_0}, 0, 0 ).
\end{gather}
One can also define the decomposition at a finite lattice spacing, for which
$w_0$ in lattice units can be additionally fixed. In doing so the isospin
symmetric part $\langle O\rangle_\mathrm{iso}$ has to be distinguished from the
value of the observable at the bare isospin-symmetric point $\left[\langle O
\rangle\right]_0$.

In this work we use the above definitions for the isospin decomposition; a
similar scheme was put forward in \cite{Horsley:2015vla}. A different scheme
would be to keep the renormalized quark masses and the strong coupling constant
fixed as QED is turned on \cite{Gasser:2003hk}. In case of light quark
observables the two schemes supposed to agree well.  This can be justified by
the smallness of the electromagnetic part of the neutral pion in the scheme of
\cite{Gasser:2003hk}. Reference \cite{Horsley:2015vla} found
$\epsilon_{\pi_0}=0.03(2)$, where $\epsilon_{\pi_0}$ is the parameter that
measures the size of the electromagnetic contribution in the neutral pion mass.
In comparison the same quantity for the charged pion was found
$\epsilon_{\pi_+}=1.03(2)$ in \cite{Horsley:2015vla}.

    \section{Isospin breaking: dynamical QED}
\label{se:obs_dynqed}

\begin{table}[t]
    \centering
    \begin{tabular}{C|C|C|C|C|R}
	\beta & a[\text{fm}] & L\times T & m_s & m_s/m_l & \# \text{conf} \\
	\hline
	\hline
	3.7000 & 0.1315 & 24\times 48 & 0.057291 & 27.899 &  716 \\ 
	       &        & 48\times 64 & 0.057291 & 27.899 &  300 \\ 
	\hline
	3.7753 & 0.1116 & 28\times 56 & 0.047615 & 27.843 &  887 \\ 
	\hline
	3.8400 & 0.0952 & 32\times 64 & 0.043194 & 28.500 & 1110 \\ 
	       &        &             & 0.043194 & 30.205 & 1072 \\ 
	       &        &             & 0.040750 & 28.007 & 1036 \\ 
	       &        &             & 0.039130 & 26.893 & 1035 \\ 
    \end{tabular}
    \caption{\label{ta:dynqed}List of {\tt 4stout} ensembles used for computing
    dynamical QED effects with gauge coupling, lattice spacing at the physical
    point, lattice size, quark masses and number of configurations.}
\end{table}

In the isospin expansion of an observable $\langle O \rangle$ (see Equation
\eqref{eq:exp}) we refer to the $e_s$ dependent terms as dynamical QED
contributions.

The sea-valence contribution is given by Equation \eqref{eq:ixx} as
\begin{gather}
    \langle O \rangle''_{11}= \left\langle \left\langle O'_1 \frac{\mathrm{dets}'_1}{\mathrm{dets}_0} \right\rangle_A \right\rangle_{U}.
\end{gather}
Here we made explicit that the path integral is carried out over two gauge
fields: the index $A$ of the expectation value means averaging over free photon
fields with the action $S_\gamma$. The rest of the path integral weight is
contained in the gluon expectation value, labeled with index $U$. The trace
over coordinate and color space in the first derivative of the fermion
determinant (see Equation \eqref{eq:detsd}) is computed exactly in the low-lying
eigenmode space of the Dirac operator and with random vectors in the
complement.  According to our findings, the noise in this term overwhelmingly
stems from the random sources, and not from the gauge fields.  For each $U$
field we generate one $A$ field and on this $(U,A)$ gauge field pair we use
about $10^4$ random vectors to estimate the first derivative
$\mathrm{dets}'_1/\mathrm{dets}_0$. The first derivative of the observable is
estimated by a finite difference $O'_1\approx \tfrac{1}{2e_v} (O_+ - O_-)$, see
eg. Section \ref{se:obs_ibmass}.

The sea-sea contribution is given by
\begin{gather}
    \langle O \rangle''_{02}=
    \left\langle \left[O_0-\left \langle O_0 \right\rangle_U\right] \left\langle \frac{\mathrm{dets}''_2}{\mathrm{dets}_0} \right\rangle_A \right\rangle_U,
\end{gather}
where the $A$-average of the second derivative of the determinant can be done
independently from the observable. This is especially useful, since the noise
in this term is dominated by fluctuations in the photon field. On each $U$
configuration we use about $2000$ photon fields, and on each photon field, 12
random sources to estimate the second derivative
$\mathrm{dets}''_2/\mathrm{dets}_0$.

For both contributions we apply the Truncated Solver Method
\cite{Bali:2009hu,Blum:2012uh}: the matrix inverters are run with a reduced
precision most of the time, and the resulting small bias is corrected using
occasional, high-precision inversions.

In this work we compute dynamical QED effects on a dedicated set of ensembles
using the {\tt 4stout} action. We have three lattice spacings, with box sizes
around $L=3$~fm with $T=2L$. Additionally, on the coarsest lattice there is
also an ensemble with an $L=6$~fm box. Table \ref{ta:dynqed} gives the
parameters of these ensembles, together with the number of configurations. The
chosen parameters in these dedicated runs match the parameters of selected {\tt
4stout} ensembles. For the observables that we consider, we see no significant
difference in the size of dynamical QED contributions between the two different
volumes. For the volume dependence of the hadron masses, see Section
\ref{se:obs_ibmass} and Figure \ref{fi:qedv}. The $L=3$~fm volume simulations
need about an order of magnitude less computer time for the same precision.
Therefore, on the finer lattices we performed simulations in the smaller volume
only.

    \section{Isospin-breaking: $w_0$-scale}
\label{se:obs_ibw0}

In this section we derive a formula that gives the electromagnetic correction
of the $w_0$-scale \cite{Borsanyi:2012zs}. The starting point is the operator
$W_\tau[U]$, which is the logarithmic derivative of the gauge-action
density along the gradient flow \cite{Luscher:2010iy}:
\begin{gather}
    W_\tau[U]\equiv \frac{d (\tau^2 E[U,\tau])}{d\log \tau},
\end{gather}
where $\tau$ is the gradient flow time and $E$ is a suitable discretization of
the gluonic gauge action density. The expectation value of this operator defines the $w_0$-scale via
\begin{gather}
    \label{eq:w0def}
    \left\langle W_{\tau= w_0^2(e)} \right\rangle = 0.3\ .
\end{gather}
Since $W_\tau[U]$ is a pure-gauge observable, it neither depends on the
valence charge nor on fermion masses: the only isospin-breaking dependence in
Equation \eqref{eq:w0def} comes from the electric sea charge. The derivatives
with respect to $\delta m$ and the valence electric charge are zero.
The expansion of
the expectation value is given by Equations \eqref{eq:exp} and \eqref{eq:ixx}:
\begin{gather}
    \langle W_\tau \rangle= \langle W_\tau \rangle_0 + e_s^2
	\left\langle \left( W_\tau -\left\langle W_\tau \right\rangle_0 \right) \frac{\mathrm{dets}_2''}{\mathrm{dets}_0} \right\rangle_0
\end{gather}
We also have to expand the $w_0$-scale:
\begin{gather}
    w_0(e_s)= w_0 + e_s^2\delta w_0
\end{gather}
and the $W$ operator:
\begin{gather}
    W_{\tau=w_0^2(e_s)}= W_{\tau=w_0^2} + e_s^2\cdot 2w_0\delta w_0\ \cdot \left.\frac{d W}{d\tau} \right|_{\tau=w_0^2}
\end{gather}
Here, $w_0$ denotes the value of the $w_0$-scale at the isospin-symmetric point, which of course
satisfies
\begin{gather}
    \label{eq:w0iso}
    \left\langle W_{\tau= w_0^2} \right\rangle_0 = 0.3
\end{gather}
From these we obtain the following formula for the electromagnetic correction:
\begin{gather}
    \delta w_0= -\left[\frac{1}{2\sqrt{\tau}}
    \left\langle\frac{d W_\tau}{d\tau}\right\rangle_0^{-1}
	\left\langle \left( W_\tau -\left\langle W_\tau \right\rangle_0 \right) \frac{\mathrm{dets}_2''}{\mathrm{dets}_0} \right\rangle_0
	\right]_{\tau=w_0^2}
\end{gather}
Section \ref{se:obs_dynqed} gives details on the fermion-determinant derivative
computations.  On our coarsest lattice spacing we computed the electromagnetic
correction in two different volumes, $L=3$~fm and $L=6$~fm (see Table
\ref{ta:dynqed} for the ensemble parameters). We obtained for this correction
$\delta w_0=-0.018(2)$ on the small and $\delta w_0=-0.018(3)$ on the large
lattice, which are in perfect agreement. Even the isospin-symmetric values show
no significant finite-size effect: we have $w_0=1.2899(9)$ on the small and
$w_0=1.2908(2)$ on the large lattice.  In our analyses, we use the isospin
symmetric $w_0$ measured on the large lattices listed in Table \ref{ta:4stout}; whereas
for $\delta w_0$ we use the small volume ensembles listed in Table
\ref{ta:dynqed}.

    \section{Isospin breaking: hadron masses}
\label{se:obs_ibmass}

\begin{figure}[t]
    \centering
    \includegraphics[width=0.7\textwidth]{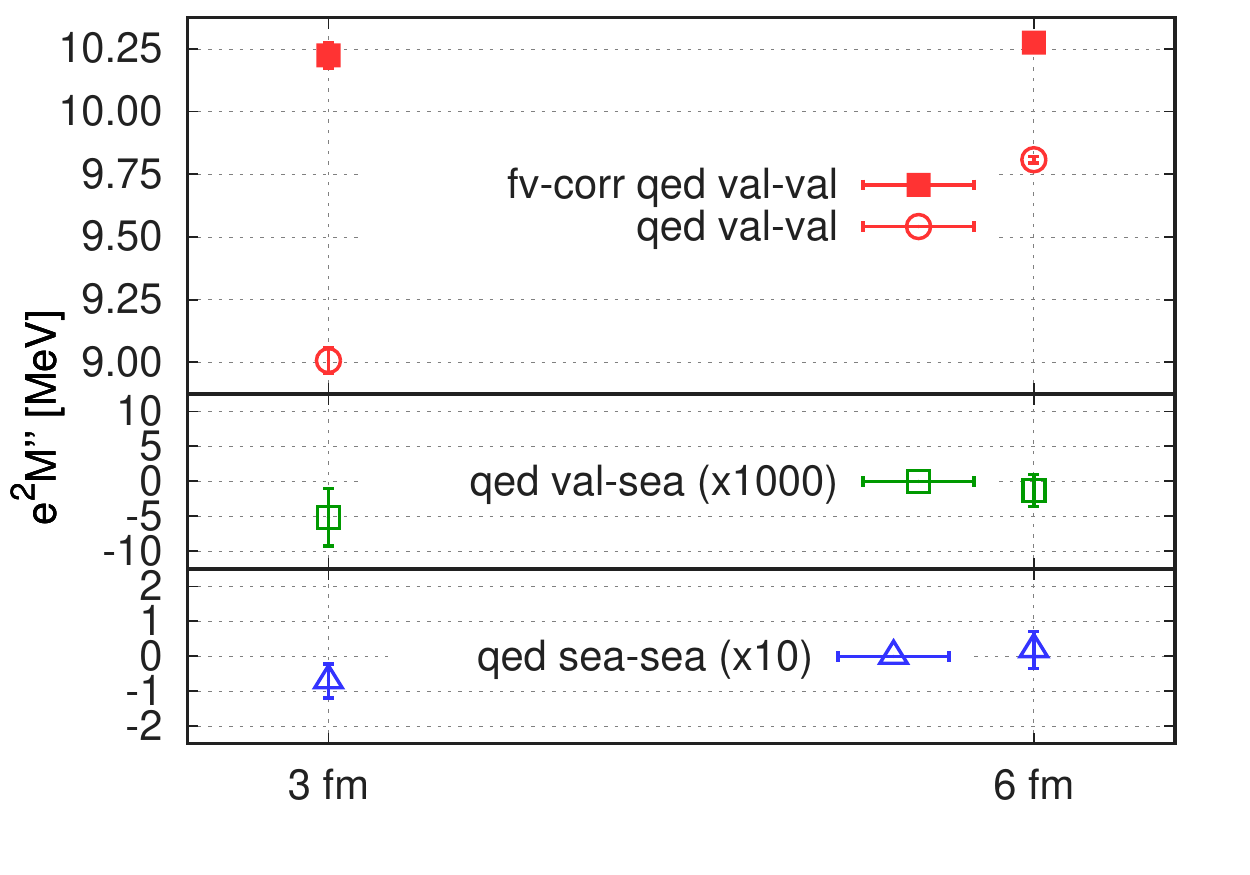}
    \caption
    {
	\label{fi:qedv}Volume dependence of various electromagnetic
	contributions to the $\pi_+$ mass. For the valence-valence contribution
	$M''_{20}$ we apply an infinite-volume correction given by Equation
	\eqref{eq:ivmass}. The valence-sea $M''_{11}$ and sea-sea $M''_{02}$
	contributions are multiplied by $1000$ and $10$ on the plot,
	respectively. The results are obtained with the {\tt 4stout} action at
	$\beta=3.7000$.
    }
\end{figure}

\begin{table}[t]
    \centering
    \begin{tabular}{L||C|C||C|C}
	\text{isospin component}    & \text{meson \#1 [fm]} & \text{meson \#2 [fm]} & \text{omega \#1 [fm]} & \text{omega \#2 [fm]} \\
	\hline
	M_0                &  2.0\dots3.5  &  2.5\dots4.0  & 1.4\dots2.0 & 1.5\dots2.1\\
	M''_{20}           &  1.5\dots3.0  &  2.0\dots3.5  & 1.2\dots2.0 & 1.3\dots2.1\\
	M'_m               &  1.5\dots3.0  &  2.0\dots3.5  & - & - \\
	M''_{11},M''_{02}  &  1.0\dots2.5  &  1.5\dots3.0  & 0.3\dots1.5 & 0.6\dots1.5\\
    \end{tabular}
    \caption{
	\label{ta:ibmass}Plateau fit ranges for different isospin-breaking
	components, for mesons and for the $\Omega$ baryon.
	}
\end{table}

In this section we describe the procedure to obtain the isospin-breaking
derivatives of a hadron mass $M$.  On certain ensembles we measure the hadron
propagator $H$ at four different values of isospin breaking:
\begin{gather}
    H_0,\quad
    H_+,\quad
    H_-,\quad
    H_{\delta m}.
\end{gather}
The first is measured at the isospin-symmetric point, the second/third
with valence electric charge $e_v=\pm\sqrt{4\pi \alpha_*}$ and zero
quark mass difference $\delta m=0$, and the fourth with $e_v=0$ and $\delta m=
2m_l \left.\tfrac{1-r}{1+r}\right|_{r=0.485}$.  These allow to calculate finite
differences with respect to $e_v$ and $\delta m$, whereas the $e_s$ derivatives
can be calculated exactly using the formulae in Equation \eqref{eq:ixx}.  In
these measurements both gluon and photon fields are fixed to Coulomb gauge, for
the former the gauge fixing procedure is applied after smearing. 

We use the notation $\mathcal{M}[\langle H \rangle]$ for the mass that is
extracted from the hadron-propagator expectation value $\langle H \rangle$. At
the isospin-symmetric point we have
\begin{gather}
    M_0= \mathcal{M}[\langle H_0 \rangle_0].
\end{gather}
The QED, sea-sea, isospin-breaking component of the propagator, $\langle H
\rangle''_{02}$, is given by Equation \eqref{eq:ixx}. Then the derivative of
the mass can be obtained by application of the chain rule:
\begin{gather}
    \label{eq:m02}
    M''_{02}=
    \left.\frac{\delta \mathcal{M}[H]}{\delta H}\right|_{\langle H_0 \rangle_0} \langle H\rangle''_{02}=
    \left.\frac{\delta \mathcal{M}[H]}{\delta H}\right|_{\langle H_0 \rangle_0}
    \left\langle (H_0 - \langle H_0 \rangle_0) \frac{\mathrm{dets}''_2}{\mathrm{dets}_0}\right\rangle_0,
\end{gather}
where we use $\partial \langle H \rangle/\partial e_s=0$ at the
isospin-symmetric point. In the case of the valence-valence QED component we
can build the derivative as a finite difference:
\begin{gather}
    \label{eq:m20}
    M''_{20}\approx \frac{1}{2e_v^2}
    \left( \mathcal{M} [\langle H_+ \rangle_0]+
    \mathcal{M} [\langle H_- \rangle_0]-
    2 \mathcal{M} [\langle H_0 \rangle_0] \right)=
    \frac{1}{e_v^2}\left( \mathcal{M} [\tfrac{1}{2}\langle H_+ + H_-\rangle_0] - \mathcal{M} [\langle H_0 \rangle_0]\right).
\end{gather}
Here we used $\langle H_+ \rangle_0= \langle H_- \rangle_0= \tfrac{1}{2}\langle
H_+ + H_- \rangle_0$. Working with the average propagator has the advantage
that the $O(e_v)$ noise is absent in its expectation value \cite{Blum:2007cy}.
Equation \eqref{eq:m20} gives an approximation that is valid up to, and
including, $O(e_v^2)$ terms. We have confirmed that these terms are on the
order of a few percent relative to $M''_{20}$ with our choice of $e_v$. There
is an analogous formula for the strong-isospin-breaking component:
\begin{gather}
    M'_m\approx
    \frac{m_l}{\delta m} \left( \mathcal{M} [\langle H_{\delta m} \rangle_0] - \mathcal{M} [\langle H_0 \rangle_0]\right).
\end{gather}
The QED valence-sea derivative is available in a mixed form: the derivation
with respect to $e_s$ is exact, but with respect to $e_v$, it is a finite
difference. Applying these to $\mathcal{M}[\langle H\rangle]$ we get:
\begin{gather}
    M''_{11}\approx
    \frac{1}{2e_v}\left[ \frac{\delta\mathcal{M}[H]}{\delta H} \left.\frac{\partial \langle H \rangle}{\partial e_s}\right|_{e_s=0}- (e_v \to -e_v)\right]=
    \left.\frac{\delta\mathcal{M}[H]}{\delta H}\right|_{\langle H_+ + H_- \rangle_0}
    \left\langle \frac{H_+-H_-}{2e_v}\frac{\mathrm{dets}'_1}{\mathrm{dets}_0} \right\rangle_0.
\end{gather}

We now specify $\mathcal{M}[H]$, i.e. the way to extract the mass from the
hadron propagator. Since we are interested in the small isospin breaking
effects here, the choice of the mass extraction procedure is not crucial. We
utilize a procedure based on effective masses instead of fitting the propagator
to multiple exponentials. This has the advantage over the standard fitting
procedure that the derivatives $\tfrac{\delta \mathcal{M}}{\delta H}$ can be
computed easily, in particular they can be given in closed analytic form.
Using two/four propagator points, an effective-mass value and its differential
can be given in analytical form for mesons/baryons \cite{Fleming:2009wb}. Then
we fit a constant to the plateau of the effective mass.  We choose two plateau
ranges, so that a systematic error can be associated with finding the plateau.
The ranges are given in Table \ref{ta:ibmass}. 

In this work the strong-isospin-breaking $M'_m$ and valence-valence
contributions $M''_{20}$ are evaluated on {\tt 4stout} configurations with box
sizes around $L=6$~fm. For the sea contributions, $M''_{11}$ and $M''_{02}$, we
use also the {\tt 4stout} action, but on smaller volumes, $L=3$~fm (see Section
\ref{se:obs_dynqed} for details of these ensembles).

In the presence of the electromagnetic interaction, hadron masses have a
finite-size effect that is power-like in the size of the box. In some cases it
can be much larger than the exponentially-suppressed, finite-size effect
related to the strong interaction. For the $\mathrm{QED}_\mathrm{L}$ photon
action, the effect in the first two orders depends on the hadron only through
its electric charge $Q$ and mass $M$ and is known analytically
\cite{Davoudi:2014qua,Borsanyi:2014jba}:
\begin{gather}
    \label{eq:ivmass}
    M(L)-M= -\frac{(Qe)^2 c}{8\pi} \left[ \frac{1}{L} + \frac{2}{ML^2} + O(L^{-3})\right]\quad
    \text{with}\quad
    c=2.837297\dots .
\end{gather}
The first two orders of this formula can be used to correct for electromagnetic
finite-size effects. Remaining $O(L^{-3})$ effects are beyond the precision of
this work and are neglected. Since, for charged hadrons, sea effects are
typically much smaller than the valence-valence contribution, we use the
universal finite-size formula \eqref{eq:ivmass} to correct the valence-valence
component and apply no correction to the rest.  We can corroborate this choice
by looking at the different isospin breaking components of the charged pion
mass on two different volumes on our coarsest lattice, as shown in Figure
\ref{fi:qedv}. The corrected $M''_{20}$ values almost agree on the two volumes.
On finer lattices we use the results of the $L=6$~fm runs, correcting them with
Equation \eqref{eq:ivmass}.  In the case of the sea contributions the
uncorrected $M''_{11}$ and $M''_{02}$ data are consistent on the two volumes.
On finer lattices we use the small-volume runs to estimate the electromagnetic
sea effects without correcting for finite-volume effects.

    \section{Current propagator $\langle JJ\rangle$}
\label{se:obs_jj}

\renewcommand{\arraystretch}{1.8}
\begin{table}
    \centering
    \begin{tabular}{C|R|R|R}
	X \longrightarrow & \multicolumn{1}{C|}{C^\mathrm{light}} & \multicolumn{1}{C|}{C^\mathrm{strange}} & \multicolumn{1}{C}{C^\mathrm{charm}}\\
	\hline
	X_0  & \tfrac{5}{9} C^\mathrm{conn}(m_l,0) & \tfrac{1}{9} C^\mathrm{conn}(m_s,0) & \tfrac{4}{9} C^\mathrm{conn}(m_c,0) \\
	X'_1
	&  \tfrac{7}{27} \left[\tfrac{\partial}{\partial e} C^\mathrm{conn}\right](m_l,0)
	& -\tfrac{1}{27} \left[\tfrac{\partial}{\partial e} C^\mathrm{conn}\right](m_s,0)
	&  \tfrac{8}{27} \left[\tfrac{\partial}{\partial e} C^\mathrm{conn}\right](m_c,0)
	\\
	X'_2
	& \tfrac{17}{81}\left[\tfrac{1}{2} \tfrac{\partial^2}{\partial e^2} C^\mathrm{conn}\right](m_l,0)
	& \tfrac{ 1}{81}\left[\tfrac{1}{2} \tfrac{\partial^2}{\partial e^2} C^\mathrm{conn}\right](m_s,0)
	& \tfrac{16}{81}\left[\tfrac{1}{2} \tfrac{\partial^2}{\partial e^2} C^\mathrm{conn}\right](m_c,0)
	\\
	X'_m
	& -\tfrac{m_l}{6}\left[\tfrac{\partial}{\partial m_l} C^\mathrm{conn}\right](m_l,0)
	& -
	& -
    \end{tabular}
    \caption
    {
	\label{ta:ibjjc}Isospin symmetric value $X_0$ and isospin-breaking
	derivatives $X'_1,X'_2,X'_m$ of various observables $X$, namely the
	light, strange and charm connected contractions of the current
	propagator, in terms of the connected vector meson contraction and its
	derivatives. See Equation \eqref{eq:conn} and \eqref{eq:clsc} for the definitions.
    }
\end{table}
\renewcommand{\arraystretch}{1.0}

In this section we consider in detail the definition and decomposition of the
current propagator:
\begin{gather}
    \langle  J_{\mu,x} J_{\bar{\mu},\bar{x}} \rangle,
\end{gather}
where $J_\mu$ is the quark electromagnetic current. In the continuum limit this
propagator can be obtained by coupling the quarks to an external photon field
$A_\mu^\mathrm{ext}$ and building the second differential with respect to this
field.  In our lattice regularization, we use this prescription to define the current propagator.
Specifically the partition function in the presence of an external photon field is given by:
\begin{gather}
    Z[A^\mathrm{ext}]\equiv \int \dots\ \mathrm{dets}[U,A+A^\mathrm{ext}; \{q_f\}, \{m_f\}, e].
\end{gather}
The current propagator is then defined as the following second differential:
\begin{gather}
    \langle J_{\mu,x} J_{\bar{\mu},\bar{x}} \rangle\equiv
    \left.\frac{\delta^2 \log Z}{\delta A_{\mu,x}^\mathrm{ext}\delta A_{\bar{\mu},\bar{x}}^\mathrm{ext}}\right|_{A^\mathrm{ext}=0}.
\end{gather}
The so defined propagator satisfies current conservation on both the source and sink sides.
To compute it, we need the first and second derivatives of the fermion matrix at zero external field:
\begin{gather}
    \begin{aligned}
	\left.\frac{\delta}{\delta A^\mathrm{ext}_{\mu,x}}\right|_0 M_f&= eq_f\cdot D_\mu[i P_x V_Ue^{ieq_fA}],\\
	\left.\frac{\delta^2}{\delta A_{\mu,x}^\mathrm{ext}\delta A_{\bar{\mu},\bar{x}}^\mathrm{ext}}\right|_0 M_f&= -e^2q_f^2 \cdot \delta_{\mu\bar{\mu}}\cdot
	D_\mu[P_x P_{\bar{x}} V_Ue^{ieq_fA}].
    \end{aligned}
\end{gather}
From these we get the current propagator as follows:
\begin{gather}
    \label{eq:jj}
    \langle J_{\mu,x} J_{\bar{\mu},\bar{x}} \rangle/e^2=
    \left\langle
	\sum_f q_f^2 C^\mathrm{conn}_{\mu,x,\bar{\mu},\bar{x}}(m_f,eq_f) + C^\mathrm{disc}_{\mu,x,\bar{\mu},\bar{x}} 
	-\sum_f \frac{q_f^2}{4} \mathrm{Tr} \left(
	M^{-1}_f\ D_\mu[P_x P_{\bar{x}}V_U e^{ieq_fA}] \delta_{\mu\bar{\mu}} \right)
	\right\rangle
\end{gather}
where the connected vector meson contraction is defined as
\begin{gather}
    \label{eq:conn}
    C^\mathrm{conn}_{\mu,x,\bar{\mu},\bar{x}}(m,e)\equiv -\frac{1}{4}\mathrm{Tr} \left( M^{-1}\ D_\mu[iP_xV_Ue^{ieA}]\  M^{-1}\ D_{\bar{\mu}}[iP_{\bar{x}}V_Ue^{ieA}]\right)
\end{gather}
and the disconnected contraction as
\begin{gather}
    \label{eq:disc}
    C^\mathrm{disc}_{\mu,x,\bar\mu,\bar x} \equiv \sum_{f,\bar f} q_f q_{\bar f} I_{\mu,x}(m_f,eq_f) I_{\bar \mu,\bar x}(m_{\bar f}, eq_{\bar f}) \quad \text{with} \quad
    I_{\mu,x}(m,e)\equiv \frac{1}{4}\mathrm{Tr}\left( M^{-1} D_\mu[i P_xV_Ue^{ieA}] \right).
\end{gather}
In these formulas, $\mathrm{Tr}$ is the trace over color and spacetime indices
and the $P_x$ projection operator clears the components of a vector on all
sites except for $x$. Here, the fermion matrix $M$ is understood with mass $m$
and on a gauge background $V_Ue^{ieA}$. The $M_f$ notation, defined in Equation
\eqref{eq:dets}, stands for the fermion matrix with $m_f$ mass and $eq_f$
charge.  Due to gauge invariance, $\langle I_{\mu,x} \rangle=0$.  Equation
\eqref{eq:jj} is our master formula for the current propagator.  In the
following we decompose it into several pieces. There are three terms.  First is
the connected contribution, second is the disconnected contribution and the
third is a contact term. This last one gives no contribution to the observables
that we are interested in and it will be omitted from now on.  To obtain the
expansion in Equation \eqref{eq:exp}, we have to calculate the
Wick-contractions $C^\mathrm{conn}$ and $C^\mathrm{disc}$ at the
isospin-symmetric point and also their isospin-breaking derivatives.

It is common to split the connected part, $\sum_f q_f^2
C^\mathrm{conn}(m_f,eq_f)$, into the contributions of individual flavors:
\begin{gather}
    \label{eq:clsc}
    \begin{aligned}
	C^\mathrm{light}  &\equiv \tfrac{4}{9} C^\mathrm{conn}(m_u,\tfrac{2}{3}e) + \tfrac{1}{9} C^\mathrm{conn}(m_d,-\tfrac{1}{3}e),\\
	C^\mathrm{strange}&\equiv \tfrac{1}{9} C^\mathrm{conn}(m_s,-\tfrac{1}{3}e),\\
	C^\mathrm{charm}  &\equiv \tfrac{4}{9} C^\mathrm{conn}(m_c,\tfrac{2}{3}e),
    \end{aligned}
\end{gather}
where we suppressed Lorentz indices and coordinates for simplicity.  The
isospin-limit components of these, as defined in Section \ref{se:obs_gen}, in
terms of $C^\mathrm{conn}$ and their derivatives are given in Table
\ref{ta:ibjjc}.  For the disconnected contribution we give here the formulas
for the isospin-symmetric point and for the strong-isospin-breaking term:
\begin{gather}
    \label{eq:ibjjd}
    \begin{aligned}[]
	C^\mathrm{disc}_0  &= \tfrac{1}{9} \left[ I_{\mu,x}(m_l,0) - I_{\mu,x}(m_s,0) + 2I_{\mu,x}(m_c,0) \right] \left[I_{\mu,x}...\to I_{\bar{\mu},\bar{x}}...\right],\\
	[C^\mathrm{disc}]'_m &= -\tfrac{3m_l}{2} \tfrac{\partial}{\partial m_l} C^\mathrm{disc}_0
    \end{aligned}
\end{gather}
The detailed implementation of these quantities will be given in Sections
\ref{se:obs_lma} and \ref{se:obs_ibjj}. From these we then
construct the total expectation value as shown in Equations \eqref{eq:exp} and
\eqref{eq:ixx}.

It is also common to split the propagator at the isospin-symmetric point into
isospin singlet and triplet parts: $\left[\langle JJ\rangle\right]_0= \langle JJ
\rangle_{I=0} + \langle JJ \rangle_{I=1}$.  These are given by
\begin{align}
    \label{eq:jji1i0}
    \begin{split}
	\langle JJ \rangle_{I=1}/e^2&\equiv \left[\langle \tfrac{9}{10}C^\mathrm{light} \rangle\right]_0= \left\langle \tfrac{1}{2} C^\mathrm{conn}(m_l,0)\right\rangle_0\ ,\\
	\langle JJ \rangle_{I=0}/e^2&\equiv \left[\langle \tfrac{1}{10}C^\mathrm{light} + C^\mathrm{strange} + C^\mathrm{charm} + C^\mathrm{disc}\rangle\right]_0=\\
	&=\left\langle \tfrac{1}{18} C^\mathrm{conn}(m_l,0) + \tfrac{1}{9} C^\mathrm{conn}(m_s,0) + \tfrac{4}{9}C^\mathrm{conn}(m_c,0) + C^\mathrm{disc}_0 \right\rangle_0\ .
    \end{split}
\end{align}

Finally, we introduce the notation $G(t)$ for the zero-momentum timelike current
propagator with averaged Lorentz indices:
\begin{gather}
    \label{eq:gdeflat}
    G(t)\equiv 
    \frac{1}{3e^2}\sum_{\vec{x},\mu=1,2,3} \frac{1}{2} \left\langle J_{\mu,t,\vec{x}} J_{\mu,0} + J_{\mu,T-t,\vec{x}} J_{\mu,0} \right\rangle.
\end{gather}
This is the lattice version of the propagator given in Equation~\labelmaingdef{}
of the main text. $G(t)$ can also be decomposed into connected terms of
different flavors and a disconnected part. Note that the imaginary parts of
these quantities are zero due to the gauge averaging.

    \section{Anomalous magnetic moment $a_\mu$}
\label{se:obs_hvp}

In this section we provide the definition for the central observable of the
paper: the leading-order hadronic vacuum polarization (LO-HVP) contribution to
the anomalous magnetic moment of the muon, $a_\mu^\mathrm{LO-HVP}$.
Furthermore, we detail the decomposition of $a_\mu^\mathrm{LO-HVP}$. Since we
consider only the LO-HVP contribution, we drop the superscript and multiply the
result by $10^{10}$, ie. $a_\mu$ stands for $a_{\mu}^\mathrm{LO-HVP} \times
10^{10}$ throughout this work.

The renormalized scalar hadronic vacuum polarization function (HVP) can be
extracted from the zero momentum current propagator $G(t)$ as
\cite{Bernecker:2011gh}:
\begin{gather}
    \label{eq:hvp}
    \hat\Pi(Q^2)\equiv \sum_{t=0}^{T/2} \left[ t^2 -
      \frac4{(aQ)^2}\sin^2\left(\frac{aQt}{2}\right)\right] G(t)
\end{gather}
where $t$ and $G(t)$ are given in lattice units here and $a$ is the lattice
spacing. This formula corresponds to a Fourier transformation followed by a
division by $Q^2$, including an explicit removal of: 1.\ a pure finite-volume
effect and 2.\ the ultraviolet divergence. Renormalization is performed on
shell such that $\hat\Pi(0)=0$.  While, in a finite volume, $\hat\Pi(Q^2)$ is
only formally defined at momenta with components that are integer multiples of
either $2\pi/L$ or $2\pi/T$, we use \eqref{eq:hvp} to analytically continue
$\hat\Pi(Q^2)$ to any real values of $Q^2$. It is worth noting that this
approach is related to the time-moment approach of \cite{Chakraborty:2014mwa}.
Time moments can also be used as input in various approximants that were put
forward in \cite{deRafael:2014gxa,Charles:2017snx}. These are based on the
application of Mellin–Barnes techniques. They converge rapidly with the number
of moments retained \cite{Charles:2017snx} and also allow for a systematic
matching to perturbation theory at short distances. 

In this work we compute all hadronic $O(e^2)$ effects in the vacuum
polarization, including ones that are reducible under cutting a photon line.
Like $G(t)$, $\hat\Pi(Q^2)$ can also be decomposed into the connected contributions
of various quark flavors and a disconnected contribution (Section
\ref{se:obs_jj}).

The LO-HVP contribution to the anomalous magnetic moment of the muon is
computed from the one-photon irreducible part of $\hat\Pi(Q^2)$, denoted by
$\hat\Pi_\mathrm{1\gamma I}(Q^2)$, using the following integral \cite{Blum:2002ii}:
\begin{gather}
    \label{eq:amu}
    a_\mu=
    10^{10}\alpha^2 \int_{0}^\infty \frac{dQ^2}{m_\mu^2}\ \omega\left(\frac{Q^2}{m_\mu^2}\right) \hat\Pi_{1\gamma I}(Q^2)
\end{gather}
where $\omega(r)$ is given after Equation~\labelmainkdef{} of the main
paper, $\alpha$ is the fine structure constant renormalized in the
Thomson-limit, and $m_\mu$ is the mass of the muon.  The difference between
having $\hat\Pi(Q^2)$ and $\hat\Pi_{1\gamma I}(Q^2)$ in the integral in
\eqref{eq:amu} is the one-photon-reducible ($1\gamma R$) contribution denoted
by $a_\mu^\mathrm{1\gamma R}$. This contribution is finite in
$\mathrm{QED}_\mathrm{L}$ with our setup. It is an $O(e^2)$ effect that is
included in the higher-order hadronic vacuum polarization (HO-HVP)
contribution. This effect is tiny compared to $a_\mu$ and has already been
computed on the lattice, as discussed in our final result section, Section
\ref{se:res_amu}.

We partition the momentum integral in Equation \eqref{eq:amu} by cutting it
into two contributions at a momentum $Q_\mathrm{max}$. Below $Q_\mathrm{max}$
we use the lattice and above that perturbation theory. In the two parts the
vacuum polarization is renormalized differently: it is renormalized to zero at
$Q=0$ on the lattice and at $Q=Q_\mathrm{max}$ in perturbation theory.  This
requires the introduction of an extra term, that accounts for this difference.
The lattice part is then splitup into the contributions of different flavors.
In detail, our partitioning takes the following form:
\begin{gather}
    \label{eq:amuterms}
    a_\mu=
    a_\mu^\mathrm{light} + a_\mu^\mathrm{strange} + a_\mu^\mathrm{charm} + a_\mu^\mathrm{disc} +
    a_\mu^\mathrm{pert} - a_{\mu}^\mathrm{1\gamma R}\ .
\end{gather}
Here, the connected light contribution is given as
\begin{gather}
    \label{eq:amusplit}
    a_\mu^\mathrm{light}=
    10^{10}\alpha^2\left[\int_0^{Q_\mathrm{max}^2} \frac{dQ^2}{m_\mu^2}\ \omega\left(\frac{Q^2}{m_\mu^2}\right) \hat\Pi^\mathrm{light}(Q^2) +
    \hat\Pi^\mathrm{light}(Q^2_\mathrm{max})\int_{Q_\mathrm{max}^2}^\infty \frac{dQ^2}{m_\mu^2}\ \omega\left(\frac{Q^2}{m_\mu^2}\right)\right] ,
\end{gather}
and similarly for the other flavors\footnote{The contributions of the bottom
and the top are not shown explicitely here, they will be added in our final
result section, Section \ref{se:res_amu}.} and the disconnected contributions.
The second term accounts for the difference in the lattice and perturbative
renormalization points, as explained above. Using Equation \eqref{eq:hvp} we
can express all of these contributions as a weighted sum of the corresponding
current propagator. For the connected light term we have, for instance:
\begin{gather}
    \label{eq:kdef}
    a_\mu^\mathrm{light}=
    10^{10}\alpha^2
    \sum_{t=0}^{T/2} K(t;aQ_\mathrm{max},am_\mu)\ G^\mathrm{light}(t)\ ,
\end{gather}
where the kernel $K(t;aQ_\mathrm{max},am_\mu)$ is given by Equations
\eqref{eq:hvp} and \eqref{eq:amusplit}. It depends on the gauge ensemble only
through the lattice spacing. 
The perturbative contribution is given by
\begin{gather}
    a_\mu^\mathrm{pert}= 
    10^{10}\alpha^2\int_{Q_\mathrm{max}^2}^{\infty} \frac{dQ^2}{m_\mu^2}\ \omega\left(\frac{Q^2}{m_\mu^2}\right)
    \left[ \hat\Pi^\mathrm{pert}(Q^2)-\hat\Pi^\mathrm{pert}(Q_\mathrm{max}^2) \right]\ .
\end{gather}
In Reference \cite{Borsanyi:2017zdw} we demonstrated on our {\tt 4stout} data set that
switching to the perturbative calculation can be safely done for
$Q_\mathrm{max}^2\gtrsim2$~GeV$^2$, ie. from this point on $a_\mu$ does not
depend on the choice of $Q_\mathrm{max}$.  In this work we use
$Q_\mathrm{max}^2= 3$~GeV$^2$.  The perturbative part for this choice was
computed in \cite{Borsanyi:2017zdw} and is given in Section \ref{se:res_amu},
where the final result for $a_\mu$ is put together. 

We also consider a modification of Equation \eqref{eq:hvp}, in which
the current propagator is restricted to a certain region in time, from $t_1$ to
$t_2$. To achieve this, we multiply the propagator by a smooth window function
\cite{Blum:2018mom}
\begin{gather}
    \label{eq:win} W(t;t_1,t_2)\equiv
    \Theta(t;t_1,\Delta) - \Theta(t;t_2,\Delta) \quad\text{with}\quad
    \Theta(t;t',\Delta)\equiv \tfrac{1}{2} + \tfrac{1}{2}\tanh[ (t-t')/\Delta ]
\end{gather}
or equivalently we replace the weight factor as $K(t)\to K(t)W(t)$.  We will
focus on a particular window defined in Reference \cite{Blum:2018mom}, with
parameters $t_1=0.4$~fm, $t_2=1.0$~fm and $\Delta=0.15$~fm. The corresponding
contribution to the magnetic moment of the muon is denoted by
$a_{\mu,\mathrm{win}}^\mathrm{LO-HVP}$ and for brevity we use
$a_{\mu,\mathrm{win}}=a_{\mu,\mathrm{win}}^\mathrm{LO-HVP}\times 10^{10}$.  We
can do the same partitioning as we did with $a_\mu$ in Equation
\eqref{eq:amuterms}. We will use those notations extended by a $\mathrm{win}$
subscript. 

    \section{Noise reduction techniques}
\label{se:obs_lma}

\begin{figure}[t]
    \centering
    \includegraphics[width=0.7\textwidth]{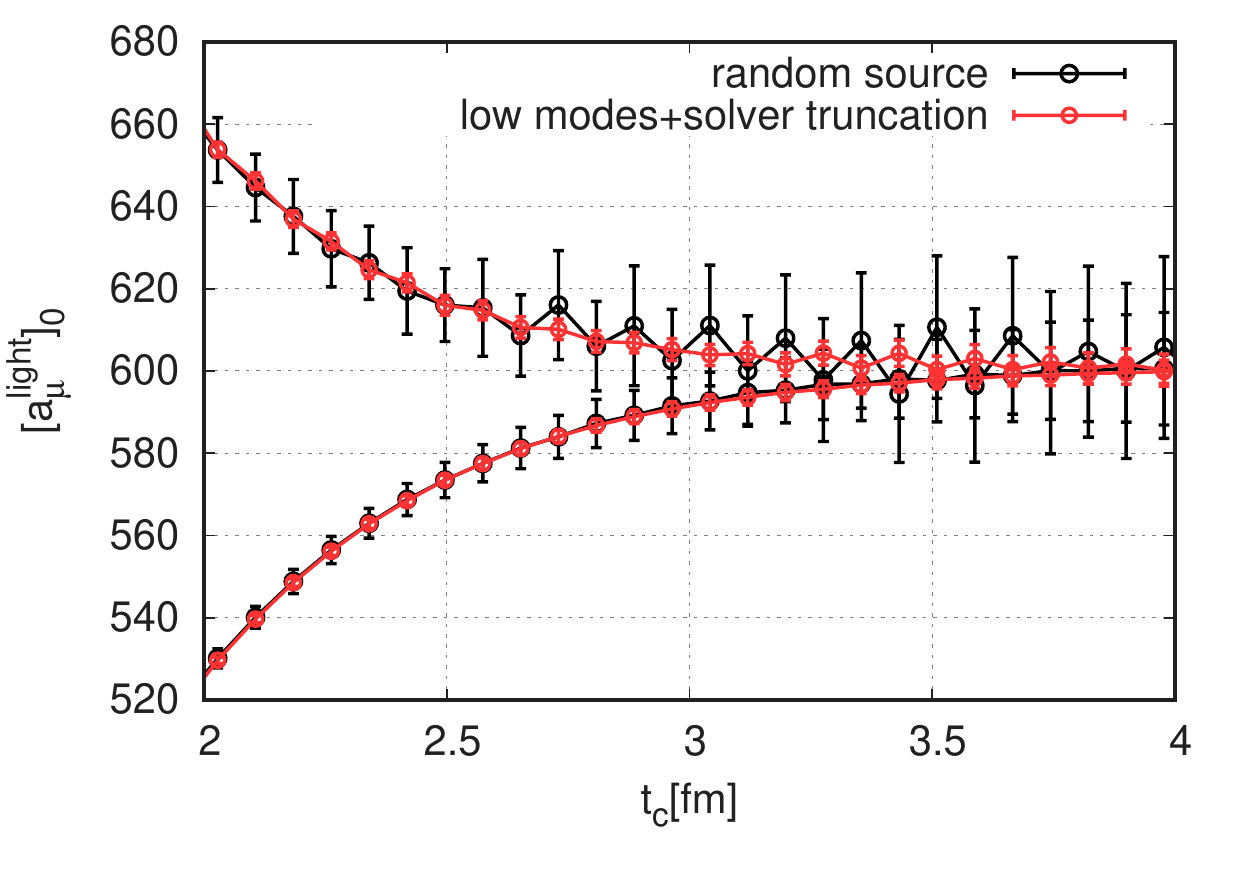}
    \caption
    {
	\label{fi:lma}Comparison of a conventional random source based
	technique, as we applied it in our earlier work
	\cite{Borsanyi:2017zdw}, and a low mode utilizing technique of this
	work on a $\beta=3.9200$ {\tt 4stout} ensemble for the case of
	$[a_{\mu}^\mathrm{light}]_0$ upper and lower bounds (see Section
	\ref{se:obs_bound}).
    }
\end{figure}

In this section we consider quantities at the isospin-symmetric point; noise
reduction techniques for the isospin-breaking part are discussed in Section
\ref{se:obs_ibjj}.  For the strange and charm connected contributions,
$C^\mathrm{strange}_0$ and $C^\mathrm{charm}_0$, and for the disconnected
contribution $C^\mathrm{disc}_0$ we use the same measurements that are
presented in our previous work \cite{Borsanyi:2017zdw}. A new measurement
procedure is implemented for the light connected component
$C^\mathrm{light}_0$. It is used to reanalyze the old configurations and make
measurements on new ensembles. This plays a key role in reducing the final
statistical error in $a_\mu$.

The technique utilizes the lowest eigenmodes of the fermion matrix; for an
early work with low eigenmodes, see \cite{Neff:2001zr}. The way in which we use
these modes here is essentially the same as in \cite{Li:2010pw}, where it is
called Low Mode Substitution. In the space orthogonal to these modes, the
computational effort is reduced considerably by applying imprecise (aka.
sloppy) matrix inversions. This is called the Truncated Solver Method
\cite{Bali:2009hu} or All Mode Averaging \cite{Blum:2012uh}. Here we describe
the technique for the connected part of the current propagator. The same
technique was applied recently for magnetic moment computations in
\cite{Aubin:2019usy} also.

We consider the connected propagator of Equation \eqref{eq:conn} for timelike
separation, and perform an averaging over the source positions, together with a
zero spatial-momentum projection at the sink:
\begin{gather}
    \label{eq:ctr}
    C(t,\bar{t}) \equiv
    \frac{1}{3L^3} \sum_{\vec{\bar{x}},\vec{x},\mu=1,2,3} C^\mathrm{conn}_{\mu,x,\mu,\bar{x}}(m_l,0)=
    -\frac{1}{12L^3} \sum_{\mu=1,2,3}\mathrm{ReTr} \left[ \mathcal{D}_{\mu,t} M^{-1} \mathcal{D}_{\mu,\bar{t}} M^{-1}\right],
\end{gather}
where $\mathcal{D}_{\mu,t}=\sum_{\vec{x}} D_\mu[iP_xU]$ is an operator that performs a symmetric, gauge-covariant
shift on a vector $v_x$:
\begin{gather}
    [\mathcal{D}_{\mu,t} v]_x =
	    \begin{cases}
		x_4=t:     & i\eta_{\mu,x} \left( U_{\mu,x}v_{x+\mu} + U^\dagger_{\mu,x-\mu}v_{x-\mu} \right)\\
		x_4\neq t: & 0
	    \end{cases}
\end{gather}
where $\eta_{\mu,x}$ are the Dirac-gamma matrices in the staggered
representation. We use the simplifying notation
$\mathcal{D}=\mathcal{D}_{\mu,t}$ and
$\overline{\mathcal{D}}=\mathcal{D}_{\mu,\bar{t}}$ in the following. In
Equation \eqref{eq:ctr}, we apply the real part because the imaginary part
vanishes anyway after averaging over gauge configurations.

Using the lowest eigenmodes of $M$ we split the quark propagator into an
eigenvector part and into its orthogonal complement, denoted by ``e'' and ''r'', respectively:
\begin{gather}
    M^{-1}= M^{-1}_e + M^{-1}_r,\\
    M^{-1}_e= \sum_i \frac{1}{\lambda_i} v_i v_i^\dagger
    \qquad \mathrm{and} \qquad
    M^{-1}_r= M^{-1} \left( 1 - \sum_i v_i v_i^\dagger \right),
\end{gather}
where $v_i/\lambda_i$ is the $i$-th eigenvector/eigenvalue of the operator $M$.
For the projection we used a modified version of the symmetric Krylov-Schur
algorithm described in \cite{Hernandez:2005:SSF}.  Correspondingly, $C$ splits
into three terms:
\begin{gather}
    C= C_{ee} + C_{re} + C_{rr},
\end{gather}
with eigen-eigen, rest-eigen and rest-rest contributions:
\begin{align}
    C_{ee}&= -\frac{1}{4L^3} \mathrm{ReTr} \left[ \mathcal{D} M^{-1}_e \overline{\mathcal{D}} M^{-1}_e\right],\\
    C_{re}&= -\frac{1}{4L^3} \mathrm{ReTr} \left[ \mathcal{D} M^{-1}_r \overline{\mathcal{D}} M^{-1}_e + \mathcal{D} M^{-1}_e \overline{\mathcal{D}} M^{-1}_r\right],\\
    C_{rr}&= -\frac{1}{4L^3} \mathrm{ReTr} \left[ \mathcal{D} M^{-1}_r \overline{\mathcal{D}} M^{-1}_r\right],
\end{align}
where an average over $\mu$ is assumed but not shown explicitely.  The benefit
of this decomposition is that the trace in the {\bf eigen-eigen part} can be
calculated exactly, and is thus equivalent to calculating the propagator with
all possible sources in position space. This is the main ingredient for the
noise reduction. Though no extra inversions are needed in this part, it has to
be optimized carefully, since there is a double sum over the eigenmodes, where
each term is a scalar product $v^\dagger_i \mathcal{D} v_j$.  In the {\bf
rest-eigen part} we have terms $v^\dagger_i \mathcal{D} M^{-1}_r
\overline{\mathcal{D}} v_i$ and also terms where $\mathcal{D}$ and
$\overline{\mathcal{D}}$ are exchanged. Therefore, this part is only a single
sum over the eigenmodes, and each term involves one matrix inversion.  Note
that these inversions are preconditioned by the eigenvectors, so they need much
less iterations than standard inversions.  Additionally, we speed up the
inversions by running them with a reduced precision, and for some randomly
selected eigenvectors we correct for the small bias by adding the difference
between a high precision solver and the reduced precision one
\cite{Bali:2009hu,Blum:2012uh}.  Finally, the {\bf rest-rest part} is evaluated
using random source vectors $\xi$: we calculate $\xi^\dagger \mathcal{D}
M^{-1}_r \overline{\mathcal{D}} M^{-1}_r \xi$, which requires two inversions
per random source. The reduced precision inverter technique is used here too.

As an example we give here the algorithm parameters for one of the {\tt 4stout}
ensembles at $\beta=3.9200$: $1032$ modes of the even-odd preconditioned Dirac
operator are projected; the high precision inversion has $10^{-8}$ accuracy;
the reduced precision inverter is capped at $400$ conjugate gradient
iterations; the bias correction is calculated with a frequency of $1/32$ and
$384$ random sources are chosen for the rest-rest term. With these choices, the
eigen-eigen part is the dominant source of the error, and since it is already
evaluated with all possible sources, we have reached the limit where the noise
comes from the fluctuations between gauge configurations.  Using this technique
we achieve a factor of five improvement in the statistical error compared to
our previous work \cite{Borsanyi:2017zdw}. There we applied a random source
technique similar to the one that we now use for the rest-rest term. The number
of random sources was $768$ per configuration. The comparison of the result
with the old and new techniques is shown in Figure \ref{fi:lma}. 

The number of projected eigenmodes is around $1000$ for all lattice spacings
in the $L=6$~fm boxes. The number of projected modes has to be scaled with the
physical four-volume, to keep the magnitude of the eigen-eigen part constant.
On the large {\tt 4HEX} lattice with $L=11$~fm we project $6048$ eigenmodes.

    \section{Upper and lower bounds on $\langle JJ\rangle$}
\label{se:obs_bound}

\begin{figure}[t]
    \centering
    \includegraphics[width=0.7\textwidth]{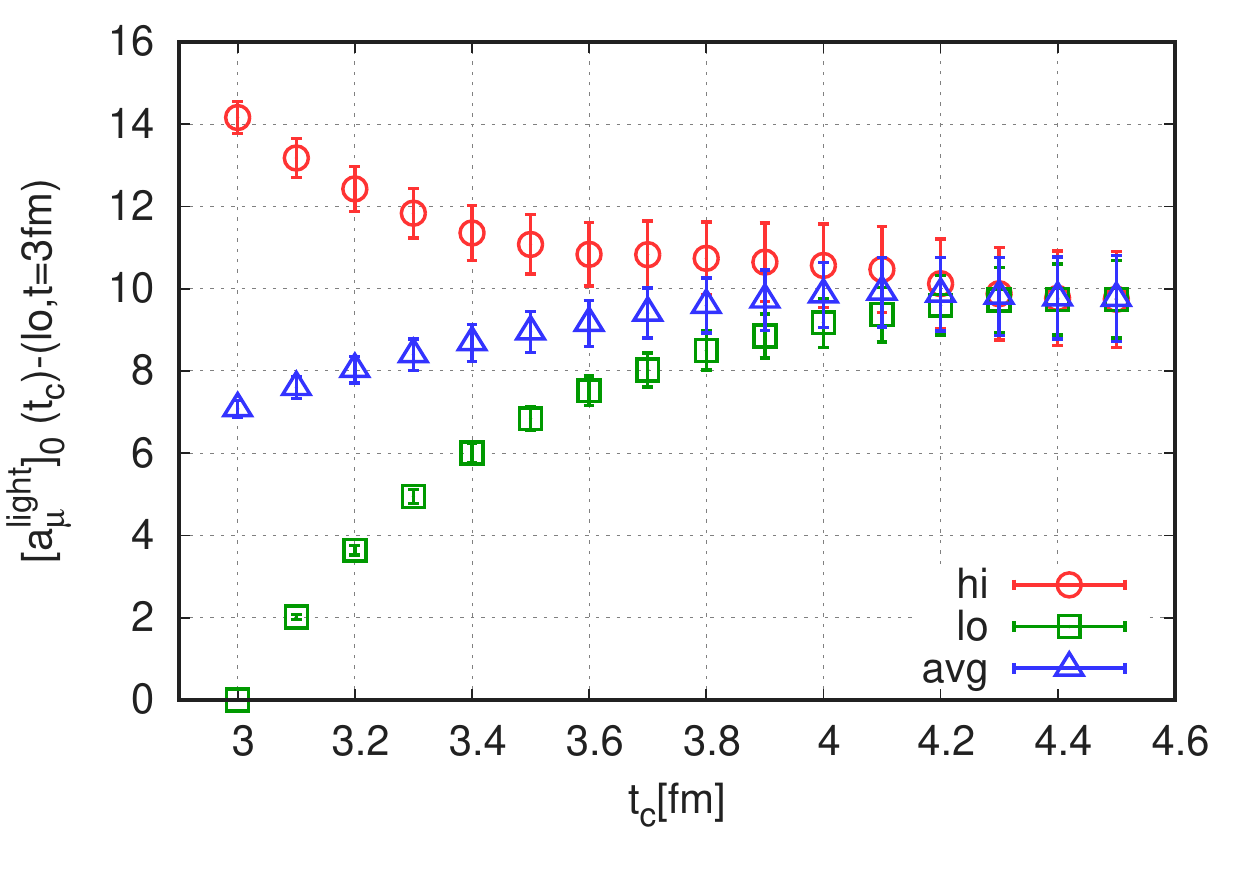}
    \caption
    {
	\label{fi:bound}Upper and lower bounds on $[a_{\mu}^\mathrm{light}]_0$
	as a function of $t_c$, ie.\ the upper limit of the time integration in
	Equation \eqref{eq:kdef}. The lower bound value at $3.0$~fm is
	subtracted and the plot shows the result of a combined fit of all
	ensembles that is evaluated at $a=0.064$~fm (see text). Such plots are
	used to set the value of $t_c$ on our ensembles. We use $t_c=4.0$~fm
	everywhere.
    }
\end{figure}

In the case of the light and disconnected contributions to the current
propagator, the signal deteriorates quickly as distance is increased. To
calculate the HVP, a sum over time of the propagator has to be performed, as
Equation \eqref{eq:kdef} shows. As was suggested in
\cite{Lehner:2016,Borsanyi:2016lpl}, we introduce a cut in time $t_c$, beyond
which the propagator is replaced by upper and lower bounds, thereby reducing
the statistical noise. Our estimate is given by the average of the bounds at a
$t_c$ where the two bounds meet. In this section only isospin-symmetric
quantities are considered.

Bounds are derived from the assumption that the current propagator is
a sum of exponentials with positive coefficients. In the case of staggered
fermions, where opposite parity states with oscillating coefficients give also
a contribution, the assumption is only satisfied after some distance and within
the statistical error. On our ensembles this is usually the case beyond about
$t\sim2.5$~fm.

For the {\bf light connected propagator} at the isospin-symmetric point the
bounds express the positivity (lower bound) and that the propagator should
decay faster than the exponential of two pions (upper bound). 
They are given as
\begin{gather}
    \label{eq:hibound}
    0 \le G^\mathrm{light}(t)\le G^\mathrm{light}(t_c) \frac{\varphi(t)}{\varphi(t_c)},
\end{gather}
where $\varphi(t)= \exp(-E_{2\pi}t)$. For $E_{2\pi}$ we use the energy
of two non-interacting pions with the smallest non-zero lattice momentum
$2\pi/L$.  The larger the $t_c$ the better the upper bound, but it comes with
more statistical noise.

The exponential decay above assumes an infinite time extent, $T=\infty$.  We
incorporate the effects of a finite-$T$ using next-to-leading-order chiral
perturbation theory. There the exponential decay with the two-pion energy gets
replaced by the following $\cosh$-type form:
\begin{gather}
    \label{eq:coshT}
    \exp(-E_{2\pi}t) \longrightarrow \frac{\cosh[E_{2\pi}(t-T/2)]+1}{\cosh(E_{2\pi}T/2)-1}\ .
\end{gather}
This is shown in detail in Section \ref{se:obs_xpt}, where the above
replacement corresponds to $\epsilon \to \epsilon_T$ in Equations
\eqref{eq:epsxpt} and \eqref{eq:epsxpt2}. In this case an appropriate upper
bound on the propagator is given by using the right hand side of Equation
\eqref{eq:coshT} as the bounding function $\varphi(t)$. We use this form of
$\varphi(t)$ to compute the upper bounds in this work.

To obtain a suitable $t_c$ on the {\tt 4stout} ensembles, we combine the
propagators of all ensembles in a single analysis. The reason for this is to
use all available statistics to analyze the behavior of the bounds on $t_c$.
The $a_\mu$ results of the ensembles depend on the pion and kaon masses and the
lattice spacing. The first two dependencies can be safely eliminated if we
consider only the tail of the propagator. (Remember, we are working close to
the physical point.) For this we subtracted the value of the lower bound at
$3.0$~fm from both bounds on each ensemble. The lattice-spacing dependence was
taken into account by making a continuum extrapolation for this ``subtracted''
$a_\mu$ at each value of $t_c$.  The result of the continuum fit for the upper
and lower bounds on $[a_{\mu}^\mathrm{light}]_0$ is shown in Figure
\ref{fi:bound}, in which the $a^2$ function used in the continuum extrapolation
is evaluated at our finest lattice spacing, $a=0.064$~fm.  The two bounds meet
around $4.0$~fm. Here the statistical error of the average covers the central
value of both bounds.  We will use this average at $t_c=4.0$~fm as our estimate
for $[a_{\mu}^\mathrm{light}]_0$ on the {\tt 4stout} ensembles. Variation of
$t_c$ in the plateau region had negligible effect on the result. Note, that
this value of $t_c$ is much larger than the one that we used in our previous
work, $t_c=3.0$~fm. This improvement is made possible by the noise reduction
technique of Section \ref{se:obs_lma}.

In the case of the isospin-symmetric {\bf disconnected propagator} the bounds
are
\begin{gather}
    0 \le -G^\mathrm{disc}(t) \le \tfrac{1}{10} G^\mathrm{light}(t_c)\frac{\varphi(t)}{\varphi(t_c)} + G^\mathrm{strange}(t) + G^\mathrm{charm}(t).
\end{gather}
Since the strange and charm terms fall off much faster than the light and
disconnected one, their contribution does not change the value of $t_c$
obtained. We use the same measurements for $G^\mathrm{disc}_0$ as in our
previous work \cite{Borsanyi:2017zdw}, and take the average of the bounds at a
single cut value: $t_c=2.5$~fm. This choice is in accordance with our findings
in \cite{Borsanyi:2017zdw}, that the variation in $t_c$ within the plateau of
the average bound has a negligible effect on the result.

    \section{Isospin-breaking effects in $\langle JJ\rangle$}
\label{se:obs_ibjj}

\begin{figure}[t]
    \centering
    \includegraphics[width=0.7\textwidth]{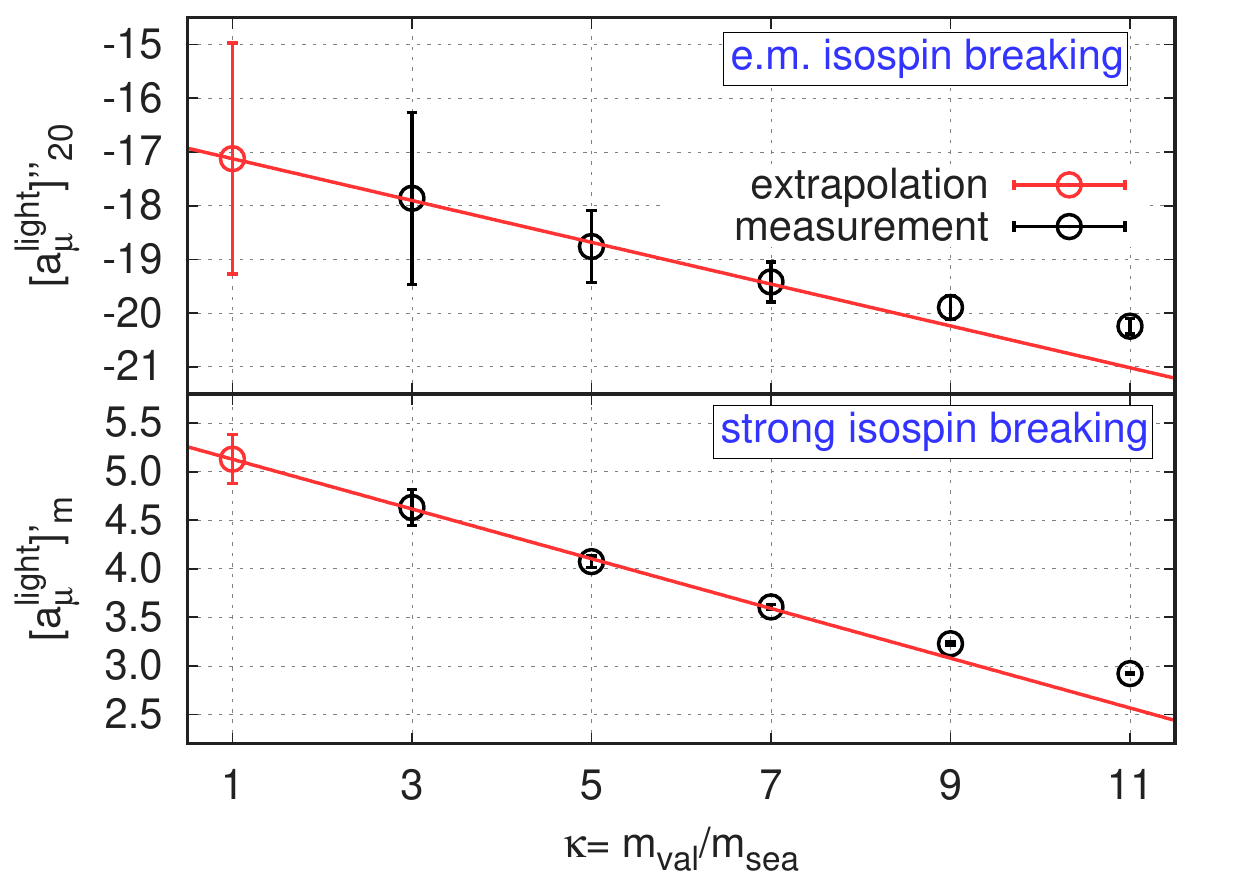}
    \caption
    {
	\label{fi:ibjj}Extrapolation procedure to obtain the electromagnetic
	and strong-isospin-breaking corrections to the light connected
	contribution.  Measurements are performed with valence over sea quark
	mass ratios of $\kappa=\{3,5,7,9,11\}$. We use the three lightest
	$\kappa$ values to obtain the value at $\kappa=1$ with a linear
	extrapolation shown in the Figure.  The data on the plot is taken from
	our coarsest lattice, corresponding to $\beta=3.7000$.
    }
\end{figure}

\begin{table}[t]
    \centering
    \begin{tabular}{L|C|C}
	\text{isospin component}& t_c[\text{fm}]\ X=G^\mathrm{light} & t_c[\text{fm}]\ X=G^\mathrm{disc} \\
	\hline
	X'_m     & 2.5 & 2.5 \\
	X''_{20} & 2.5 & 2.0 \\
	X''_{11}, X''_{02} & 1.0 & 1.5
    \end{tabular}
    \caption
    {
	\label{ta:tcfm}Cuts in time for different isospin breaking components
	of the light and disconnected propagator. For each component we use two
	different cuts: the one that is given in the Table, and another that is
	$0.5$~fm larger.
    }
\end{table}

In this section we describe the procedure that we use to compute the
isospin-breaking corrections to the current propagator. We consider the
contribution of the light and strange quarks only. The charm quark contribution
was computed on the lattice in \cite{Giusti:2019xct}.

We start with the {\bf connected contributions} (see Table \ref{ta:ibjjc}). The
electric derivatives in those formulas, $X_{1}'$ and $X_{2}''$, are measured by
finite differences, as in the case of the isospin-breaking in hadron masses.
Specifically, we compute the following contractions
\begin{gather}
    C^\mathrm{conn}(m_s,0),\quad
    C^\mathrm{conn}(m_s,+\tfrac{1}{3}e_*),\quad
    C^\mathrm{conn}(m_s,-\tfrac{1}{3}e_*)
\end{gather}
for the strange quark and
\begin{gather}
    \label{eq:connkappa}
    C^\mathrm{conn}(\kappa m_l,0),\quad
    C^\mathrm{conn}(\kappa m_l,+\tfrac{1}{3}e_*),\quad
    C^\mathrm{conn}(\kappa m_l,-\tfrac{1}{3}e_*)
\end{gather}
for the light quark, where $e_*$ is the physical value of the electric
coupling. From these the finite-difference approximators of the first and
second derivatives can be built in the standard way. In the light quark case,
the $G^\mathrm{conn}(m_l)$ propagators are noisy. Instead of the low-mode
technique of Section \ref{se:obs_lma}, we use a simpler approach to reduce the
noise, which is sufficiently accurate.  We perform computations with valence
quark masses that are some multiple $\kappa$ of the sea quark mass, $m_l$, and
then implement a chiral extrapolation to the target point at $\kappa=1$. We
measure the contractions in Equation \eqref{eq:connkappa} at five different
values of $\kappa$ and use the three lightest of these, $\kappa=3,5,7$, to
perform a linear extrapolation gauge-configuration by gauge-configuration.
Figure \ref{fi:ibjj} shows the result of this chiral extrapolation procedure on
our coarsest ensemble for the case of $[a_{\mu}^\mathrm{light}]_{20}''$.
A quadratic extrapolation including all five $\kappa$ values gives a result
that is consistent with the $\kappa=3,5,7$ linear extrapolation within
statistical uncertainty.

For the strong-isospin derivative of the connected contraction
$[C^\mathrm{light}]'_{m}$ (see last line of Table \ref{ta:ibjjc}) we implement
directly the operator corresponding to the mass derivative. Again we use a
chiral extrapolation from non-physical valence quark masses with
$\kappa=3,5,7$, similar to the case of the electromagnetic derivative. This
extrapolation procedure is also plotted in Figure \ref{fi:ibjj} for
$[a_{\mu}^\mathrm{light}]_m'$.

Finally the sea-valence and sea-sea electromagnetic derivatives of the
connected part, $[a_{\mu}^\mathrm{light}]_{11/02}''$ and
$[a_{\mu}^\mathrm{strange}]_{11/02}''$, are measured on the set of ensembles
that are dedicated to dynamical QED effects (see Table \ref{ta:dynqed}). There
we combine the low-mode, noise-reduction technique with the mass extrapolation
procedure above to reach sufficient accuracy.

In order to further reduce the noise in the light quark isospin-breaking
corrections, when computing quantities like $a_\mu$, we set the propagator to
zero after some cut $t_c$. The value of $t_c$ is obtained by looking for a
plateau in the derivative of $a_\mu^\mathrm{light}$ as a function of $t_c$.
For the different isospin-breaking derivatives these values are given in Table
\ref{ta:tcfm}. To assess the systematic error of the procedure, we also use a
second value of $t_c$ that is $0.5$~fm larger.  Let us note here that, when
computing various derivatives of $a_\mu$, not only do we have to consider the
derivatives of the propagator, but also those of the lattice scale, which enters
in the weighting function $K(t;Q_\mathrm{max}a,m_\mu a)$ (see Equation
\eqref{eq:kdef}).

Now let us turn to the {\bf disconnected contribution}. The basic operator for
this measurement is the trace of the quark propagator, $I(m,e)$ of Equation
\eqref{eq:disc}. This is computed with the help of the low-lying eigenmodes of
the Dirac operator, in a similar way to the calculation of the connected
diagram $C^\mathrm{conn}(m,e)$, described in Section \ref{se:obs_lma}. In this
case, the computation is technically much simpler, since in $I(m,e)$ there is
only one quark propagator under the trace, whereas in $C^\mathrm{conn}(m,e)$
there are two. The eigenvectors depend on the electric coupling, and we compute
them for each value of $e$ that we need.  For the electromagnetic derivative it
is useful to rewrite the current
\begin{gather}
    \sum_{f=\{u,d,s\}} q_f I(m_f,q_fe)=
    \tfrac{2}{3}I(m_l,\tfrac{2}{3}e) - \tfrac{1}{3}I(m_l,-\tfrac{1}{3}e) - \tfrac{1}{3}I(m_s,-\tfrac{1}{3}e)
\end{gather}
using a Taylor expansion around $e=0$ in the following way:
\begin{gather}
    \sum_{f=\{u,d,s\}} q_f I(m_f,q_fe)=
    -2I(m_l,0) + 2 I(m_l,\tfrac{1}{3}e) + \tfrac{1}{3}I(m_l,-\tfrac{1}{3}e) - \tfrac{1}{3}I(m_s,-\tfrac{1}{3}e) + O(e^3).
\end{gather}
The advantage of this form is that we can compute the first and second
derivatives without having to compute the traces with $\pm\tfrac{2}{3}e$
charge.  For the strong isospin derivative, where only the first derivative is
needed, it suffices to compute only one additional trace at a slightly
different mass than $m_l$. We use $0.9\cdot m_l$.  Altogether we measure
traces at the following masses and electric couplings:
\begin{gather}
    \begin{gathered}
	I(m_l,0),\qquad
	I(m_s,0),\\
	I(0.9\cdot m_l,0),\\
	I(m_l,+\tfrac{1}{3}e_*),\qquad
	I(m_l,-\tfrac{1}{3}e_*),\qquad
	I(m_s,+\tfrac{1}{3}e_*),\qquad
	I(m_s,-\tfrac{1}{3}e_*),
    \end{gathered}
\end{gather}
from which all the isospin-breaking derivatives can be constructed using finite
differences.  All these traces are measured using the same set of random
vectors, so that the random noise does not wash out the signal in the
differences.

    \section{Staggered chiral perturbation theory}
\label{se:obs_xpt}

\def\SXPT{SXPT}
\def\LO{LO}
\def\NLO{NLO}
\def\NNLO{NNLO}

\renewcommand{\arraystretch}{1.2}
\begin{table}[t]
    \centering
    \begin{tabular}{L|C|C|C|C|L}
	D & N_L & N_4 & N_5 & N_6 & \\
	\hline
	\hline
	1   & 0 & 0 & 0 & 0 & \multirow{ 1}{*}{LO}\\
	\hline
	2   & 0 & 1 & 0 & 0 & \multirow{ 2}{*}{NLO} \\
	    & 1 & 0 & 0 & 0 & \\
	\hline
	5/2 & 0 & 0 & 1 & 0 & \multirow{ 1}{*}{N$\sqrt{\text{N}}$LO} \\
	\hline
	3   & 0 & 2 & 0 & 0 & \multirow{ 4}{*}{NNLO}\\
	    & 0 & 0 & 0 & 1 & \\
	    & 1 & 1 & 0 & 0 & \\
	    & 2 & 0 & 0 & 0 & \\
	\hline
	\hline
    \end{tabular}
    \caption
    {
	\label{ta:powcnt}Solutions to the power counting formula of
	Reference~\cite{SWME:2011aa} for the current-current correlator.  For
	an \NNLO{} calculation, diagrams of dimension \(D \le 3\) will
	contribute.  Each solution describes diagrams of dimension \(D\) having
	\(N_L\) loops, along with some specified number of vertices from the
	Lagrangians of each order denoted by $N_{4,5,6}$. The number of
	vertices from the leading order Lagrangians $N_2$ is only limited by
	the number of loops.
    }
\end{table}
\renewcommand{\arraystretch}{1.0}

In this section we consider the current propagator in staggered chiral
perturbation theory (\SXPT{}) to next-to-next-to-leading order (\NNLO{}). The
goal of this effort is to describe the taste violation and finite-size
effects in our lattice simulations. We work in the isospin-symmetric limit
throughout this section.

In continuum chiral perturbation theory and in momentum space, the \NNLO{}
contribution was computed in \cite{Golowich:1995kd,Amoros:1999dp}, and the
finite-volume corrections are given in \cite{Bijnens:2017esv}. A coordinate
space computation including finite-volume effects was given recently in
\cite{Aubin:2019usy}.  In staggered chiral perturbation theory, only the
next-to-leading order (\NLO{}) has been computed \cite{Aubin:2006xv}. Here we
work out the following order and give the final result in the coordinate space
representation. This order requires the \NLO{} contributions to the staggered
chiral Lagrangian, which is given in \cite{Sharpe:2004is}.  This Lagrangian has
already been used to compute pseudoscalar meson masses to \NLO{}
\cite{SWME:2011aa}. This result will be an important ingredient here: just as
in continuum chiral perturbation theory, the current propagator to \NNLO{} can
be considerably simplified if one writes it in terms of masses including \NLO{}
corrections.

The \SXPT{} Lagrangian relevant to our computation
is given by
\begin{gather}
    \mathcal{L}=
    \mathcal{L}_2 + \mathcal{L}_{2,LS} +
    \mathcal{L}_4 + \mathcal{L}_{4,SV} +
    \mathcal{L}_5 +
    \mathcal{L}_6\ ,
\end{gather}
with $\mathcal{L}_{2,4,6}$ the standard continuum \LO{}, \NLO{} and \NNLO{}
Lagrangians of Gasser and Leutwyler \cite{Gasser:1984gg} and Bijnens, Colangelo
and Ecker \cite{Bijnens:1999sh}. We denote $\mathcal{S}_i=\int dx\
\mathcal{L}_i$ the corresponding actions, $i\in \{2;2,LS;4;4,SV;5;6\}$.  We use
the standard \SXPT{} power counting scheme of Reference~\cite{Lee:1999zxa},
whereby the \LO{} contributions are $O(p^2) \approx O(m_q) \approx O(a^2)$,
where $p$ stands for a derivative operation, $m_q$ is the light quark mass and
$a$ is the lattice spacing. According to this counting there is a \LO{}
staggered correction $\mathcal{L}_{2,LS}$ given by the Lagrangian of Lee and
Sharpe \cite{Lee:1999zxa} and a \NLO{} staggered correction
$\mathcal{L}_{4,SV}$ given by Sharpe and van de Water \cite{Sharpe:2004is}.
There is also a staggered specific contribution between \NLO{} and \NNLO{},
$\mathcal{L}_5$. More details on these terms are given below.  There are
further $O(a^2)$ terms in the Lagrangian, which --differently from the
staggered corrections-- are invariant under the continuum taste symmetry. We
are not giving them explicitly here; their effect is to change the low-energy
constants of the theory by $O(a^2)$ amounts -- at least to the order that we
consider here.

A general power counting formula is provided in the Appendix of
\cite{SWME:2011aa}. Under this scheme, we say that a diagram $\mathcal{M}(p,
m_q, a^2)$ has dimension $D$ if it scales as $\mathcal{M}(p, m_q, a^2) \to
\lambda^D \mathcal{M}(p, m_q, a^2)$ under a rescaling of the external momenta,
quark masses and lattice spacing by $p \to \sqrt{\lambda} p$, $m_q \to
\lambda m_q$, and $a^2 \to \lambda a^2$.  In Table~\ref{ta:powcnt}, we
enumerate the contributions that are required for our \NNLO{} computation under
this counting scheme. Note that the \LO{} contribution is zero.

The field variables are denoted by $\phi$. They describe all the flavors and
tastes of staggered pions. They are expressed as linear combinations of the
product of the Hermitian generators of the \(U(4)\) taste group and of the
\(U(N)\) replicated flavor group, ie.  $\phi \equiv \sum_{\alpha a}
\phi_{\alpha a} \xi_\alpha T_a$.  Here the taste index $\alpha$ runs over the
16 element set $\{5,\mu5,\mu\nu,\mu,I\}$ with $\mu<\nu$. A possible
representation for the taste generators can be built from the $\xi_\mu$
Dirac-matrices and the $4\times4$ identity matrix as:
\begin{gather}
    \xi_\alpha\in \{\quad
	\xi_5,\quad
	\xi_{\mu5}=i\xi_\mu\xi_5,\quad
	\xi_{\mu\nu}= i\xi_\mu\xi_\nu,\quad
	\xi_\mu,\quad
	\xi_I= 1\quad
	\}
\end{gather}
For the $U(N)$ generators $T_a$ we use the generalized Gell-Mann matrices and
the $a$ index runs from $0$ to $N^2-1$. We work with $N_f$ degenerate flavors
and rooting is implemented with the replica trick, ie. $N=N_fN_r$, with
$N_r\to\frac{1}{4}$ at the end of the computation.  In most cases, $\phi$
appears in an exponential form in the Lagrangian, $U=\exp(i\phi/F)$, where $F$
is the pion decay constant in the chiral limit. Traces are taken
over both the taste and replica-flavor indices. Generators are normalized as
$\mathrm{tr}(\xi_\alpha T_a\cdot \xi_\beta T_b)= 2\delta_{ab}\delta_{\alpha\beta}$.

The current propagator is obtained by incorporating an external Hermitian
vector field \(v_\mu = v^\dagger_\mu\) in the Lagrangian, setting \(v_\mu = Q
A_\mu\) (where $Q$ is the charge matrix) and taking the second order functional
derivative of the partition function with respect to $A_\mu$:
\begin{align}
    \label{eq:jjxpt_gen}
    \Braket{J_{\mu}(x) J_{\bar{\mu}}(\bar x)}/e^2
    &\equiv {\left. \frac{\delta^2 \log Z\!\left[v_\nu = Q A_\nu\right]}
                         {\delta A_\mu(x) \, \delta A_{\bar{\mu}}(\bar x)} \right|}_{A=0}\nonumber\\
    &= \Braket{\frac{\delta\mathcal{S}\left[v_\nu = Q A_\nu\right]}{\delta A_\mu(x)}
               \frac{\delta\mathcal{S}\left[v_\nu = Q A_\nu\right]}{\delta A_{\bar{\mu}}(\bar x)}
             - \frac{\delta^2\mathcal{S}\left[v_\nu = Q A_\nu\right]}{\delta A_\mu(x) \, \delta A_{\bar{\mu}}(\bar x)}}\ 
\end{align}
up to vanishing disconnected terms. In the following we take the current to
operate as a taste singlet, so \(Q\) is proportional to the identity matrix in
taste space, that is $Q= Q_a \xi_{I} T_a$.  This restriction is discussed in
more detail at the end of the computation. In flavor space, \(Q\) is diagonal
but non-singlet. The field-strength tensor of the electromagnetic vector
potential is denoted by $F_{\mu\nu}= \partial_\mu A_\nu - \partial_\nu A_\mu$.

In the first term of Equation \eqref{eq:jjxpt_gen}, we will refer to the
derivative $\frac{\delta\mathcal{S}}{\delta A_\mu}$ as current term.  The
second term in Equation \eqref{eq:jjxpt_gen} gives rise to contact terms
proportional to $\delta(x-\bar x)$ and derivatives thereof.  The anomalous
magnetic moment considered in this work is obtained by integrating the
propagator with a kernel function, that behaves as $(x-\bar x)^4$ for small
differences. This eliminates all contact terms with less than four derivatives.
The smallest order they can enter is therefore $O(p^6)$, since the two photon
fields add an extra $p^2$ to the counting.

\subsection*{Leading-order contributions}

The \(O(p^2)\) and \(O(m_q)\) terms are given by the standard Euclidean
chiral Lagrangian, with an additional mass term for the taste-flavor singlet
\begin{align}
    \mathcal{L}_2 =\frac{F^2}{4} \mathrm{tr}\left( D_\mu U D_\mu U^\dagger \right)
		       - \frac{F^2}{4} M^2 \mathrm{tr} \left( U + U^\dagger \right)
		       + \frac{m_0^2}{12} (\mathrm{tr} \phi)^2 \,,
\end{align}
where $M^2$ is the tree-level Goldstone-boson mass and $D_\mu= \partial_\mu U - i
[v_\mu,U]$ is the covariant derivative including the vector field.  Note that
the external vector field is $O(p)$ in the chiral power counting scheme.  The
singlet meson mass has to be sent to infinity at the end of the computation
(\(m_0 \to \infty\)).  The functional derivatives described above give the
current couplings
\begin{align}
    \frac{\delta\mathcal{S}_2}{\delta A_\mu(x)}
    &= \frac{i}{2} \mathrm{tr}\left(Q \left[\partial_\mu \phi, \phi\right]\right)
    - \frac{i}{24 F^2} \mathrm{tr}\left( \left[Q, \partial_\mu \phi\right] \phi \phi \phi
                            - 3 Q \phi \left[\partial_\mu \phi, \phi\right] \phi \right)
    + O(\phi^6)\,.
\end{align}

The leading order staggered terms have been described by Lee and
Sharpe~\cite{Lee:1999zxa} and generalized to multiple flavors by Aubin and
Bernard~\cite{Aubin:2003mg}. We write these terms as \(\mathcal{L}_{2,LS}\).
Since these are \(O(a^2)\), they can contribute to our \NNLO{}
calculation in diagrams with up to two loops, so when expanding in \(\phi\), we
consider terms up to and including \(O(\phi^4)\). The \(O(\phi^2)\) terms can
be absorbed into the \(O(\phi^2)\) mass term from \(\mathcal{L}_2\), providing
taste-dependent corrections to the tree-level mass. There are also extra terms
for the flavor singlets of each taste, shifting these masses separately from
the other flavor components.
Since \(\mathcal{L}_{2,LS}\) does not depend on the external field, there are
no contributions to the current terms.

\subsection*{Next-to-leading order contributions}

The \(O(p^4)\), \(O(p^2 m_q)\), and \(O(m_q^2)\) vertices have been described
by Gasser and Leutwyler~\cite{Gasser:1984gg}. We denote this Lagrangian by
\(\mathcal{L}_4\), and use the standard notation for the low-energy
constants.  According to Table~\ref{ta:powcnt} they contribute to the
calculation in diagrams with up to one loop, so when expanding in \(\phi\), we
only consider terms of \(O(\phi^2)\). These terms can mostly be absorbed into
redefinitions of the coefficients for the mass term, and a
rescaling of the field variable $\phi$. Such a rescaling cannot affect the
final result, since we are computing the correlation function of an external field.
The rescaling of the field variable absorbs the contributions to the current
term from the terms in the Lagrangian that are proportional to the low-energy constants
\(L_4\) and \(L_5\) and only the following term remains:
\begin{align}
    \frac{\delta\mathcal{S}_4}{\delta A_\mu(x)}
    &= \frac{2i}{F^2}\ L_9\ \mathrm{tr}\left( Q \partial_\nu \left[\partial_\mu \phi, \partial_\nu \phi\right] \right)
    + O(\phi^4).
\end{align}

The staggered specific \NLO{} contributions involve \(O(a^2 p^2)\), \(O(a^2
m_q)\), and \(O(a^4)\) terms, which have been described by Sharpe \& van de
Water~\cite{Sharpe:2004is}.  As \NLO{} terms, these can only contribute at one
loop, so we only consider their expansion up to \(O(\phi^2)\). Similar to
\(\mathcal{L}_{2,LS}\), most of the terms of $\mathcal{L}_{4,SV}$ can be
absorbed into the \LO{} mass as taste-dependent corrections and by field
rescaling. There are three groups of terms that cannot be treated this way:
\begin{enumerate}

    \item In the first group we have $O(a^2p^2)$ terms that violate the
	remnant $SO(4)$ taste symmetry of the \LO{} action, for example
	$\sum_\mu \langle \partial_\mu \phi \xi_\mu \partial_\mu \phi \xi_\mu
	\rangle$. The net effect of these terms on the current
	propagator is only to change the mass that appears in the pion propagators
	by $SO(4)$ violating terms. In our lattice simulations these $SO(4)$
	violations are tiny: on our coarsest lattice the pion mass splittings,
	within the $SO(4)$ multiplets, are about $50$ times smaller than the
	splittings between different multiplets. We will set them to zero in
	our \SXPT{} computation.

    \item In the second group we have $O(a^2p^2)$ terms that depend
	explicitly on the external field, and not through the covariant
	derivatives. Reference~\cite{Sharpe:2004is} calls these ``extra source-terms''.
	In principle, these can contribute to the current terms.
	However, they are all proportional to the commutator of the
	vector field with a taste matrix $\xi_\alpha$, and since we take our
	vector current to be a taste singlet, such commutators vanish.
	An example for such a term is $\sum_{\mu,\nu}\langle [A_\mu,\xi_\nu]
	\phi \xi_\nu \partial_\mu \phi\rangle$.

    \item Finally, there might be taste-symmetry-breaking terms containing the
	field-strength tensor $F_{\mu\nu}$. A spurion analysis, similar to the
	one in Section III of Reference \cite{Aubin:2006xv}, indicates that no
	such terms are allowed in the $\mathcal{L}_{4,SV}$ Lagrangian.

\end{enumerate}

\subsection*{Contributions from beyond next-to-leading order}

In \SXPT{}, it is possible to construct a chiral Lagrangian between \NLO{} and
\NNLO{}, as discussed by Bailey, Kim and Lee~\cite{SWME:2011aa}. We denote this
Lagrangian $\mathcal{L}_5$. In it, terms might arise from the dimension-8 or
dimension-9 Lagrangians in the Symanzik effective theory, that are either
$O(a^4p)$ or $O(a^5)$.  In order to contribute to our calculation, vertices at
this order would need to appear in a tree-level diagram (see
Table~\ref{ta:powcnt}, where we denote it N$\sqrt{\text{N}}$LO). As a result,
the only terms that could contribute to the current propagator would be contact
terms. However, such terms require the square of the vector field $A_\mu\approx
O(p)$ in order to have a non-zero second functional derivative and therefore
they must be $O(p^2)$.  We conclude that $\mathcal{L}_5$ gives no contribution
in this calculation.

Finally, we come to the \NNLO{} Lagrangian. Similarly to the Bailey-Kim-Lee
Lagrangian discussed above, \NNLO{} terms can only contribute to tree-level
diagrams through contact terms at \(O(\phi^0)\). The continuum terms are
described by Bijnens, Colangelo and Ecker~\cite{Bijnens:1999sh}, and we write
them as \(\mathcal{L}_6\).  As we already mentioned, in our observables only
contact terms that are at least $O(p^6)$ can contribute, and the
only such term is:
\begin{align}
    \frac{\delta^2 \mathcal{S}_6}{\delta A_\mu(x)\delta A_{\bar{\mu}}(\bar{x})}
    &= C_{93}\ \mathrm{tr}({Q^2})\ (\delta_{\mu\bar{\mu}} \partial^2 - \partial_\mu \partial_{\bar{\mu}}) \partial^2 \delta(x-\bar{x}) + O(\phi^2)\,.
\end{align}

The $O(a^2)$ taste-violating contributions at \NNLO{} must similarly be contact
terms at \(O(\phi^0)\). As mentioned above, such terms must be at least
$O(p^6)$ in order to contribute to the $a_\mu$. However, $O(a^2 p^6)$ terms can
enter only beyond NNLO, so they are not relevant in our calculation.

\subsection*{Infinite-$L$ and $T$ result}

Performing the computations with the Lagrangians from the previous subsections
we obtain the full current propagator. As we mentioned in the beginning, it is
also necessary to compute the \NLO{} mass shift \(\delta M^2_\alpha\) for each
taste arising from the terms not absorbed into the tree-level mass
\(M^2_\alpha\).

To arrive to the propagator that is used in our computation, we take Equation
\eqref{eq:jjxpt_gen}, apply a spatial integral over \(x\) and assume that
\(\mu,\bar{\mu}\) are spatial. With all the terms enumerated above we end up with:
\begin{align}
    \label{eq:jjxpt_raw}
    \begin{aligned}
	&\frac{1}{e^2}\int d^3x \Braket{J_{\mu}(\vec x,t) J_{\bar{\mu}}(0)}= \\
	&\qquad= \sum_\alpha \int \frac{d^3p}{{(2\pi)}^3} 2 N Q_\mathrm{ns}^2
	\left[1 + \frac{16}{F^2} L_9 E_{p,\alpha}^2 - \frac{N}{4F^2} \sum_\beta \mathcal{G}_{0,\beta}
	+ \delta M^2_\alpha \frac{d}{dM_\alpha^2} \right]
	\frac{e^{-2E_{p,\alpha}t}}{E_{p,\alpha}^2} p_\mu p_{\bar\mu} + \\
	&\qquad+ \sum_{\alpha,\beta} \int \frac{d^3p}{{(2\pi)}^3} \frac{d^3r}{{(2\pi)}^3}
        \frac{N^2Q_\mathrm{ns}^2}{8F^2}
	\frac{e^{-2E_{r,\beta}t}E_{p,\alpha} - e^{-2E_{p,\alpha}t}E_{r,\beta}}{E_{p,\alpha}^2E_{r,\beta}^2 (E_{p,\alpha}^2-E_{r,\beta}^2)}
	p_\mu r_{\bar{\mu}} (\vec{p}\cdot \vec{r})
	+ \text{contact terms},
    \end{aligned}
\end{align}
where we define the relativistic energy of a free particle with $\alpha$ taste
as \(E_{p,\alpha} = \sqrt{M_\alpha^2 + \vec{p}^2}\). The  summations mean a sum
over sixteen tastes of the flavor non-singlet pions, $\alpha\in
\{5,\mu5,\mu\nu,\mu,I\}$. The flavor-singlet pseudoscalars only contribute to
the \(\delta M^2_\alpha\) mass-shift terms. These can be transformed away by
switching to the \NLO{} mass everywhere in the formula, ie. applying the shift
\(M_\alpha^2 \to M_\alpha^2 + \delta M_\alpha^2\). This changes the result by
effects that are higher order than the \NNLO{} considered here. The non-singlet
charge squared is defined as \(Q^2_\mathrm{ns} = Q_a Q_a - Q_0 Q_0\). Since the
result is proportional to $Q^2_\mathrm{ns}$, the flavor-singlet part of the
current gives no contribution at this order.

We use dimensional regularization at scale $\mu$ and the
$\overline{\mathrm{MS}}$ scheme to work with the ultraviolet divergent loop
integrals.  Specifically, the one-loop integral $\mathcal{G}_{0,\beta}$ is
given by:
\begin{align}
    \mathcal{G}_{0,\beta} = \int \frac{d^3p}{{(2\pi)}^3}\ \frac{1}{2E_{p,\beta}}=
    \frac{M_\beta^2}{16\pi^2}\left( \log\frac{M_\beta^2}{\mu^2} + R \right) 
\end{align}
where $R$ contains the divergence isolated by the $\overline{\mathrm{MS}}$
prescription, see eg. \cite{Scherer:2002tk}.  Seemingly, there is also a
singularity in the integrand of the double loop integral in the second line of
Equation \eqref{eq:jjxpt_raw}, when $\vec{p}=\vec{r}$. This singularity is
superficial, since there is a zero in the numerator which cancels it. It is
useful to work with the terms of the numerator separately, in which case the
separated terms are singular. These singularities have to be regulated. We do
this by adding an $+i\eta$ into the denominator. The full expression has to be
smooth as $\eta\to0$, and we take this limit at the end of the computation.

The only low-energy constant, that appears in the result, is $L_9$ from the
continuum Gasser-Leutwyler Lagrangian $\mathcal{L}_{4,GL}$. The ultraviolet
divergences coming from the single loop integral $\mathcal{G}_{0,\beta}$ and the double
loop integral on the second line of Equation~\eqref{eq:jjxpt_raw} can all be
absorbed into $L_9$. This procedure defines a renormalized $L_9^r$, and also a
scheme-independent $\overline{L}_9$ in the standard way \cite{Scherer:2002tk}.

The only contact term that affects our observables is the one proportional to
$C_{93}$; we will ignore this term here, since it has no effect on the finite
volume and taste-splitting corrections.

We now move to our specific case of two degenerate light quarks with rooting,
so we set \(N_f = 2\), \(N_r = \frac{1}{4}\), \(L_9 =-\frac{1}{2} l_6\) and
\(Q = (\frac{1}{6} + \frac{1}{2} \sigma_3) \otimes 1\) in flavor-replica
space, which give \(Q^2_\mathrm{ns} = N_r/4\). One of the two momentum
integrals in Equation \eqref{eq:jjxpt_raw} can be performed analytically, and
we also average over the Lorentz indices to arrive to our final expression:
\begin{align}
    G(t)\equiv \frac{1}{3e^2}\sum_{\mu=1,2,3} \int d^3x \Braket{J_{\mu}(\vec x, t) J_{\mu}(0)} = \frac{1}{3}\sum'_\alpha \int \frac{d^3p}{{(2\pi)}^3} \frac{e^{-2E_{p,\alpha}t}}{E_{p,\alpha}^2} p^2
    \left[1 + \frac{1}{F^2}\sum'_\beta \Gamma(p^2;M_\alpha,M_\beta)
	\right],
    \label{eq:jjxpt_ren}
\end{align}
where $p^2=\vec{p}^{\ 2}$ and the summation symbol
$\sum'_{\alpha}=\tfrac{1}{16}\sum_\alpha$ stands for averaging over the taste
index $\alpha$.  The first part in the square-bracket is the well-known \NLO{}
expression of \cite{Aubin:2006xv}, ie.\ the taste average of the continuum
\NLO{} result. The second part of the square-bracket contains the \NNLO{}
correction and is given explicitly as
\begin{gather}
	\nonumber
	\Gamma(p^2;M_\alpha,M_\beta)=
	\frac{p^2+M_\alpha^2}{12\pi^2}\left( \overline{l}_6 - \log\frac{M_\beta^2}{M_5^2} \right)
	+\frac{5(p^2+M^2_\alpha)}{36\pi^2}
	- \frac{M_\beta^2}{6\pi^2}+
	\\
	- \frac{p^2+M^2_\alpha-M^2_\beta}{6\pi^2}
	\begin{cases}
	    \sqrt{\frac{1-x}{x}}\arcsin\sqrt{x}                & \text{if}\quad x<1\\
	    \sqrt{\frac{x-1}{x}}\log ( \sqrt{x} + \sqrt{x-1} ) & \text{if}\quad x>1
	\end{cases}
    \quad\text{with}\quad
    x=\frac{p^2+M_\alpha^2}{M_\beta^2}.
    \label{eq:jjxpt_gam}
\end{gather}
The scheme independent $\overline{l}_6$ is introduced with the Goldstone pion
mass $M_5$ as
\begin{gather}
  \label{eq:l6}
    l_6= -\frac{1}{96\pi^2}\left( \overline{l}_6 + \log \frac{M_5^2}{\mu^2} + R \right).
\end{gather}
Equations \eqref{eq:jjxpt_ren} and \eqref{eq:jjxpt_gam} reproduce the continuum
\NNLO{} result of \cite{Aubin:2019usy}, if we set the masses of all tastes to
$M_5$. 

When we use the above result later in this paper, we take the flavor
non-singlet pion masses from the simulations and fix the remaining two
parameters as: $F=92.21$~MeV and $\overline{l}_6=16$ as in
\cite{Aubin:2019usy}.

\subsection*{Finite-$L$ effects}

The above computation can also be carried out in finite volume. In the
continuum case, this was already done in \cite{Aubin:2019usy}. Here we
generalize those formulas in the presence of taste violations. We use the same
techniques that were applied there, and we also correct that
computation. A detailed derivation will not be given here, we just briefly
describe the main strategy and give the results.

The current propagator $G(t)$ in finite volume can be obtained from the
one in infinite volume by replacing the momentum integrals with sums. In the
infinite-volume expression, Equation \eqref{eq:jjxpt_raw}, there are terms with
single and double integrals. The finite-size correction for those with a
single integral can be obtained from a Poisson summation formula, formally:
\begin{gather}
    \frac{1}{L^3}\sum_p - \int \frac{d^3p}{(2\pi)^3}=
    \int \frac{d^3p}{(2\pi)^3} \sum_{n\neq 0} e^{i\vec n\vec pL}\equiv
    \sumint_{p,n\neq0},
\end{gather}
whereas for double integrals the correction involves more terms:
\begin{gather}
    \frac{1}{L^3}\sum_p \frac{1}{L^3}\sum_r- \int\frac{d^3p}{(2\pi)^3}\int\frac{d^3r}{(2\pi)^3}=
    \sumint_{p,n\neq0} \sumint_{r,m\neq0}\ +\ \sumint_{p,n\neq0} \int\limits_r\ +\ \int\limits_p^{} \sumint_{r,m\neq0}\ ,
\end{gather}
where we also introduced the notation $\int_p \equiv \int \tfrac{d^3
p}{(2\pi)^3}$.  In some cases the momentum integrals can be more easily
performed if the line of integration is deformed into the complex plane.
Specifically we need the following integrals:
\begin{equation}
    \label{eq:fvint}
    \begin{aligned}
	I_1[f]&\equiv\sumint_{p,n\neq0} f(p^2)=
	\frac{1}{2\pi^2} \sum_{n^2=1}^\infty \frac{\nu_n}{nL} \int_0^\infty dp\ p\sin(npL)  f(p^2)\\
	I_2(M_\alpha)&\equiv -\sumint_{p,n\neq0} \frac{1}{E_{p,\alpha}}=
	-\frac{M_\alpha}{2\pi^2} \sum_{n^2=1}^\infty \frac{\nu_n}{nL} \int_1^\infty \frac{e^{-ynM_\alpha L} y dy}{\sqrt{y^2-1}}\\
	I_3(p^2;M_\alpha,M_\beta)&\equiv\frac{1}{6}
	\sum_{s=\pm}\sumint_{r,m\neq0} \frac{r^2}{E_{r,\beta}(E_{r,\beta}^2-E_{p,\alpha}^2 + is\eta)}=
	\frac{M_\beta}{6\pi^2} \sum_{m^2=1}^\infty \frac{\nu_m}{mL}
	\left[\int_1^\infty \frac{e^{-ymM_\beta L} y^3 dy}{\sqrt{y^2-1}(y^2+x-1)} + \right.\\
	&+\left.\frac{\pi}{2}
	\frac{x-1}{\sqrt{x}}
	\begin{cases}
	    \exp(-m\sqrt{1-x}M_\beta L)  & \text{if}\quad x<1\\
	    \cos{(m\sqrt{x-1}M_\beta L)} & \text{if}\quad x>1
	\end{cases}
	\right]
	\quad\text{with}\quad
	x=\frac{p^2+M_\alpha^2}{M_\beta^2}.
    \end{aligned}
\end{equation}
Here $f(p^2)$ is an arbitrary integrable function, $\nu_\xi=\sum_{\vec n^2=\xi}
1$ and the $\eta>0$ regulator was introduced according to our earlier
discussion.  In the $I_3$ integral, the second term in the square-bracket comes
from poles at $r^2=(p^2+M_\alpha^2-M_\beta^2)/M_\beta^2$.  This pole term was
dropped in \cite{Aubin:2019usy} by saying that the pole can be shifted outside
a complex contour.  We give the result as the finite- minus infinite-volume
difference and we split it into five terms, one \NLO{} and four \NNLO{}:
\begin{gather}
    G(t;L)-G(t;\infty)= \Delta G(t)_\mathrm{NLO} + \sum_{i=1}^4 \Delta G(t)_{\mathrm{NNLO},i}.
\end{gather}
To keep the formulas simple we
introduce the notation
\begin{gather}
    \label{eq:epsxpt}
    \epsilon(p^2;M_\alpha,t) \equiv
    \frac{e^{-2E_{p,\alpha}t}}{E_{p,\alpha}^2}\ p^2\ .
\end{gather}
Our result for the finite-volume correction is then:
\begin{align}
    \label{eq:fvxpt}
    \begin{aligned}
	\Delta G(t)_\mathrm{NLO}&= \frac{1}{3}\sum'_\alpha I_1[ \epsilon(M_\alpha,t) ]\\
	F^2\Delta G(t)_\mathrm{NNLO,1}&= \frac{1}{3} \sum'_{\alpha,\beta}
	I_1[\epsilon(M_\alpha,t) \Gamma(M_\alpha,M_\beta)]\\
	F^2\Delta G(t)_\mathrm{NNLO,2}&= \frac{1}{3} \sum'_{\alpha,\beta}
	I_1[\epsilon(M_\alpha,t)]\cdot I_2(M_\beta)\\
	F^2\Delta G(t)_\mathrm{NNLO,3}&= \frac{1}{3} \sum'_{\alpha,\beta}
	\int_p\epsilon(p^2;M_\alpha,t) \left[ I_2(M_\beta) + I_3(p^2; M_\alpha,M_\beta) \right]\\
	F^2\Delta G(t)_\mathrm{NNLO,4}&= \frac{1}{3} \sum'_{\alpha,\beta}
	I_1[\epsilon(M_\alpha,t) I_3(M_\alpha,M_\beta)].
    \end{aligned}
\end{align}
The $\tfrac{1}{3}$ prefactors arise from the Lorentz-index averaging and
$\sum'_{\alpha,\beta}=\tfrac{1}{256}\sum_{\alpha,\beta}$.  In the continuum
these formulas agree with the ones in \cite{Aubin:2019usy}, up to the pole
contribution that affects the $\Delta G(t)_\mathrm{NNLO,3}$ and $\Delta
G(t)_\mathrm{NNLO,4}$ terms. The largest \NNLO{} term by far is $\Delta
G(t)_\mathrm{NNLO,1}$, and has the same order of magnitude as the \NLO{} term
on our lattices.

\subsection*{Finite-$T$ effects}

Until now we assumed that the temporal extent of the lattice is infinite,
$T=\infty$.  The corrections introduced by a finite $T$ can also be computed in
\SXPT{}. The integrals over the time component of the momenta become
sums, which are then related to integrals via Poisson's summation
formula. Formally:
\begin{gather}
    \int \frac{dp_4}{2\pi} \longrightarrow \frac{1}{T} \sum_{p_4} = \int \frac{dp_4}{2\pi} \sum_{n_4} e^{ip_4 n_4 T}\ .
\end{gather}
We give here the result of this procedure for three integrals that appear in
the current-current correlator, in a one-loop and in a two-loop diagram, respectively:
\begin{gather}
    \label{eq:fintxpt}
\begin{aligned}
    &\int \frac{dp_4}{2\pi} \frac{e^{ip_4t}}{p_4^2+E_p^2}
    \longrightarrow \frac{\cosh\left[ E_p\left(t-T/2\right)\right]}{2E_p\sinh(E_pT/2)}\\
    &\int
    \frac{dp_4}{2\pi}
    \frac{dq_4}{2\pi}
    \frac{e^{ip_4t}}{p_4^2+E^2}
    \frac{e^{iq_4t}}{q_4^2+E^2}
    \longrightarrow \left( \frac{\cosh\left[ E_p\left(t-T/2\right)\right]}{2E_p\sinh(E_pT/2)} \right)^2=
    \frac{\cosh\left[ 2E_p\left(t-T/2\right)\right] +1 }{4E_p^2\left[\cosh(E_pT)-1\right]}\\
    &\int dt'
    \int
    \frac{dp_4}{2\pi} 
    \frac{dq_4}{2\pi} 
    \frac{dr_4}{2\pi} 
    \frac{ds_4}{2\pi} 
    \frac{e^{ip_4(t-t')}}{p_4^2+E_p^2}
    \frac{e^{iq_4(t-t')}}{q_4^2+E_p^2}
    \frac{e^{ir_4t'    }}{r_4^2+E_r^2}
    \frac{e^{is_4t'    }}{s_4^2+E_r^2}
    \longrightarrow
\end{aligned}\\
    \nonumber
    \frac{
	E_rE_p^2\cosh\left[2E_r(t-T/2)\right]\sinh(E_pT) - \{ p \leftrightarrow r \} +
	(E_p^2-E_r^2)\left[E_r\sinh(E_pT) + E_p\sinh(E_rT) + E_pE_rT\right]
    }
    { 16E_p^3E_r^3(E_p^2-E_r^2) \left[ \cosh(E_pT)-1 \right] \left[ \cosh(E_rT)-1 \right]}
\end{gather}
The computation proceeds as in the case of finite-$L$. As in
Equation \eqref{eq:epsxpt}, we introduce new
notations to keep the formulas simple:
\begin{gather}
    \label{eq:epsxpt2}
    \epsilon_{T}(p^2;M_\alpha,t)\equiv \frac{\cosh\left[2E_{p,\alpha}(t-T/2)\right]+1}{E_{p,\alpha}^2\left[ \cosh(E_{p,\alpha}T)-1\right]}\ p^2\ ,\qquad
    \sigma_{T}(p^2;M_\alpha)\equiv \frac{p^2}{E_{p,\alpha}^2\left[ \cosh(E_{p,\alpha}T)-1\right]}\ .
\end{gather}
For the correction due to finite $L$ and $T$ we get:
\begin{gather}
    G(t;L,T) - G(t;\infty,\infty)=\\
    \nonumber
    \\
    \nonumber
	\begin{aligned}
	    &\left[ \frac{1}{3}\sum'_\alpha \int_p \epsilon(p^2;M_\alpha,t) \right]_{\epsilon \to \epsilon_T-\epsilon} +
	    \left[ \frac{1}{3}\sum'_\alpha \int_p \epsilon(p^2;M_\alpha,t) \frac{1}{F^2} \sum'_{\beta} \Gamma(p^2;M_\alpha,M_\beta)\right]_{\epsilon \to \epsilon_T-\sigma_T-\epsilon}\\
	    &+\Bigg[ \Delta G(t)_\mathrm{NLO} \Bigg]_{\epsilon\to \epsilon_T}+ \left[\sum_{i=1}^4 \Delta G(t)_{\mathrm{NNLO},i}\right]_{\epsilon \to \epsilon_T-\sigma_T}\\
	    &+\frac{T}{18F^2} \left[ \sum'_\alpha \sumint_{p,n} \sigma_T(p^2;M_\alpha) \right]^2\\
	    &+\frac{1}{9F^2} \left[\sum'_\alpha \sumint_{p,n} \sigma_T(p^2;M_\alpha)\right]\left[ \sum'_\alpha \sumint_{p,n} \sigma_T(p^2;M_\alpha) \frac{1-e^{-E_{p,\alpha}T}}{E_{p,\alpha}}\right]\\
	    &-\frac{2}{3F^2} \left[ \sum'_\alpha \sumint_{p,n} \epsilon_T(p^2;M_\alpha,t) \right] \left[ \sum'_\alpha \sumint_{p,n} \frac{1}{E_{p,\alpha}\left( e^{E_{p,\alpha}T}-1\right)} \right]\\
	    &+\frac{1}{18F^2} \sum'_{\alpha} \sumint_{p,n} \left[ \epsilon_{T}(p^2;M_\alpha,t) - \sigma_T(p^2,M_\alpha)\right]
	    \sum'_{\beta}\sum_{s=\pm}\sumint_{r,m} \frac{r^2}{E_{r,\beta}(E^2_{r,\beta}-E^2_{p,\alpha}+is\eta)}\frac{1-e^{-E_{r,\beta}T}}{\cosh(E_{r,\beta}T)-1}
	\end{aligned}
\end{gather}
The first line is the correction due to finite $T$ at infinite $L$, ie.
$G(t;\infty,T)-G(t;\infty,\infty)$. The rest of the formula gives the
correction due to finite $L$ at finite $T$, ie. $G(t;L,T)-G(t;\infty,T)$. The
terms in the second line are obtained by replacing $\epsilon$ by $\epsilon_T$
or by $\epsilon_T-\sigma_T$ in Equations \eqref{eq:fvxpt} from the previous
subsection.  The first and second lines give most of the contribution to the
finite-size effect for the cases of interest in this work.  The rest, from the
third to the sixth line, are genuine finite-$T$ corrections in that they vanish
at infinite $T$. They are all NNLO and are negligible for the lattices
considered here.

\subsection*{Taste non-singlet contributions to the current}

Here we explore briefly to what extent our previous assumption, that the vector
current is a taste singlet, is justified. For this purpose it is useful to work
with the valence staggered fermion fields $\chi$. The lattice current propagator,
that we introduced in Section \ref{se:obs_jj}, can be derived from the
following zero-spatial-momentum operator:
\begin{gather}
    \label{eq:stjj}
    \sum_{\vec x}\left( \overline{\chi}_{x} U_{\mu,x} \eta_{\mu,x}\chi_{x+\mu} +
    \overline{\chi}_{x+\mu} U^\dagger_{\mu,x} \eta_{\mu,x}\chi_{x}\right)\ ,
\end{gather}
where the $\eta_{\mu,x}$ phase is the staggered representation of the
Dirac matrix and we drop the flavor index for simplicity. Equation
\eqref{eq:stjj} is just the conserved vector current of the staggered fermion
action.

Staggered bilinears can be assigned with a spin$\otimes$taste structure, see
\cite{Follana:2006rc} for a modern treatment. In the case of the operator in
\eqref{eq:stjj}, two such assignments can be given:
\begin{gather}
    \label{eq:spta}
    \gamma_\mu \otimes 1
    \qquad\text{and}\qquad
    \gamma_\mu \gamma_4 \gamma_5\otimes \xi_4 \xi_5\ .
\end{gather}
The conserved current couples to states of both types. The first corresponds to
the taste-singlet vector current, the case which we have fully covered
previously.  The second is a taste non-singlet
pseudovector\footnote{Pseudovectors are to be distinguished from
axial-vectors, the latter have $\gamma_{\mu}\gamma_5$ spin structure.}. Its
correlator gives the characteristic contribution to the staggered propagator
which oscillates in time with a factor $(-1)^{x_4}$.  The observables that we
consider in this paper are obtained by integrating the propagator over the
whole time range, or at least over some physical distance.  In the continuum
limit such an integration completely eliminates the oscillating contribution.
At finite lattice spacing it gives an $O(a^2)$ suppression compared to the
taste singlet vector contribution.

It is possible to compute the pseudovector propagator in chiral perturbation
theory. This requires the inclusion of an external antisymmetric tensor source
$t_{\mu\nu}$. This has been done to \NNLO{} in \cite{Cata:2007ns}, which has
chosen $t_{\mu\nu}\approx O(p^2)$ for the chiral counting of the tensor field.
With this choice a vertex with the tensor field first appears in the \NLO{}
Lagrangian. The one-loop contribution to the tensor-tensor propagator is zero,
due to charge conjugation and parity invariance. The two-loop contribution has
two \NLO{} vertices and is thus $O(p^{10})$. This, combined with the
suppression explained in the previous paragraph, gives $O(a^2p^{10})$, which is
well beyond the order to which we work.

As explained in \cite{Cata:2007ns}, there is an ambiguity in the
chiral-counting assignment of $t_{\mu\nu}$. A closely related fact is that the
tensor-tensor propagator is renormalization scheme dependent, since there is no
conservation law for the tensor current. Even if we used a counting
$t_{\mu\nu}\approx O(1)$, the oscillating contribution would still be beyond
the \NNLO{} to which we work.

Based on these arguments, we ignore the taste non-singlet contribution to the
current in our \SXPT{} computation.

    \section{Meyer-Lellouch-Lüscher-Gounaris-Sakurai model}
\label{se:obs_llgs}

In this section we describe a phenomenological model that we use to make
predictions for finite-volume corrections in Section \ref{se:obs_fv}. We also
use it to correct for taste-breaking effects in the $I=1$ contribution to the
Euclidean, current-current correlation function, in Section \ref{se:obs_taimp}.
We work in the isospin-symmetric limit throughout this section.

\subsection*{Model for the pion form factor}

As shown in \cite{Bernecker:2011gh}, in infinite volume the Euclidean,
current-current correlation function is a Laplace transform of the
corresponding spectral function, $\rho(E)$:
\begin{equation}
    \label{eq:Ctspecdec}
    G(t)= \int_0^\infty dE E^2 \ e^{-E|t|}\rho(E)\ ,
\end{equation}
for $t\ne 0$. Here, $G(t)$ is defined in Equation~\labelmaingdef{} of the main
paper and $E$ is the center-of-mass energy.

For large $|t|$, $G(t)$ is dominated by the low end of the spectrum, which is
governed by two-pion, scattering states. In addition, phenomenology indicates
that two-pion states, up to
$E=1.8$~GeV~\cite{Davier:2019can,Keshavarzi:2018mgv}, are responsible for over
70\% of the total $a_\mu$ and over 85\% of the $I=1$ contribution computed in
this work. Thus, the contribution of two-pion states should not only provide a
good description of the long-distance behavior of the current-current
correlator, important for understanding finite-volume effects, but also a
reasonable model for this correlator at all distances relevant for the
determination of $a_\mu$. Now, the two-pion contribution to the spectral
function is given by (see eg. \cite{Jegerlehner:2009ry})
\begin{equation}
    \label{eq:2pispec}
    \rho(E)\vert_{\pi\pi}=
    \frac{1}{6\pi^2}\left(\frac{k}{E}\right)^3|F_\pi(k)|^2
    \ ,
\end{equation}
where $E=2\sqrt{M_\pi^2+k^2}$, with $k$ the magnitude of the pions'
back-to-back momenta in the center-of-mass frame, and $F_\pi(k)$ is
the timelike, pion, electromagnetic form factor.

A good phenomenological description of the pion form factor and the
corresponding $\pi$-$\pi$ scattering phase shift is given by the
Gounaris-Sakurai (GS) parametrization \cite{Gounaris:1968mw}. In the context of
estimating finite-volume effects in $g_\mu-2$, it was used first in
\cite{Francis:2013qna}.  The GS parametrization describes well the experimental
spectral function in the $I=1$ channel from threshold to $E$ around $1$~GeV,
thus covering the very important $\rho$-resonance contribution.  This
parametrization is given by:
\begin{equation}
    \label{eq:GSFpi}
    F_\pi(k) = \frac{M^2(0)}{M^2(E)-E^2-iM_\rho\Gamma(E)}
\ ,
\end{equation}
with the energy-dependent width
\begin{equation}
    \Gamma(E) = \Gamma_\rho\left(\frac{k}{k_\rho}\right)^3\left(\frac{M_\rho}{E}\right)
    \ ,
\end{equation}
where $k_\rho=\sqrt{\frac{M_\rho^2}4-M_\pi^2}$, and the
energy-dependent mass squared:
\begin{equation}
    M^2(E) = M_\rho^2 + \frac{\Gamma_\rho M^2_\rho}{k_\rho^3}\left[k^2\left[h(E)-h(M_\rho)\right]
    -\left[E^2-M_\rho^2\right]\frac{k_\rho^2}{2M_\rho} h'(M_\rho)\right]
\ .
\end{equation}
Here, the pion-loop function is
\begin{equation}
    h(E)=\frac{2k}{\pi E}\log\frac{E+2k}{2M_\pi}
\,
\end{equation}
and $h'(E)$ is its derivative with respect to $E$. To complete
the model, we determine the width of the $\rho$ in terms of the
$\rho$-$\pi\pi$ coupling, $g$, to leading order in an effective theory
where the $\rho$ and $\pi$ are pointlike particles. It is
straightforward to show that
\begin{equation}
    \label{eq:gammarho}
    \Gamma_\rho = \frac{g^2}{6\pi}\frac{k_\rho^3}{M_\rho^2}
\ .
\end{equation}
From these
expressions for $F_\pi(k)$, the phase-shift is simply obtained using
Watson's theorem:
\begin{equation}
    \label{eq:delta11GS}
    \delta_{11}(k) = \arg F_\pi(k)
    \ .
\end{equation}
$G^\mathrm{GS}(t)$ will denote the infinite-volume correlator given by Equations
\eqref{eq:Ctspecdec}, \eqref{eq:2pispec}, \eqref{eq:GSFpi} and
\eqref{eq:gammarho}.

The GS model has two free parameters: $g$ and $M_\rho$. Since our simulations
are performed very near the physical mass point, we fix these parameters to
their physical value, neglecting their sub-percent uncertainties:
$M_\rho=775$~MeV is the mass of the $\rho_0$ meson from
\cite{Tanabashi:2018oca}, and $g=5.95$ is obtained from Equation
\eqref{eq:gammarho}, using the width of $\rho_0$ and the mass of $\pi_\pm$,
also from \cite{Tanabashi:2018oca}.

\subsection*{Model for finite-volume effects}

In a finite spatial volume of size $L\times L\times L$, the two-pion
spectrum is discrete because of momentum quantization, and the
spectral representation of the current-current correlator becomes a
sum, instead of an integral, over two-pion states. Thus, the large-t
behavior of the corresponding, finite-volume correlation function,
$G(t;L)$, can be written as:
\begin{equation}
    \label{eq:CtspecdecL}
    G(t;L)\overset{|t|\to\infty}{\longrightarrow}\frac{1}{3}\sum_{n>0} |\vec{A}_n|^2 e^{-E_n|t|}\ ,
\end{equation}
where $n$ labels the energy eigenstates, in order of
increasing energy. Below the four-pion, inelastic threshold, the
energy of state $n$ is given by $E_n=2\sqrt{M_\pi^2+k_n^2}$, with
$k_n$ determined by the infinite-volume, $I^G(J^{PC})=1^+(1^{--})$,
$\pi$-$\pi$ scattering phase shift, $\delta_{11}(k)$, through
Lüscher's formula \cite{Luscher:1991cf,Luscher:1990ux}:
\begin{equation}
    \label{eq:lusch}
    \phi(q_n)+\delta_{11}(k_n) = n\pi, \qquad n=1,\,2,\,\ldots\ ,
\end{equation}
where $q=kL/2\pi$ and $\phi(q)$ is given in \cite{Luscher:1991cf} (see also
\cite{Lellouch:2011qw}).  The amplitude of the $n$-th state $\vec{A}_n$ is
proportional to $\langle 0\vert \vec{J}_{I=1}(0) \vert n\rangle$, where
$\vec{J}_{I=1}$ collects the spatial components of the isospin $I=1$
contribution to the quark electromagnetic current defined after Equation~\labelmaingdef{}
of the main paper. This amplitude is determined by the
phase-shift and by the timelike, pion, electromagnetic form factor, through a
Lellouch-Lüscher (LL) equation \cite{Lellouch:2000pv,Lin:2001ek,Meyer:2011um}
\begin{equation}
    \label{eq:lellusch}
    |\vec{A}_{n}|^2 =
    \left[q\phi'(q)+k\delta_{11}'(k)\right]^{-1}_{k=k_n}\frac{2k_n^5}{\pi
      E_n^2}|F_\pi(k_n)|^2 \ ,
\end{equation}
where the primes indicate a derivative of the function with respect to its
argument. We assume that Equations \eqref{eq:lusch} and \eqref{eq:lellusch} are
also approximately true above the inelastic threshold, because the $\rho$
decays almost exclusively into two pions \cite{Tanabashi:2018oca}.  Equations
\eqref{eq:delta11GS} and \eqref{eq:CtspecdecL} then define the
Meyer-Lellouch-Lüscher-Gounaris-Sakurai (MLLGS) model for the finite-volume current
correlator. It will be denoted $G^\mathrm{MLLGS}(t;L)$.

It is important to note that the GS parametrization for $\rho(E)|_{\pi\pi}$,
obtained from Equations \eqref{eq:GSFpi} and \eqref{eq:gammarho}, decreases as
$(E\log{E})^{-2}$ for large $E$ and becomes smaller than that of free pions
after the $\rho$ peak, when $E\gtrsim 1.1$~GeV. Thus, in the sums over two-pion
states in finite-volume, one can reasonably neglect terms for which $E_n$ is
greater than $1.2$~GeV. In the reference volume with
$L=L_\mathrm{ref}=6.272$~fm, used in Section \ref{se:obs_fv}, this corresponds
to $n=8$ for the Goldstone. 

In Section \ref{se:obs_fv}, we use this MLLGS model to compute the finite-volume
correction to the $I=1$ contribution to $a_\mu$. It is determined in the
continuum limit for the reference volume $L_\mathrm{ref}$ and is given by
integrating the difference of the infinite and finite-volume correlators:
\begin{equation}
  \label{eq:LLGSFV}
    a_{\mu}^\mathrm{GS}(L=\infty)-a_{\mu}^\mathrm{MLLGS}(L_\mathrm{ref})=
    10^{10}\alpha^2
    \int_0^\infty dt\, K(t)
    \left[G^\mathrm{GS}(t)-G^\mathrm{MLLGS}(t;L_\mathrm{ref})\right]
\ .
\end{equation}
In Section \ref{se:obs_fv}, this difference is compared to the results of a
dedicated lattice study of finite-volume effects in $a_\mu$.  The good
agreement represents a strong validation of the model.

\subsection*{Model for taste violations}

Here we generalize the MLLGS model, for the finite-volume correlation function,
to include the lattice spacing effects arising from taste breaking. Indeed, the
dominant, taste-breaking effects in $a_\mu$ are expected to be those associated
with the two-pion spectrum: these states give the dominant contribution to
$a_\mu$ and the masses of pions are significantly affected by taste breaking on
coarser lattices. Our conserved, quark electromagnetic current couples, not
only to two-Goldstone-pion states, but also to fifteen additional pairs of more
massive taste partners of the pion. We label the masses of these states with
$M_{\tau}$, where the $\tau$ index runs over the 16 element set
$\tau\in\{5,\mu5,\mu\nu,\mu,I\}$ with $\mu<\nu$.

In a description where the pions are free, corresponding to NLO staggered XPT,
the two-pion states have energies, $E_{n,\tau}^{(0)}=2\sqrt{M_{\tau}^2+k_n^2}$,
with $k_n=|\vec{n}|(2\pi/L)$, $n=\vec{n}^2$ and $\vec{n}\in\mathbb{Z}^3$. In
the interacting case, we make the assumption that two-pion-state energies have
the same set of taste copies, but with the momentum, $k_{n,\tau}$, given by the
L\"uscher quantization condition of Equation \eqref{eq:lusch}. We further
assume that a similar conclusion holds for the amplitudes, i.e.\ that they
satisfy Equation \eqref{eq:lellusch} with $k_{n,\tau}$ given by Equation
\eqref{eq:lusch}. Thus we model the long-distance behavior of the correlator in
a finite spatial volume and at finite lattice spacing as:
\begin{equation}
    \label{eq:CtspecdecSL}
    G(t;L,a)\overset{|t|\to\infty}{\longrightarrow}
    \frac{1}{3}\sum_{n>0}
    \frac{1}{16}\sum_{\tau}|\vec{A}_{n,\tau}|^2 e^{-E_{n,\tau}|t|}\ ,
\end{equation}
which we denote $G^\mathrm{SMLLGS}(t;L,a)$. In the formula, the lattice spacing
dependence arises from the splittings in the pion spectrum. In implementing
this model, we assume that all taste-partner pairs of pions couple to the
physical, lightest, $\rho$ taste, with the same coupling $g$. Thus, we
implicitly assume that the dependence of $\Gamma_\rho$ and of $\delta_{11}(k)$
on $M_\tau$ is mostly kinematic, an assumption which is borne out by
simulations \cite{Metivet:2014bga}. All of these choices guarantee that our
model has the correct continuum limit. Also, we keep states up to $n=8$, even
for the more massive taste partners. However, we only apply the model to
simulations in which the physical $\rho$ has sufficient phase space to decay
into all sixteen, taste-partner pion pairs. This excludes only our coarsest
simulation, with $\beta=3.7000$.

We use the SMLLGS model to correct taste-breaking effects in the $I=1$
contribution to $a_\mu$, simulation by simulation. This significantly reduces
the $a$-dependence of the dominant contribution to $a_\mu$, allowing for a more
precise determination of its continuum limit.  As shown in Section
\ref{se:obs_taimp}, the optimal way in which to apply taste-breaking corrections
is to consider these corrections in different time-windows. Thus, for each
simulation we compute,
\begin{equation}
  \label{eq:SLLGSFV}
    a_{\mu,\mathrm{win}}^\mathrm{MLLGS}(L_\mathrm{ref})-a_{\mu,\mathrm{win}}^\mathrm{SMLLGS}(L,a)
    =10^{10}\alpha^2
    \int_0^\infty dt\, K(t)\, W(t;t_1,t_2)
    \left[G^\mathrm{MLLGS}(t;L_\mathrm{ref})-G^\mathrm{SMLLGS}(t;L,a)\right]\ ,
\end{equation}
where the window function, $W(t;t_1,t_2)$, is defined in Equation
\eqref{eq:win} and $L$ is the size of the lattice for that simulation. The
lattice spacing dependence enters in SMLLGS from the taste breaking in the pion
masses $M_{\tau}$ and these are also taken from the simulation. This correction
is applied as an additive shift on the measured $a_\mu$.

\subsection*{Inclusion of finite-$T$ effects}

In this section, we generalize the MLLGS and SMLLGS models to include the effects
of the finite-time extent, $T$, of the lattice. In our simulations, gauge-boson
fields obey periodic boundary conditions in time and quark fields, antiperiodic
ones. Finite-$T$ effects are expected to be smaller than those due to finite
$L$, because $T\ge L$ in our simulations and because they occur only in a
single spacetime direction.

The MLLGS model describes the contributions to the current-current correlator of
two-pion states, in the presence of interactions. Thus, in studying finite-$T$
effects, we consider terms in which each current couples only to two
pions\footnote{There are also terms in which, four, six, eight, etc. pions
couple to the currents. Compared to the two pion terms, these are exponentially
suppressed in time but also suppressed by the small coupling of four or more
pions to the $\rho$, which dominates the energy range that we are modeling
here.}. The three largest contributions are:
\begin{enumerate}

    \item The Euclidean propagation of interacting $\pi^+\pi^-$ states between
	times $0$ and $t>0$, with energies $E_n$. These contributions fall-off
	exponentially as $e^{-E_nt}$.
	
    \item The Euclidean propagation of interacting $\pi^-\pi^+$ states from $t$
	to $T$, which increases as $e^{-E_n(T-t)}$ as $t$ approaches $T/2$.
	
    \item The contribution of a single pion that wraps around the time
	direction. This is at the same order in decreasing exponentials of $T$
	as the previous contribution. At any given time, there is only a single
	pion on the lattice. Therefore, the energy of this state is that of a
	free pion with at least one unit of momentum, i.e.
	$E_n^{(0)}/2=\sqrt{n(2\pi/L)^2+M_\pi^2}$, with $n\ge 1$ the
	norm-squared of a three vector in $\mathbb{Z}^3$. This contribution is
	constant in $t$ and proportional to $e^{-E_n^{(0)}T/2}$. Moreover,
	since the pion couples to the current without recoiling, the matrix
	element describing this coupling is also that of the free pion theory.
	This contribution appears with a factor of $2$, because it can be
	caused by the propagation of either a $\pi^+$ or a $\pi^-$.
	Altogether, this contribution has the same form as it does at NLO in
	XPT.

\end{enumerate}
The next order in decreasing exponentials of $T$ is a term of three pions
propagating from $0$ to $t$ and a single pion from $t$ to $T$. We have
estimated this contribution and find that it contributes to $a_\mu$ at a level
that is orders of magnitude smaller than our statistical error. Thus, we
neglect it, as well as all higher-order winding terms and, to describe
finite-$T$ effects, we replace $G^\mathrm{MLLGS}(t;L)$ by
\begin{equation}
\label{eq:CtspecdecLT}
G^\mathrm{MLLGS}(t;L,T) \equiv G^\mathrm{MLLGS}(t;L) + \frac13\sum_{n> 0}|\vec{A}_n|^2\ e^{-E_n(T-t)}
+ \frac23\sum_{n> 0}|\vec{A}_n^{(0)}|^2\ e^{-E_n^{(0)}T/2},
\end{equation}
with the free pion amplitude squared, $L^3|\vec{A}_n^{(0)}|^2=4\nu_n
n(2\pi/LE_n^{(0)})^2$, and keeping only states up to $n=8$, as above. Here,
$\nu_n$ counts the number of vectors of $\mathbb{Z}^3$ that have norm squared
$n$.

Using similar arguments, the inclusion of taste-violations is straightforward.
One merely performs the replacement $E_n \to E_{n,\tau}$ and averages over the
sixteen tastes. This yields $G^\mathrm{SMLLGS}(t;L,T,a)$.

Now, using the above correlators, it is straightforward to generalize Equation
\eqref{eq:LLGSFV} to also include finite-$T$ corrections. We obtain:
$$
a_{\mu}^\mathrm{GS}(L=\infty,T=\infty)-a_{\mu}^\mathrm{MLLGS}(L_\mathrm{ref},T_\mathrm{ref})
$$
\begin{equation}
  \label{eq:LLGSFVT}=
    10^{10}\alpha^2
    \left[ \int_0^\infty dt\, K(t)G^\mathrm{GS}(t)- \int_0^{T_\mathrm{ref}/2} dt\, K(t)G^\mathrm{MLLGS}(t;L_\mathrm{ref},T_\mathrm{ref})\right]
\ .
\end{equation}
Similarly, for each simulation, we can estimate taste-breaking effects on an
$T\times L^3$ lattice, through:
$$
a_{\mu,\mathrm{win}}^\mathrm{MLLGS}(L_\mathrm{ref},T_\mathrm{ref})-a_{\mu,\mathrm{win}}^\mathrm{SMLLGS}(L,T,a)=10^{10}\alpha^2\left[
    \int_0^{T_\mathrm{ref}/2} dt\, K(t)\, W(t;t_1,t_2)G^\mathrm{MLLGS}(t;L_\mathrm{ref},T_\mathrm{ref})\right.
$$
\begin{equation}
  \label{eq:SLLGSFVT}
      \left.-\int_0^{T/2} dt\, K(t)\, W(t;t_1,t_2)G^\mathrm{SMLLGS}(t;L,T,a)\right]\ ,
\end{equation}
thus generalizing Equation \eqref{eq:SLLGSFV} to finite $T$.

    \section{Rho-pion-gamma model}
\label{se:obs_rho}

In this section we briefly review a model that we use to predict finite-size
effects in Section \ref{se:obs_fv} and perform the taste improvement in Section
\ref{se:obs_taimp}. The model is an effective field theory of the rho, the pion
and the photon; we call it the rho-pion-gamma model and abbreviate it RHO. It
was proposed long ago by Sakurai \cite{Sakurai:1960ju}, has been used by
Jegerlehner and Szafron to describe $\rho-\gamma$ mixing
\cite{Jegerlehner:2011ti} and has been used recently by HPQCD to remove
taste-breaking effects in $a_\mu$ \cite{Chakraborty:2016mwy}.

The Lagrangian of the model is given by
\begin{gather}
    \mathcal{L}=
    -\frac{1}{4}F^2_{\mu\nu}(A)
    -\frac{1}{4}F^2_{\mu\nu}(\rho)
    +\frac{1}{2}m_\rho^2 \rho_\mu\rho^\mu
    -\frac{e}{2g_\gamma} F_{\mu\nu}(\rho)F^{\mu\nu}(A)
    + (D_\mu \pi)^\dagger (D^\mu \pi) - m_\pi^2\pi^\dagger\pi\ ,
\end{gather}
where the $\rho_\mu$, $\pi$ and $A_\mu$ are the rho, pion and photon fields,
the covariant derivative is $D_\mu= \partial_\mu -ieA_\mu -ig\rho_\mu$ and the
field-strength tensor is $F_{\mu\nu}(V)= \partial_\mu V_\nu - \partial_\nu
V_\mu$. The parameters of the model are the coupling constants $e,g,g_\gamma$
and the masses $m_\pi$ and $m_\rho$. We work with a Minkowski metric in this
Section, unless stated otherwise.

In the following we compute the photon propagator including effects up to
$O(g^2,g_\gamma^2,gg_\gamma)$.  The only momentum-dependent loop in diagrams is
the two-pion loop, whose renormalized contribution we denote $\Sigma(q^2)$ --
we use dimensional regularization here. The relevant diagrams can be re-summed
thanks to a Dyson equation, and the resulting photon propagator in momentum
space is given by:
\begin{gather}
    \label{eq:gprop}
    iG_{\gamma\gamma}^\mathrm{RHO}(q)=
    \frac{1}{q^2} \frac{1}{1-e^2\Sigma(q^2)} +
    \frac{e^2}{g_\gamma^2} \frac{\left[1-gg_\gamma\Sigma(q^2)\right]^2}{q^2\left[1-g^2\Sigma(q^2)\right]-m_\rho^2}\ .
\end{gather}
At the rho pole, this expression can be matched to a Breit-Wigner propagator:
\begin{gather}
    iG_{\gamma\gamma}^\mathrm{RHO}(q)\to \frac{F_\rho^2/(2M_\rho^2)}{q^2-M_\rho^2+iM_\rho\Gamma_\rho}\ ,
\end{gather}
determined by the physical mass, width and electromagnetic decay constant of
the rho meson: $M_\rho$, $\Gamma_\rho$ and $F_\rho$. In the matching one can
neglect the imaginary part of the complex residue at the pole because it
contributes at $O(g^4)$. The matching of the model parameters to the physical
ones is given by:
\begin{gather}
    \label{eq:match}
    \begin{aligned}
	m_\rho^2&= M_\rho^2 \left[ 1-g^2\mathrm{Re}\Sigma_\rho\right]\ ,\\
	g^2&=-\frac{\Gamma_\rho}{M_\rho}\frac{1}{\mathrm{Im}\Sigma_\rho}\ ,\\
	\frac{1}{g_\gamma^2}&= \frac{F_\rho^2}{2M_\rho^2}\mathrm{Re}
	\left[ 1 + gg_\gamma\Sigma_\rho - \tfrac{1}{2}g^2\Sigma_\rho -\tfrac{1}{2}g^2M_\rho^2\Sigma'_\rho\right]^2\ ,
    \end{aligned}
\end{gather}
where $\Sigma_\rho=\Sigma(M_\rho^2)$ and $\Sigma'=d\Sigma/dq^2$. In this paper
we use $M_\rho=775$~MeV, $g_\rho=5.95$ and $F_\rho=210$~MeV. When correcting
the light contribution to $a_\mu$ in the isospin limit the pion-mass parameter
is set to the pion mass measured on the ensemble.
The vacuum-polarization $\Pi^\mathrm{RHO}(q^2)$ can then be obtained from the
photon propagator in Equation \eqref{eq:gprop}. After applying a
zero-momentum-squared subtraction, and dropping of $O(g^4)$ terms, we get:
\begin{gather}
    \label{eq:pihat}
    \hat{\Pi}^\mathrm{RHO}(q^2)= \hat{\Sigma}(q^2)
    + \frac{1}{g_\gamma^2} \left[1-gg_\gamma \Sigma(0) + \tfrac{1}{2}g^2\Sigma(0)\right]^2
    \cdot \frac{q^2[1-gg_\gamma \hat{\Sigma}(q^2)]^2}{q^2[1-g^2\hat\Sigma(q^2)] - m_\rho^2/[1-g^2\Sigma(0)]}\ ,
\end{gather}
where a hat on a symbol denotes on-shell subtraction, i.e. $\hat{\Pi}(q^2)=
\Pi(q^2) - \Pi(0)$. 

Finite-size effects are obtained by replacing $\Sigma(q^2)$ with its
finite-volume counterpart. Moreover, on the lattice we perform the subtraction
of the HVP function directly at the values of the lattice parameters, in finite
volume. This means that $\Sigma(0)$ must also be replaced by its finite-volume
value. In addition, to isolate the finite-size effects proper to HVP, the
Lagrangian must be renormalized in a short-distance scheme, as above. Thus, the
parameters of the model in finite volume are those of the infinite-volume
model, i.e. $m_\rho$, $g$ and $g_\gamma$ obtained in Equations
\eqref{eq:match}, with the infinite-volume $\Sigma$.

In this model staggered, taste-violation effects enter through the mass
splitting of the pions, while all other parameters are kept fixed.  As a result
$\Sigma(q^2;M_\pi^2)$ is replaced by the taste-averaged,
pion-vacuum-polarization function, $\frac{1}{16}\sum_\alpha
\Sigma(q^2;M_{\pi,\alpha}^2)$ in Equation \eqref{eq:pihat}, while we keep using
the continuum $\Sigma$ in the matching expressions given by Equations
\eqref{eq:match}.

This yields the following expression for HVP, which includes finite-size and taste-violation effects:
\begin{gather}
    \begin{gathered}
	\hat{\Pi}^\mathrm{SRHO}(q^2)= \hat{\Sigma}_\mathrm{latt}(q^2) +\\
	+ \frac{F_\rho^2}{2M_\rho^2}\cdot
	\left[1+gg_\gamma \hat\Sigma_\mathrm{rho} -\tfrac{1}{2}g^2\hat\Sigma_\mathrm{rho} -\tfrac{1}{2} g^2 M_\rho^2 \hat\Sigma'_\mathrm{rho}\right]^2
	\cdot \frac{q^2[1-gg_\gamma \hat\Sigma_\mathrm{latt}(q^2)]^2}{q^2[1-g^2\hat\Sigma_\mathrm{latt}(q^2)] - M_\rho^2[1-g^2\hat\Sigma_\mathrm{rho}]}\ ,
    \end{gathered}
\end{gather}
The label SRHO denotes the staggered version of the rho-pion-gamma model. Here,
$\hat\Sigma_\mathrm{latt}$ is the taste-averaged, pion-vacuum-polarization
function computed in finite volume and at finite lattice spacing, and
$\hat\Sigma_\mathrm{rho}= \mathrm{Re}\hat\Sigma_\rho +
\Sigma(0)-\Sigma_\mathrm{latt}(0)$. The difference of the $\Sigma$ at zero
momentum squared is scheme independent.

For correcting the lattice results, we need the above function for Euclidean
momenta, $Q^2=-q^2$, where the pion vacuum polarization is a real function.
Additionally, we need a vacuum polarization that corresponds to the coordinate
space current propagator $G(t)$, multiplied by a window-function $W(t)$, see
Section \ref{se:obs_hvp}. In particular we need a recipe to construct
\begin{gather}
    \hat\Pi_\mathrm{win}(Q^2)= 2\int_0^\infty dt\ \frac{1}{Q^2}\left[ \cos(Qt)-1+\tfrac{1}{2}(Qt)^2 \right] W(t) G(t) 
\end{gather}
from the available $\hat\Pi(Q^2)$ function, which is obtained by using $W(t)=1$
in the above equation. For a smooth and symmetric window function, $W(t)=W(-t)$,
the following formula holds:
\begin{gather}
    \hat\Pi_\mathrm{win}(Q^2)=
    \int_{-\infty}^{\infty} \frac{dP}{2\pi}
    \frac{1}{Q^2}\left[ \widetilde{W}(P-Q) - \widetilde{W}(P) - \frac{1}{2}Q^2\widetilde{W}''(P) \right]
    P^2 \hat\Pi(P^2)\ ,
\end{gather}
where $\widetilde{W}$ is the Fourier-transform of $W$, i.e. $\widetilde{W}(Q)= \int_{-\infty}^\infty dt \cos(Qt) W(t)$.

    \section{Finite-size effects in $a_\mu$}
\label{se:obs_fv}

\begin{figure}[t]
    \centering
    \includegraphics[width=0.7\textwidth]{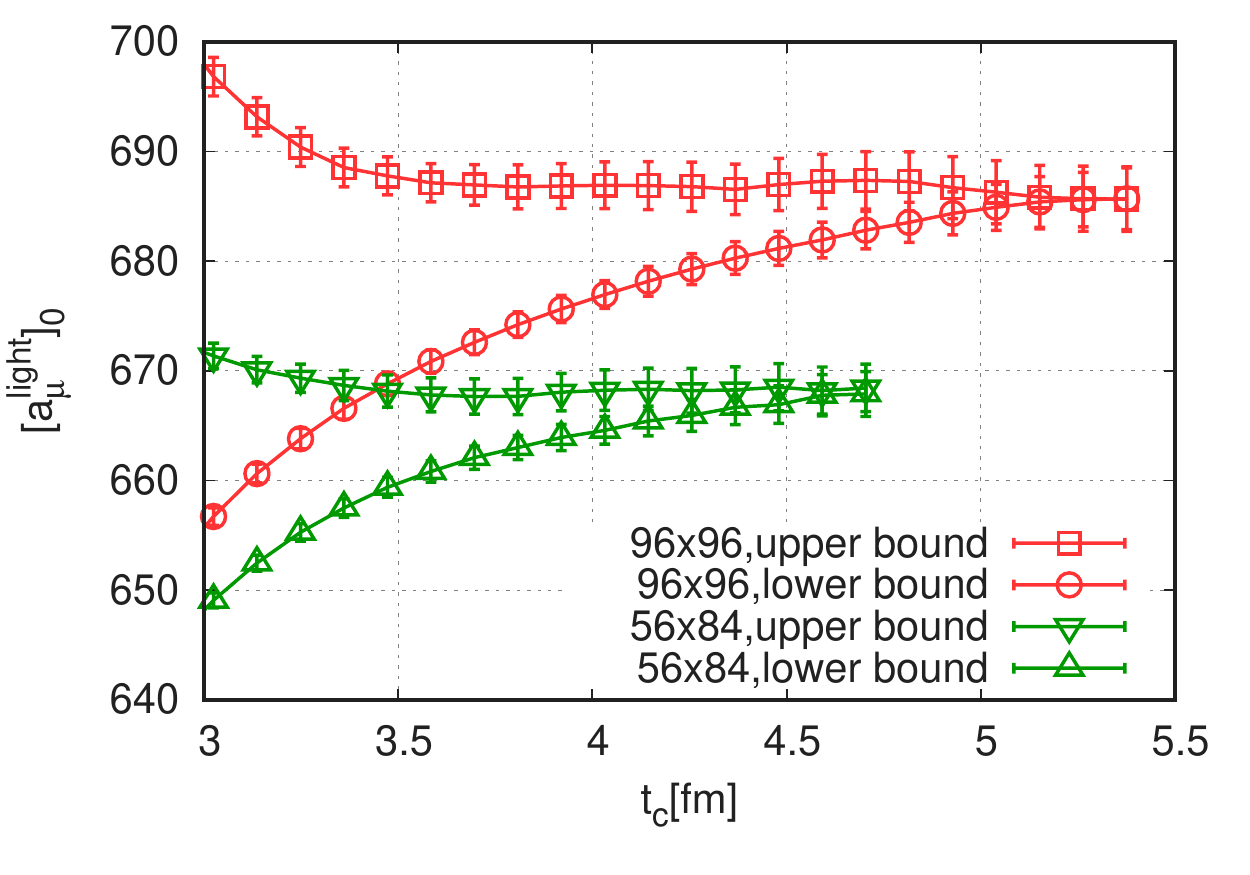}
    \caption
    {
	\label{fi:fv4hex}Upper and lower bounds on the light isospin-symmetric
	component of $a_\mu$. The results shown here are obtained with the {\tt
	4HEX} action on two different volumes at $a=0.112$~fm lattice spacing
	and $M_\pi=121$~MeV Goldstone-pion mass. We also have another
	simulation with $M_\pi=104$~MeV mass. From these two we interpolate to
	$M_\pi=110$~MeV. This value ensures that a particular average of pion
	tastes is fixed to the physical value of the pion mass (see text).
    }
\end{figure}

Finite-size effects on $a_\mu$ were the largest source of uncertainty in our
previous work \cite{Borsanyi:2017zdw}. In this section we present the
computation of these effects in a systematic way, which includes dedicated
lattice simulations, chiral perturbation theory and phenomenological models.
The concrete goal of this section is to provide a single number that is to be
added to the continuum-extrapolated lattice result obtained in a reference box,
which is defined by a spatial extent of $L_\mathrm{ref}=6.272$~fm and a
temporal extent of $T_\mathrm{ref}=\tfrac{3}{2}L_\mathrm{ref}$.

First we concentrate on the finite-size effect of the isospin-symmetric part.
Section \ref{se:obs_split} details our isospin decomposition. The
isospin-breaking part will be discussed later in the last subsection. The
isospin-symmetric part can be further decomposed into an $I=0$ and an $I=1$
channel. From these the $I=1$ is supposed to give the majority of the
finite-size effect. We focus on the $I=1$ first, and give an estimate of the
$I=0$ contribution later. According to Equation \eqref{eq:jji1i0} the $I=1$
result is given by the $\left(\frac{9}{10}\right)$'th of the connected light
contribution.

We perform dedicated lattice simulations with two different lattice geometries:
one on a $56\times84$ lattice with the reference box size and another on a
large $96\times96$ lattice with box size $L=L_\mathrm{big}=10.752$~fm and
$T=T_\mathrm{big}=L_\mathrm{big}$.  Since taste violations severely distort the
finite-size behavior, we designed a new action with highly-suppressed taste
breaking for these computations. The details of the {\tt 4HEX} action and the
simulation parameters are given in Section \ref{se:act_4hex}. Our strategy is
then to compute the finite-size correction as the following sum:
\begin{gather}
    \label{eq:telescope}
    \begin{gathered}
	a_\mu(\infty,\infty)-a_\mu(L_\mathrm{ref},T_\mathrm{ref})=\\
	=[a_\mu(L_\mathrm{big},T_\mathrm{big})-a_\mu(L_\mathrm{ref},T_\mathrm{ref})]_{\tt 4HEX} +
	[a_\mu(\infty,\infty)-a_\mu(L_\mathrm{big},T_\mathrm{big})]_\mathrm{XPT}.
    \end{gathered}
\end{gather}
The first difference on the right hand side is taken from the dedicated {\tt
4HEX} simulations.  The second difference is expected to be much smaller than
the first and is taken from a non-lattice approach: chiral perturbation theory.

We consider four non-lattice approaches for both differences on the right hand
side of Equation \eqref{eq:telescope}. In case of the first difference, they
will be compared to our {\tt 4HEX} simulations. The first is chiral
perturbation theory (XPT), discussed in detail in Section \ref{se:obs_xpt}.
The second is the Meyer-Lellouch-Luscher-Gounaris-Sakurai model (MLLGS), with details
in Section \ref{se:obs_llgs}. In this approach we compute values for the
reference box only, and not for the large box. This is because $L_\mathrm{big}$
is relatively large and one would have to deal with a large number of states,
which is not practical in that approach.  The third approach is that of Hansen
and Patella (HP) \cite{Hansen:2019rbh}, who use a generic field theory
framework to relate the finite-size effect to the electromagnetic form factor
of the pion, the latter being determined on the lattice. Note that their first
published result does not include effects that are of order $e^{-\sqrt{2}M_\pi
L}$. These can be significant and have been added later \cite{Hansen:2020whp}.
Though the latter version also includes finite-$T$ effects, it does so with
assumptions that are not applicable to our lattices, where $T<2L$. Therefore we
use the HP approach here in the infinite-$T$ limit. The fourth non-lattice
approach is the rho-pion-gamma model of \cite{Chakraborty:2016mwy}, which we
abbreviate as RHO here, and is reviewed in Section \ref{se:obs_rho}. This model
is also considered in the infinite-$T$ limit only.

\subsection*{Results with the {\tt 4HEX} action}

We compute the first difference in Equation \eqref{eq:telescope} using
dedicated simulations with the {\tt 4HEX} action.  First we describe the way in
which we fix the physical point in these simulations. For this purpose, it is
instructive to look at the influence of taste violations on the finite-size
effect in NNLO staggered chiral perturbation theory (SXPT).  The necessary
formulas are given in Section \ref{se:obs_xpt}. We apply them to various cases
that are described below.  The following numbers are obtained for the
finite-size effect:
\begin{center}
    \begin{tabular}{C|C|C|C|C}
	\text{NNLO SXPT results for} \to & \text{continuum} & \text{\tt 4stout} & \text{\tt 4HEX} & \text{\tt 4HEX@110MeV}\\
	\hline
	a_\mu(L_\mathrm{big},T_\mathrm{big})-a_\mu(L_\mathrm{ref},T_\mathrm{ref}) &  15.7 & 1.6 & 8.1 & 15.8 \\
    \end{tabular}
\end{center}
The first number gives the continuum prediction, which is about 2\% of the
total $a_\mu$.  The second number stands for the {\tt 4stout} action at a
lattice spacing of $a=0.112$~fm. Here, most of the pion tastes are too heavy to
play any role in the finite-size behavior. According to SXPT the finite-size
effect is practically non-existent there. The {\tt 4HEX} action has much
suppressed taste violations, and the corresponding number, the third in the
table, is already much closer to the continuum. Until now the Goldstone pion
mass is set to the physical value $M_\pi= M_{\pi_0,*}$. This pion is the
lightest of the sixteen pions in the taste multiplet. We can get much closer to
the size of the continuum finite-size effect if we use Goldstone-pion masses
below $M_{\pi_0,*}$. For example one can set a taste-averaged pion mass to the physical value.

In NLO SXPT the slope of the hadronic vacuum polarization is proportional to
$\sum_\alpha M^{-2}_{\pi,\alpha}$. This motivates to use the
harmonic-mean-square (HMS), defined by
\begin{gather*}
    M^{-2}_{\pi,\mathrm{HMS}}\equiv \frac{1}{16} \sum_\alpha M^{-2}_{\pi,\alpha}\ ,
\end{gather*}
to average over the tastes.  Setting $M_{\pi,\mathrm{HMS}}= M_{\pi_0,*}$
requires lowering the Goldstone-pion mass to $M_\pi=110$~MeV. With this choice,
the finite-size effect is of the same size on the lattice and in the continuum
in NNLO SXPT. This gives the fourth number in the table. This choice results in
much smaller lattice artefacts than the usual setting with the Goldstone-pion,
at least for an observable like the finite-size effect. 

To generate the {\tt 4HEX} data set, we performed simulations with two
different Goldstone pion masses: $M_\pi=104$~MeV and $121$~MeV. To set the
physical point as described above, we perform an interpolation from these two
pion masses to $M_\pi=110$~MeV. 

To compute $a_{\mu}^\mathrm{light}$ from the current propagator in our {\tt
4HEX} simulations we use the upper and lower bounds described in Section
\ref{se:obs_bound}. The results are plotted in Figure \ref{fi:fv4hex} for the
$M_\pi=121$~MeV simulation point. The bounds meet at around $4.2$~fm and
$4.7$~fm on the small and large volumes, respectively. At these distances we
take the average of the two bounds as an estimate for $a_{\mu}^\mathrm{light}$.
The results are given in the table below:
\begin{center}
    \begin{tabular}{C|C|C|C}
	M_\pi \text{ in {\tt 4HEX}}\to & 104\text{ MeV} & 121\text{ MeV} & 110\text{ MeV} \\
	\hline
	a_{\mu}^\mathrm{light}(56\times 84) &685.9(2.7)& 668.3(2.0)& 679.5(1.9)\\
	a_{\mu}^\mathrm{light}(96\times 96) &710.7(1.9)& 684.3(1.7)& 701.1(1.3)
    \end{tabular}
\end{center}
In the last column we also give the interpolated value at the physical point,
using the HMS averaged pion-mass prescription defined above.

We only have one lattice spacing with the {\tt 4HEX} action, so no proper
continuum extrapolation of the finite-size effect can be done. We estimate the
cutoff effect of the result by comparing the total $a_\mu$ with the {\tt 4HEX}
action at this single lattice spacing to the continuum extrapolated {\tt
4stout} lattice result, both in the $L_\mathrm{ref}$ volume. The {\tt 4HEX}
result is about 7\% larger than the continuum value. Therefore we reduce the
measured finite-size effect by 7\%, and assign a 7\% uncertainty to this
correction step. For the difference we get
\begin{gather}
    \label{eq:fvref}
    a_\mu(L_\mathrm{big},T_\mathrm{big})-a_\mu(L_\mathrm{ref},T_\mathrm{ref})= 18.1(2.0)_\mathrm{stat}(1.4)_\mathrm{cont}\ .
\end{gather}
The result is obtained from the $a_\mu^\mathrm{light}$ numbers from above
including a multiplication by the $\left(\frac{9}{10}\right)$ charge factor.
The first error is statistical, the second is an estimate of the cutoff effect. 

\subsection*{Results from non-lattice approaches}

The table below collects the finite-size effect computed in various non-lattice
approaches:
\begin{center}
    \begin{tabular}{L|C|C|C|C|C}
	& \text{NLO XPT} & \text{NNLO XPT} & \text{MLLGS} & \text{HP} & \text{RHO} \\
	\hline
	a_\mu(L_\mathrm{big},T_\mathrm{big})-a_\mu(L_\mathrm{ref},T_\mathrm{ref}) & 11.6 & 15.7 & 17.8 &    - &    - \\
	a_\mu(L_\mathrm{big},\infty)-a_\mu(L_\mathrm{ref},\infty)                 & 11.2 & 15.3 & 17.4 & 16.3 & 14.8 \\
    \end{tabular}
\end{center}
As we mentioned before, the MLLGS approach was not used in the large box. The
MLLGS numbers in the table are actually a difference of the MLLGS prediction for
$a_\mu(\infty,\infty)-a_\mu(L_\mathrm{ref},T_\mathrm{ref})$ and the residual
finite-size effects of the big lattice taken from NNLO XPT. We also give
results for the case of infinite time extent. We see that, according
to the models, the finite-$T$ effect is much smaller than the finite-$L$ effect.

The different models give a finite-size effect of similar size that agrees well
with the lattice determination of Equation \eqref{eq:fvref}. Only the NLO
result differs by about $3\sigma$'s. The fact that NLO chiral perturbation
theory underestimates the finite-size effect was already shown in
\cite{Giusti:2018mdh}, at a non-physical pion mass. Using physical pion mass, a
dedicated finite-volume study was carried out in \cite{Shintani:2019wai}. It
reaches the same conclusion as we do, albeit with larger errors.

The good agreement for the finite-size effect of the reference box,
between the models and the lattice, gives us confidence that the
models can be used to reliably compute the very small, residual,
finite-size effect of the large box. We get:
\begin{center}
    \begin{tabular}{L|C|C|C|C}
	& \text{NLO XPT} & \text{NNLO XPT} & \text{HP} & \text{RHO} \\
	\hline
	a_\mu(\infty,\infty)-a_\mu(L_\mathrm{big},T_\mathrm{big}) & 0.3 & 0.6 &   - &   - \\
	a_\mu(\infty,\infty)-a_\mu(L_\mathrm{big},\infty)         & 1.2 & 1.4 & 1.4 & 1.4 \\
    \end{tabular}
\end{center}
For an infinite-time extent the NNLO XPT, the HP and RHO approaches agree nicely. As a
final value for the large box finite-size effect we take the NNLO XPT result including
the finite-$T$ effects:
\begin{gather}
    \label{eq:fvbig}
    a_\mu(\infty,\infty)-a_\mu(L_\mathrm{big},T_\mathrm{big})= 0.6(0.3)_\mathrm{big}\ ,
\end{gather}
where the uncertainty is an estimate of higher-order effects, given here by the
difference of the NNLO and NLO values.

Until now we have been discussing the $I=1$ finite-size effects. Here we make
an estimation of the $I=0$ channel using XPT. The long-distance behavior of the
$I=0$ channel is dominated by three pions. A necessary photon-pion-pion-pion
vertex arises from the Wess-Zumino-Witten term in the chiral Lagrangian, which
is NLO \cite{Scherer:2002tk}. The lowest order diagram involves two such
vertices and two loops, thus the contribution is N$^4$LO. From the NLO
and NNLO values for
$a_\mu(\infty,\infty)-a_\mu(L_\mathrm{ref},T_\mathrm{ref})$ we
estimate the size of the N$^4$LO term, from which we take
$0.0(0.6)_\mathrm{I=0}$ as our estimate for the $I=0$ finite-size effect.

\subsection*{Finite-size effect in the isospin-breaking contributions}

\begin{figure}[t]
    \centering
    \includegraphics[width=0.7\textwidth]{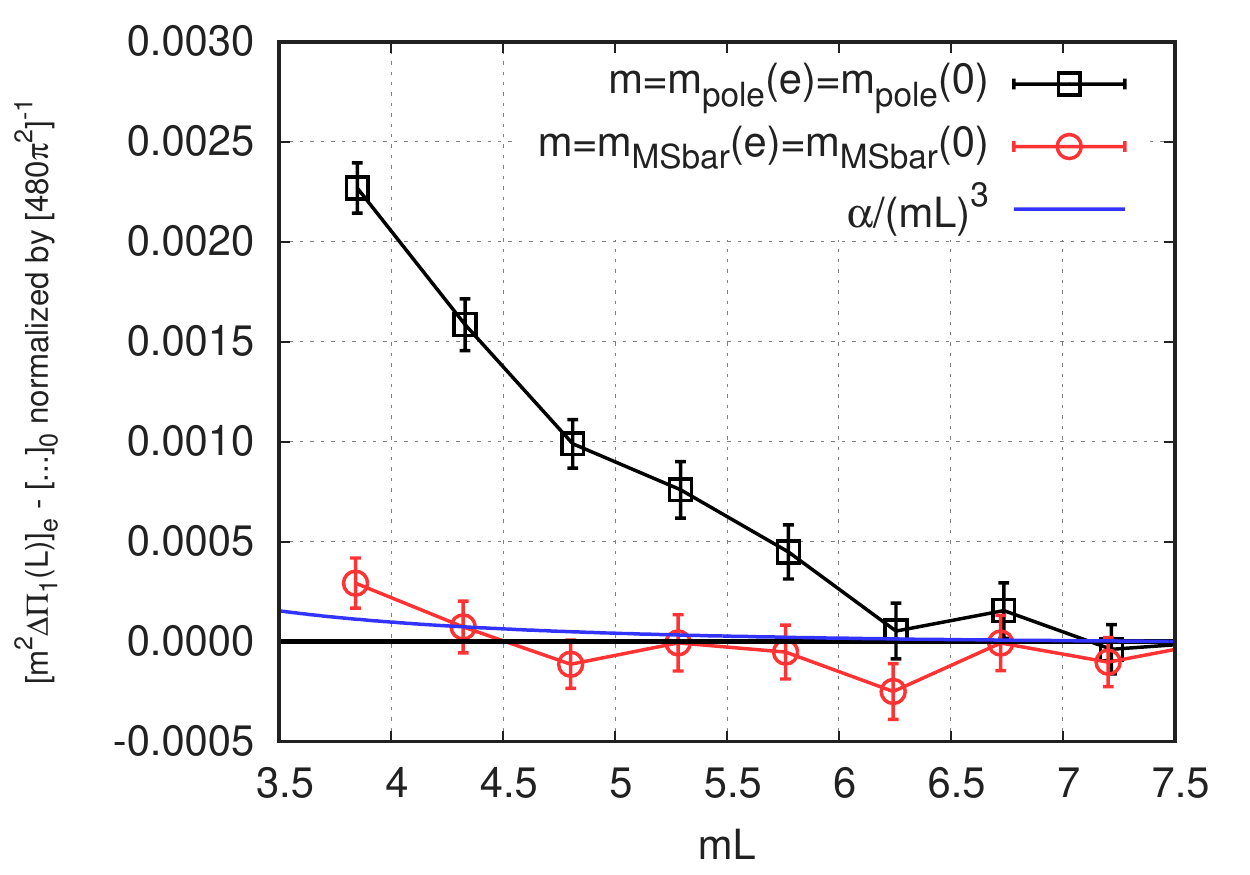}
    \caption
    {
	\label{fi:sqed}Electromagnetic part of the finite-size effect for
	scalar QED. Shown is the dimensionless double difference for the slope
	of the vacuum polarization, ie. $m^2\left[ \Delta \Pi_1(L,e)-\Delta
	\Pi_1(L,0) \right]$, normalized by the free field value
	$m^2\Pi_1(\infty,0)=[480\pi^2]^{-1}$ (see text). Two different
	definitions for the electromagnetic part are shown, defined by matching
	with two different masses, $m_\text{pole}$ and
	$m_{\overline{\text{MS}}}$. A blue curve shows an order of magnitude
	estimate of the finite-size effect from \cite{Bijnens:2019ejw}.
    }
\end{figure}

A comprehensive study of the electromagnetic finite-size effects on the current
propagator has appeared recently \cite{Bijnens:2019ejw}. The authors conclude
that if all particles, except the photon, are treated in infinite volume, then
the finite-size effects are of order $\alpha/(M_{\pi}L)^3$.  In practice,
however, when all particles reside in the finite box, the usual exponential
finite-size effects become dominant over their electromagnetic counterpart
suppressed by $\alpha$. In this case it is useful to separate the
electromagnetic contributions from the isospin-symmetric part. The QED part
exhibits an $\alpha/(M_{\pi}L)^3$ behavior. The isospin-symmetric part will
have an exponential suppression governed by the neutral-pion mass,
$\exp(-M_{\pi_0}L)$. These isospin-symmetric effects are sizeable and discussed
earlier in this Section. A subtle point here is the definition of the
electromagnetic contribution or equivalently the matching of QCD+QED to QCD.

It is instructive to study, in a simple model, the role of matching in the size
of finite-volume effects. For this purpose we carry out lattice simulations in
scalar QED.  Only quenched QED is implemented, since dynamical QED effects
enter at order $O(e^4)$. We perform two sets of simulations:
\begin{enumerate}

    \item First we perform simulations in QED with a bare scalar mass
	$m_0=0.1210$, a coupling $\alpha=1/137$ and in $L^4$ boxes in the range
	$L=16\dots32$. The mass of the charged scalar boson extracted from the
	propagator and extrapolated to infinite volume is
	$m_\text{pole}(e)=0.2406$. One can also define a renormalized mass
	using the $\overline{\text{MS}}$ prescription \cite{Bijnens:2019ejw},
	and we find $m_{\overline{\text{MS}}}(e)=0.2401$.
	
    \item We also perform simulations without QED, which is just the free
	scalar field theory, in the same box sizes and at two bare values of
	the mass, $m_0=0.2405$ and $0.2415$. We use these two values to perform
	the interpolations that are necessary for the different matching
	conditions. Note that even in the free case, the pole mass is slightly
	different from the bare mass due to lattice artefacts
	\cite{Montvay:1994cy}. We use
	$m_{\overline{\text{MS}}}(0)=m_\text{pole}(0)$.
	
\end{enumerate}
We choose a simple observable, the slope of the vacuum polarization function
$\Pi_1(L,e)\equiv \left.\frac{d\Pi(Q^2)}{dQ^2}\right|_{Q^2=0}$. This is built
from the conserved current of the scalar field theory, which is given as
\begin{gather}
    j_{\mu,x}/i= \phi^\dagger_{x+\mu} e^{ieA_{\mu,x}} \phi_x - \phi_{x}^\dagger e^{-ieA_{\mu,x}} \phi_{x+\mu}\ .
\end{gather}
From the measured $\Pi_1$, we build the difference $\Delta \Pi_1(L,e)\equiv
\Pi_1(32,e)-\Pi_1(L,e)$ and investigate its $L$ dependence. 
To define the QED part, we compute the difference between $\Delta \Pi_1(L,e)$
and $\Delta \Pi_1(L,0)$. This can be done in different ways, depending on how
the two theories, with and without QED, are matched. One way is to use
$m_\text{pole}$ to match the theories, another is to do the same with the
$m_{\overline{\text{MS}}}$. The results are shown in Figure \ref{fi:sqed}.
As one can see, the $\overline{\text{MS}}$ matching leads to the
expected $\alpha/(m L)^3$ behavior, but pole-mass matching leaves the
electromagnetic part with a much larger finite-size effect.

In QCD+QED we can define matching schemes similar to the pole and
$\overline{\text{MS}}$ mass matchings of scalar QED. The pole mass is the
measured mass of the charged particle, whereas the $\overline{\text{MS}}$ mass
is the parameter that appears in the renormalized Lagrangian. In QCD+QED the
analogue of $m_\mathrm{pole}$ is the charged-pion mass, and that of
$m_{\overline{\text{MS}}}$ is the renormalized quark mass. If we were to base
our matching on $M_{\pi_+}$, the finite-size effects would be much larger than
the expected $\alpha/(M_\pi L)^3$. Whereas if we use the renormalized quark
masses in the matching, we expect to see the $\alpha/(M_\pi L)^3$ finite-volume
behavior in the electromagnetic part. In our scheme, as introduced in Section
\ref{se:obs_split}, we keep the neutral-pion mass, $M_{\pi_0}$, fixed instead
of the charged-pion mass $M_{\pi_+}$. This should be very close to a
scheme where the renormalized quark masses are kept fixed. As such, we expect
the QED corrections to $a_\mu$ to exhibit an $\alpha/(M_\pi L)^3$ behavior in
our scheme too.

The finite-size effect of $O[\alpha/(M_\pi L)^3]$ from the electromagnetic part
is very small compared to the precision of our study. The finite-size effect of
the strong-isospin-breaking part must also be small: it is actually exactly
zero in NLO chiral perturbation theory. In our reference box
$L_\mathrm{ref}=6.272$ an $\alpha/(M_\pi L_\mathrm{ref})^3$ relative correction
corresponds to a finite-size effect of $0.1$ in $a_\mu$. We will use a value of
$0.0(0.1)_\mathrm{qed}$ as an estimate of the finite-size effect of the
isospin-breaking in our reference box.

\subsection*{Final result}

For our final result for the finite-size effect of the reference box, we add
the numbers in Equations \eqref{eq:fvref} and \eqref{eq:fvbig}, as well as the
estimates from the conclusions of the preceeding two subsections, giving:
\begin{gather}
    \label{eq:fv}
    a_\mu(\infty,\infty) - a_\mu(L_\mathrm{ref},T_\mathrm{ref})= 18.7(2.0)_\mathrm{stat}(1.4)_\mathrm{cont}(0.3)_\mathrm{big}(0.6)_{I=0}(0.1)_\mathrm{qed}[2.5]\ .
\end{gather}
the first error is the statistical uncertainty of our {\tt 4HEX} computation,
the second is an estimate of the {\tt 4HEX} cutoff effects, the third is the
uncertainty of the residual finite-size effect of the ``big'' lattice, the
fourth is a XPT estimate of the $I=0$ finite size effect, the fifth is an
estimate of the isospin-breaking effects. The last, total error in the
square-brackets is the first five added in quadrature. The vast majority of the
finite-size effect is obtained using the {\tt 4HEX} lattice computation; for
the rest we apply analytic methods. These methods have been validated by the
lattice computation: for the majority contribution they give values that are
consistent with the lattice.

    \section[Taste improvement]{Taste improvement\protect\footnote{ We thank the
referees for helping us to improve this important component of our analysis
through the constructive criticism of their reports. We are also grateful to
the participants of the online workshop, ``The hadronic vacuum polarization
from lattice QCD at high precision''~\cite{HM:202011}, in particular T.~Blum,
G.~Colangelo, M.~Hoferichter, N.~Husung, C.~Lehner, H.~Meyer, R.~Sommer, R.~Van
de Water and H. Wittig, for interesting comments and discussions on these and
related issues.} }
\label{se:obs_taimp}

As is well known, some of the most important cutoff effects of staggered
fermions are taste violations. At long distances, these violations distort the
pion spectrum. Since $a_\mu$ is predominantly a long-distance observable,
dominated by a two-pion contribution, including the $\rho$ resonance, we expect
these effects to be largest in the light-quark terms. This is visible in
Figure~S4 of \cite{Borsanyi:2017zdw}, where the continuum extrapolations of the
light connected and disconnected contributions to $a_\mu$ are clearly steeper
than those of the more massive strange-quark contribution, which displays very
small discretization effects.

It is instructive to compare the lattice artefacts of $a_\mu^\mathrm{light}$ in
different fermion formulations that are found in the literature. For this
purpose we interpolate available lattice data to a common lattice spacing of
$a=0.10$~fm and compute its deviation from the quoted continuum limit.  If
necessary, we scale the lattice artefact by assuming an $a^2$ dependence.  With
our action, at a lattice spacing of $a=0.10$~fm, the result for
$a_\mu^\mathrm{light}$ is about 11\% smaller than its continuum value, as can
be seen in Figure~\ref{fi:cont} (unimproved data set). For twisted-mass
fermions, as used by the ETM collaboration, the deviation is 28\% downward,
which we read off from the left panel of Figure~14 in \cite{Giusti:2018mdh}.
The $O(a^2)$-improved Wilson fermion formulation of the Mainz group has an
$a_\mu^\mathrm{light}$ that is 18\%/6\% larger than in the continuum, as can be
seen on the left/right panel of Figure~8 in \cite{Gerardin:2019rua}. The
former/latter number is obtained without/with an improvement related to the
scale setting.  Aubin {\it et al.}\ \cite{Aubin:2019usy}, with so-called
highly-improved staggered fermions, obtain a 9\% smaller value at $a=0.10$~fm
than in the continuum, as can be determined from their Table III. The same
action with a different current is used by the the Fermilab--HPQCD--MILC
collaboration \cite{Davies:2019efs}. Here the coefficient of the lattice
spacing dependence is explicitely given and we get an $a_\mu^\mathrm{light}$ at
$a=0.10$~fm, that is 3\% below the continuum limit. Finally, in the case of
domain-wall fermions, as used by the RBC/UKQCD group, we have found no
published results on the lattice-spacing dependence of $a_\mu^\mathrm{light}$.
However, for the window observable $a_{\mu,\mathrm{win}}^\mathrm{light}$ the
deviation from the continuum is about 3.5\%, which we read off from their
Figure~8 in \cite{Blum:2018mom}. Note that, in our case, the lattice artefacts
in the window are even smaller, about 2.5\%.

In this section we investigate various physically motivated models for reducing
long-distance taste violations in our lattice results. After comparing lattice
artefacts in our lattice data with predictions of these models, we propose a
procedure to improve the approach to the continuum limit.  These improvements
will be applied on light-quark observables at the isospin-symmetric point,
whose taste violations have the largest impact on our final uncertainties. 

There is a subtle problem in connection with taste violations and finite-size
effects. On coarse lattices, finite-size effects are largely suppressed, since
most of the pion taste partners are heavy. The finite-size effects increase
gradually for finer lattices, but even on our finest lattice, only about half
of the expected effect is present, as discussed in Section \ref{se:obs_fv}. A
good correction for cutoff effects should also restore the finite-volume
dependence that is expected in the continuum limit.

To address the issue of finite-volume effects, we have considered three
techniques: NNLO staggered chiral perturbation theory in Section
\ref{se:obs_xpt}, MLLGS in Section \ref{se:obs_llgs} and RHO in Section
\ref{se:obs_rho}.  In this section we investigate and discuss the suitability
of their staggered versions for reducing the taste violations present in our
lattice data. We call the resulting corrections taste improvements, because
they improve the continuum extrapolation of our lattice data without, in
principle, modifying the continuum-limit value. Indeed, these corrections
vanish in that limit, as taste-breaking effects should.

\subsection*{Lattice artefacts: data vs models}

\begin{figure}[t]
    \centering
    \includegraphics[width=0.7\textwidth]{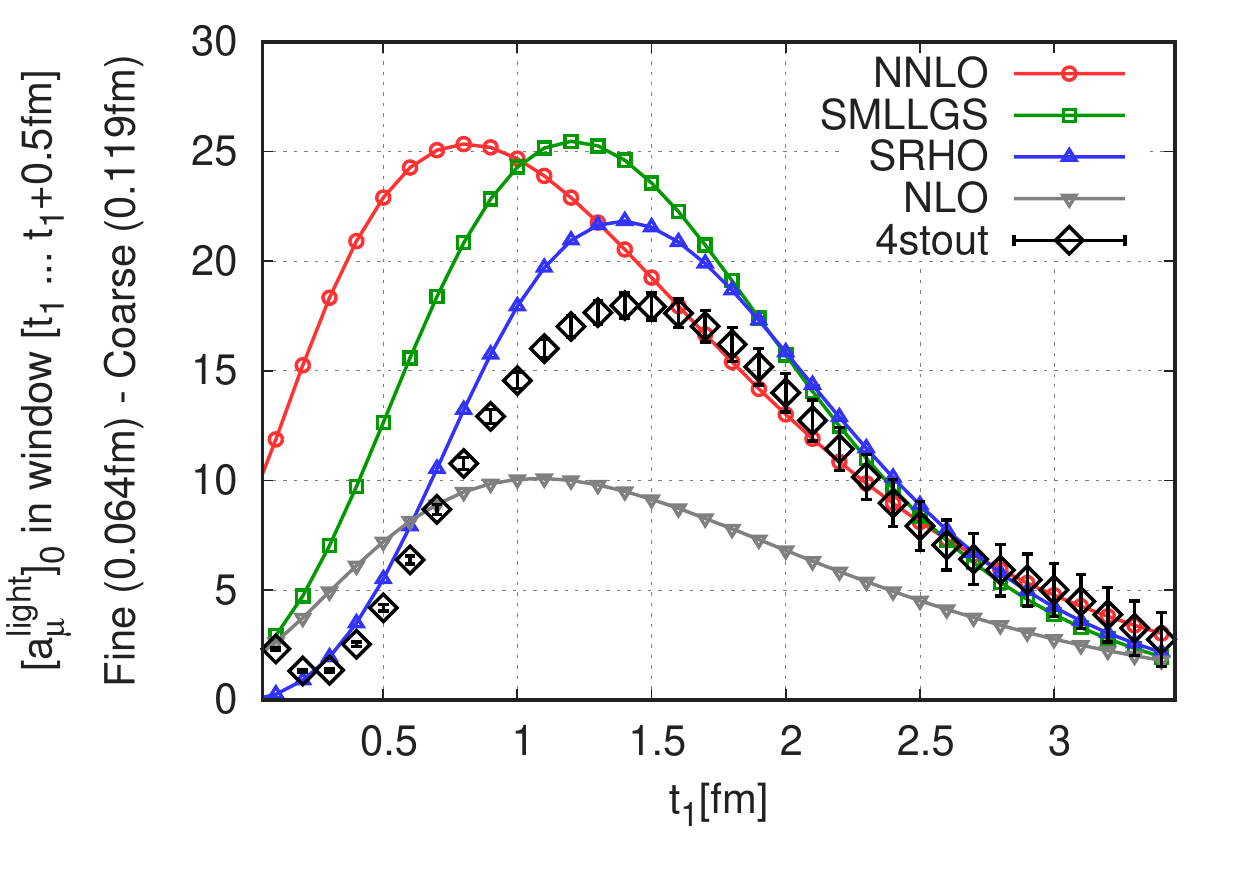}
    \caption
    {
	\label{fi:amuwsxx} $[a_\mu^\mathrm{light}]_0$ computed with a sliding
	window: the window starts at $t_1$ and ends $0.5\,\mathrm{fm}$ later. The plot
	shows the difference between a fine and a coarse lattice, the volumes
	are $L=6.14\,\mathrm{fm}$ and $L=6.67\,\mathrm{fm}$. The black squares with errors are
	obtained from the simulation. The colored curves are the predictions of
	NLO and NNLO staggered chiral perturbation theory, the SRHO and the
	SMLLGS models.  They are computed at the parameters (pion mass, taste
	violation, volume) of the simulations.
    }
\end{figure}

In Section~\ref{se:obs_fv}, we showed that the models NNLO, MLLGS and RHO
describe the long-distance physics associated with finite-volume effects, as
measured in our simulations. Here we find that they also describe the physics
associated with taste violations, at least at larger distances. This is
illustrated in Figure \ref{fi:amuwsxx}, where cutoff effects in the integrand
of $[a_\mu^\mathrm{light}]_0$ are plotted as a function of Euclidean time. More
specifically, we define the physical observable, obtained by convoluting the
integrand of $[a_\mu^\mathrm{light}]_0$ with the window function
$W(t;t_1,t_1+0.5\,\mathrm{fm})$ of Equation~\eqref{eq:win}, and consider the
difference in the value of this observable, obtained on a fine ($\beta=4.0126$)
and a coarse lattice ($\beta=3.7500$), at a sequence of $t_1$ separated by
$0.1\,\mathrm{fm}$. These are compared to the NLO SXPT, NNLO SXPT, SRHO and
SMLLGS predictions for this quantity, evaluated at the exact parameters of the
ensembles.

These hadronic models are not expected to describe short-distance, QCD
behavior: we exclude their application below $t=0.4\,\mathrm{fm}$. On the other
hand, all three models, as well as NLO SXPT, are expected to correctly describe
taste-breaking effects at large enough $t_1$, when $L$ is also large. As
Figure~\ref{fi:amuwsxx} shows, this is verified for $t_1\gsim 3.5\,\mathrm{fm}$
in our volumes. We now discuss in detail how well each model performs, in
comparison to our simulation results shown in Figure \ref{fi:amuwsxx}, as well
as the extent to which they can be trusted:
\begin{itemize}

    \item The SMLLGS, the SRHO and the NNLO taste improvements describe the
	numerical data very nicely for $t_1\gsim 2.0\,\mathrm{fm}$, fairly well
	for $t_1\gsim 1.0\,\mathrm{fm}$ and all the way down to $t_1\simeq
	0.4\,\mathrm{fm}$ in the case of SRHO. All three slightly overestimate
	the observed cutoff effects, the rho-meson based approach performing
	best, whereas NNLO displays a large deviation from the lattice results
	in the $t_1\le 0.8\,\mathrm{fm}$ region.

    \item SMLLGS and SRHO are based on similar physical input and perform nearly
	identically down to $t_1\simeq 1.8\,\mathrm{fm}$, confirming that SRHO
	correctly captures the contribution of the rho meson.  However, SMLLGS
	only accounts for the contributions of two-pion states up to the
	rho-meson mass, whereas SRHO includes the contributions of the
	full tower of two-pion states: the latter is expected to work down to
	shorter distances, which is what is observed in
	Figure~\ref{fi:amuwsxx}. Thus, in what follows we only consider SRHO,
	with the added confidence brought by its agreement with SMLLGS at
	Euclidean times $\gsim 1.8\,\mathrm{fm}$.

    \item NLO SXPT starts failing shortly below $t_1=3.5\,\mathrm{fm}$, as does
	the convergence of SXPT for determining taste violations, as measured
	by the distance between the NLO and NNLO curves. This failing of NLO
	SXPT is not surprising because, at that order, the effective theory has
	no information about the rho resonance that plays such an important
	role in the $I{=}J{=}1$ channel.  That information only enters at NNLO,
	through the low energy constant LEC $l_6$ of Equation~\eqref{eq:l6}.
	This LEC determines the short-distance contribution to the slope of the
	electromagnetic form factor of the pion at vanishing virtuality.
	Because NLO SXPT misses this important information about the rho, we
	choose not to include this improvement in any of our analyses of
	$a_\mu^\mathrm{light}$, even though this correction goes in the right
	direction, as Figure~\ref{fi:amuwsxx} shows.

    \item If SRHO is viewed as a resummation of SXPT, then the convergence of
	SXPT beyond NNLO appears to be quite good for determining taste
	violations above $t_1=1.3\,\mathrm{fm}$, as the SRHO and NNLO
	prediction agree fairly well in that region. Nevertheless, because NNLO
	XPT severely underestimates the hadronic contribution to vacuum
	polarization itself, we choose not to use it as a taste improvement in
	determining the central value of our lattice result for
	$a_\mu^\mathrm{light}$.

    \item The lattice results have a maximum at $t_1=1.4\,\mathrm{fm}$, as does
	the SRHO improvement, reinforcing our confidence that this model
	captures the relevant physics.

    \item Since our taste-improvement models are not expected to capture the
	relevant physics below $t=0.4\,\mathrm{fm}$, which is
	confirmed by Figure~\ref{fi:amuwsxx}, we use none in this region.

\end{itemize}

\subsection*{Taste-improvement procedure}

\begin{figure}[t]
    \centering
    \includegraphics[width=0.7\textwidth]{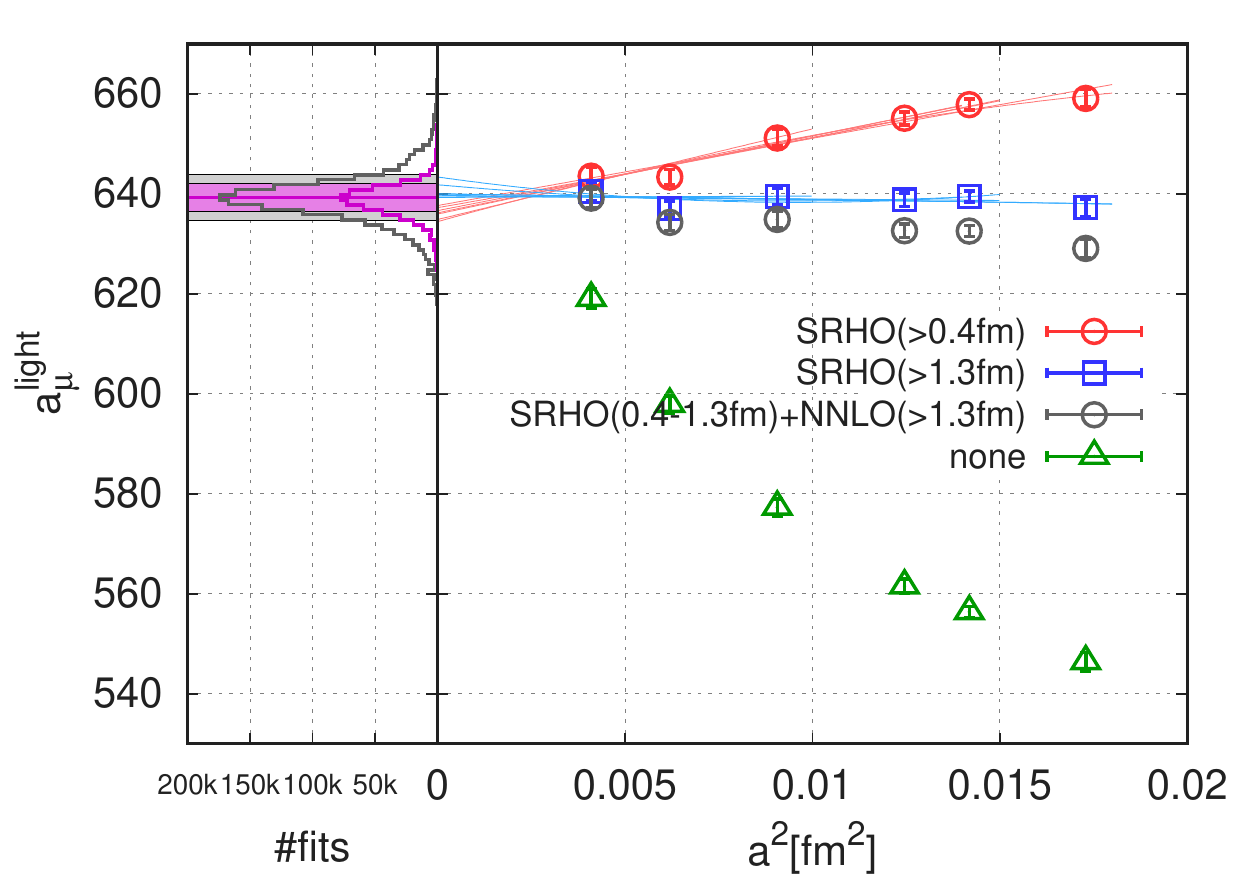}
    \caption
    {
	\label{fi:cont} Example continuum limits of $a_\mu^\mathrm{light}$.
	The light green triangles labeled none correspond to our lattice
	results with no taste improvement. The blue squares have undergone no
	taste improvement for $t< 1.3\,\mathrm{fm}$ and SRHO improvement above.
	The blue curves correspond to example continuum extrapolations of those
	improved data to polynomials in $a^2$, up to and including $a^4$. Note
	that extrapolations in $a^2\alpha_s(1/a)^n$, with $n=3$, are also
	considered in our final result. The red circles and curves are the same
	as the blue points, but correspond to SRHO taste improvement for $t\ge
	0.4\,\mathrm{fm}$ and none for smaller $t$. The darker grey circles
	correspond to results corrected with SRHO in the range
	$0.4{-}1.3\,\mathrm{fm}$, NNLO SXPT for larger $t$. The purple
	histogram results from the fits described in items 1)-5) of the text.
	The grey one from those of item 6). The purple band is the result
	described in 5) and the grey band includes the systematic error
	described in items 7) and 8).
    }
\end{figure}

The physics-based considerations and approximate agreement with the simulation
results, discussed in the previous subsection, lead us to apply the following
taste corrections to our simulations results for
$[a_\mu^\mathrm{light}]_0(L,T,a)$, before performing continuum extrapolations:
\begin{gather}
    \label{eq:cont_imp}
    \begin{aligned}[]
	[a_\mu^\mathrm{light}]_0(L,T,a) \to
	[a_\mu^\mathrm{light}]_0(L,T,a)
	&+\tfrac{10}{9}\left[a_{\mu,t \ge t_\mathrm{sep}}^\mathrm{RHO}(L_\mathrm{ref},T_\mathrm{ref})-a_{\mu,t \ge t_\mathrm{sep}}^\mathrm{SRHO}(L,T,a) \right],
    \end{aligned}
\end{gather}
with $t_\mathrm{sep}=0.4,0.7,1.0,1.3\,\mathrm{fm}$. The charge factor of
$(10/9)$ is required because the corrections computed in
Sections~\ref{se:obs_xpt}, \ref{se:obs_rho} and \ref{se:obs_llgs} correspond to
the $I{=}1$ contribution, not the light one. Note that by using
$L_\mathrm{ref}$ and $T_\mathrm{ref}$ in the above Equation, we are applying a
very small volume correction to interpolate all of our simulation results to
the same reference, four-volume so that they can be continuum extrapolated
together.  For the disconnected contribution, we apply the same improvement as
in Equation~\eqref{eq:cont_imp}, but with a charge factor of $(-1/9)$ instead
of $(10/9)$, as appropriate for this case.

The taste-improved data is then continuum-extrapolated using our standard fit
procedure, in the course of which isospin-breaking effects are also included
(Section \ref{se:res_fit}). For the error estimation we use the histogram
technique (Section \ref{se:res_err}). To estimate the systematic uncertainty of
our SRHO-model-based taste-improvement procedure we use NNLO SXPT. The
resulting error is added in quadrature to the systematic error from the
histogram. The central values and the detailed error budget of this analysis
can be found in Section \ref{se:res_amu}. 

The procedure is illustrated in Figure~\ref{fi:cont}, which shows the data sets
for $a_\mu^\mathrm{light}$ without and with taste improvements, as functions of
$a^2$. (See also Figure~\labelmaincont{} of the main paper, which zooms in on
the taste-improved, continuum extrapolations.) The SRHO improvement with
$t_\mathrm{sep}=0.4\,\mathrm{fm}$ are shown as red points, while blue points
correspond to $t_\mathrm{sep}=1.3\,\mathrm{fm}$. These plots already include
isospin-breaking contributions.

A more detailed account of the taste-improved, continuum extrapolation procedure
of $a_\mu^\mathrm{light}$ is as follows:
\begin{enumerate}

    \item For $0\le t\le 0.4\,\mathrm{fm}$, we consider only uncorrected
	lattice results.

    \item For $0.4\,\mathrm{fm}< t\le 1.3\,\mathrm{fm}$, two-pion
	taste-violating effects are clearly visible, but we do not know
	precisely where they become important. Thus, we consider
	$a_\mu^\mathrm{light}$ that is obtained without taste improvement in
	the window $[0.4\,\mathrm{fm},t_\mathrm{sep}]$ and with SRHO
	improvement in the window $[t_\mathrm{sep},1.3\,\mathrm{fm}]$, with
	$t_\mathrm{sep}=0.4, 0.7, 1.0, 1.3\,\mathrm{fm}$. 

    \item For $t>1.3\,\mathrm{fm}$, two-pion, taste violations must be present
	and are seen. It is reasonable to perform continuum extrapolations
	including these effects. Thus, in this range of Euclidean times, we
	only consider $a_\mu^\mathrm{light}$ with SRHO taste improvement.

    \item Even after taste improvement, there remain discretization effects
	that must be extrapolated away by taking a continuum limit. However,
	once nonlinear, taste-breaking effects have been corrected for, we expect
	the remaining continuum extrapolations to be mild and similar to those
	of quantities that do not suffer from these large, two-pion, taste
	violations. This is what is observed in Figure~\ref{fi:cont}. Thus, for
	these residual, continuum extrapolations, we consider lattice spacing
	behaviors that are low-order power expansions in $a^2\alpha_s(1/a)^n$,
	with $n=0,3$, as suggested in \cite{web:hvpsommer} based on the work
	\cite{Husung:2019ytz}.
  
    \item Items 1)-4) yield histograms, for $a_\mu^\mathrm{light}$ and
	$a_\mu^\mathrm{disc}$, from which we obtain our final central values,
	the final statistical errors and the systematic errors associated,
	amongst others, with the choice of the functional form for the residual
	continuum extrapolation, as explained in 4). With the other systematic
	errors that are detailed in Sections~\ref{se:res_fit} and
	\ref{se:res_err}, this represents over half-a-million different
	analysis procedures.

    \item The procedure described in 1)-5) does not take into account the
	systematic uncertainty associated with our choice of SRHO for taste
	improvement for $t>1.3\,\mathrm{fm}$. Since applying no taste
	improvement in that region is not an option, because of the
	nonlinearities introduced by two-pion, taste violations, we turn to
	NNLO SXPT, only as a means to estimate the uncertainty associated with
	this choice.  It is known that NNLO XPT describes the current-current
	correlator for $t\to\infty$ but that, as $t>1.3\,\mathrm{fm}$ is
	reduced, it provides only an unsaturated, lower bound on the
	correlator. Thus, we define this systematic uncertainty as ERR = (SRHO
	$-$ NNLO SXPT) for $t>1.3\,\mathrm{fm}$. An example of our lattice
	results with SRHO improvement between $t=0.4\,\mathrm{fm}$ and
	$t=1.3\,\mathrm{fm}$ and NNLO SXPT improvement above are shown as grey
	points in Figure~\ref{fi:cont}.  Then, we perform the same
	over-half-a-million fits as in item 4), but with SRHO, SRHO-ERR and
	SRHO+ERR improvements. This yields the grey histogram in
	Figure~\ref{fi:cont}.

    \item From this histogram we extract the contribution which comes from the
	variation in the improvement model from SRHO-ERR to SRHO+ERR. Since the
	histogram may have slightly longer tails than would a Gaussian, to be
	conservative we take the 1-sigma error, associated with choosing the
	SRHO model for taste improvement above $1.3\,\mathrm{fm}$, to be the
	largest of the 68\% confidence interval or half the 95\% confidence
	interval. For the $a_\mu^\mathrm{light}$ shown in Figure~\ref{fi:cont},
	it is the latter which is largest. For $a_\mu^\mathrm{disc}$, it is the
	former.

    \item To the systematic error determined in 5), we add the 1-sigma
	systematic error computed in 7). Note that this second error
	corresponds to the full difference of using SRHO versus SRHO $-$ ERR =
	NNLO SXPT for $t>1.3\,\mathrm{fm}$. Thus, our procedure guarantees the
	our central values for $a_\mu^\mathrm{light}$ and $a_\mu^\mathrm{disc}$
	are not biased by NNLO SXPT and that the error is given by the full
	difference between SRHO and NNLO SXPT.

\end{enumerate}

\begin{figure}[t]
    \centering
    \includegraphics[width=0.7\textwidth]{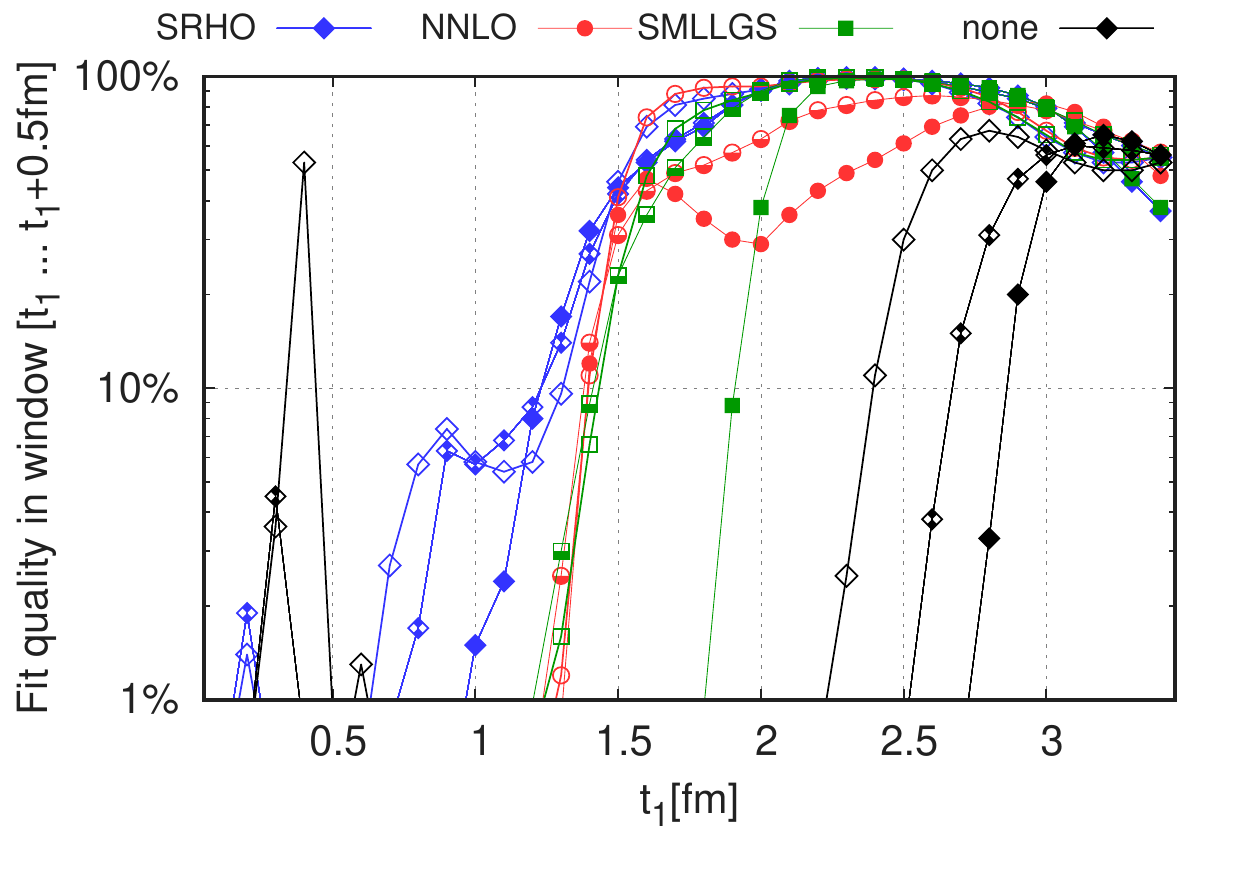}
    \caption
    {
	\label{fi:fitqsxx} Fit qualities of continuum-extrapolation fits for
	$[a_\mu^\mathrm{light}]_0$ as a function of $t_1$, where the propagator
	is restricted to a window $[t_1,t_1+0.5\,\mathrm{fm}$$]$. Different
	colors correspond to different taste-improvement procedures: none, NNLO
	staggered chiral perturbation theory and the SMLLGS and SRHO models.
	Different symbols with the same color correspond to different number of
	the coarse lattice spacings ignored in the fit:
	filled/half-filled/empty for zero/one/two.
    }
\end{figure}

We can also verify, {\em a posteriori}, that the above physics-motivated
choices are reasonable. Because long-distance, taste violations introduce
nonlinearities in $a^2$ (see Section~\ref{se:act_4stout} and
Figure~\ref{fi:tavi}, in particular), we expect that taste improvement will
reduce these nonlinearities and increase the asymptotic scaling regime by
making the $a^2$ dependence more linear. This can be monitored by looking at
the goodnesses of fit of linear continuum extrapolations. Figure
\ref{fi:fitqsxx} illustrates the situation. In order to determine where the
asymptotic scaling regime starts, we also leave out different numbers of coarse
lattice spacings from the continuum extrapolation. As the figure indicates, our
physically motivated choices for the window boundaries lead to improved
linearity, suggesting that our corrected results are in the asymptotic scaling
regime. An almost equivalent way to see this is to attach Akaike Information
Criterion (AIC) weights to the continuum extrapolations, which automatically
selects the models that have good fit qualities. Here, however, instead of
using this somewhat black-box approach, we determine the boundaries of the
windows based on the physical arguments made above and ascribe equal weights to
the results obtained by varying these boundaries. Moreover, we allow for
nonlinear continuum extrapolations of the corrected results, when such
corrections are required by the data.

\subsection*{Isoscalar contribution}

An important check of our understanding, that the leading discretization
effects in $a_\mu^\mathrm{light}$ and $a_\mu^\mathrm{disc}$ are due to
taste-breaking effects in the two-pion spectrum, is provided by the study of
the $I=0$ contribution to $a_\mu$, as suggested by arguments made in
\cite{Gerardin:2019rua}. In this subsection, that serves illustrative purposes,
we work with the isospin-symmetric data sets for simplicity and the subscript
``0'' will be omitted from the observables.  For continuum extrapolations we
use Type-II fit forms without isospin-breaking and with a fixed value of
$w_0=0.1724$~fm. As explained in Section~\ref{se:res_fit}, these fits differ
slightly from the Type-I fits used to obtain our final result for $a_\mu$.

The isoscalar contribution to $a_\mu$,
\begin{equation}
  \label{eq:amuIeq0}
        a_\mu^{I=0}\equiv 
        \frac1{10}a_\mu^\mathrm{light} +a_\mu^\mathrm{disc}+\cdots
  \ ,
\end{equation}
where the ellipsis stands for the quark-connected contributions of the more
massive $s$, $c$, \ldots\ quarks, receives no two-pion contributions: it starts
with three pions. Moreover, as shown in Section~\ref{se:obs_xpt}, the
three-pion contributions only appear at $O(p^{10})$, suggesting that their
taste-breaking effects should be very small. Thus, if our understanding of
discretization errors in $a_\mu^\mathrm{light}$ and
$a_\mu^\mathrm{disc}$ is correct, the large taste-breaking corrections
observed in those quantities must be largely absent from $a_\mu^{I=0}$. As
a consequence, we expect the continuum extrapolation of $a_\mu^{I=0}$ to be
much milder.

\begin{figure}[t]
    \centering
    \includegraphics[width=0.7\textwidth]{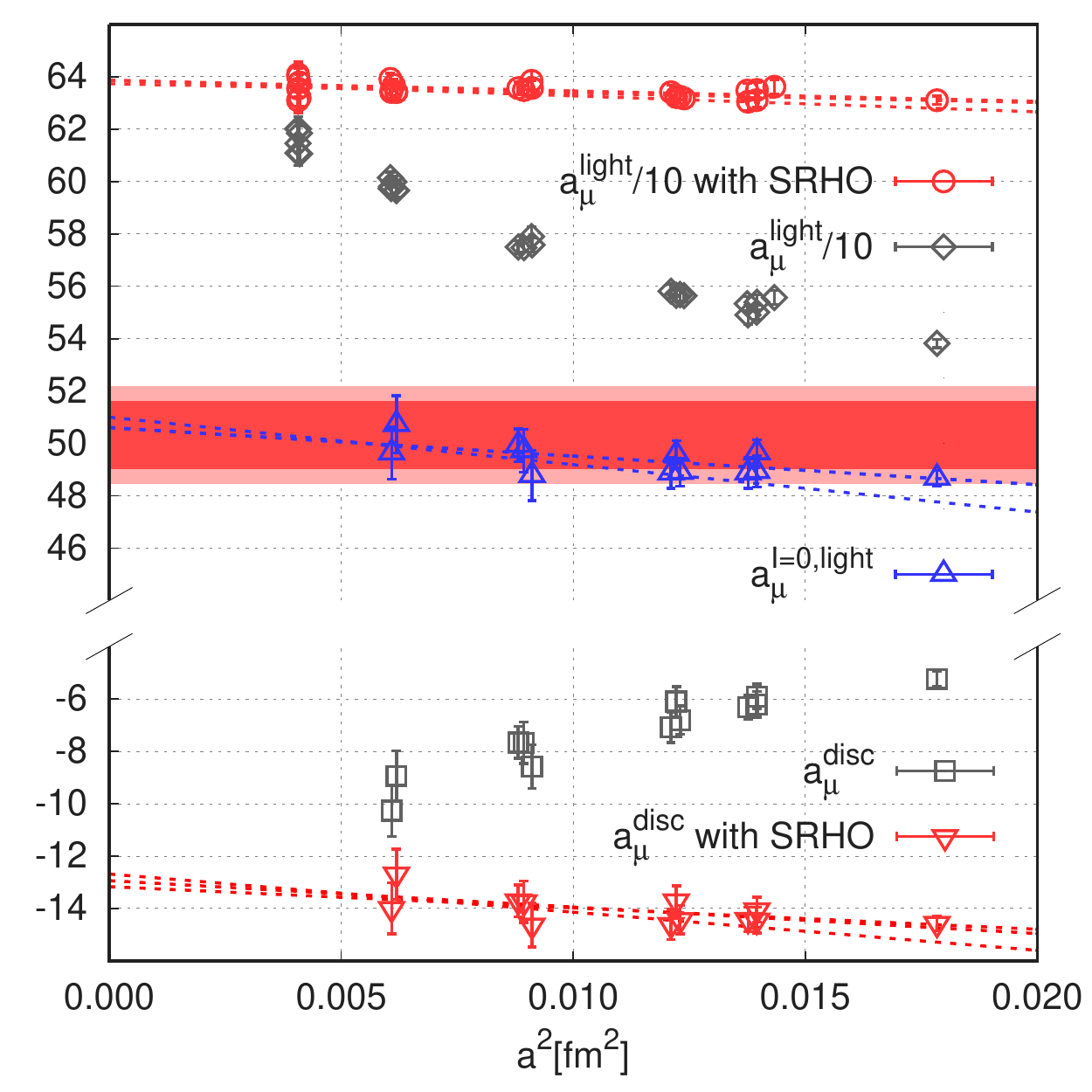}
    \caption
    {
	\label{fi:ieq0} Comparison of the continuum extrapolation of
	$a_\mu^{I=0,\mathrm{light}}$ to those of $a_\mu^\mathrm{light}$ and
	$a_\mu^\mathrm{disc}$. The upper set of grey points corresponds to our
	uncorrected results for $\frac1{10}a_\mu^\mathrm{light}$. The upper red
	ones are these same results with our standard SRHO taste improvement.
	They have a much milder continuum limit that exhibits none of the
	nonlinear behavior of the grey points. The red curves show typical
	examples of our illustrative, Type II continuum extrapolations of those
	points. The lower set of grey and red points and curves are the
	same quantities, but for $a_\mu^\mathrm{disc}$. Combining the results
	from the two, individual, continuum extrapolations of
	$\frac1{10}a_\mu^\mathrm{light}$ and $a_\mu^\mathrm{disc}$, according
	to Equation~\eqref{eq:amuIeq0}, gives the result with statistical
	errors illustrated by the red band, and with combined statistical and
	systematic errors, by the broader pink band. The blue points correspond
	to our results for $a_\mu^{I=0,\mathrm{light}}$, for each of our
	simulations, and are obtained by combining the two sets of grey points,
	according to Equation~\eqref{eq:amuIeq0}. As these blue points show,
	the resulting continuum-limit behavior of $a_\mu^{I=0,\mathrm{light}}$
	is much milder than that of either the uncorrected
	$a_\mu^\mathrm{light}$ or $a_\mu^\mathrm{disc}$, and shows none of the
	curvature exhibited by them. This behavior resembles much more that of
	the taste-improved, red points. Moreover, all of the blue points,
	including typical continuum extrapolations drawn as blue lines, lie
	within the bands.  This suggests that our taste improvements neither
	bias the central values of our continuum extrapolated
	$a_\mu^\mathrm{light}$ and $a_\mu^\mathrm{disc}$, nor do they lead to
	an underestimate of uncertainties.
    }
\end{figure}

That is exactly what is shown in Figure \ref{fi:ieq0}, where we compare the
$a^2$ behavior of $a_\mu^{I=0}$ to those of $a_\mu^\mathrm{light}$ and
$a_\mu^\mathrm{disc}$.  We focus on the sub-contribution,
$a_\mu^{I=0,\mathrm{light}}$, that is obtained by keeping only the first two
terms in Equation~\eqref{eq:amuIeq0}. This restriction only makes the check
stronger, because taste-breaking effects are larger, in relative terms, in
$a_\mu^{I=0,\mathrm{light}}$ than in $a_\mu^{I=0}$. As the figure shows,
$a_\mu^{I=0,\mathrm{light}}$ has a much milder continuum extrapolation than
either $a_\mu^\mathrm{light}$ or $a_\mu^\mathrm{disc}$, displaying none of the
curvature present in the latter two. In fact, this continuum extrapolation is
very similar to that of the connected-strange contribution, whose leading
contributions come from two-kaon states, not two-pion ones. The
connected-strange contribution can be viewed as having discretization errors
which are typical of low-energy, light-quark quantities that are not (heavily)
affected by the distortion of the pion spectrum.

In Figure \ref{fi:ieq0} we show a band that corresponds to the value of
$a_\mu^{I=0,\mathrm{light}}$ obtained by combining the individual, continuum
limits of $a_\mu^\mathrm{light}$ and $a_\mu^\mathrm{disc}$. The latter are
extrapolated to the continuum limit after implementing our standard taste
improvements. The fact that all of the $a_\mu^{I=0,\mathrm{light}}$ data points
are contained within the band is yet another confirmation that our treatment of
taste-breaking effects is correct and that our taste-improved, continuum
extrapolations are unbiased.

    \clearpage
    \newpage
    \section{Global fit procedure}
\label{se:res_fit}

In this section we describe the procedure that is used to obtain the physical
values of $a_\mu$. Two types of fit functions are introduced, Type-I and
Type-II, which differ in their input parameters. In Type-I fits these are
experimentally measurable quantities. In Type-II fits the inputs are
observables that are not directly accessible in experiments. Type-II fits are
needed to implement the separation of observables into isospin-symmetric
contribution and isospin-breaking corrections that is described in Section
\ref{se:obs_split}. We close the section by presenting an alternative fit
procedure.

\subsection*{Type-I fits}

In the case of Type-I fits we parameterize the quark-mass and electric-charge
dependence of an observable $Y$ around the physical point and for small isospin
breaking with a linear function $f$:
\begin{gather}
    \label{eq:fitI}
    Y= f( \{X\}; A, B, \dots )\equiv A + B X_l + C X_s + D X_{\delta m} + E X_{vv} + F X_{vs} + G X_{ss}\ .
\end{gather}
The $X_l, X_s,\dots$ are called independent variables of the fit function,
though they can be (statistically) correlated. The $A,B,\dots$ are called the
fit coefficients.  The Type-I fits have the feature that their independent
variables $\{X\}$ are quantities that are experimentally measurable.  Here the
$X_l$ and $X_s$ variables describe the deviation from the physical light and
strange mass
\begin{gather}
    \label{eq:fitx1}
	X_l= \frac{M_{\pi_0}^2}{M_{\Omega}^2} - \left[\frac{M_{\pi_0}^2}{M_{\Omega}^2}\right]_*\ ,\qquad
	X_s= \frac{M_{K_\chi}^2 }{M_{\Omega}^2} - \left[\frac{M_{K_\chi}^2 }{M_{\Omega}^2}\right]_*
\end{gather}
with $*$ denoting the experimental value. No higher orders in $X_l$ or $X_s$
are needed, since we work close to the physical point.  The remaining $X$
variables measure the distance from the isospin-symmetric limit
\begin{gather}
    X_{\delta m}= \frac{\Delta M_K^2 }{M_{\Omega}^2}\ ,\qquad
    X_{vv}= e_v^2\ , \qquad
    X_{vs}= e_ve_s\ , \qquad
    X_{ss}= e_s^2\ ,
\end{gather}
where $e_v$ and $e_s$ are the valence and sea electric charges, respectively.
Higher-order isospin-breaking terms are not considered in this work. The meson
masses are defined as
\begin{align}
    \begin{aligned}
	M_{K_\chi}^2    &\equiv \tfrac{1}{2} \left(M_{K_0}^2 + M_{K_+}^2 - M_{\pi_+}^2 \right)\ ,\\
	\Delta M_K^2    &\equiv M_{K_0}^2 - M_{K_+}^2\ .
    \end{aligned}
\end{align}
In case of the neutral pion we use the combination
\begin{align}
    M_{\pi_\chi}^2  &\equiv \tfrac{1}{2} \left(M_{uu}^2 + M_{dd}^2\right)\ ,
\end{align}
where the masses of mesons $uu$ and $dd$ are obtained from contractions
involving connected diagrams only. It can be shown in partially-quenched
chiral perturbation theory coupled to photons \cite{Bijnens:2006mk}, that
$M_{\pi_0} = M_{\pi_\chi}$ up to terms that are second order in isospin
breaking.

The coefficients $A,B,\dots$ in Equation \eqref{eq:fitI} are specific to the
observable $Y$. They can depend on the lattice spacing, and also on the $X$
variables defined above, in particular we use:
\begin{equation}
    \label{eq:abcdefg}
    \begin{aligned}
	A&= A_0 + A_2 \left[a^2\alpha_s(1/a)^n\right] + A_4 \left[ a^2\alpha_s(1/a)^n\right]^2 + A_6 \left[ a^2\alpha_s(1/a)^n\right]^3,\\
	B&= B_0 + B_2 a^2,\\
	C&= C_0 + C_2 a^2,\\
	D&= D_0 + D_2 a^2 + D_4 a^4 + D_l X_l + D_s X_s,\\
	E&= E_0 + E_2 a^2 + E_4 a^4 + E_l X_l + E_s X_s,\\
	F&= F_0 + F_2 a^2,\\
	G&= G_0 + G_2 a^2.
    \end{aligned}
\end{equation}

The lattice spacing $a$ is defined in a so-called mass-dependent, scale-setting
scheme: for any ensemble, $a$ is given as the ratio of the $\Omega$ mass
measured in lattice units divided by its experimental value. For the strong
coupling, we use its four-flavor, $\overline{\mathrm{MS}}$ value at scale
$1/a$, ie.\ $\alpha_s(1/a)$. We determine this value from the world average
value of $\alpha_s(M_Z)$ \cite{Zyla:2020zbs}, by running the latter down from
$M_Z$ to $1/a$ in five-loop perturbation theory~\cite{Herzog:2017ohr}, taking
into account four-loop threshold corrections \cite{Schroder:2005hy} at the
$b$-quark mass given in \cite{Zyla:2020zbs}.

For observables, which give a large contribution to the final result, we
change the power of the strong coupling constant in the lattice-spacing
dependence of the $A$ coefficient from the commonly used $n=0$ to $n=3$, a
value suggested in \cite{web:hvpsommer} based on the work
\cite{Husung:2019ytz}. The $n=0$ case corresponds to the usual polynomial
expansion in $a^2$. The $n=3$ case, with an $a^2\alpha_s(1/a)^3$ behavior,
describes well the lattice-spacing dependence of the taste-violation in the
pion spectrum in our range of lattice spacings. The change related to the
choice of $n$ is part of our systematic error.  In the $A$ coefficient, the
dependence on $a^2\alpha_s(1/a)^n$ is taken to be linear or quadratic and, in
some cases, even cubic, if required by the data.  In all other coefficients the
lattice spacing dependence is assumed to be a function of $a^2$ only.  In the
$D$ and $E$ coefficients, up to quadratic dependencies are used, in all other
cases only a linear one is needed.  Depending on the fit qualities, some of
these parameters will be set to zero.

The parameters $A_0,A_2,A_4,B_0,\dots$ can be determined by performing a fit
for sufficiently many ensembles that scatter around the physical point.  The
physical value of $Y$ can then be obtained from this fit as
\begin{gather}
    Y_*= A_0 + D_0 \left[X_{\delta m}\right]_* + (E_0+F_0+G_0)\cdot e_*^2\ ,
\end{gather}
ie. by setting the independent variables $X$ to their physical values,
including setting the valence and sea electric charges to the physical value of the
coupling $e_*$. The value $e_*$ is related to the experimental value of the
fine structure constant as $e_*=\sqrt{4\pi\alpha_*}$. This choice is valid up
to second order in isospin-breaking.

As described in Section \ref{se:obs_gen}, isospin-breaking corrections are
obtained by measuring derivatives with respect to the $\delta m$, $e_s$ and
$e_v$ parameters. These can be incorporated into the above procedure by
deriving a system of coupled equations: one by taking Equation \eqref{eq:fitI}
at the isospin-symmetric point, and the other four by applying the isospin
breaking derivatives, see Equations \eqref{eq:ibder1} and \eqref{eq:ibder2}. We
then find the following five equations:
\begin{align}
    \label{eq:fiveeq}
    \begin{aligned}[]
	[Y]_0&= [A + BX_l +C X_s]_0\\
	[Y]'_m&= [D X_{\delta m}]'_m\\
	[Y]''_{20}&= \left[A + BX_l + CX_s + DX_{\delta m}\right]''_{20} + [E]_0\\
	[Y]''_{11}&= \left[A + BX_l + CX_s + DX_{\delta m}\right]''_{11} + [F]_0\\
	[Y]''_{02}&= \left[A + BX_l + CX_s + DX_{\delta m}\right]''_{02} + [G]_0
    \end{aligned}
\end{align}
where various isospin components of the coefficients $A,B,\dots$ have to be included, eg. the isospin symmetric value of $E$ is given by:
\begin{gather}
    [E]_0= E_0 + E_2 [a^2]_0 + E_4 [a^4]_0 + E_l [X_l]_0 + E_s [X_s]_0\ .
\end{gather}
The first line in \eqref{eq:fiveeq} parameterizes the isospin-symmetric data,
and is the only equation that depends on the $A_0$ parameter. The next equation
describes strong-isospin-breaking, where the electromagnetic coefficients
$E,F,G$ trivially drop out. $B$ and $C$ are also absent here, since they depend
symmetrically on the $u$ and $d$ quarks.  This equation is the main constraint
for $D$. The final three equations are the electric derivatives; they constrain
the $E,F$ and $G$ coefficients.

Note that the derivatives in Equation \eqref{eq:fiveeq} are with
respect to the bare parameters. The strong-isospin-breaking derivative
$[\dots]'_m$ defines a renormalized observable, but the electric charge
derivatives do not. This is due to the fact that the electric charge changes the
running of the quark masses and the lattice spacing. However, differences like
\begin{gather}
    \label{eq:q20ren}
    \left[ Y \right]''_{20}- \left[A + BX_l +CX_s +DX_{\delta m}\right]''_{20},
\end{gather}
which actually appear in \eqref{eq:fiveeq}, are free of divergences. When
preparing plots to illustrate the continuum extrapolation, the electric
derivatives will always refer to such renormalized combinations.

\subsection*{Type-II fits}

We introduce a second type of parametrization, called Type-II, in order to
obtain the isospin decomposition described in Section \ref{se:obs_split}.
Type-II fits use the $w_0$-scale for scale setting and are defined through:
\begin{gather}
    \label{eq:fitII}
	Y= f(\{\tilde{X}\}; \tilde{A},\tilde{B},\dots)\equiv
	\tilde{A} +
	\tilde{B} \tilde{X}_l +
	\tilde{C} \tilde{X}_s +
	\tilde{D} \tilde{X}_{\delta m} +
	\tilde{E} \tilde{X}_{vv} +
	\tilde{F} \tilde{X}_{vs} +
	\tilde{G} \tilde{X}_{ss}\ ,
\end{gather}
where the independent variables of the fit function are defined as
\begin{gather}
    \begin{gathered}
	\tilde{X}_l= M_{\pi_\chi}^2 w_0^2 - [M_{\pi_\chi}^2 w_0^2]_*\ ,\qquad
	\tilde{X}_s= M_{ss}^2 w_0^2 - [M_{ss}^2 w_0^2]_*\ ,\\
	\tilde{X}_{\delta m}= \Delta M^2 w_0^2\ ,\qquad
	\tilde{X}_{vv}= X_{vv}\ ,\qquad
	\tilde{X}_{vs}= X_{vs}\ ,\qquad
	\tilde{X}_{ss}= X_{ss}\ ,
    \end{gathered}
\end{gather}
with $\Delta M^2=M_{dd}^2-M_{uu}^2$. Some of the $\tilde{X}$ variables contain
$w_0$, $M_{ss}$ and $\Delta M^2$, that cannot be measured experimentally. The
physical values of these quantities have to be determined from a Type-I fit of
Equation \eqref{eq:fitI} first.  $\tilde{A}$, $\tilde{B}$,
\dots in general depend on hadron masses and on the lattice spacing,
analogously to the dependencies in Equation \eqref{eq:abcdefg}. Here the
lattice spacing is defined through $w_0$: it is the physical value of $w_0$
divided by the one measured in lattice units. The fit procedure is also
completely analogous to the one described above, including the coupled
equations for the different isospin components.  The isospin decomposition can
be obtained from the Type-II fit coefficients as
\begin{gather}
    \label{eq:xiso}
    [Y]_\mathrm{iso}= \tilde{A}_0,\quad
    [Y]_\mathrm{sib}= \tilde{D}_0 [\Delta M^2 w_0^2]_*,\quad
    [Y]_\mathrm{qed}= (\tilde{E}_0+\tilde{F}_0+\tilde{G}_0)\cdot e_*^2.
\end{gather}
One can also decompose the electromagnetic contribution further to valence-valence,
valence-sea and sea-sea parts:
\begin{gather}
    \label{eq:xiso2}
    [Y]_\mathrm{qed-vv}= \tilde{E}_0e_*^2,\quad
    [Y]_\mathrm{qed-sv}= \tilde{F}_0e_*^2,\quad
    [Y]_\mathrm{qed-ss}= \tilde{G}_0e_*^2.
\end{gather}
The two fit types, Type-I in Equation \eqref{eq:fitI} and Type-II in Equation
\eqref{eq:fitII} have to yield the same physical value $Y_*$ within error bars.
This was always the case for the observables considered here.  Later, when we
discuss the fits, it will be obvious from the text which parametrization we are
working with, so we drop the $\ \tilde{}\ $ from the coefficients of the
Type-II fits for simplicity.

\subsection*{Correlations}

In both parametrizations we have to work with a system of equations such as
\eqref{eq:fiveeq}, where the unknown parameters are contained in $A,B,\dots$.
To obtain these we perform a fit taking $[Y]_0$ and the isospin derivatives
from several ensembles. The $[Y]_0,[Y]'_m$ and $[Y]''_{20}$ components are
measured on the same $L\approx 6$~fm ensembles of Table \ref{ta:4stout}, and
they are therefore correlated.  One also has to take into account the
correlation between the sea quark derivatives $[Y]''_{11}$ and $[Y]''_{02}$
that are measured on the $L\approx 3$~fm ensembles of Table \ref{ta:dynqed}.
These correlations have to be properly included in the fit.  Also, we have to
take into account the correlation of $Y$ and the independent variables $\{X\}$,
including the lattice spacing. Specifically we compute and minimize the
following function to determine the fit parameters $A_0, \dots$:
\begin{gather}
    \chi^2= \sum_{i,j} ( Y_i - f_i )\ {\mathrm{Cov}}^{-1}_{ij}\ ( Y_j - f_j ).
\end{gather}
Here the sums run over all ensembles and $Y_i$ ($f_i$) are the values of the
observable (function) on ensemble $i$. The matrix $\mathrm{Cov}_{ij}$ is the
statistical covariance of the residuals $Y_i-f_i$, computed as
\begin{gather}
    \mathrm{Cov}_{ij}=
    \overline
    {
	\left[ (Y_i - f_i) - \overline{(Y_i-f_i)} \right]
	\left[ (Y_j - f_j) - \overline{(Y_j-f_j)} \right]
    }\ ,
\end{gather}
where we denote the statistical average with an overline. Using
the jackknife samples this can be obtained as:
\begin{gather}
    \label{eq:covij}
    \mathrm{Cov}_{ij}= \frac{N_J-1}{N_J}\sum_{J=1}^{N_J} \left[
	\left( Y^{(J)}_i - f_i^{(J)}\right) -
	\left( Y^{(0)}_i - f_i^{(0)}\right)
	\right]
    \Big[ \dots i\to j\dots \Big],
\end{gather}
where an upper index $(J)$ means that the quantity is computed on the $J$-th
jackknife sample and $J=0$ stands for the average over all jackknife samples.
The minimization of the $\chi^2$-function yields non-linear equations for the
parameters, since the $\mathrm{Cov}$ matrix depends on them too.  To solve the
minimization problem numerically, we first guess the minimum by ignoring the
parameter dependence of the $\mathrm{Cov}$ matrix. In all cases this was
already a good starting point, which is related to the fact, that the errors on
$Y$ are typically much larger than on $X$.  This guessing can be iterated and
after a few iterations we switch to Newton's method to accelerate the
convergence.

\subsection*{Alternative fit procedure}

In addition to the previously described fit procedure we also use an
alternative approach, in which isospin corrections are included in a
different way. The idea is to use the fit function, eg. the Type-I function in
Equation \eqref{eq:fitI}, directly without working with the isospin breaking
derivatives of that function. For this purpose we create new, ``virtual''
ensembles in addition to the already existing isospin-symmetric ones.  These
virtual ensembles have an isospin breaking with one or more of the $e_s$,
$e_v$ and $(m_d-m_u)/m_l$ parameters set to non-vanishing values, close but
not necessarily exactly to their physical values. The observables on the
virtual ensembles are computed using the isospin-symmetric values and isospin
breaking derivatives measured on the original ensembles. For the global fit we
use the original ensembles together with these newly created ones. Since the
virtual ensembles were created from the original isospin-symmetric ensembles,
there are strong correlations between them.  Computing the covariance matrix
is similar to Equation \eqref{eq:covij}, but now the indices $i,j$ run over
all ensembles, including the newly created ones.

We used this technique, with the Type-I fit function, to compute the quantities
$w_0$, $M_{ss}^2$ and $\Delta M^2$. In all three cases the results had similar
uncertainties as the original approach, presented earlier in this section, and
for which the results can be found in Section \ref{se:res_w0etal}. Also, the
central values agreed within their systematic uncertainty in the two
approaches.

    \section{Uncertainty estimation}
\label{se:res_err}

\begin{figure}[t]
    \centering
    \includegraphics[width=0.7\textwidth]{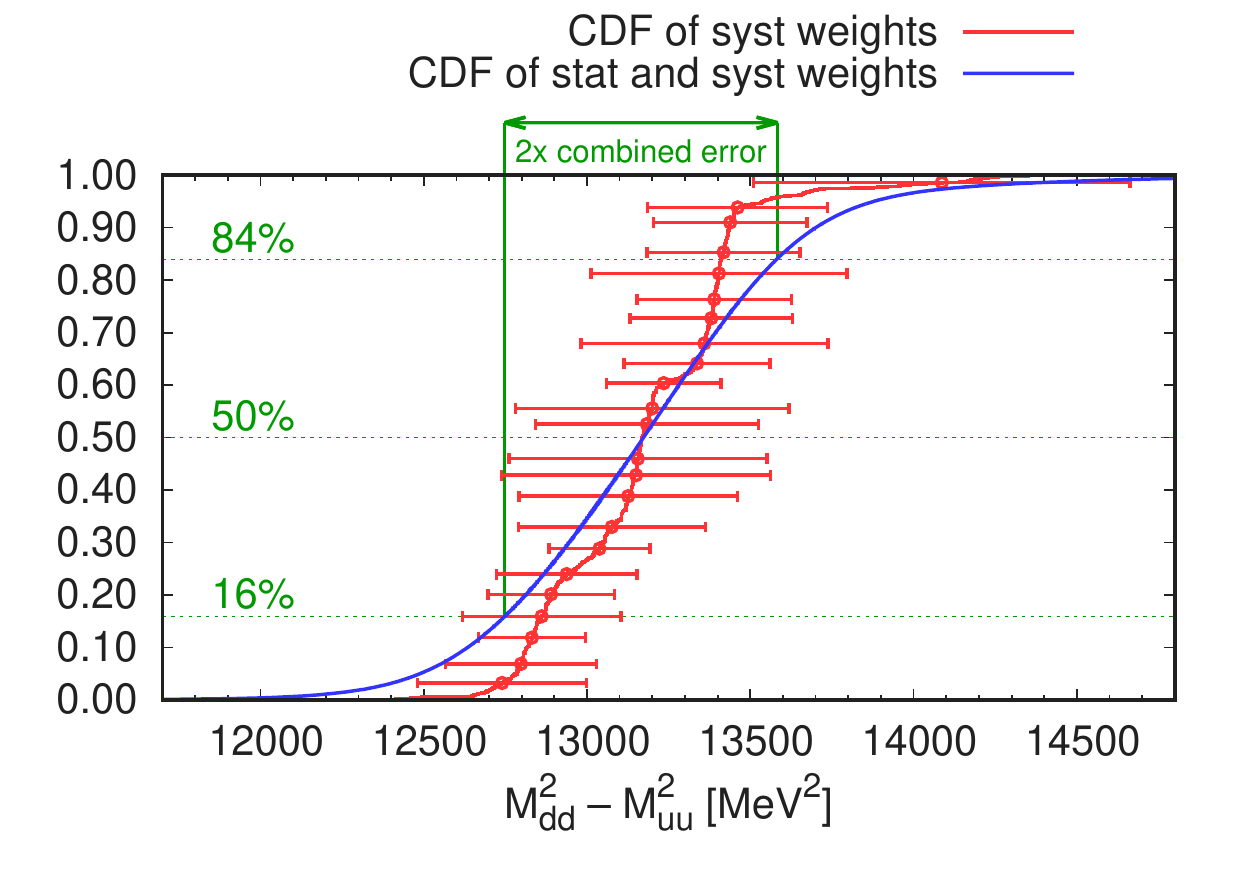}
    \caption{
	\label{fi:error} Cumulative distribution functions (CDF) of $\Delta
	M^2=M_{dd}^2-M_{uu}^2$ values. The red curve shows the CDF of about 3k
	different analyses obtained from their corresponding AIC weights. At a
	given point along the curve the horizontal band is the statistical
	error of the analysis at that point. The blue curve shows a CDF
	corresponding to a combined distribution of the AIC weights and the
	Gaussian distributions of the statistical errors. This latter curve is obtained from Equation
	\eqref{eq:cdf} with $\lambda=1$. We use the median and the width of this curve to
	define the central value and the total error. For the separation of the total
	error into a statistical and a systematic part we also use the CDF with $\lambda=2$.
    }
\end{figure}

\subsection*{Calculation of statistical errors}

We use the jackknife method to calculate the statistical errors. To suppress
the auto-correlation between data from subsequent configurations we introduce a
blocking procedure. It is very convenient to use an equal number of blocks for
all ensembles. In this work we use $N_J=48$ blocks. With this choice we have
typically 100 trajectories or more in a block, which is much larger than the
autocorrelation time of the topological charge (around $20$ on our finest
ensembles). For the blocks we apply the delete-one principle, resulting in
$N_J$ jackknife samples plus the full sample.

We keep the correlation between all quantities calculated from the same
ensemble. For simplicity, we match the jackknife samples between ensembles,
too. This means that each global fit using all ensembles at the same time is
performed $N_J+1$ times. The covariance matrix is calculated only for the main
sample, there is no need for the errors on the correlations here.

\subsection*{Estimation of systematic errors}

Systematic errors are notoriously challenging to estimate. Here we are
fortunate to have, at our disposal, thirty-one, large-scale simulations
performed with quark masses straddling their physical values, at six values of
the lattice spacing ranging from $0.064$ to $0.1315\,\mathrm{fm}$, in large
volumes ranging from $9\times 6^3\,\mathrm{fm}^4$ to $(11\,\mathrm{fm})^4$,
including strong-isospin-breaking and electromagnetic effects. This allows us
to eliminate the leading, sytematic uncertainties of previous calculations. It
also allows us to obtain a reasonable estimate of the remaining systematics,
associated mainly with the neglected higher-order terms in the expansions or
models used to perform the necessary continuum and infinite-volume
extrapolations. The goal here is not to have an error that covers even very
unlikely scenarios that only marginally agree with our lattice data. Rather, we
aim to investigate all reasonable descriptions of this data and, as long as the
ensuing results are close to being Gaussian distributed, to determine the
standard, 68\% confidence interval of these results. This guarantees that
meaningful comparisons, with experimental results for instance, can be made in
terms of differences measured in units of combined standard deviations.

Throughout the chain of analyses many choices are made, ranging from fit
windows and mass extractions, through the various Ans\"atze for the large-time
behavior of the $JJ$ correlator, to the various parametrizations of a global
fit. We call the global fit with a specific set of such choices an
analysis.  Each choice of $k$ possible options introduces a factor $k$ in the
total number of analyses, which already includes a factor of $N_J+1$
corresponding to the statistical sampling. Here we describe the procedure to
derive a systematic error coming from the ambiguity of these choices. We follow
closely the strategy introduced by us in \cite{Borsanyi:2014jba} and also
extend it by a new method to separate statistical and systematic errors.

For a target observable $y$ we build a histogram from the different analyses.
Each analysis gets a weight assigned. This weight is given by the Akaike
Information Criterion (AIC). The AIC is derived from the Kullback-Leibler
divergence, which measures the distance of the fit function from the true
distribution of the points. A derivation of the formula can be found in Section
11 of \cite{Borsanyi:2014jba}. Here we use a slightly modified version of the
AIC:
\begin{gather}
    \label{eq:aic}
    \mathrm{AIC} \sim \exp\left[-\tfrac{1}{2}\left(\chi^2 + 2 n_{\rm par} - n_{\rm data}\right)\right],
\end{gather}
where the $\chi^2$, the number of fit parameters $n_{\rm par}$ and the number
of data points $n_{\rm data}$ describe the global fit.  The first two terms
in the exponent correspond to the standard AIC, the last term is introduced to
weight fits with different number of ensembles; this happens when we apply
cuts in the lattice spacing.  It can be derived from the Kullback-Leibler
divergence and arises from the first term in Equation (S41) of
\cite{Borsanyi:2014jba}.  In case of normally distributed errors this term can
be computed and one obtains $\frac{1}{2}n_{\rm data}$, which then leads to Equation
\eqref{eq:aic} that we use in this paper.

The analyses differing only in the parametrization of the fit function, or in
a cut in the lattice spacing, are weighted with their AIC weights; in the
directions corresponding to other systematic variations, a flat weighting is
applied.  Finally, the weights are normalized in such a way, that their sum
over all analyses equals 1.

Let $w_i$ denote the weight of the $i$-th analysis for a quantity $y$, with
$\sum_i w_i=1$. We interpret this weight as a probability. The statistical
uncertainties can be included by noting that, due to the central limit theorem,
they follow a Gaussian distribution $N(y;m_i,\sigma_i)$ with a central value
$m_i$ and a standard deviation $\sigma_i$. These parameters are given by the
jackknife average and the jackknife error calculated from the jackknife samples
in the $i$-th analysis.  We then define a joint probability distribution
function of $y$, including both statistical and systematic uncertainties, as:
\begin{gather}
    \sum_i w_i N(y; m_i, \sigma_i).
\end{gather}
In the following we work with the cumulative distribution function (CDF):
\begin{gather}
    \label{eq:cdf}
    P(y;\lambda)= \int^y_{-\infty} dy'\ \sum_i  w_i N(y'; m_i, \sigma_i \sqrt{\lambda}).
\end{gather}
Here, for later use, we introduce a parameter $\lambda$ that rescales the
statistical error.

The median of the CDF is our choice for the central value of $y$ and its
total error is given by the 16\% and 84\% percentiles of the CDF:
\begin{gather}
    \label{eq:err0}
    \sigma_{\rm total}^2\equiv \left[\tfrac{1}{2}(y_{84}-y_{16})\right]^2\qquad
    \text{with}\qquad
    P(y_{16}; 1) = 0.16,\qquad
    P(y_{84}; 1) = 0.84.
\end{gather}
One could define a systematic error by evaluating the 16\% and 84\% percentiles
of the $P(y;0)$ function, since here the choice $\lambda=0$ erases the
statistical contribution to the distribution. However, $P(y;0)$ is a sum of
step functions (shown in red in Figure \ref{fi:error}), making the percentiles
a function that has jumps,
which makes the definition of the systematic error highly sensitive to the
value of the percentile chosen. Here we make a more robust choice for the
systematic error.  First we demand, that:
\begin{gather}
    \label{eq:err1}
    \sigma_{\rm stat}^2 + \sigma_{\rm sys}^2 \equiv \sigma_{\rm total}^2.
\end{gather}
Now, let us note that the rescaling of each jackknife error $\sigma_i^2$ with
a factor $\lambda$ is expected to increase the total squared statistical error
with the same factor:
\begin{gather}
    \label{eq:err2}
    \lambda\sigma_{\rm stat}^2 + \sigma_{\rm sys}^2\equiv \left[\tfrac{1}{2}(\tilde{y}_{84}-\tilde{y}_{16})\right]^2\qquad
    \text{with}\qquad
    P(\tilde{y}_{16}; \lambda) = 0.16\ ,\qquad
    P(\tilde{y}_{84}; \lambda) = 0.84\ .
\end{gather}
Equations \eqref{eq:err0}, \eqref{eq:err1} and \eqref{eq:err2} then provide a
definition for separate statistical and systematic errors.  If the $\lambda$ is
not too small, then the joint CDF is smooth and has no sudden jumps, see Figure
\ref{fi:error}, and the procedure is insensitive to the choice of $\lambda$. We
use $\lambda=2$ in our error estimations. 

To understand the composition of the systematic error we calculate the {\bf
error budget} for all important quantities in the following way. Imagine that
the full analysis uses 9 values of lattice spacing cuts, and we are interested
in the corresponding systematic error. We first determine 9 total errors for
each possible cuts.  From these we construct a second CDF, which is a sum of 9
Gaussians as in Equation \eqref{eq:cdf}, with $i=1\dots9$, the $\sigma_i$ being
the total error and $m_i$ the average of the 16 and 84 percentiles of the fits
with the $i$-th cut, and the $w_i$ the sum of the weights of those fits.  From
this CDF we derive the systematic error as done above for the original CDF,
which is our result for the systematic error corresponding to the 9 cuts. We
remark that the systematic errors are correlated within one error budget,
distorting the quadratic sum of the components, that ought to sum up to the
full systematic error.

\subsection*{Error propagation}

Here we describe the way to propagate errors to consecutive analysis steps.
Such a case occurs when we perform a Type-II fit using the physical values of
$w_0$, $M_{ss}$ and $\Delta M^2$ that were determined in a Type-I fit. 

The statistical errors are taken into account by keeping the jackknife samples
throughout the whole analysis and computing the statistical error only in the
end, ie. after the Type-II fit.  For the systematic error there are certain
analysis choices, like hadron mass fit ranges, that are shared between the
Type-I and Type-II fits. We carry these over as we do with the jackknife
samples.

There are also systematics that are independent in the two types of fits. For
those, one would like to combine all of the corresponding analyses of the
Type-I fit with all of those of the Type-II fit. The number of individual
analyses can already be several thousand for each type of fit, and by mixing
each analysis in the first step with each analysis in the second step, the
total number of analyses would easily reach a million. These many combinations
are unnecessary, since they include many bad fits with tiny weights. 

In our approach we select $N_I$ results from the Type-I fit by an ``importance
sampling'': we uniformly split the probability interval $[0,1]$ into $N_I$ bins
and, for each bin, take whichever individual fit corresponds to the midpoint of
that bin. This produces a list of $N_I$ Type-I analyses, sampled according to
their importance.  This selection is then the input into the Type-II fit, and
the total number of analyses in the second step will only get multiplied by a
factor of $N_I$ instead of several thousands.  We choose $N_I=8$ in our
analyses. We ascertained that when using these $N_I=8$ fits, both the
statistical and the systematic errors are approximately the same as when
considering all fits.

    \clearpage
    \section{Results for $w_0$, $M_{ss}$ and $\Delta M^2$}
\label{se:res_w0etal}

\begin{table}[p]
    \centering
    \begin{tabular}{l|rr|rr|rr}
	& \multicolumn{2}{c|}{$w_0$[fm]} & \multicolumn{2}{c|}{$M_{ss}$[MeV]} & \multicolumn{2}{c}{$\Delta M^2$[MeV$^2$]}\\
	\hline\hline
	median                         &   0.17236 &          &    689.89 &           &   13170 \\
	\hline
	total error                    &        70 & (0.4\%)  &        49 & (0.07\%) &      420 & (3.2\%) \\
	statistical error              &        29 &          &        28 &          &      320 &         \\
	systematic error               &        63 &          &        40 &          &      270 &         \\
	\hline
	$M_\pi/M_K/M_{ss}$ fit         &      $<$1 &          &        24 &           &       0 \\
	$M_\pi/M_K/M_{ss}$ fit QED     &         4 &          &         2 &           &     140 \\
	$M_\Omega$ fit                 &        16 &          &         4 &           &       0 \\
	$M_\Omega$ fit QED             &         9 &          &      $<$1 &           &      10 \\
	\hdashline
	$M_\Omega$ experimental        &         5 &          &         1 &           &   $<$10 \\
	Continuum limit (beta cuts)    &        62 &          &        24 &           &     120 \\
	$A_0$ on/off                   &        on &          &        on &           &     off \\
	$A_2$ on/off                   &        on &          &        on &           &     off \\
	$A_4$ on/off                   &        60 &          &        38 &           &     off \\
	$B_0$ on/off                   &         2 &          &         3 &           &     off \\
	$B_2$ on/off                   &       off &          &       off &           &     off \\
	$C_0$ on/off                   &        on &          &        on &           &     off \\
	$C_2$ on/off                   &         9 &          &         5 &           &     off \\
	$D_0$ on/off                   &       off &          &       off &           &      on \\
	$D_2$ on/off                   &       off &          &       off &           &      on \\
	$D_4$ on/off                   &       off &          &       off &           &      40 \\
	$D_l$ on/off                   &       off &          &       off &           &      40 \\
	$D_s$ on/off                   &       off &          &       off &           &      40 \\
	$E_0$ on/off                   &        on &          &        on &           &      on \\
	$E_2$ on/off                   &        18 &          &         2 &           &      on \\
	$E_4$ on/off                   &       off &          &       off &           &      40 \\
	$E_l$ on/off                   &         7 &          &      $<$1 &           &      20 \\
	$E_s$ on/off                   &        10 &          &         2 &           &      70 \\
	$F_0$ on/off                   &        on &          &        on &           &      on \\
	$F_2$ on/off                   &      $<$1 &          &         1 &           &   $<$10 \\
	$G_0$ on/off                   &        on &          &        on &           &     off \\
	$G_2$ on/off                   &         2 &          &         3 &           &     off \\
	\hline\hline
    \end{tabular}
    \caption
    {
	\label{ta:w0etal}Physical values and error budgets for $w_0$, $M_{ss}$
	and $\Delta M^2$. The errors are to be understood on the last digits of
	the central value, as usual. Both statistical and systematic
	uncertainties of these Type-I fits are propagated to the Type-II fits.
	The systematic uncertainties below the dashed line are propagated by
	choosing $N_I=8$ representative fits, as described in the text.
    }
\end{table}

\begin{figure}[p]
    \centering
    \includegraphics[width=0.7\textwidth]{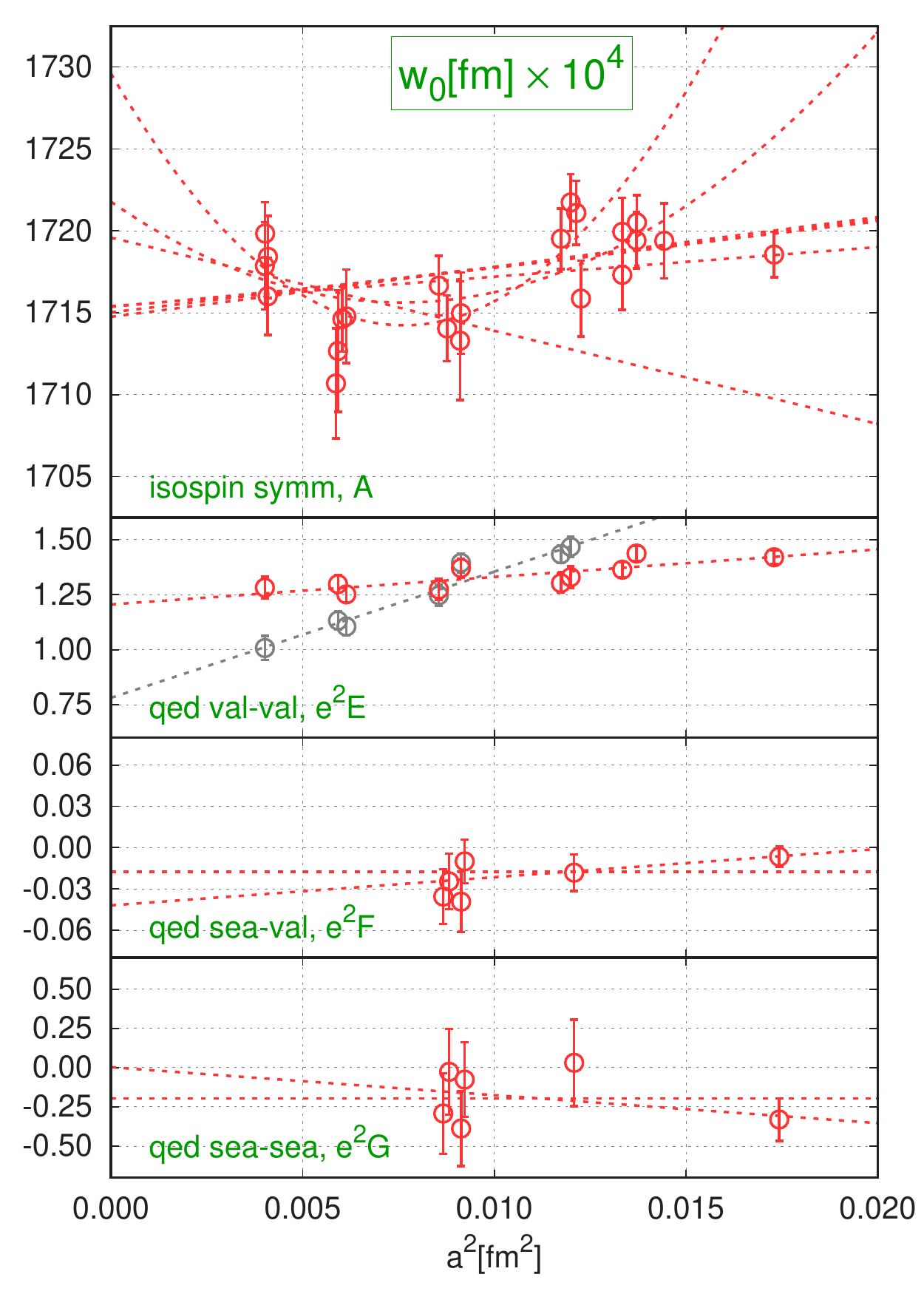}
    \caption
    {
	\label{fi:w0} Continuum extrapolations of the contributions to $w_0
	M_\Omega$. From top to bottom: isospin-symmetric, electromagnetic
	valence-valence, sea-valence and sea-sea component. The results are
	multiplied by $10^4/[M_\Omega]_*$.  For the definitions of the
	components see Equations \eqref{eq:fiveeq} and \eqref{eq:q20ren}. The
	electric derivatives are multiplied by $e_*^2$.  Dashed lines are
	continuum extrapolations corresponding to the lattice spacing dependent
	part of the $A$,$E$,$F$ and $G$ coefficients. They are illustrative
	examples from our several thousand fits.  Only the lattice spacing
	dependence is shown: the data points are moved to the physical light
	and strange quark mass using the $X_l$ and $X_s$ dependent terms in the
	fit. This adjustment varies from fit to fit, the red datapoints are
	obtained in an $a^2$-linear fit to all ensembles. If in a fit the
	adjusted points differed significantly from the red points, we show
	them with grey color. The final result is obtained from a weighted
	histogram of the several thousand fits.
    }

\end{figure}

\begin{figure}[p]
    \centering
    \includegraphics[width=0.7\textwidth]{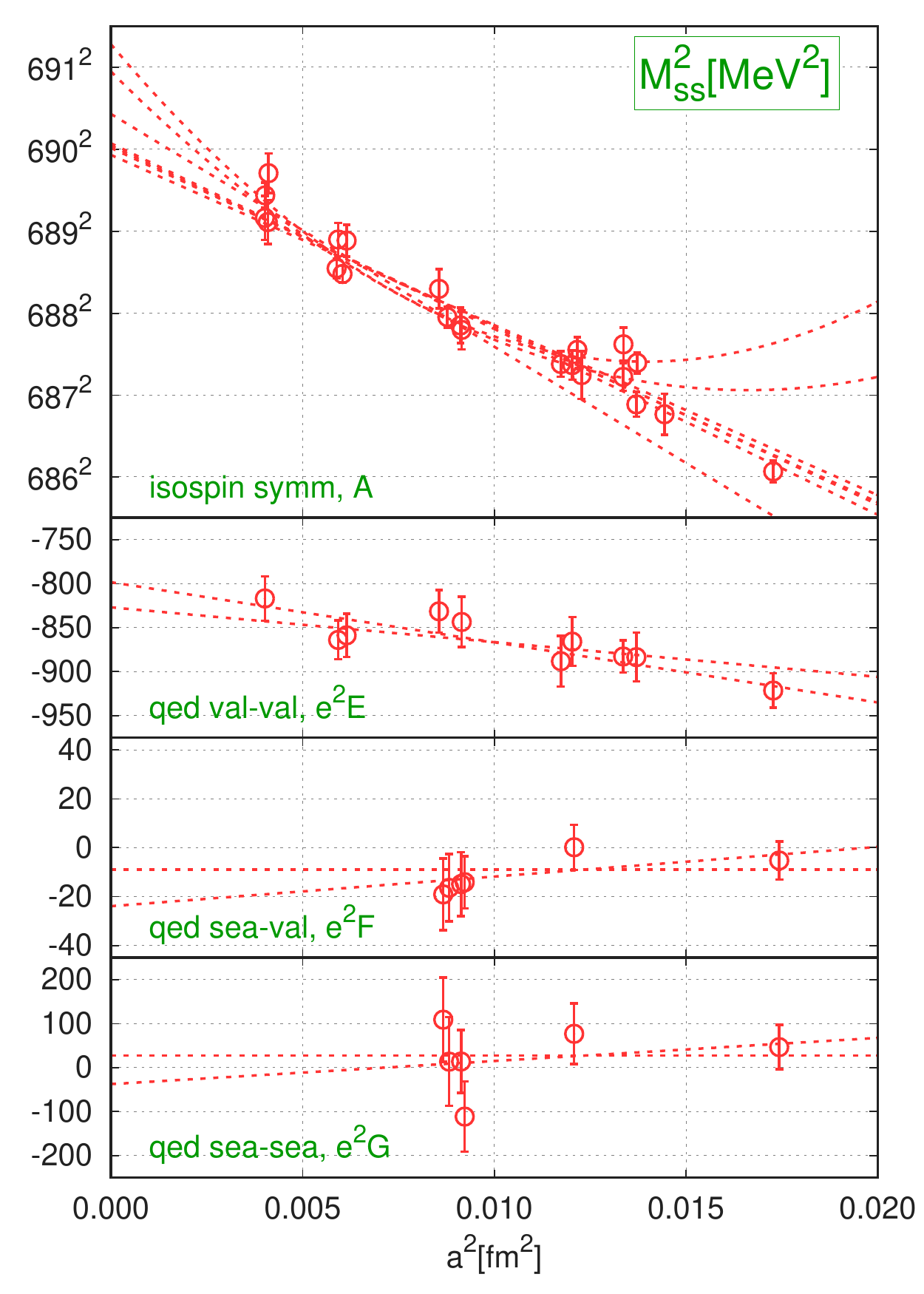}
    \caption
    {
	\label{fi:mss} Continuum extrapolations of the contributions to
	$M_{ss}^2$. Plotted is the ratio $M_{ss}^2/M_\Omega^2$ multiplied by
	$[M_\Omega]_*^2$. Other details as in Figure \ref{fi:w0}.
    }
\end{figure}

\begin{figure}[t]
    \centering
    \includegraphics[width=0.7\textwidth]{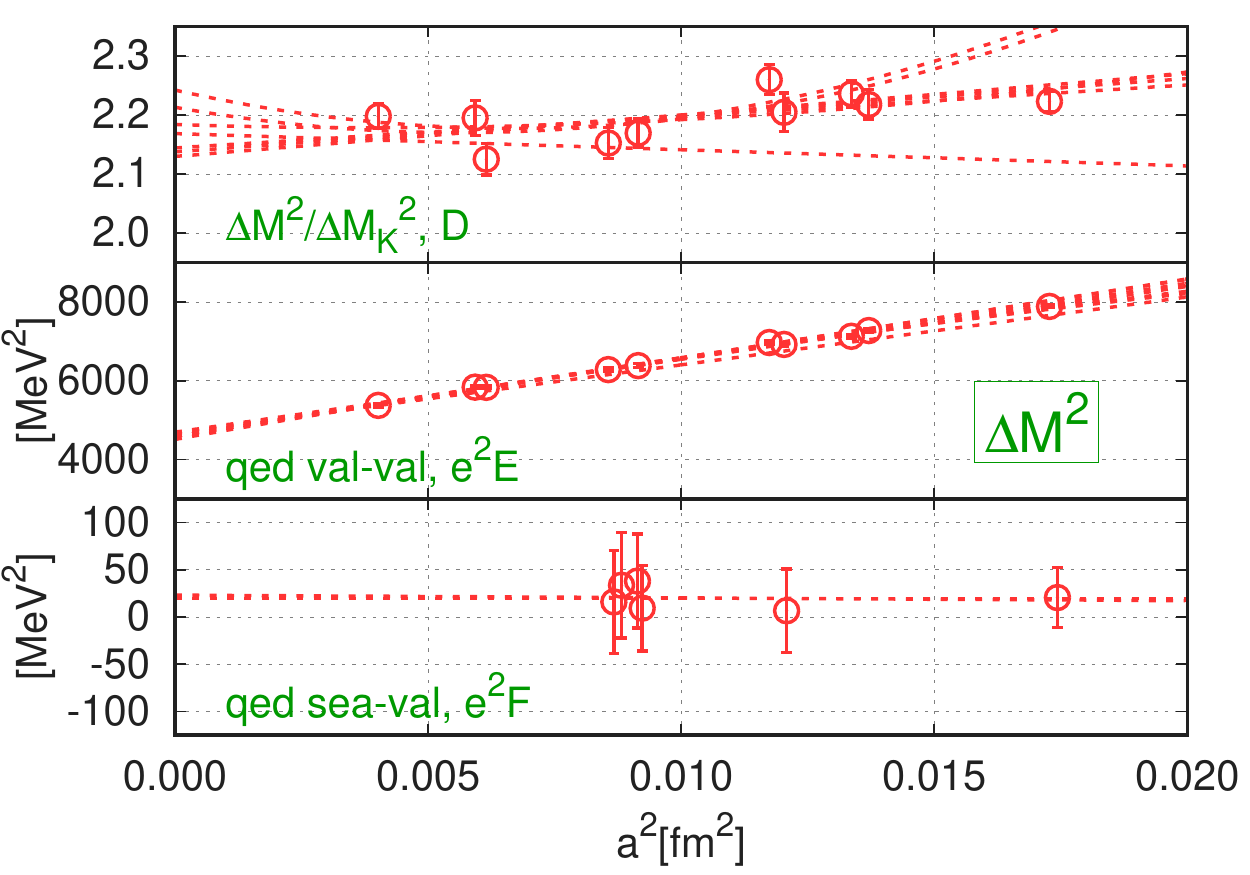}
    \caption
    {
	\label{fi:dmsq} Continuum extrapolations of the contributions to
	$\Delta M^2$.  From top to bottom: $[\Delta M^2]'_m/[\Delta M_K^2]'_m$,
	electromagnetic valence-valence and sea-valence components of $\Delta
	M^2/M_\Omega^2$.  The electric derivatives are multiplied by
	$[e^2M_\Omega^2]_*$.  Other details as in Figure \ref{fi:w0}.
    }
\end{figure}

In this section we describe briefly the details of the global fits that are
used to obtain the physical values of $w_0$, $M_{ss}$ and $\Delta M^2$ from the
experimental values of hadron masses, including the mass of the $\Omega$
baryon. In all three cases we use the Type-I fit function of Equation
\eqref{eq:fitI}, which can be related to isospin-breaking derivatives as
described in Equation \eqref{eq:fiveeq}. The set of parameters, that are used
in these fits, can be read off from Table \ref{ta:w0etal}. For a given
observable some of the parameters are included in all fits, some never, and there are
also some that are either included or excluded. A systematic error is
associated with the latter and is given in the Table.

In the case of $w_0$, the observable we fit is $Y=w_0M_\Omega$. Since $w_0
M_\Omega$ is symmetric under $u\leftrightarrow d$ exchange, no leading order
strong-isospin-breaking terms can appear. Thus we can set the
strong-isospin-breaking coefficient ($D$) to zero. 

To account for the systematic error due to the different continuum
extrapolations we apply both linear and quadratic functions in the
isospin-symmetric component, also we skip zero/one/two/three of the coarsest
lattice spacings in the linear and zero/one/two lattice spacings in the
quadratic fits. For the tiny valence QED component only linear fits are
applied, with zero/one/two skips; for the even smaller sea QED contributions we
have either constant or linear fit with all lattice spacings.

The systematic error of the hadron mass fits is taken into account by 24
different combinations of the fit ranges: three for the $M_\Omega$ mass, two for
the pseudoscalars, two for the isospin breaking of the $M_\Omega$ and two for
the isospin breaking of the pseudoscalars. The pseudoscalar fit ranges are
given in Table \ref{ta:meson}.  For the $\Omega$ mass we use two fit ranges
from the four-state fit and one from the GEVP procedure that are given in Table
\ref{ta:omega}. The fit ranges for the isospin breaking components can be found
in Table \ref{ta:ibmass}. To account for the experimental error on $M_\Omega$
we carry out the analysis with two different experimental values: one that
corresponds to the central value plus the experimental error; the other with
this error subtracted. 

Altogether, these yield a total of 129024 fits. When the different analyses
are combined into a histogram to determine the systematic error, the results from
different fit functions or lattice spacing cuts are weighted with the Akaike
Information Criterion, the rest with flat weighting. We obtain
\begin{equation}
    \label{eq:w0}
    [w_0]_* = 0.17236(29)(63)[70]\ \text{fm},
\end{equation}
where the first error is statistical, and the second is systematic, the third
is the total error; we reach a relative precision of $0.4$\%.  The split up of
the error into different sources can be found in Table \ref{ta:w0etal}.  In
Figure \ref{fi:w0} we show the various isospin components of $w_0M_\Omega$
against the lattice spacing squared together with the different continuum
extrapolations. For the electric derivatives we took the definition in Equation
\eqref{eq:q20ren}.  Our result \eqref{eq:w0} is in good agreement with earlier
four-flavor determinations: $w_0=0.1715(9)$~fm of \cite{Dowdall:2013rya} and
$w_0=0.1714\left(\begin{smallmatrix}+15 \\-12\end{smallmatrix}\right)$~fm of
\cite{Bazavov:2015yea}. In these works the isospin-breaking effects were
only estimated, whereas in our case they are fully accounted for.

The same procedure is used for $M_{ss}$ as for $w_0$. We actually work with
$Y=(M_{ss}/M_{\Omega})^2$ instead of $M_{ss}/M_{\Omega}$, since the fit
qualities are much better in the first case.  The 129024 different fits give
\begin{equation}
    \label{eq:mss}
    [M_{ss}]_*= 689.89(28)(40)[49]\ \text{MeV},
\end{equation}
with statistical, systematic and total errors as above. The error budget can be found
in the second column of Table \ref{ta:w0etal} and the continuum extrapolations
for $M_{ss}^2$ are shown in Figure \ref{fi:mss}.

Finally we also carry out the analysis for $Y=\Delta M^2/M_\Omega^2$ with $\Delta
M^2=M_{dd}^2-M_{uu}^2$. Since this observable has no isospin-symmetric part,
the $A$, $B$ and $C$ coefficients are set to zero. Also, since this is an
isospin splitting effect, no electromagnetic sea-sea effects can contribute, so
the fit function becomes:
\begin{gather}
    \frac{\Delta M^2}{M_\Omega^2}= D \left(\frac{\Delta M_K^2}{M_\Omega^2}\right) + E e_v^2 + F e_ve_s
\end{gather}
Differently from the fits earlier we use $a^2$-quadratic fits also in the $D$
and $E$ coefficients.  We apply four lattice spacing cuts by skipping
zero/one/two/three of the coarsest lattices for linear fits and three cuts for
quadratic fits with zero/one/two skips.  Other systematics were treated as in
the above fits.  Altogether we have 3328 fits, which give a central value with
statistical, systematic and total errors as:
\begin{equation}
    [\Delta M^2]_*= 13170(320)(270)[420]\ \text{MeV}^2.
\end{equation}
The corresponding error budget can be found in Table \ref{ta:w0etal}.  In
Figure \ref{fi:dmsq} we show continuum extrapolations for the $\Delta
M^2/\Delta M_K^2$ ratio and the valence-valence and sea-valence electric
derivatives; these correspond to the $D$, $E$ and $F$ coefficients in the fit
function.

    \section{Alternatives to the $M_\Omega$ scale-setting}
\label{se:res_w0ext}

\begin{figure}[t]
    \centering
    \includegraphics[width=0.7\textwidth]{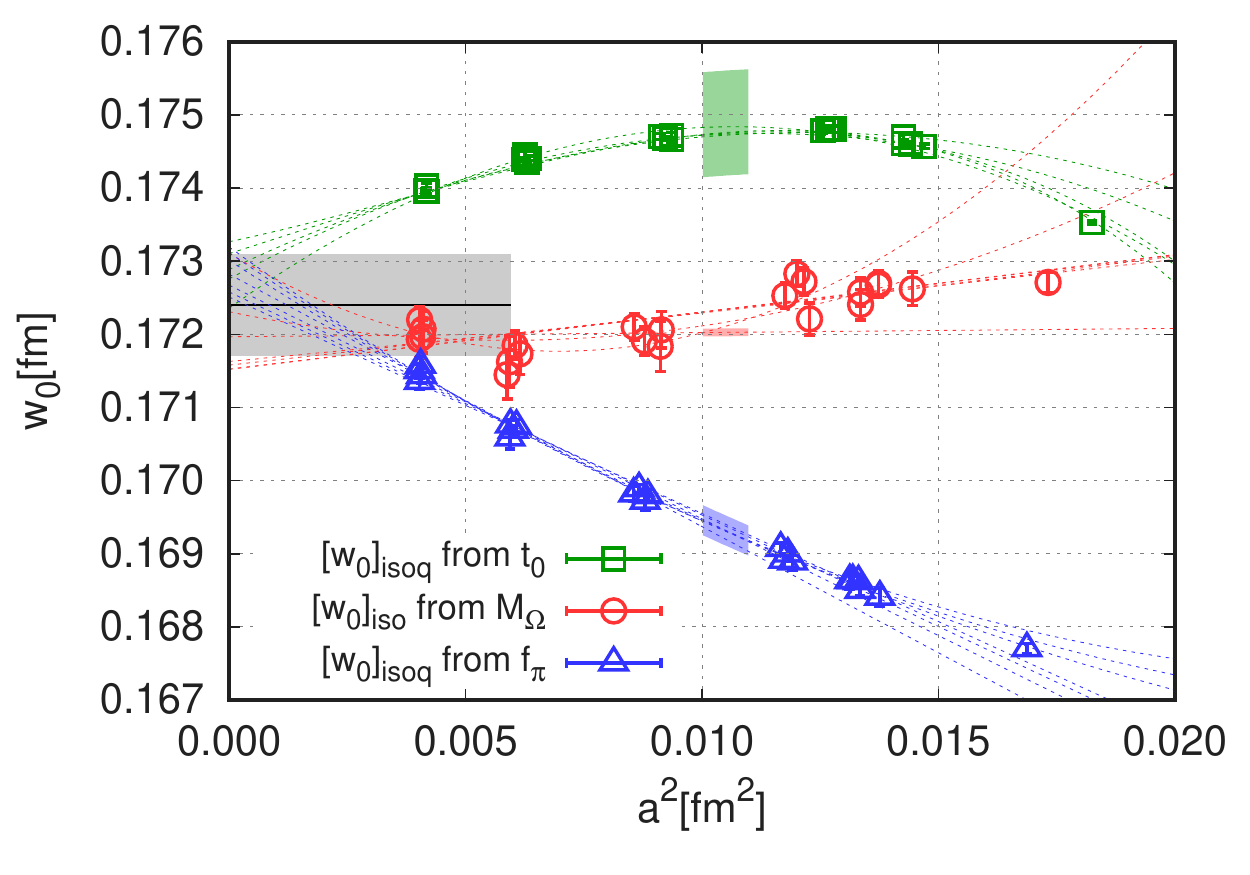}
    \caption
    {
	\label{fi:w0other} Continuum extrapolation of the isospin-symmetric
	value of $w_0$ using three different inputs: $t_0$ from the lattice
	work \cite{Bazavov:2015yea}, $M_\Omega$ from experiment
	\cite{Tanabashi:2018oca} and $f_\pi$ from a combination of chiral perturbation theory and experiment
	\cite{Tanabashi:2018oca}. The dashed lines are quadratic and cubic functions
	of $a^2$ in case of $t_0$, and linear and quadratic otherwise. The colored shaded regions
	around $a^2=0.010$~fm$^2$ correspond to the uncertainty in the input quantity. The horizontal
	grey shaded region is our final $w_0$ determination from Equation \eqref{eq:w0}. Note,
	there is a difference in the definition of the isospin-symmetric point in the different inputs.
    }
\end{figure}

To check our determination of the physical value of $w_0$, we consider two
other scale-setting quantities: the pion decay constant $f_\pi$ and the
Wilson-flow scale $t_0$.  Both approaches are independent from the systematics
of the $M_\Omega$ mass determination. We also investigate $a^4$ effects by
changing the definition of $w_0$ by lattice artefacts. We do not consider
$f_\pi$ and $t_0$ in our final analysis, because their relation to experiments
is indirect. We work in the isospin-symmetric limit throughout this section. 

In current lattice simulations it is common to use the pion decay constant for
scale setting. This observable however is well defined only in the absence of
electromagnetism, and thus useful only in simulations in the isospin-symmetric
point. It is possible to connect the experimental decay rate of the pion to an
isospin-symmetric pion decay constant, $f_\pi$. Current state-of-the-art uses a
chiral perturbation theory based approach, which yields $f_\pi=130.50(14)$~MeV
\cite{Tanabashi:2018oca}.  There are also computations underway to determine
$f_\pi$ \cite{Carrasco:2015xwa,Giusti:2017dwk,DiCarlo:2019thl} on the lattice.
In these approaches the isospin-symmetric point is defined using renormalized
quark masses, which is different from our hadronic scheme in Section
\ref{se:obs_split}. When turning on the electromagnetic interaction, our scheme
keeps certain neutral hadron masses and $w_0$ constant, contrary to the one
used in the $f_\pi$-based scheme, where the renormalized quark masses and the
strong coupling are fixed.

Here we carry out an analysis to determine an isospin-symmetric value of
$[w_0]_\mathrm{isoq}$ using the above $f_\pi$ as input. We introduce the
notation $\mathrm{isoq}$ to emphasize the difference from our definition of the
isospin-symmetric value $[w_0]_\mathrm{iso}= [w_0]_*$. To obtain
$[w_0]_\mathrm{isoq}$ we also need a pion and kaon mass that is purified from
isospin-breaking effects. For these we take $M_\pi=134.8(3)$~MeV and
$M_K=494.2(3)$~MeV \cite{Aoki:2016frl}. 

The fit procedure is similar to the Type-I fits that we performed before for
$w_0M_\Omega$. The physical point is given by the $f_\pi,M_\pi$ and $M_K$
values above. Since we work with the isospin-symmetric component, only the $A$,
$B$ and $C$ coefficients of Equation \eqref{eq:fitI} are kept. We apply both
linear and quadratic fits in $a^2$, with the usual cuts in the lattice
spacing. Figure \ref{fi:w0other} shows representative fits from this analysis,
with good fit qualities. The continuum extrapolated values are consistent with
our $[w_0]_*$ from Equation \eqref{eq:w0}. However the spread between
the different continuum extrapolations is smaller, since the curvature
of $w_0f_\pi$ in $a^2$ is smaller than in $w_0M_\Omega$.

Another way to determine $w_0$ is to take the $t_0$-scale, also
defined from the Wilson-flow, as input. This determination basically computes
the $w_0/t_0$ ratio. For the physical value of $t_0$ we use
$[t_0]_\mathrm{isoq}=0.1416\left(\begin{smallmatrix}+8
\\-5\end{smallmatrix}\right)$~fm from \cite{Bazavov:2015yea}, which has a
precision of about $0.5\%$. The same analysis is carried out as before, with
the difference that now we also include cubic fits in $a^2$, since the data
shows a very strong curvature and the linear fits have a bad quality. Figure
\ref{fi:w0other} shows representative fits, giving continuum values consistent
with using $M_\Omega$ as input, Equation \eqref{eq:w0}.

\begin{figure}[t]
    \centering
    \includegraphics[width=0.7\textwidth]{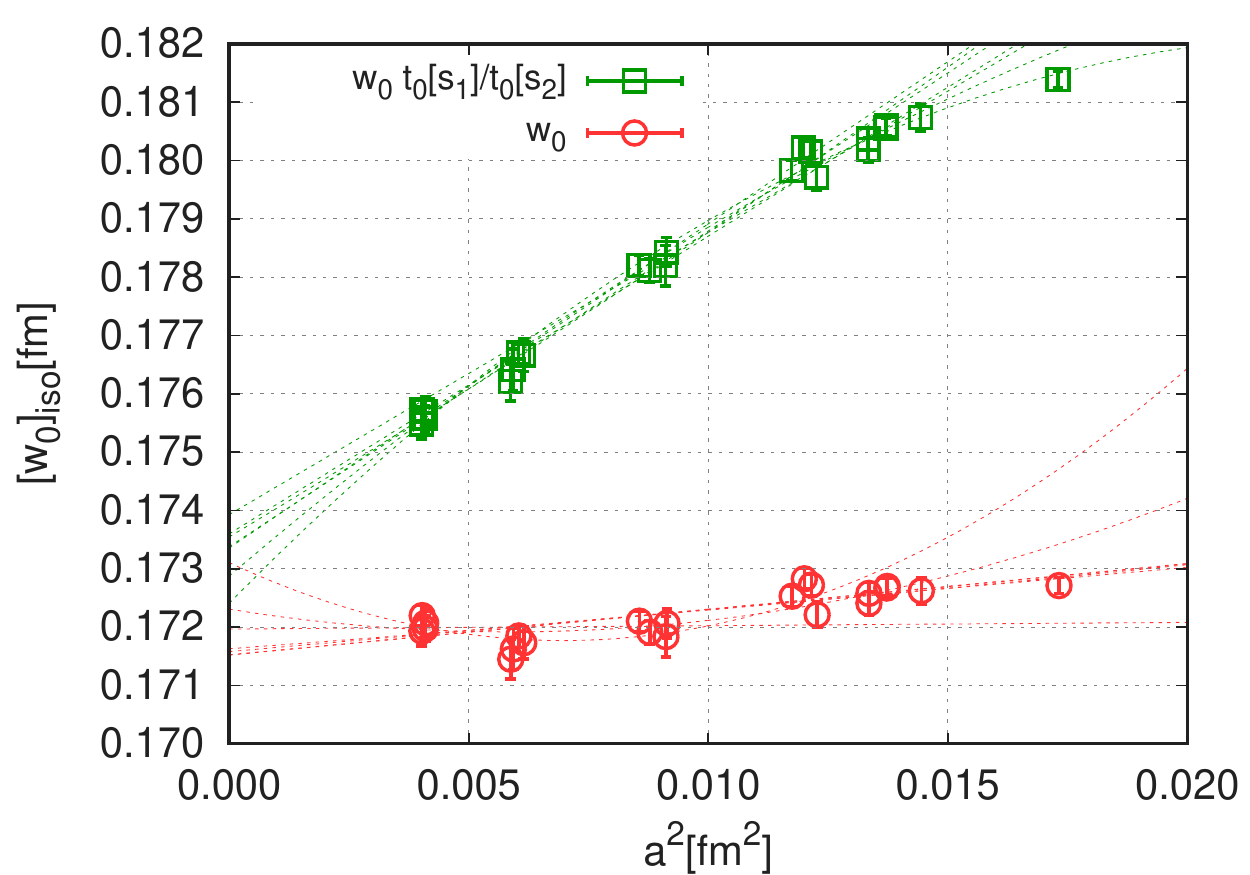}
    \caption
    {
	\label{fi:tratio} Continuum extrapolation of the isospin symmetric
	value of $w_0$ using $M_\Omega$. Two $w_0$ definitions, the standard
	one (red circles), and another modified by the $t_0[s_1]/t_0[s_2]$ ratio
	(green squares) are shown. The ratio approaches 1 in the continuum limit.
    }
\end{figure}

Finally we show here a method to determine $[w_0]_\mathrm{iso}$, which is also
based on $M_\Omega$ as an input parameter, but uses the idea of a t-shift in
the Wilson flow \cite{Cheng:2014jba}. The main reason for this analysis is to
determine whether the strong quadratic upward trend in $w_0$ for small lattice
spacings, see top panel of Figure \ref{fi:w0}, is a genuine cutoff effect?
Indeed, the Wilson flow is known to have a transient for small flow times.
Although the affected region shrinks as one approaches the continuum limit, the
effect might be sizable particularly if we want to reach an accuracy on the few
per-mil level.

The t-shift in the Wilson-flow replaces $\langle t^2E(t) \rangle$ with $\langle
t^2E(t+s a^2) \rangle$, which is essentially applying the flow on a smeared
gauge field (pre-smearing). It can be interpreted as an improved operator for
the energy density. Obviously, in the continuum limit flows with or without
t-shifts are the same.  We measured a combination, $w_0\cdot
t_0(s_1)/t_0(s_2)$, which obviously gives back $w_0$ in the continuum limit.
Clearly, this combination has a different lattice spacing dependence
than the original $w_0$, determined in the previous Section.
There are several $s_1,s_2$ choices, which eliminate the strong $a^4$
behavior, the upward turning of $w_0$ for small lattice spacings. They do so without changing the result
in the continuum limit within errors. This finding
indicates that the upward trend is indeed a cutoff effect related to the Wilson
flow and can be removed by modified operators of the Wilson flow. As an
illustration we show $s_1=0.35$ and $s_2=0.21$ in Figure~\ref{fi:tratio}.

The latter procedure reduces the coefficient of the $a^4$ term in the continuum
extrapolation and, as a consequence, could also reduce the error on $w_0$.
Since the t-shift method is a somewhat unconventional way to determine $w_0$,
we leave it as an illustration of how cutoff effects can play a role and we
quote our original $w_0$, Equation \eqref{eq:w0}, with the larger error as our
final result.

    \section{Results for $a_{\mu}$ and its various contributions}
\label{se:res_amu}

\def\amus{53.379(89)(67)}
\def\amul{639.3(2.0)(4.2)}
\def\amud{-18.61(1.03)(1.17)}
\def\amuserr{[111]}
\def\amulerr{[4.6]}
\def\amuderr{[1.56]}
\def\amutot{707.5(2.3)(5.0)[5.5]}

In this section we present results for the strange, light and disconnected
components of $a_\mu$ in the continuum and infinite-volume limits.  We perform
the two limits in two separate steps.  We introduce a reference box with
spatial extent $L_\mathrm{ref}=6.272$~fm and time extent
$T_\mathrm{ref}=\tfrac{3}{2}L_\mathrm{ref}$.  These correspond approximately to
the size of our boxes in the {\tt 4stout} ensemble set. In this reference box
we perform the continuum extrapolation for each flavor component. The
finite-size effect of the reference box is then added in the second step. For
this we prepared dedicated lattice simulations, including a large box of size
$L_\mathrm{big}=T_\mathrm{big}=10.752$~fm, as discussed in Section
\ref{se:obs_fv}. The simulations give the difference
$a_\mu(L_\mathrm{big},T_\mathrm{big}) - a_\mu(L_\mathrm{ref},T_\mathrm{ref})$,
which are in good agreement with non-lattice estimates. For the tiny residual
finite-size effect, $a_\mu(\infty,\infty) -
a_\mu(L_\mathrm{big},T_\mathrm{big})$, the predictions of the non-lattice
approaches are taken.

In this section we use both Type-I and Type-II parametrizations from Section
\ref{se:res_fit} to perform the global fits. They give compatible results for
the observables, however in most cases the Type-I results are more precise.
This can be partly explained by the relatively large error on the $w_0$ value,
Equation \eqref{eq:w0}, which is caused by the strong curvature of
$w_0M_\Omega$ at small lattice spacings. We will take the final result from the
Type-I fit, and use the Type-II fit to perform the isospin decomposition. To get
the isospin-symmetric value we take the total result from the Type-I fit and
subtract the isospin-breaking contributions obtained from the Type-II fit.

Type-II fits require the physical values of $w_0$, $M_{ss}$ and $\Delta
M^2$ as input. These were determined in the previous Section. From those fits
we keep the 24 different possibilities related to the hadron mass
determinations. The remaining systematic variations of these are represented by
$N_I=8$ suitably chosen fit combinations, as discussed in Section
\ref{se:res_err}.

Some of the fit parameters are included in all fits, some never used, and there
are also ones that are included in half of the fits and excluded in the other
half. Which of these options is applied for a given parameter is decided by
looking at the influence of the parameter on the fit result. The options chosen
can be read off from Table \ref{ta:amuerr}.

\begin{table}[p]
    \centering
    \begin{tabular}{l|rr|rr|rr}
	& \multicolumn{2}{c|}{$a_{\mu}^\mathrm{strange}(L_\mathrm{ref},T_\mathrm{ref})$} & \multicolumn{2}{c|}{$a_{\mu}^\mathrm{light}(L_\mathrm{ref},T_\mathrm{ref})$} & \multicolumn{2}{c}{$a_{\mu}^\mathrm{disc}(L_\mathrm{ref},T_\mathrm{ref})$}\\
	\hline\hline
	median                           &   53.379  &         &    639.3 &          &    -18.61  & \\
	\hline                                 
	total error                      &      111  & (0.2\%) &      4.6 & (0.7\%)  &      1.56  & (8.3\%)\\
	statistical error                &       89  &         &      2.0 &          &      1.03  &\\
	systematic error                 &       67  &         &      1.9 &          &      1.11  &\\
	difference to NNLO improvement   &       --  &         &      3.7 &          &      0.36  &\\
	\hline                                                                               
	$M_\pi/M_K/M_{ss}$ fit           &        5  &         &   $<$0.1 &          &   $<$0.01  &\\
	$M_\pi/M_K/M_{ss}$ fit QED       &        3  &         &      0.1 &          &   $<$0.01  &\\
	$M_\Omega$ fit                   &       56  &         &      0.3 &          &      0.04  &\\
	$M_\Omega$ fit QED               &        2  &         &      0.1 &          &   $<$0.01  &\\
	$M_\Omega$ experimental          &        5  &         &      0.1 &          &      0.01  &\\
	Continuum limit (beta cuts)      &       47  &         &      0.3 &          &      0.68  &\\
	$a^2\alpha_s^n$ with $n=0$ or $3$&       --  &         &      1.1 &          &      0.57  &\\
	taste improvement ranges         &       --  &         &      0.7 &          &      0.11  &\\
	$t_c$ in Table \ref{ta:tcfm}     &       --  &         &      0.2 &          &      0.23  &\\
	$A_0$ on/off                     &       on  &         &       on &          &        on  &\\
	$A_2$ on/off                     &       on  &         &       on &          &        on  &\\
	$A_4$ on/off                     &       26  &         &   $<$0.1 &          &       off  &\\
	$B_0$ on/off                     &       11  &         &       on &          &      0.06  &\\
	$B_2$ on/off                     &      off  &         &      off &          &       off  &\\
	$C_0$ on/off                     &       on  &         &       on &          &        on  &\\
	$C_2$ on/off                     &      off  &         &      off &          &       off  &\\
	$D_0$ on/off                     &      off  &         &       on &          &        on  &\\
	$D_2$ on/off                     &      off  &         &       on &          &        on  &\\
	$D_4$ on/off                     &      off  &         &      off &          &       off  &\\
	$D_l$ on/off                     &      off  &         &      0.2 &          &      0.01  &\\
	$D_s$ on/off                     &      off  &         &       on &          &       off  &\\
	$E_0$ on/off                     &       on  &         &       on &          &        on  &\\
	$E_2$ on/off                     &       on  &         &   $<$0.1 &          &      0.58  &\\
	$E_4$ on/off                     &      off  &         &      off &          &       off  &\\
	$E_l$ on/off                     &        3  &         &   $<$0.1 &          &       off  &\\
	$E_s$ on/off                     &     $<$1  &         &      off &          &       off  &\\
	$F_0$ on/off                     &       on  &         &       on &          &        on  &\\
	$F_2$ on/off                     &        1  &         &   $<$0.1 &          &       off  &\\
	$G_0$ on/off                     &       on  &         &       on &          &        on  &\\
	$G_2$ on/off                     &        3  &         &      0.1 &          &       off  &\\
	\hline\hline
    \end{tabular}
    \caption
    {
	\label{ta:amuerr}Continuum extrapolated results and error budget for
	the strange, light and disconnected contributions to $a_\mu$. The
	errors are to be understood on the last digits of the central value, as
	usual. The results correspond to a box size $L_\mathrm{ref}=6.272$~fm
	and $T_\mathrm{ref}=\tfrac{3}{2}L_\mathrm{ref}$.
    }
\end{table}

\begin{table}[p]
    \centering
    \begin{tabular}{l|r|r|r}
	& \multicolumn{1}{c|}{$a_{\mu}^\mathrm{strange}(L_\mathrm{ref},T_\mathrm{ref})$} & \multicolumn{1}{c|}{$a_{\mu}^\mathrm{light}(L_\mathrm{ref},T_\mathrm{ref})$} & \multicolumn{1}{c}{$a_{\mu}^\mathrm{disc}(L_\mathrm{ref},T_\mathrm{ref})$}\\
	\hline\hline
	total     & \amus   & \amul & \amud\\
	\hline
	iso & 53.393(89)(68)& 633.7(2.1)(4.2)& -13.36(1.18)(1.36)\\
	qed & -0.0136(86)(76)& -0.93(35)(47)& -0.58(14)(10)\\
	qed-vv & -0.0086(42)(41)& -1.24(40)(31)& -0.55(15)(10)\\
	qed-sv & -0.0014(11)(14)& -0.0079(86)(94)& 0.011(24)(14)\\
	qed-ss & -0.0031(76)(69)& 0.37(21)(24)& -0.040(33)(21)\\
	sib& --& 6.60(63)(53)& -4.67(54)(69)\\
	\hline\hline
    \end{tabular}
    \caption
    {
	\label{ta:amu}Continuum extrapolated results for the different
	components of the strange, light and disconnected contributions to
	$a_\mu$. The results correspond to a box size
	$L_\mathrm{ref}=6.272$~fm and
	$T_\mathrm{ref}=\tfrac{3}{2}L_\mathrm{ref}$.
    }
\end{table}

\subsection*{Connected strange contribution}

\begin{figure}[p]
    \centering
    \includegraphics[width=0.7\textwidth]{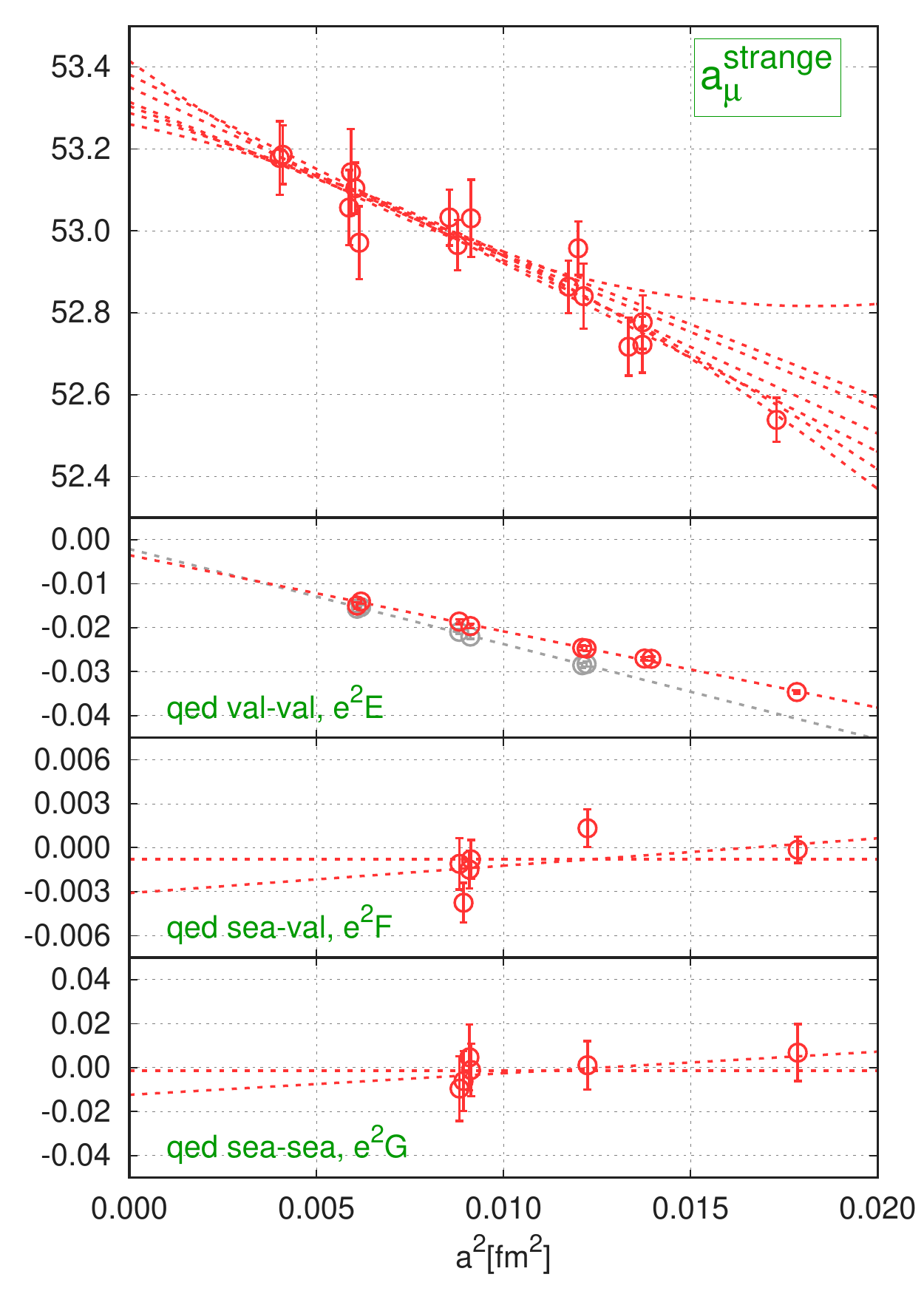}
    \caption
    {
	\label{fi:amus} Continuum extrapolations of the contributions to
	$a_{\mu}^\mathrm{strange}(L_\mathrm{ref},T_\mathrm{ref})$. The top
	panel shows the total value obtained in a Type-I fit, the lower panels are
	the isospin breaking contributions from Type-II fits.  For more
	explanations see the caption of Figure \ref{fi:w0}.
    }
\end{figure}

The strange contribution to the connected component of $a_\mu$, denoted by
$a_{\mu}^\mathrm{strange}$, is obtained from the strange flavor term of the
connected contractions $C^\mathrm{strange}$ given in Equations \eqref{eq:conn}
and \eqref{eq:clsc}. Its isospin-symmetric component, as well as its
electromagnetic isospin-breaking derivatives are given in Table \ref{ta:ibjjc}.
The propagator is then summed over space to project to zero momentum and in
time with a weight factor:
\begin{gather}
    \label{eq:amus}
    a_{\mu}^\mathrm{strange}=
    10^{10}\alpha^2
    \sum_{t=0}^{T/2} K(t;aQ_\mathrm{max},am_\mu)\ 
    \frac{1}{6}
    \sum_{\vec{x},\mu=1,2,3}
     \left\langle C^\mathrm{strange}_{\mu,t,\vec{x};\mu,0} + C^\mathrm{strange}_{\mu,T-t,\vec{x};\mu,0}\right\rangle\ ,
\end{gather}
see Equations \eqref{eq:gdeflat} and \eqref{eq:kdef}.  Strong-isospin-breaking
does not enter in $a_{\mu}^\mathrm{strange}$, so the $D$ coefficient can be set
to zero in the Type-I fit function. The systematic error estimation was carried
out as in the cases of $w_0$ and $M_{ss}$ in Section \ref{se:res_w0etal}. The
differences are, that $E_2$ is always kept and $C_2$ is not used.  Altogether
we have 32256 fits: the continuum limit and fit-form-related variations are
weighted with AIC, the rest with a flat distribution.  In the continuum limit
we get
\begin{gather}
    \label{eq:res_s}
    a_{\mu}^\mathrm{strange}(L_\mathrm{ref},T_\mathrm{ref})= \amus\amuserr\ ,
\end{gather}
with statistical, systematic and total errors. The result is
obtained in a finite box and the finite-size correction term will be added in a
later step.  The error budget for the total $a_{\mu}^\mathrm{strange}$ is given
in Table \ref{ta:amuerr}. We also perform a Type-II fit, from which we obtain
the different isospin contributions in Table \ref{ta:amu}. The corresponding
continuum extrapolations are shown in Figure \ref{fi:amus}. 

\subsection*{Connected light contribution}

The contribution of the light flavors to the connected part of $a_\mu$, denoted
by $a_{\mu}^\mathrm{light}$, is given by replacing the strange contraction
$C^\mathrm{strange}$ with the connected light quark contraction
$C^\mathrm{light}$ in Equation \eqref{eq:amus}.  In the isospin-symmetric part
a bounding procedure is applied on the propagator to reduce the noise as
discussed in Section \ref{se:obs_bound}.  In the isospin-breaking parts we
apply a cut in time, beyond which the propagator is set to zero, see Section
\ref{se:obs_ibjj}. Two different cuts, given in Table \ref{ta:tcfm}, are used
to estimate the corresponding systematic error. 

As explained in detail in Section \ref{se:obs_taimp}, the continuum
extrapolation is carried out by first applying an improvement on the
$a_{\mu}^\mathrm{light}$ observable on each ensemble. This is necessary in
order to remove large cutoff effects related to taste violations. We use the
rho-pion-gamma model (SRHO) of Section \ref{se:obs_rho} for this purpose. The
improvement is applied from a distance of $t_\mathrm{sep}$, for which we use
four different choices: $0.4$, $0.7$, $1.0$ and $1.3$~fm. The variation
corresponding to the choice of $t_\mathrm{sep}$ is part of our error budget,
which is given in Table \ref{ta:amuerr}. The above improvement procedure is
only applied to the isospin-symmetric component.

In addition, we consider cases in which we replace SRHO improvement by NNLO
SXPT in the previous fits, for Euclidean times larger than $1.3$~fm. These
results are only used to compute an extra systematic error related to the
choice of the SRHO model for taste improvement. Thus, we build a histogram of
the fit results obtained using SRHO, SRHO + ERR and SRHO - ERR improvements,
where ERR means the difference between the NNLO SXPT and SRHO fits. Then we
compute the systematic error corresponding to the variation across the SRHO and
SRHO$\pm$ERR with our usual procedure, see Section \ref{se:res_err}. The NNLO
SXPT distribution has longer tails than a Gaussian, so we take the central 95\%
region and quote the half of it as a systematic error (the usual 68\% would
give a smaller value). The corresponding value can be found in Table
\ref{ta:amuerr} and the full histogram is shown in grey in Figure~\labelmaincont{}
of the main paper. Note, that this procedure does not change
the central value of our result, which is obtained without any reference to
SXPT.

There is a small variation in the size of the lattices between different
ensembles. Our taste improvements are set up in a way, that the improved
results correspond to the same box size $L=L_\mathrm{ref}$. Finite-$T$ effects
are corrected for using NNLO SXPT.

\begin{figure}[p]
    \centering
    \includegraphics[width=0.7\textwidth]{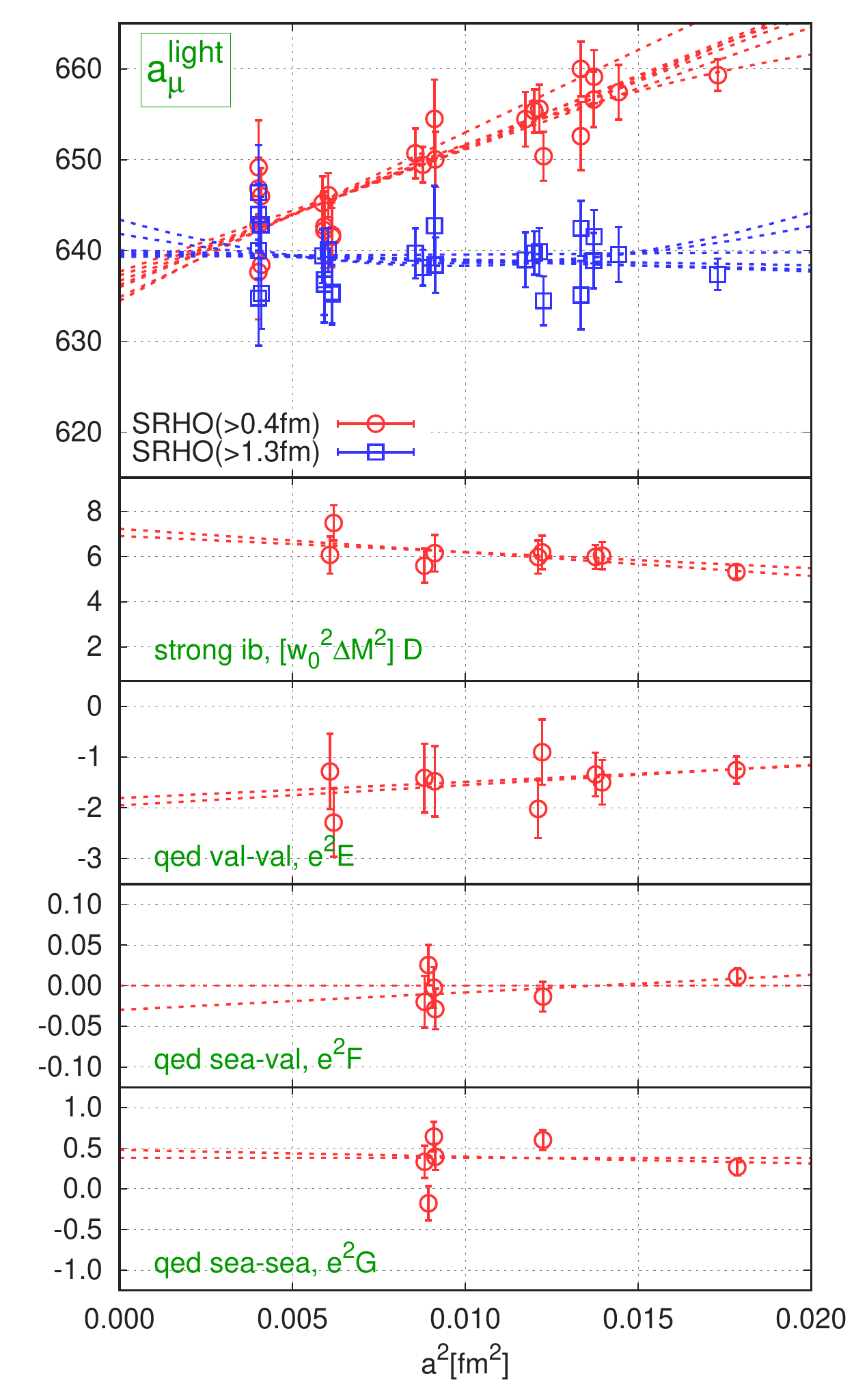}
    \caption
    {
	\label{fi:amul} Continuum extrapolations of the contributions to of
	$a_{\mu}^\mathrm{light}(L_\mathrm{ref},T_\mathrm{ref})$. For more
	explanations see the caption of Figure \ref{fi:amus}. To the total
	observable (upper panel) we apply taste improvement: we plot the
	SRHO-improved data sets with improvement applied for Euclidean times
	larger than $0.4$/$1.3$~fm, as red/blue points.
    }
\end{figure}

The above variations in the analysis procedure yield 516096 Type-I fits. We
apply the usual weighting, in particular the different choices of
$t_\mathrm{sep}$ and $n$ are given a flat-weight. For the total light connected
contribution we get
\begin{gather}
    \label{eq:res_l}
    a_{\mu}^\mathrm{light}(L_\mathrm{ref},T_\mathrm{ref})= \amul\amulerr\ ,
\end{gather}
with statistical, systematic and total errors. The used fit parameters and the
error budget is given in Table \ref{ta:amuerr}.  Performing analogous Type-II
fits we get the breakup into individual isospin contributions, given in Table
\ref{ta:amu}. The corresponding continuum extrapolations are shown in Figure
\ref{fi:amul}.

\subsection*{Disconnected contribution}

The disconnected contribution, denoted by $a_{\mu}^\mathrm{disc}$, is obtained
using Equation \eqref{eq:amus} with $C^\mathrm{disc}$, given in Equation
\eqref{eq:disc}, instead of $C^\mathrm{strange}$. In our previous work, at the
isospin-symmetric point \cite{Borsanyi:2017zdw}, we computed the effect of
charm quarks on $a_\mu^\mathrm{disc}$ at the coarsest lattice spacing, and
found that it changes the result by a value much smaller than the statistical
error.  Thus, we perform the current disconnected analysis without taking into
account valence charm quarks.

\begin{figure}[p]
    \centering
    \includegraphics[width=0.7\textwidth]{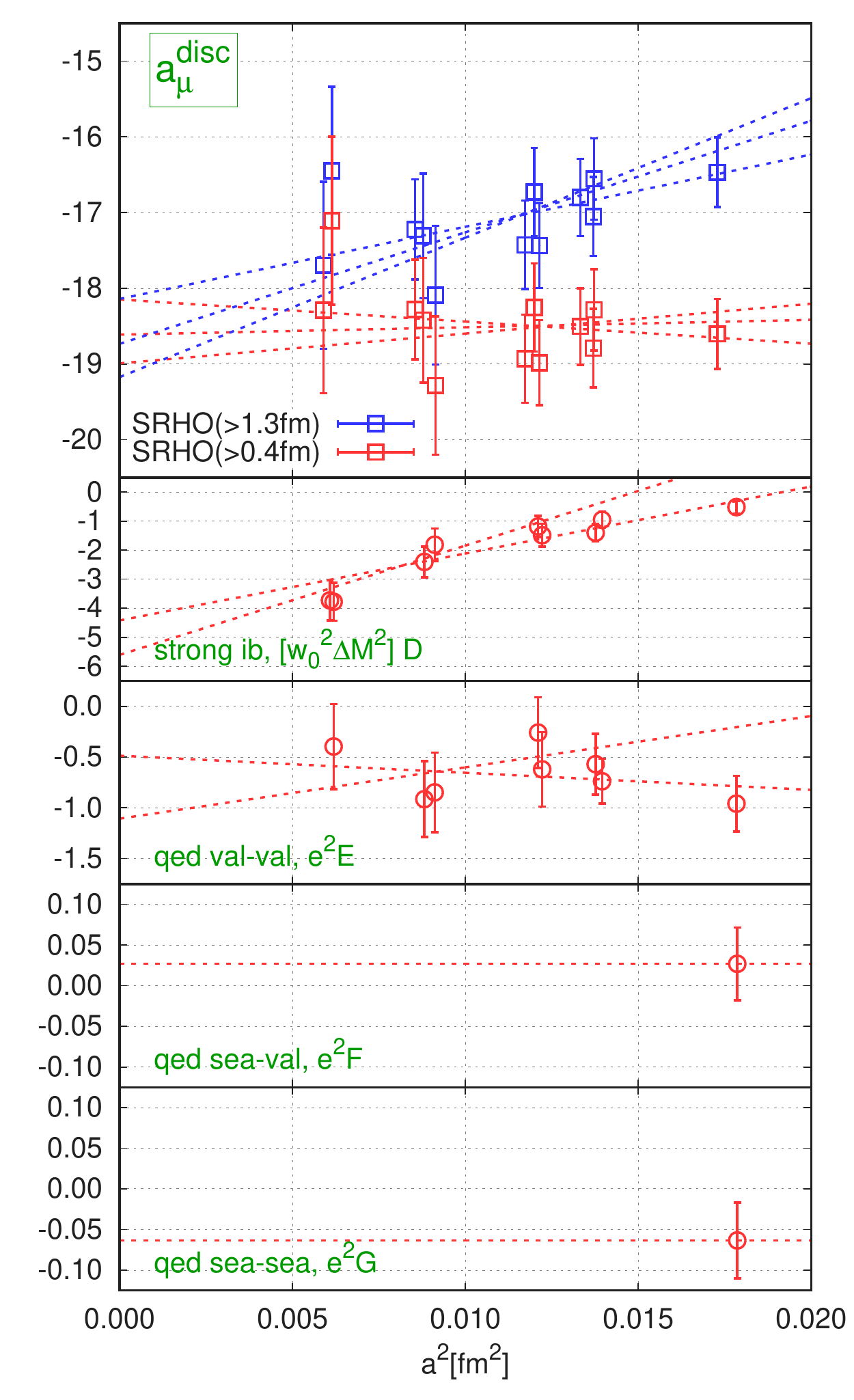}
    \caption
    {
	\label{fi:amudi} Continuum extrapolations of the contributions to of
	$a_{\mu}^\mathrm{disc}(L_\mathrm{ref},T_\mathrm{ref})$.  For more
	explanations see the caption of Figure \ref{fi:amul}.
    }
\end{figure}

The same analysis procedure is applied as in the case of
$a_\mu^\mathrm{light}$: we use upper and lower bounds for the isospin-symmetric
part, a cut in time for the isospin breaking components, two types of lattice
spacing dependence $n=0,3$ and we improve the continuum limit with the
same taste-improvement procedures (SRHO and NNLO SXPT). Several of these
improvements are completely flat as a function of $a^2\alpha_s(1/a)^n$, as
such the quadratic fits give a curvature that is consistent with zero. We
therefore use only linear continuum extrapolations.  A further difference is
that we have one less lattice spacing as in the case of $a_\mu^\mathrm{light}$.
We end up with 55296 fits in total and a result of
\begin{gather}
    \label{eq:res_di}
    a_{\mu}^\mathrm{disc}(L_\mathrm{ref},T_\mathrm{ref})=\amud\amuderr\ .
\end{gather}
The error budget is given in Table \ref{ta:amuerr}. The Type-II fit gives the
individual contributions in Table \ref{ta:amu}, and the continuum
extrapolations are shown in Figure \ref{fi:amudi}.

\subsection*{Finite-size effects and other contributions}

\begin{table}[t]
    \centering
    \begin{tabular}{l|r|l}
	\hline
	strange, $a_\mu^\mathrm{strange}(L_\mathrm{ref},T_\mathrm{ref})$          &           \amus & this work, Equation \eqref{eq:res_s}\\
	light, $a_\mu^\mathrm{light}(L_\mathrm{ref},T_\mathrm{ref})$              &           \amul & this work, Equation \eqref{eq:res_l}\\
	disconnected, $a_\mu^\mathrm{disc}(L_\mathrm{ref},T_\mathrm{ref})$        &           \amud & this work, Equation \eqref{eq:res_di}\\
	finite-size, $a_\mu(\infty,\infty)-a_\mu(L_\mathrm{ref},T_\mathrm{ref})$  &       18.7(2.5) & this work, Equation \eqref{eq:fv}\\ 
	charm iso, $[a_\mu^\mathrm{charm}]_\mathrm{iso}$                          &  14.6(0.0)(0.1) & \cite{Borsanyi:2017zdw}, Table S2\\ 
	charm qed, $[a_\mu^\mathrm{charm}]_\mathrm{qed}$                          &      0.0182(36) & \cite{Giusti:2019xct}\\
	charm effect on $a_\mu^\mathrm{disc}$                                     &          $<$0.1 & \cite{Borsanyi:2017zdw}, Section 4 in Supp. Mat.\\
	bottom, $a_\mu^\mathrm{bottom}$                                           &       0.271(37) & \cite{Colquhoun:2014ica}\\
	perturbative, $a_\mu^\mathrm{pert}$                                       &            0.16 & \cite{Borsanyi:2017zdw}, Table S5\\
	one-photon-reducible subtraction, $-a_\mu^\mathrm{1\gamma R}$             &      -0.321(11) & \cite{Chakraborty:2018iyb}, Table II\\
	\hline
    \end{tabular}
    \caption
    {
	\label{ta:amuall}List of contributions to $a_\mu$, ie. the leading
	order hadronic vacuum polarization contribution to the muon anomalous
	magnetic moment multiplied by $10^{10}$.
    }
\end{table}

Beside the three contributions that we have presented until now, there are a
number of smallers ones that have to be added to get the final value on
$a_\mu$. These are listed in Table \ref{ta:amuall} together with source
indications. Several of these have a size that is much smaller than our
accuracy. They are given here for completeness. We discuss them, now, one by
one.

The previously presented results correspond to the reference box. Its
finite-size effect is computed in Section \ref{se:obs_fv}. This is the fourth
entry in Table \ref{ta:amuall}, which already includes the tiny contribution of
the electromagnetic finite-size effects.

The contribution of the connected charm quark, $a_\mu^\mathrm{charm}$, was
computed in the isospin-symmetric limit by many groups. Here we use our own
result from \cite{Borsanyi:2017zdw}.  An upper bound on the small effect of the
charm on $a_\mu^\mathrm{disc}$ was given in \cite{Borsanyi:2017zdw}. We use the
value as an error here.  The result was obtained using a single lattice
spacing. The even smaller isospin-breaking corrections on $a_\mu^\mathrm{charm}$
was computed in \cite{Giusti:2019xct}.

Until now we have considered four quark flavors. Obviously, the contributions of
the remaining flavors have to be added.  For the bottom quark contribution
there is a lattice determination available, \cite{Colquhoun:2014ica}, whose
value we use here. The top can be safely neglected at our level of precision.

The $a_\mu$ in this work involves an integration in momentum up to
$Q_\mathrm{max}^2=3 \text{ GeV}^2$, as discussed in Section \ref{se:obs_hvp}. The
integration from this value to infinity can be computed in perturbation theory.
We use the value given in \cite{Borsanyi:2017zdw}.

As mentioned in Section \ref{se:obs_hvp}, we compute the current propagator
with all $O(e^2)$ effects included. In this result the one-photon-reducible
($1\gamma R$) contribution belongs to the higher order HVP. This term has to be
subtracted if we are interested in the leading order HVP. A recent lattice
determination of this term can be found in \cite{Chakraborty:2018iyb}. The
corresponding diagram was labeled by the letter 'c' in their Figure 2.

Summing all the contributions gives
\begin{gather}
    a_\mu=\amutot\ ,
\end{gather}
which is our final result for the LO-HVP contribution. The first error is
statistical. It includes the statistical errors of the strange, light and
disconnected contributions. The latter two are the dominant ones. All other
uncertainties are added in quadrature, and this is given as the second,
systematic error.  Its major sources are the finite-size effect estimation, the
taste improvement and the continuum extrapolation. The last error in brackets
is the combined error, which corresponds to a relative precision of $0.8\%$.

    \clearpage
    \section{Result for $a_{\mu,\mathrm{win}}$}
\label{se:res_win}

\begin{table}[p]
    \centering
    \begin{tabular}{l|rr|rr|rr}
	& \multicolumn{2}{c|}{$a_{\mu,\mathrm{win}}^\mathrm{strange}(L_\mathrm{ref},T_\mathrm{ref})$} & \multicolumn{2}{c|}{$a_{\mu,\mathrm{win}}^\mathrm{light}(L_\mathrm{ref},T_\mathrm{ref})$} & \multicolumn{2}{c}{$a_{\mu,\mathrm{win}}^\mathrm{disc}(L_\mathrm{ref},T_\mathrm{ref})$}\\
	\hline\hline
	median                           &   27.170 &           &   207.56 &              &   -1.207 &\\
	\hline
	total error                      &       30 & (0.1\%)   &     1.39 & (0.7\%)      &    0.060 & (5.0\%) \\
	statistical error                &       28 &           &     0.39 &              &    0.028 &\\
	systematic error                 &       13 &           &     1.33 &              &    0.053 &\\
	\hline
	$M_\pi/M_K/M_{ss}$ fit           &        1 &           &     0.03 &              & $<$0.001 &\\
	$M_\pi/M_K/M_{ss}$ fit QED       &        1 &           &  $<$0.01 &              & $<$0.001 &\\
	$M_\Omega$ fit                   &        8 &           &     0.15 &              &    0.005 &\\
	$M_\Omega$ fit QED               &     $<$1 &           &     0.01 &              & $<$0.001 &\\
	$M_\Omega$ experimental          &        2 &           &     0.01 &              & $<$0.001 &\\
	Continuum limit (beta cuts)      &        5 &           &     0.15 &              &    0.007 &\\
	$a^2\alpha_s^n$ with $n=0$ or $3$&       -- &           &     1.21 &              &    0.005 &\\
	taste improvement ranges         &       -- &           &     0.07 &              &    0.019 &\\
	$A_0$ on/off                     &       on &           &       on &              &       on &\\
	$A_2$ on/off                     &       on &           &       on &              &       on &\\
	$A_4$ on/off                     &     $<$1 &           &     0.02 & (also $A_6$) &      off &\\
	$B_0$ on/off                     &        3 &           &       on &              &       on &\\
	$B_2$ on/off                     &      off &           &      off &              &    0.003 &\\
	$C_0$ on/off                     &       on &           &       on &              &       on &\\
	$C_2$ on/off                     &        1 &           &     0.01 &              &    0.005 &\\
	$D_0$ on/off                     &      off &           &       on &              &       on &\\
	$D_2$ on/off                     &      off &           &       on &              &       on &\\
	$D_4$ on/off                     &      off &           &      off &              &      off &\\
	$D_l$ on/off                     &      off &           &  $<$0.01 &              &      off &\\
	$D_s$ on/off                     &      off &           &       on &              &      off &\\
	$E_0$ on/off                     &       on &           &       on &              &       on &\\
	$E_2$ on/off                     &       on &           &       on &              &       on &\\
	$E_4$ on/off                     &      off &           &      off &              &      off &\\
	$E_l$ on/off                     &     $<$1 &           &     0.03 &              &    0.002 &\\
	$E_s$ on/off                     &       on &           &      off &              &      off &\\
	$F_0$ on/off                     &       on &           &       on &              &       on &\\
	$F_2$ on/off                     &        1 &           &     0.02 &              &      off &\\
	$G_0$ on/off                     &       on &           &       on &              &       on &\\
	$G_2$ on/off                     &        2 &           &     0.03 &              &      off &\\
	\hline\hline
    \end{tabular}
    \caption
    {
	\label{ta:amuwerr}Continuum extrapolated results and error budget for
	the strange, light and disconnected contributions to
	$a_{\mu,\mathrm{win}}$. The errors are to be understood on the last
	digits of the central value, as usual. The results correspond to a box
	size of $L_\mathrm{ref}=6.272$~fm and
	$T_\mathrm{ref}=\tfrac{3}{2}L_\mathrm{ref}$.
    }
\end{table}

\begin{table}[p]
    \centering
    \begin{tabular}{l|r|r|r}
	& \multicolumn{1}{c|}{$a_{\mu,\mathrm{win}}^\mathrm{strange}(L_\mathrm{ref},T_\mathrm{ref})$} & \multicolumn{1}{c|}{$a_{\mu,\mathrm{win}}^\mathrm{light}(L_\mathrm{ref},T_\mathrm{ref})$} & \multicolumn{1}{c}{$a_{\mu,\mathrm{win}}^\mathrm{disc}(L_\mathrm{ref},T_\mathrm{ref})$}\\
	\hline
	\hline
	total  &   27.170(28)(13)& 207.56(0.39)(1.33)&  -1.207(28)(53)\\
	\hline
	iso    &   27.175(28)(13)& 206.77(0.40)(1.34)&  -0.853(35)(54)\\
	qed    &  -0.0050(35)(37)&      0.035(40)(44)&   -0.117(17)(6)\\
	qed-vv &  -0.0018(14)(10)&      0.044(26)(16)&   -0.115(18)(5)\\
	qed-sv & -0.00142(60)(30)&    -0.0219(91)(50)&   0.0013(23)(0)\\
	qed-ss &  -0.0018(31)(38)&      0.007(26)(36)& -0.0026(25)(18)\\
	sib    &               --&      0.753(40)(16)&    -0.237(9)(6)\\
	\hline\hline
    \end{tabular}
    \caption
    {
	\label{ta:amuw}Continuum extrapolated results for the different isospin
	components of the strange, light and disconnected contributions to
	$a_{\mu,\mathrm{win}}$. The results correspond to a box size of
	$L_\mathrm{ref}=6.272$~fm and
	$T_\mathrm{ref}=\tfrac{3}{2}L_\mathrm{ref}$.
    }
\end{table}

\begin{figure}[t]
    \centering
    \includegraphics[width=0.7\textwidth]{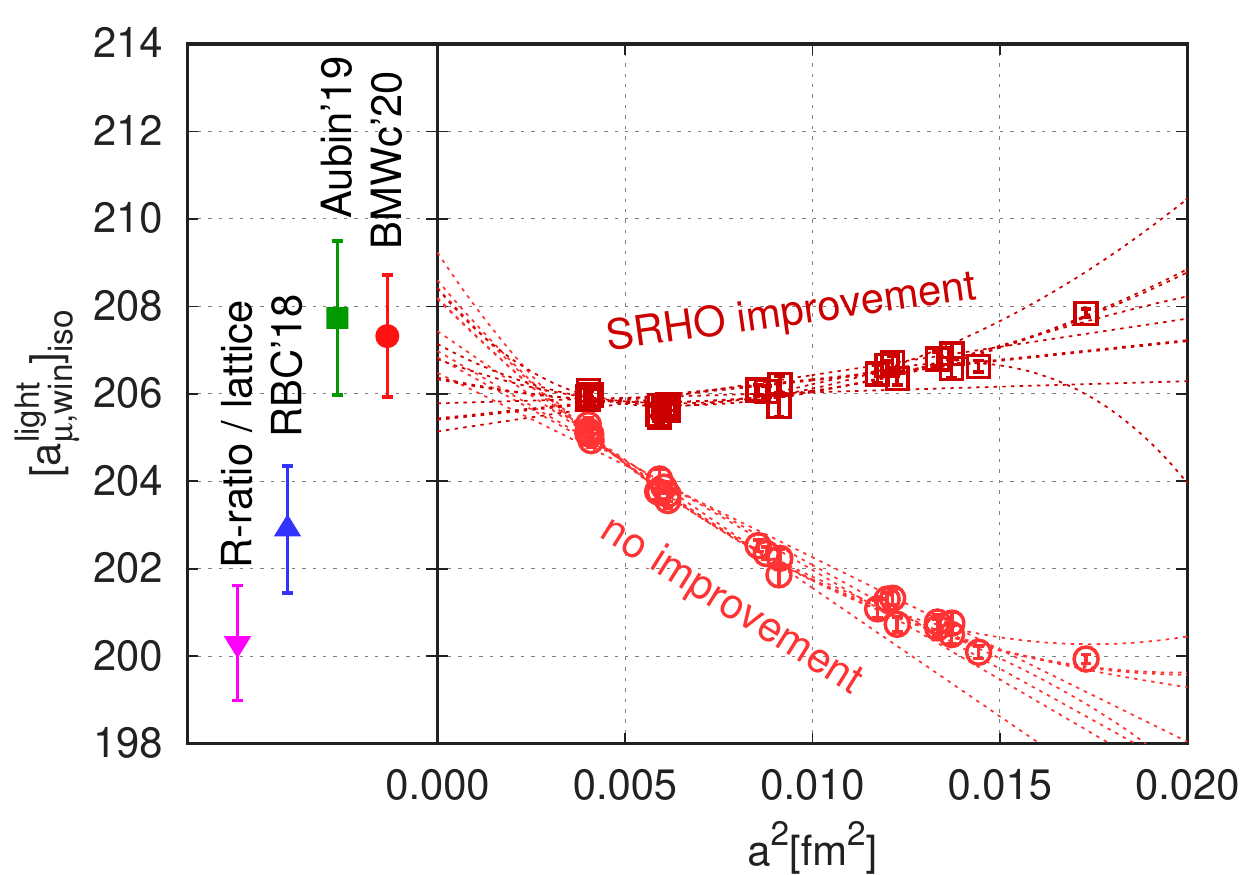}
    \caption
    {
	\label{fi:amuw} Continuum extrapolation of
	$[a_{\mu,\mathrm{win}}^\mathrm{light}]_\mathrm{iso}$. Two types of
	improvements are shown: one where the SRHO model is used to
	improve the lattice result and another where no improvement is
	performed. For each, some continuum extrapolations are shown as
	illustration with dashed lines. They include fits linear, quadratic and
	cubic in $a^2$ and also where different numbers of coarse lattices are
	skipped in the fit. The data points on the plot are corrected for
	light and strange quark mass effects, this adjustment is different from
	fit to fit. Our final value in the continuum limit comes from a
	histogram of about 250,000 fits and is given by the filled red circle
	in the left panel. This histogram also includes fits with a lattice
	spacing dependence of $a^2\alpha_s^n$ with $n=3$.  The results are
	corrected for finite-size effects using Equation \eqref{eq:amuw_fv}.
	Other lattice computations are shown with a green box
	\cite{Aubin:2019usy} and a blue triangle \cite{Blum:2018mom}. A value
	computed from the R-ratio method is also given (see text for details).
    }
\end{figure}

The work \cite{Blum:2018mom} defined a particularly useful observable
$a_{\mu,\mathrm{win}}$, in which the current propagator is restricted to a time
window $[t_1,t_2]$, using a smooth weight function $W(t;t_1,t_2)$. See Section
\ref{se:obs_hvp} for the definition of $W$. The advantage of
$a_{\mu,\mathrm{win}}$ over $a_{\mu}$ is that, by choosing an appropriate
window, the calculation can be made much less challenging on the lattice than
for the full $a_\mu$. Here we will be interested in the window between
$t_1=0.4$~fm and $t_2=1.0$~fm, ie. in an intermediate time range. By this
choice we eliminate both the short-distance region, where large cutoff effects
are present, and the long-distance region, where the statistical uncertainties,
taste violations and finite-size effects are large. Because the determination
of $a_{\mu,\mathrm{win}}$ does not require overcoming many of the challenges
described in the main paper, other lattice groups have obtained this quantity
with errors comparable to ours \cite{Blum:2018mom,Aubin:2019usy}. This allows
for a sharper benchmarking of our calculation. At the same time
$a_{\mu,\mathrm{win}}$ can also be computed using the phenomenological
approach. This is done in Section \ref{se:res_pheno}. Therefore,
$a_{\mu,\mathrm{win}}$ is also a powerful tool to compare the results of
lattice and phenomenological computations.

To compute $a_{\mu,\mathrm{win}}$ on the lattice we perform similar global fits
that were used to get $a_\mu$ in Section \ref{se:res_amu}. In case of the light
and disconnected contributions we use the SRHO taste improvement. The starting
point of the improvement can take three different values: $0.4$, $0.7$
and $1.0$~fm; the latter corresponds to applying no improvement at all. Beside
the usual $a^2$ dependence, we also perform continuum extrapolations with
$a^2\alpha_s(1/a)^n$ with $n=3$, just as in the case of the total
$a_\mu$.  For the light contribution the variation in the $n$ from $0$ to
$3$ gives the largest systematic uncertainty. In case of the light contribution
we can even resolve cubic terms in $a^2\alpha_s(1/a)^n$, these are also
included in our fits. A difference compared to the $a_\mu$ fit procedure is
that no cuts are applied on the propagator in time; the window function
suppresses the propagator for distances beyond $t_2=1.0$~fm.

The results for the strange, light and disconnected contributions and different
isospin breaking corrections are summarized in Table \ref{ta:amuw} and the
error budget for the total values and the fit parameters used, in Table
\ref{ta:amuwerr}. The largest source of error is the continuum extrapolation of
the light connected component. However, it is still much smaller than the
typical size of uncertainties in the full $a_\mu$ determination. In Figure
\ref{fi:amuw} we plot the continuum extrapolation of the isospin-symmetric
component of $a_{\mu,\mathrm{win}}$.

The comparison of lattice results for
$[a_{\mu,\mathrm{win}}^\mathrm{light}]_\mathrm{iso}$ is particularly
interesting, because it allows to benchmark the leading, light-quark
contribution to $a_\mu$ through a quantity that can be computed precisely
without resorting to highly-advanced techniques. Using our result obtained in
the reference box from Table~\ref{ta:amuw} and correcting for finite-size
effects (see later) we get:
\begin{gather}
    [a_{\mu,\mathrm{win}}^\mathrm{light}]_\mathrm{iso}= 207.3(0.4)(1.3)[1.4]\ ,
\end{gather}
with statistical, systematic and total uncertainties.  This result is
$0.2\sigma$ smaller than
$[a_{\mu,\mathrm{win}}^\mathrm{light}]_\mathrm{iso}=207.7(1.8)$ of
Aubin'19~\cite{Aubin:2019usy}. Here we use the continuum-extrapolated value,
which Aubin'19 obtain from their two finest lattices in the upper panel of
their Figure~7, because its errors bars cover the results of the other
continuum extrapolations that they consider. Compared with
$[a_{\mu,\mathrm{win}}^\mathrm{light}]_\mathrm{iso}=202.9(1.4)$ of
RBC'18~\cite{Blum:2018mom} our result is $2.2\sigma$ larger. These two
comparisons yield an average deviation of 1.2$\sigma$. These two lattice
results are also shown in Figure \ref{fi:amuw}.

Additionally, we also made an analysis of the charm quark contribution. The
total $a_\mu^\mathrm{charm}$ was obtained in our previous work
\cite{Borsanyi:2017zdw}. Here we perform a Type-II fit for
$a_{\mu,\mathrm{win}}^\mathrm{charm}$. Only the isospin-symmetric component is
used and we obtain the following result:
\begin{gather}
    [a_{\mu,\mathrm{win}}^\mathrm{charm}]_\mathrm{iso}= 2.7(1)\ .
\end{gather}
Here the error is the systematic uncertainty: the statistical is an order of
magnitude smaller. The isospin breaking of the charm should be well below the
uncertainties of the fit. See Table \ref{ta:amuall} for the case of
$a_\mu^\mathrm{charm}$.  Furthermore, in our dedicated finite-size study with
the ${\tt 4HEX}$ action we compute the difference of the light contribution
between the ``big'' and reference boxes and obtain
\begin{gather}
    a_{\mu,\mathrm{win}}^\mathrm{light}(L_\mathrm{big},T_\mathrm{big})- a_{\mu,\mathrm{win}}^\mathrm{light}(L_\mathrm{ref},T_\mathrm{ref})= 0.57(2)\ ,
\end{gather}
where the error is statistical. Applying the same procedure for
$a_{\mu,\mathrm{win}}$ as we did for $a_\mu$ in the finite-size study of
Section \ref{se:obs_fv}, we get for the finite-size effect:
\begin{gather}
    \label{eq:amuw_fv}
    a_{\mu,\mathrm{win}}(\infty,\infty)- a_{\mu,\mathrm{win}}(L_\mathrm{ref},T_\mathrm{ref})= 0.49(2)(4)\ .
\end{gather}
The first error is statistical, the second is an estimate of the cutoff effect
of the ${\tt 4HEX}$ action. We find that the finite-$T$ effects are even less
important than in the case of the total $a_\mu$, where they were already much
smaller than the finite-$L$ effects.

Summing up these contributions we get
\begin{gather}
    \label{eq:amuwL}
    a_{\mu,\mathrm{win}}= 236.7(0.4)(1.3)[1.4]\ \text{(lattice)}\ ,
\end{gather}
where the first error is statistical, the second is systematic and the third in
the square brackets is the first two added in quadrature. Further
contributions, that are listed in Table \ref{ta:amuall}, should have an effect
much smaller than the uncertainties of this result.

The value can be directly compared to the one obtained from the R-ratio method
in Section \ref{se:res_pheno}:
\begin{gather}
    \label{eq:amuwR}
    a_{\mu,\mathrm{win}}=  229.7(1.3)\ \text{(R-ratio)}\ ,
\end{gather}
which is smaller than the lattice result by $3.7\sigma$ or $3.1$\%. We can also
derive an R-ratio result for the isospin-symmetric light contribution.  From
the value in Equation \eqref{eq:amuwR} we subtract the lattice results for all
contributions, except for
$[a_{\mu,\mathrm{win}}^\mathrm{light}(L_\mathrm{ref},T_\mathrm{ref})]_\mathrm{iso}$
and its finite-size correction. We get:
\begin{gather}
    \label{eq:amuwR-light}
    [a_{\mu,\mathrm{win}}^\mathrm{light}]_\mathrm{iso}=  200.3(1.3)\ \text{(R-ratio \& lattice)}\ .
\end{gather}
This value is compared in Figure \ref{fi:amuw} to continuum and infinite-volume
extrapolated lattice results from this work and from other lattice groups.

\subsection*{Crosscheck with overlap fermions}

\begin{figure}[t]
    \centering
    \includegraphics[width=0.7\textwidth]{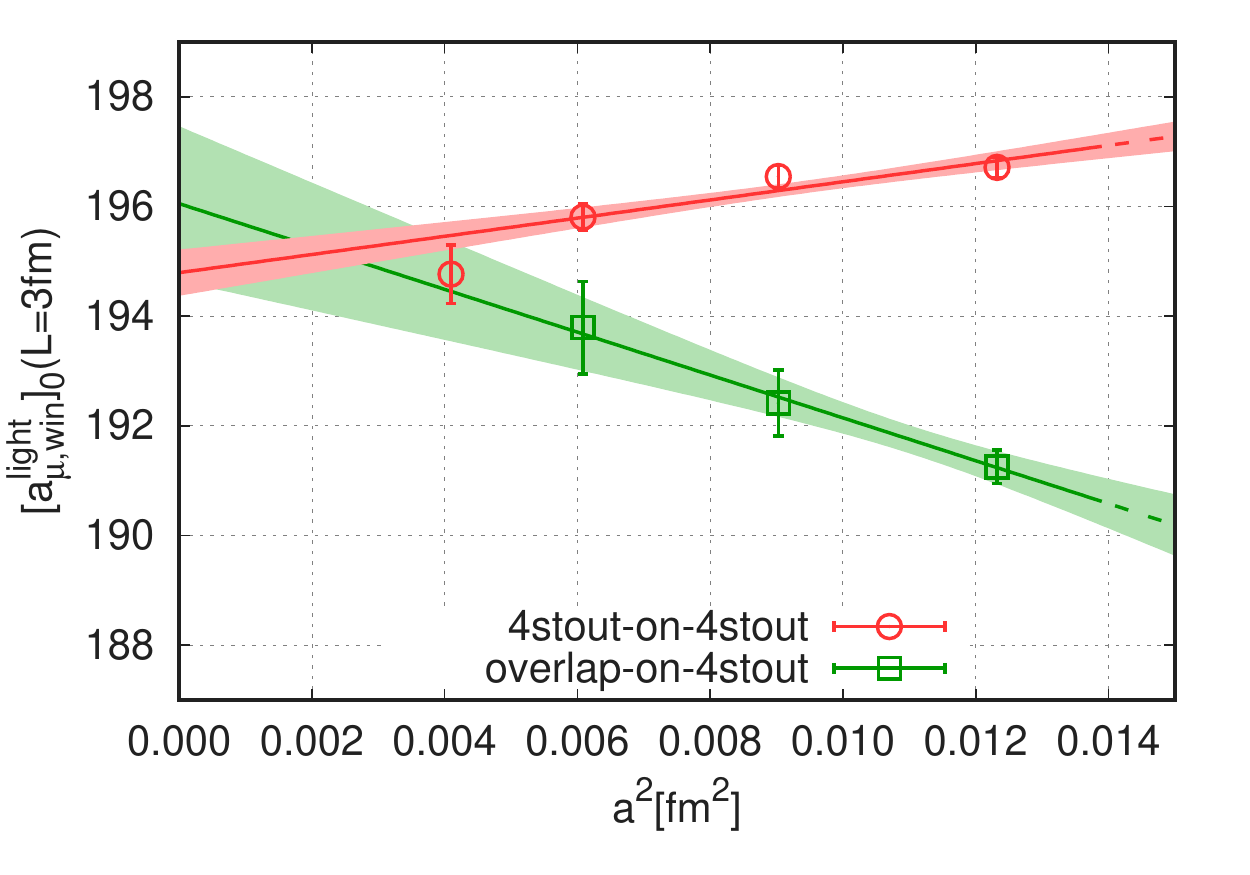}
    \caption
    {
	\label{fi:ov3fm} Continuum extrapolation of
	$[a_{\mu,\mathrm{win}}^\mathrm{light}]_0$. The two datasets correspond
	to the staggered-on-staggered and the overlap-on-staggered simulations.
    }
\end{figure}

We perform a crosscheck of the above results with a mixed action formulation:
overlap valence and {\tt 4stout} staggered sea quarks. Our goal is to provide
more evidence that the continuum extrapolation and the current renormalization
are done correctly in the {\tt 4stout} case. The overlap fermion action, the
matching and the current renormalization are described at length in Section
\ref{se:act_overlap}. Our target is the isospin symmetric value of the light
connected window observable $[a_{\mu,\mathrm{win}}^\mathrm{light}]_0$. We use
lattices of size $L\approx3$~fm in this crosscheck, the related finite-volume
effects are about five percent.

For the measurement of the overlap current propagator we use 512 random wall
sources per configuration, the trace over color indices is performed randomly.
For the staggered current propagator we use the same noise reduction technique
as on the $L\approx6$~fm lattices. In the overlap case we find, that the
coarsest lattice, corresponding to $\beta=3.7000$, is outside the $a^2$-scaling
region. Figure \ref{fi:ov3fm} shows the results together with continuum
extrapolations, they are in good agreement for the two formulations. The
scenario, in which replacing staggered by overlap fermions removes the
discrepancy between the R-ratio and the lattice result, seems improbable. In
order for this to happen, the overlap result would have to get more than
$4\sigma$ smaller than we determine it here.

    \section{Phenomenological determination of $a_{\mu,\mathrm{win}}$}
\label{se:res_pheno}

\begin{table}[t]
    \centering
    \begin{tabular}{lr}
      \hline\hline
      $\quad$ Ref. & $a_\mu\qquad$ \\
	\hline
	KNT18 \cite{Keshavarzi:2018mgv} & 693.26(2.46) \\
	KNT19 \cite{Keshavarzi:2019abf} & 692.78(2.42) \\
	DHMZ17 \cite{Davier:2017zfy}    & 693.1(3.4) \\
	DHMZ19 \cite{Davier:2019can}    & 693.9(4.0) \\
	CHHKS19 \cite{Colangelo:2018mtw,Hoferichter:2019mqg} & 692.3(3.3) \\
	\hline
	\hline
    \end{tabular}
    \caption
    {
	\label{ta:pheno} Recent phenomenological determinations of $a_\mu$.
	KNT stands for Keshavarzi, Nomura and Teubner; DHMZ for Davier,
	Hoecker, Malaescu and Zhang; CHHKS for Colangelo, Hoferichter, Hoid,
	Kubis and Stoffer.
    }
\end{table}

The purpose of this section is to describe the computation of the
phenomenological result for $a_{\mu,\mathrm{win}}$, which we compare
with the corresponding lattice result in Section \ref{se:res_win}.
For this we use the R-ratio from $e^+e^-$ collision experiments and the
corresponding covariance matrix from the work of KNT18
\cite{Keshavarzi:2018mgv}, courteously given to us by Keshavarzi, Nomura
and Teubner.

As Table~\ref{ta:pheno} shows, there is a sizeable difference in the
uncertainties on $a_\mu$ in the recent phenomenological
literature. While the KNT19 result has an error of 2.42, DHMZ19 gives
a 65\% larger error of 4.0. Since we do not want to risk overstating
possible differences between the phenomenological and lattice
approaches, we extend the error estimates of KNT18
\cite{Keshavarzi:2018mgv} by two additional sources, bringing them
much closer to those of DHMZ.

{\bf a.} The first source of uncertainty is related to a tension
between the two $e^+e^-$ experiments, KLOE and BaBar, which have the
smallest uncertainties in the window from $0.6$ to $0.9$~GeV
center-of-mass energy. These two experiments exhibit, for the
pion-pion channel in that window, a close to $3\sigma$ discrepancy or
a $2.7$\% relative difference \cite{Anastasi:2017eio}.  Note that the
overlapping energy region of the pion-pion channel for KLOE and BaBar
($0.324-0.972$~GeV) provides about 70\% of the total $a_\mu$. The
discrepancy, fully accounted for in DHMZ19~\cite{Davier:2019can}, has a
strong impact on the discrepancy between the measurement of
$g_\mu-2$ and theory predictions based on the R-ratio.

In order to address this discrepancy, we follow the prescription of
the Particle Data Group (PDG) for similar tensions between
experimental results \cite{Tanabashi:2018oca}.  After calculating the
weighted average of all the experimental results in the $0.6-0.9$~GeV energy
range \cite{Anastasi:2017eio}, the PDG prescription tells us to
adjust the error by a factor of $S=[\chi^2/(N-1)]^{1/2}$, where $N$ is
the number of experiments (in our case $N=5$). This yields an
uncertainty of $1.97$ instead of the $1.32$ of
\cite{Keshavarzi:2018mgv}. This uncertainty is far less than half the
difference between BaBar and KLOE, because the other, less precise
experiments dilute the discrepancy. We include this increased error
estimate as
$(1.97^2-1.32^2)^{1/2}=(1.46)_{\pi\pi}$, where ``${\pi\pi}$''
denotes the uncertainty coming from the tension between experiments in
the pion-pion channel. Note that the other experimental channels
may have similar uncertainties, which would increase the error
further.

It is worth pointing out that hadronic $\tau$ decays can be used, in
principle, to provide an independent measurement of the spectral
function in this important low-energy region, as first proposed in
\cite{Alemany:1997tn} and updated in
\cite{Davier:2009ag,Davier:2010nc,Davier:2013sfa}. However, this
requires controlling isospin-breaking corrections
\cite{Cirigliano:2001er,Cirigliano:2002pv,Jegerlehner:2011ti,Bruno:2018ono,Miranda:2020wdg},
which is a challenge and has led to putting this approach aside in the
last few years.

{\bf b.} Another possible source of uncertainty comes from the way in
which the dispersive integral of the experimental data for the
various, final-state channels is performed, including correlations. KNT
\cite{Keshavarzi:2018mgv,Keshavarzi:2019abf} use a trapezoidal rule
and argue that the error resulting from this choice is
negligible. They also take into account correlations in systematic
uncertainties within the same experiment and between different
experiments, as well as within and between different channels, over
extended ranges of center-of-mass energy. On the other hand, DHMZ
\cite{Davier:2019can} limit the effects of these correlations to small
energy bins and use splines for integrating the data, correcting for
biases if necessary. The end result is that, despite using the same
experimental input, the two teams find results for the various
channels which differ, more often outside the error bars of
\cite{Keshavarzi:2018mgv} than of \cite{Davier:2019can}. To account
for this when using the correlation matrices from
\cite{Keshavarzi:2018mgv}, we follow a suggestion put forward at the
last ``Muon g-2 Theory Initiative'' meeting \cite{web:white}. We add,
to results obtained with these correlations, an uncertainty obtained
by summing, in quadrature, half the differences of the individual
channels.  This gives an additional error of $(2.26)_\mathrm{int}$,
where ``$\mathrm{int}$'' stands for the uncertainty related to the
integration and correlation procedures.

Thus, using the well known dispersive integral (see eg. Section 5 of
\cite{Jegerlehner:2017gek})
\begin{gather}
    \label{eq:amuR}
    a_\mu= 10^{10}\left(\frac{\alpha m_\mu}{3\pi}\right)^2
    \int_{s_\mathrm{th}}^{\infty}\frac{ds}{s^2}\ R(s)\hat{K}(s)\ ,
\end{gather}
the experimental R-ratio data set of \cite{Keshavarzi:2018mgv} and the
perturbative R-ratio from the {\tt rhad} package \cite{Harlander:2002ur}, we
obtain,
$a_\mu=693.27(2.46)_\mathrm{stat}(1.46)_{\pi\pi}(2.26)_\mathrm{int}[3.65]$,
where the first error reproduces the one given in \cite{Keshavarzi:2018mgv},
while the second and third errors are computed above. The last error, in
brackets, is the quadratically combined error of the first three. It is larger
than the one of KNT and is closer to the error of DHMZ17, but still a bit
smaller than that of DHMZ19. This enlarged error, as well as the most recent
value of $4.0$ from DHMZ, reduce a bit the strong tension between the
measurement of $g_\mu-2$ and the theory predictions based on the R-ratio
method.

Having checked that we are able to reproduce well known R-ratio results, we
repeat the whole procedure for $a_{\mu,\mathrm{win}}$ of Equation
\eqref{eq:win}. For this observable, high statistical precision is easier to
reach on the lattice and the continuum and infinite-volume extrapolations are less difficult.
In addition, we expect that the R-ratio method yields a similar relative error
for $a_{\mu,\mathrm{win}}$ as for $a_\mu$. These facts make
$a_{\mu,\mathrm{win}}$ a convenient observable to compare the two approaches
and, eventually, to combine them for improved overall precision
\cite{Blum:2018mom}.

To determine $a_{\mu,\mathrm{win}}$ from R-ratio data, we transform
the latter to Euclidean coordinate space by a Laplace transform
\cite{Bernecker:2011gh}, where a weighted integral with weight
function $K(t)W(t)$ has to be performed, as described in Section
\ref{se:obs_hvp}. One ends up with an integral as in Equation
\eqref{eq:amuR}, but the kernel $\hat{K}$ replaced by
$\hat{K}_\mathrm{win}$:
\begin{gather}
    \label{eq:khat_window}
    \begin{aligned}
	\hat{K}_\mathrm{win}(s)
	=\frac{3s^{5/2}}{8m_\mu^2}
	\int_0^\infty dt\
	e^{-\sqrt{s}t} K(t)
	W(t;t_1,t_2),
    \end{aligned}
\end{gather}
where $K(t)$ is given by Equation~\labelmainkdef{} of the main paper. The
window parameters $t_1=0.4$~fm, $t_2=1.0$~fm and $\Delta=0.15$~fm are the same
as in Section \ref{se:res_win}. We proceed with the computation of the
$s$-integral as in the case of $a_\mu$. In particular, we include the
${\pi\pi}$ and $\mathrm{int}$ errors as follows:

{\bf a.} Repeating the R-ratio method, with the $\hat{K}_\mathrm{win}$
kernel, gives the same relative difference between the KLOE and BaBar
results, ie. $2.7\%$, as with the original kernel function
$\hat{K}$.  Carrying out the PDG procedure for adjusting errors, we
obtain a value of $0.5$ for the additional ``${\pi\pi}$'' error.

{\bf b.} Since we do not have the contributions of the individual
experimental channels in our window for both the KNT and the DHMZ
frameworks, we simply scale down the ``$\mathrm{int}$'' error of the
full $a_\mu$. Thus, instead of $2.26$, we obtain $0.8$ as an
``$\mathrm{int}$'' uncertainty.

Putting the above components together and also including the tiny perturbative
contribution from \cite{Harlander:2002ur}, we obtain
\begin{gather}
    \label{eq:amuwpheno}
    a_{\mu,\mathrm{win}}= 229.7(0.9)_\mathrm{stat}(0.5)_{\pi\pi}(0.8)_\mathrm{int}[1.3]\ .
\end{gather}
The last error, in brackets, is all errors added in quadrature. This value is
compared with our lattice result in Section \ref{se:res_win}.

Finally, let us compare our result for $a_\mu$ with the one in the recently
published white paper \cite{Aoyama:2020ynm}. The KNT19, DHMZ19 and the white
paper predictions for $a_\mu$ (rounded to the nearest tenth for brevity) are
692.8(2.4), 693.9(4.0) and 693.1(4.0), respectively. Our procedure using the
KNT database and following the PDG prescription gives 693.3(3.7). The white
paper number and our result are almost the same, they differ by one-twentieth
of a sigma and the error estimates differ only by less than a tenth of a sigma.
Therefore, we believe that our estimate is a fair assessment of $a_\mu$. The
same can be also said about our result for $a_{\mu,\mathrm{win}}$ in Equation
\eqref{eq:amuwpheno}, which uses the same procedure as for $a_\mu$, namely the
KNT data set and the above treatment for the integration and correlation
uncertainties.

    \section{Consequences for electroweak precision observables?}
\label{se:res_ewp}

\begin{figure}[t]
    \centering
    \includegraphics[width=0.7\textwidth]{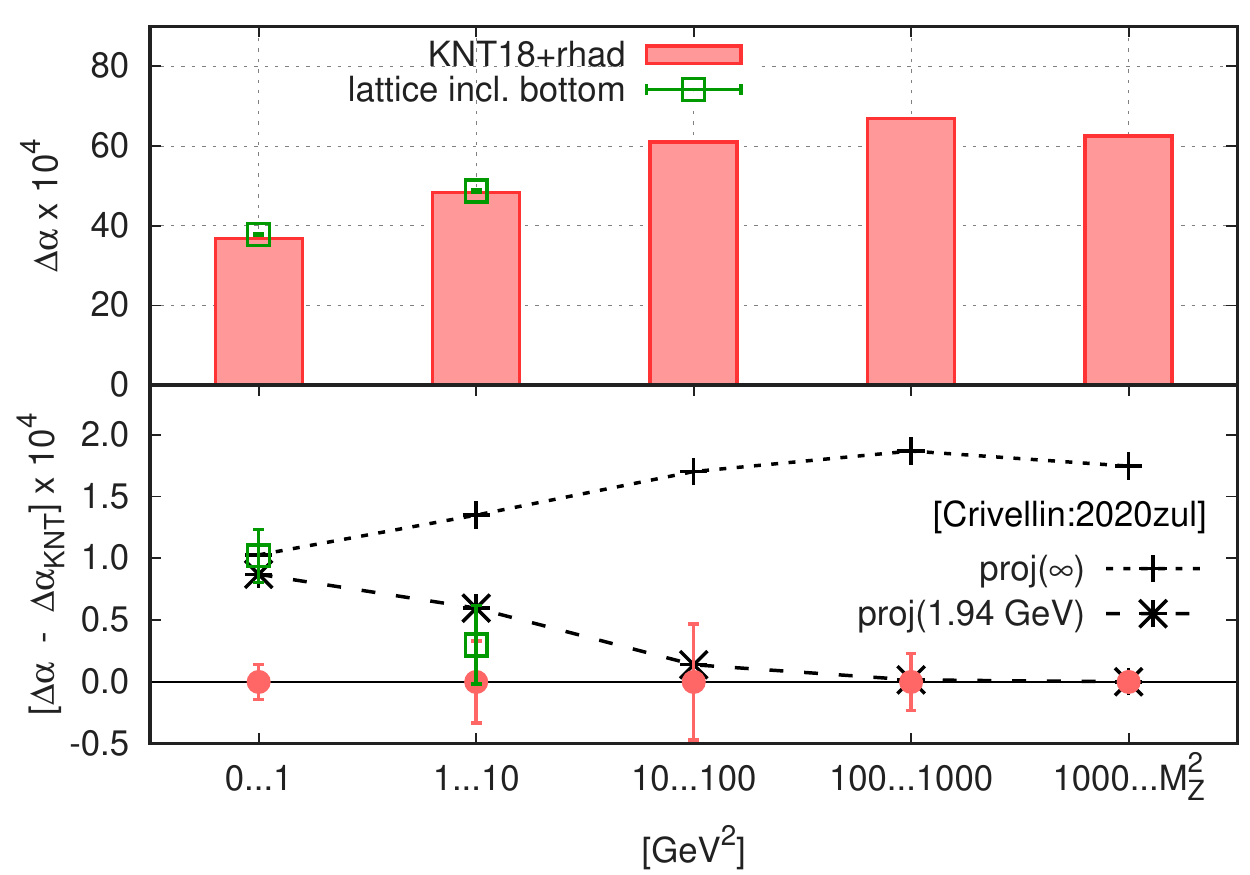}
    \caption
    {
	\label{fi:ewp} Hadronic contribution to the running of the
	electromagnetic coupling for Euclidean momenta. For each bin we show
	$\Delta\alpha^{(5)}_\mathrm{had}(-q_\mathrm{max}^2)-\Delta\alpha^{(5)}_\mathrm{had}(-q_\mathrm{min}^2)$,
	where $q_\mathrm{min/max}^2$ is the lower/upper end of the bin.  The
	conventional value of $\Delta\alpha^{(5)}_\mathrm{had}(M_Z^2)$ can be
	obtained by summing the results of all bins and performing a rotation
	to Minkowski space (see text). The upper panel shows results obtained
	from the experimental R-ratio of KNT18 \cite{Keshavarzi:2018mgv} and
	perturbation theory \cite{Harlander:2002ur}, as well as our lattice
	results.  The bottom panel compares the results with the R-ratio as a
	baseline.  The ``reference point'' scenario of CHMM
	\cite{Crivellin:2020zul} (``proj($\infty$)'') is shown with crosses.
	Their ``proj($1.94$~GeV)'' scenario is shown with bursts.
    }
\end{figure}

In this section we investigate the claim, put forward by Crivellin,
Hoferichter, Manzari and Montull (CHMM) \cite{Crivellin:2020zul}, that the
result of this paper may lead to a significant tension with electroweak
precision fits for the hadronic contribution to the running of the
electromagnetic coupling, $\Delta\alpha_\mathrm{had}^{(5)}(M_Z^2)$. Here
``$(5)$'' means that we consider five active flavors and $M_Z$ is the mass of
the $Z$-boson. CHMM's analysis is based on the earlier work of
\cite{Passera:2008jk} that studies the impact, on precision electroweak fits,
of increasing the HVP contribution to reproduce the measured value of
$(g_\mu-2)/2$. That work has been recently updated and expanded in
\cite{Keshavarzi:2020bfy}. In contrast, assuming that the difference
between the lattice and R-ratio predictions for $a_\mu$ is given solely by a
difference in the slopes of the corresponding HVP functions at $Q^2{=}0$
--which is a reasonable first approximation-- \cite{deRafael:2020uif} shows
that the resulting shift in $\Delta\alpha_\mathrm{had}^{(5)}(M_Z^2)$ is at
least nine times smaller than the error in the electroweak fit determination of
$\Delta\alpha_\mathrm{had}^{(5)}(M_Z^2)$, given in \cite{Crivellin:2020zul} and
in Equation~\eqref{eq:alfaew} below. The consequences of our result for
electroweak precision observables has also been investigated very recently in
\cite{Malaescu:2020zuc}, where the impact of correlations between $a_\mu$ and
$\Delta\alpha_\mathrm{had}^{(5)}(M_Z^2)$, in the context of R-ratio
calculations, has been analyzed in detail.

Indeed, the authors of
\cite{Crivellin:2020zul} perform a global fit to electroweak observables and
obtain
\begin{gather}
    \label{eq:alfaew}
    \Delta\alpha_\mathrm{had}^{(5)}(M_Z^2)= 270.2(3.0) \times 10^{-4}\ \text{(electroweak)}\ .
\end{gather}
Note that this value is somewhat smaller, both in value and in uncertainty, than
the latest result of the Gfitter group \cite{Haller:2018nnx}. The same
observable can also be obtained from the experimental
R-ratio~\cite{Davier:2019can,Keshavarzi:2019abf}, eg. using the KNT19
result~\cite{Keshavarzi:2019abf} as do CHMM:
\begin{gather}
\label{eq:alfaRratio}
  \Delta\alpha_\mathrm{had}^{(5)}(M_Z^2)=
    276.1(1.1) \times 10^{-4}\ \text{(R-ratio)}\ ,
\end{gather}
which is $1.8\sigma$ higher than the electroweak-fit value.

CHMM then consider a variety of scenarios to estimate the possible value of
$\Delta\alpha_\mathrm{had}^{(5)}(M_Z^2)$ from the results given in the present
paper. In their ``reference point'' scenario (``proj$(\infty)$''), they assume
that the relative difference between the R-ratio and our result for $a_\mu$
corresponds to an energy-independent rescaling in the $e^+e^-\to\mbox{hadrons}$
spectral function for all center-of-mass energies, from threshold to infinity.
Thus, they obtain
\begin{gather}
    \label{eq:chmm}
    \Delta\alpha_\mathrm{had}^{(5)}(M_Z^2)= 276.1(1.1) \times 10^{-4}\times\frac{712}{693} = 283.8(1.3)\times 10^{-4}\ \text{(CHMM)}\ ,
\end{gather}
which deviates by $4.2\sigma$ from the electroweak-fit value \footnote{We use
here 712, which was the central value of the result, $a_\mu=712.4(4.5)$, in
the arxiv-v1 of this work. The difference to our current result,
$a_\mu=707.5(5.5)$, is caused by three effects of roughly the same size:
changing the scale setting from $w_0$ to $M_\Omega$, adding finite-$T$ effects
and adding new measurements on the strong-isospin breaking of the disconnected
contribution.}. This leads CHMM to conclude that the result of the present
paper, while removing the discrepancy with the experimental determination of
$a_\mu$, may create a new one, now with electroweak precision measurements.

We now take a closer look at the claim of CHMM and point out problems with
their assumption. We will not study the running of $\alpha$ all the way up to
$M_Z^2$ here, because that would take us significantly beyond the scope of
the present study of $a_\mu$.  However, without too much effort we can
investigate the running of $\alpha$ in the Euclidean regime up to scales
accessible in our lattice computation.\footnote{If we were able to run in the
Euclidean up to $M_Z^2$, the conversion to timelike $M_Z^2$ can be computed
in perturbation theory and the resulting correction is significantly smaller
than the present error bars on $\Delta\alpha_\mathrm{had}^{(5)}(M_Z^2)$, as
shown eg. in \cite{Blondel:2019vdq}.} The Euclidean running of the coupling is
obtained from the Euclidean HVP as
\begin{gather}
    \Delta\alpha^{(5)}_\mathrm{had}(-q^2)= e^2 \hat\Pi^{(5)}(q^2)\ ,
\end{gather}
where $\hat\Pi^{(5)}$ is the five-flavor HVP given, with the notations of Section \ref{se:obs_hvp}, by:
\begin{gather}
    \hat\Pi^{(5)}= 
    \hat\Pi^\mathrm{light}+
    \hat\Pi^\mathrm{strange}+
    \hat\Pi^\mathrm{charm}+
    \hat\Pi^\mathrm{disc}+
    \hat\Pi^\mathrm{bottom}\ .
\end{gather}
Here we compute $\Pi^{(5)}$ both on the lattice and from the R-ratio.

\begin{table}[t]
    \centering
    \begin{tabular}{L|C|C|C|C|C|C|C}
	& \hat\Pi^\mathrm{light} & \hat\Pi^\mathrm{strange} & \hat\Pi^\mathrm{charm} & \hat\Pi^\mathrm{disc} & \infty-\mathrm{ref} & \hat\Pi^\mathrm{bottom} & \hat\Pi^{(5)}\\
	\hline
	\hat\Pi(1)         & 355.7(2.3) & 41.8(0.1) & 17.0(0.1) & -5.1(0.6) & 2.8(0.2) & 0.3(0.1) & 412.6(2.4) \\
	\hat\Pi(10)-\hat\Pi(1) & 363.4(2.2) & 67.7(0.3) & 96.6(2.6) & -0.2(0.2) & 0.2(0.0) & 2.9(0.5) & 530.7(3.5) \\
    \end{tabular}
    \caption
    {
	\label{ta:ewp} Continuum extrapolated lattice results for the HVP. We
	give separately the light, strange, charm and disconnected
	contributions and also the finite-size effects as computed from the
	{\tt 4HEX} simulations. The bottom quark contribution was obtained from
	the work \cite{Colquhoun:2014ica}. The last column is the sum of all.
    }
\end{table}

To compute the HVP on the lattice we apply the same Type-II fit procedure as we
use for the determination of $a_\mu$ and $a_{\mu,\mathrm{win}}$ in the previous
Sections. We correct for finite-size effects by computing them in our {\tt
4HEX} simulations. For the bottom quark we combine the first four moments of
$\hat\Pi^\mathrm{bottom}$, determined by the HPQCD collaboration
\cite{Colquhoun:2014ica}, into a Pad\'e approximation. For the purpose of the
present discussion it suffices to consider two observables: $\hat\Pi^{(5)}(1)$ the
value of $\hat\Pi^{(5)}$ at $q^2=1\text{ GeV}^2$ and the difference in $\hat\Pi^{(5)}$
at $q^2=10\text{ GeV}^2$ and $1\text{ GeV}^2$, denoted by
$\hat\Pi^{(5)}(10)-\hat\Pi^{(5)}(1)$. Our final continuum extrapolated results are
given in Table \ref{ta:ewp}.

In the case of $\hat\Pi^{(5)}(10)-\hat\Pi^{(5)}(1)$ we have departed from our standard
procedure in two details. First, one must note that the full dataset gives bad
fit qualities in the continuum extrapolation, especially for the charm
contribution. This is because the precision of our results is orders of
magnitude better than eg. for $a_\mu$. Thus, we fit only a subset of about 50
configurations, maximally spaced along the simulation chain. This increases the
statistical error and leads to acceptable fit qualities. Also, we observe
lattice artefacts that are much larger than for the other observables in the
paper. In the charm case they are on the level of $100$\%. To reflect this in
the uncertainty of the continuum extrapolation, we use fit functions that are
quadratic polynomials in $a^2$ and also apply a flat weighting of the results
of different procedures in our determination of the systematic error.

To compute the HVP from the R-ratio we apply a dispersion integral
(see eg. Section 3 of \cite{Jegerlehner:2017gek})
\begin{gather}
    \hat\Pi^{(5)}(q^2)= \frac{q^2}{12\pi^2}\int_{s_\mathrm{th}}^{\infty} ds\ \frac{R(s)}{s(s+q^2)}
\end{gather}
to the R-ratio data set of \cite{Keshavarzi:2018mgv}.  Uncertainties are
computed from the covariance matrix of the data.  For energies above the range
of this data set we use the perturbative result from the {\tt rhad} package
\cite{Harlander:2002ur}.

In Figure \ref{fi:ewp} we show HVP differences corresponding to five energy
bins, starting at zero and ending at the scale $M_Z^2$. The first bin gives the
difference of the HVP between 1 and 0 GeV$^2$, the second between 10 and 1
GeV$^2$ and so on. The running to the scale $M_Z^2$ can then be obtained by
summing the values in the five bins. In the top panel we show the result
obtained from the R-ratio. For the first two bins, we also have lattice
results. The bottom panel shows our lattice results with the R-ratio result as
a baseline.

In the first bin, corresponding to an energy range from $0$ to $1$ GeV$^2$, we
see a difference between the lattice and the R-ratio determinations, that is
about $1.0$ in $\Delta\alpha^{(5)}_\mathrm{had}$. This corresponds to a
relative deviation of approximately $2.8\%$, which is about the same as we have
in the total $a_\mu$.  This fact is not surprising because over $99\%$ of
$a_\mu$ comes from this spacelike region of momenta. In the second bin,
corresponding to the energy range from $1$ to $10$~GeV$^2$, the lattice and
R-ratio results already agree.  For the bins with larger energies we show no
lattice results: discretization errors are too large to allow a controlled
continuum extrapolation.

In Figure \ref{fi:ewp} we also show two scenarios from CHMM. The first
is their ``reference point'' projection (``proj($\infty$)'') described
above. There, the R-ratio and lattice results are assumed to have the
same $2.8\%$ relative discrepancy in all energy bins up to $M_Z$, as
observed in the first bin. The difficulty with this assumption is that
CHMM extrapolates over two orders of magnitude in energy, while using
data only from the first bin. In addition, our lattice result in the
second energy bin is already in clear disagreement with their
hypothesis, and invalidates their estimate of Equation
\eqref{eq:chmm}.

The second of their scenarios considered here is the one in which they
assume that the $2.8\%$ rescaling of the spectral function only
applies to center-of-mass energies below $1.94$~GeV (``proj($1.94$~GeV)'').
This second scenario agrees much better
with our lattice results, as can be seen in the first two bins. If
future lattice calculations confirm that the agreement holds in the
remaining bins, the tension implied by our lattice calculation of
$a_\mu$ on $\Delta\alpha^{(5)}_\mathrm{had}$ would be $2.4\sigma$,
significantly smaller than the $4.2\sigma$ of the ``reference point''
scenario of \cite{Crivellin:2020zul} and only slightly larger than the
$1.8\sigma$ already observed with the R-ratio result of
Equation~\eqref{eq:alfaRratio}.

    \printbibliography[section=1]
\end{refsection}

\end{document}